\documentclass[10pt, partial]{aucklandthesis}

\usepackage[utf8]{inputenc}
\usepackage[T1]{fontenc}

\usepackage{amssymb}
\usepackage{multirow}
\usepackage{graphicx}
\usepackage[caption=false]{subfig}
\usepackage{tabularx} 
\usepackage{amsmath} %for dealing with mathematics,
\usepackage{amsthm} %for dealing with theorem environments,
\usepackage{pifont} %for check mark and cross mark
\newcommand{\cmark}{\ding{51}}% 
\newcommand{\xmark}{\ding{55}}%
\DeclareMathOperator{\Tr}{Tr} 
\usepackage{placeins}
\usepackage{cite}
\usepackage{url}
\usepackage{navigator} %For using URL in text
\usepackage{makeidx}
\usepackage{multicol} 
\usepackage{footmisc} %for footnotes

\usepackage[hidelinks]{hyperref}% Can also be used with [colorlinks=false, linktocpage=true]
\begin{document}

% ====================================================
%
% FRONTMATTER
%
% Arabic pagination, starting with the title page
% which is counted but not numbered
%
% ====================================================
\pagenumbering{roman}
% Specify the title page content
\title{Signed Network Structural Analysis and Applications with a Focus on Balance Theory}
\subtitle{}
\author{Samin Aref}
\degreesought{Doctor of Philosophy (Ph.D.)} 
\degreediscipline{Computer Science}
\degreecompletionyear{2019}

% Print the title page
\maketitle

% Abstract, up to 350 words % at the last count on 30 may it was 360 words

\cleardoublepage
\begin{abstract}
	{This research is an effort to understand small-scale properties of networks resulting in global structure in larger scales. Networks are modelled by graphs and graph-theoretic conditions are used to determine the structural properties exhibited by the network. Our focus is on \textit{signed networks} which have positive and negative signs as a property on the edges. We analyse networks from the perspective of \textit{balance theory} which predicts \textit{structural balance} as a global structure for signed social networks that represent groups of friends and enemies. The vertex set of \textit{balanced} signed networks can be partitioned into two subsets such that each negative edge joins vertices belonging to different subsets. 
		
	The scarcity of balanced networks encouraged us to define the notion of \textit{partial balance} in order to quantify the extent to which a network is balanced. We evaluate several numerical \textit{measures of partial balance} and recommend using the \textit{frustration index}, a measure that satisfies key axiomatic properties and allows us to analyse graphs based on their levels of partial balance.
	
	The exact algorithms used in the literature to compute the frustration index, also called the \textit{line index of balance}, are not scalable and cannot process graphs with a few hundred edges. We formulate computing the frustration index as a graph optimisation problem to find the minimum number of edges whose removal results in a balanced network given binary decision variables associated with graph nodes and edges. We use our first optimisation model to analyse graphs with up to 3000 edges.
	
	Reformulating the optimisation problem, we develop three more efficient \textit{binary linear programming} models. Equipping the models with \textit{valid inequalities} and \textit{prioritised branching} as speed-up techniques allows us to process graphs with 15000 edges on inexpensive hardware. Besides making exact computations possible for large graphs, we show that our models outperform heuristics and approximation algorithms suggested in the literature by orders of magnitude.
	
	We extend the concepts of balance and frustration in signed networks to applications beyond the classic friend-enemy interpretation of balance theory in social context. Using a high-performance computer, we analyse graphs with up to 100000 edges to investigate a range of applications from biology and chemistry to finance, international relations, and physics. 
	
	}
\end{abstract}

% Dedication (optional)
\thesisdedication{Dedicated to\\ M. (infatuation)\\ H. (sharp sword)\\ M. (moonlight)\\ and S. (star)}

% Preface and/or acknowledgements (optional)
\cleardoublepage
\chapter*{Acknowledgements}
	{I would like to express my very great appreciation to Dr. Mark C. Wilson for supervising this research and motivating me in the past couple of years. The experience of working with him was extremely valuable and I am deeply indebted to him for sharing his knowledge and expertise. It was proven to me numerous times that having him as a supervisor has played a key role in the success of this Ph.D. project.
		
	I also would like to thank Dr. Andrew J. Mason for co-supervising this research. I am very grateful to him not only for his valuable comments, but for sharing his mathematical modelling expertise which strengthened the contributions of this thesis.
	
	I was privileged to have Dr. Serge Gaspers and Prof. Gregory Gutin as examiners of this thesis. I am thankful for their essential comments which helped in revising and improving the thesis.
		
	This research would not be completed without the tremendous support of my partner for whom my heart is filled with gratitude. I cannot thank her enough for her selfless and pure love that has lighten up my life. The challenges we faced could have not possibly been overcome without her devotion and dedication.
	
	I am also indebted to a lifetime of love and support from my parents and my sister. Their presence has always encouraged me to accept new challenges such as a Ph.D. program in New Zealand. I am grateful for having the best father, the best mother, and the best sister I can possibly imagine.
	
	I would like to acknowledge University of Auckland for investing in these ideas. The support provided by Department of Computer Science, Centre for eResearch, and {Te P\={u}naha Matatini} was greatly appreciated.
	
	In the end, I would like to thank everyone who has taught me something; past teachers and professors as well as authors of the papers I have read and the reviewers who have commented on my works.
	}

% Contents, lists of tables and figures
\settocdepth{subsection} % choose chapter, section, subsection 
\cleardoublepage\tableofcontents
\cleardoublepage\listoffigures
\cleardoublepage\listoftables

% Glossary (optional)
%\input{glossary}

\cleardoublepage
\pagenumbering{arabic}
\setcounter{page}{1} % re-sets the page counter

\chapter{Introduction}
\cleardoublepage
	{We investigate small-scale properties of networks resulting in global structure in larger scales. Networks are modelled by graphs and graph-theoretic conditions are used to determine the structural properties exhibited by the network. Our focus is on \textit{signed networks} which have positive and negative signs as a property on the edges. We analyse networks from the perspective of \textit{balance theory} which predicts \textit{structural balance} as a global structure for signed social networks that represent groups of friends and enemies. The vertex set of \textit{balanced} signed networks can be partitioned into two subsets such that each negative edge joins vertices belonging to different subsets. 
	
	The scarcity of balanced networks encouraged us to define the notion of \textit{partial balance} in Chapter \ref{ch:1} in order to quantify the extent to which a network is balanced. We evaluate several numerical \textit{measures of partial balance} using randomly generated graphs and basic axioms. The results highlight using the \textit{frustration index}, a measure that satisfies key axiomatic properties and allows us to analyse graphs based on their levels of partial balance \cite{aref2015measuring}.
	
	Two types of random graphs that we use are Erd\H{o}s-R\'{e}nyi graphs and Barab\'{a}si-Albert graphs \cite{bollobas2001random}. Erd\H{o}s-R\'{e}nyi graphs, denoted by $G(n,M)$, are a type of random graphs generated based on a model named after Paul Erd\H{o}s and Alfr\'{e}d R\'{e}nyi in which given a fixed vertex set of size $n$, all graphs with $M$ edges are equally likely to be generated. Note that another model for generating random graphs is contemporaneously introduced by Edgar Gilbert in which each edge has a fixed probability $p$ of being present or absent in a graph with $n$ nodes. Such randomly generated graphs are denoted by $G(n,p)$, but also referred to as Erd\H{o}s-R\'{e}nyi graphs. Throughout this thesis, we used the term Erd\H{o}s-R\'{e}nyi graphs alongside the distinctive notation to clarify the type of Erd\H{o}s-R\'{e}nyi graph.
	
	Barab\'{a}si-Albert graphs are another type of random graphs that are generated based on the preferential attachment process \cite{bollobas2001random}. According to this random graph generation model, a graph is grown by attaching new nodes each with a certain number of edges that are preferentially attached to existing high-degree nodes.	Different types of random graphs can be generated using the \textit{NetworkX} package. NetworkX provides functions which take parameters such as size and order and generate graphs according to certain random graph generation models and processes such as Erd\H{o}s-R\'{e}nyi model or preferential attachment process. 
	
	The exact algorithms used in the literature to compute the frustration index, also called the \textit{line index of balance}, are not scalable and cannot process graphs with a few hundred edges. In Chapter \ref{ch:2}, we formulate computing the frustration index as a graph optimisation problem in order to find the minimum number of edges whose removal results in a balanced network given binary decision variables associated with graph nodes and edges. We use our first optimisation model to analyse graphs with up to 3000 edges. Such computations take a few seconds on an ordinary computer \cite{aref2017computing}.
	
	In Chapter \ref{ch:3}, we reformulate the optimisation problem to develop three more efficient \textit{binary linear programming} models. Equipping the models with \textit{valid inequalities} and \textit{prioritised branching} as speed-up techniques allows us to process graphs with 15000 edges. Using our more advanced models, such instances take less than a minute on inexpensive hardware. Besides making exact computations possible for large graphs, we show that our models outperform heuristics and approximation algorithms suggested in the literature by orders of magnitude \cite{aref2016exact}.
	
	In Chapter \ref{ch:4}, we extend the concepts of balance and frustration in signed networks to applications beyond the classic friend-enemy interpretation of balance theory in social context. Using a high-performance computer, we analyse graphs with up to 100000 edges to investigate a range of applications from biology and chemistry to finance, international relations, and physics. The longest solve time for these instances is 9.3 hours. We use the frustration index as a measure of distance to monotonicity in biological networks, a predictor of fullerene chemical stability, a measure of bi-polarisation in international relations, a measure of financial portfolio performance, and an indicator of ground-state energy in models of atomic magnets \cite{aref2017balance}. 
	
	Chapters \ref{ch:1} -- \ref{ch:4} of this thesis are based on the results from the following papers \cite{aref2015measuring, aref2017computing, aref2016exact, aref2017balance}. Links to publisher's verified versions of the four papers are provided in the bibliography. Each chapter is written as a self-contained paper and the readers who are interested in a specific chapter can directly jump to that chapter. Those who read this thesis as a whole may notice several preliminary definitions recurring at the beginning of each chapter. In particular, the readers may notice an overlap between Chapter \ref{ch:2} and Chapter \ref{ch:3} that both concern computing the frustration index. More introductory discussions regarding computing the frustration index are provided in Chapter \ref{ch:2}, while Chapter \ref{ch:3} concerns more advanced discussions about the efficiency of such computations.
}

% ====================================================
%
% MAINMATTER
%
% Include external chapter files here using
% the \input{} command
%
% If you run out of memory during compilation,
% switch some or all chapters to \include{} instead of \input{}, 
% but watch out for pagination problems.
%
% ====================================================

\cleardoublepage

\chapter{Measuring Partial Balance in Signed Networks}\label{ch:1}

\maketitle

%\cleardoublepage
\section*{Abstract}
{Is the enemy of an enemy necessarily a friend? If not, to what extent does this tend to hold? Such questions were formulated in terms of signed (social) networks and necessary and sufficient conditions for a network to be ``balanced" were obtained around 1960. Since then the idea that signed networks tend over time to become more balanced has been widely used in several application areas. However, investigation of this hypothesis has been complicated by the lack of a standard measure of partial balance, since complete balance is almost never achieved in practice. We formalise the concept of a measure of partial balance, discuss various measures, compare the measures on synthetic datasets, and investigate their axiomatic properties. The synthetic data involves Erd\H{o}s-R\'{e}nyi and specially structured random graphs. We show that some measures behave better than others in terms of axioms and ability to differentiate between graphs. We also use well-known datasets from the sociology and biology literature, such as Read's New Guinean tribes, gene regulatory networks related to two organisms, and a network involving senate bill co-sponsorship. Our results show that substantially different levels of partial balance is observed under cycle-based, eigenvalue-based, and frustration-based measures. We make some recommendations for measures to be used in future work.}
%\textbf{Keywords:}
%{structural analysis, signed networks, balance theory, axiom, frustration index, algebraic conflict}
%

%2000 Math Subject Classification: 05C22, 05C38, 91D30, 90B10

\cleardoublepage
\section{Introduction to Chapter \ref{ch:1}} \label{1s:intro}

Transitivity of relationships has a pivotal role in analysing social interactions. Is the enemy of an enemy a friend? What about the friend of an enemy or the enemy of a friend? Network science is a key instrument in the quantitative analysis of such questions. Researchers in the field are interested in knowing the extent of transitivity of ties and its impact on the global structure and dynamics in communities with positive and negative relationships. Whether the application involves international relationships among states, friendships and enmities between people, or ties of trust and distrust formed among shareholders, relationship to a third entity tends to be influenced by immediate ties.

There is a growing body of literature that aims to connect theories of social structure with network science tools and techniques to study local behaviours and global structures in signed graphs that come up naturally in many unrelated areas. The building block of structural balance is a work by Heider \cite{heider_social_1944} that was expanded into a set of graph-theoretic concepts by Cartwright and Harary \cite{cartwright_structural_1956} to handle a social psychology problem a decade later. The relationship under study has an antonym or dual to be expressed by the opposite sign \cite{harary_structural_1957}. In a setting where the opposite of a negative relationship is a positive relationship, a tie to a distant neighbour can be expressed by the product of signs reaching him. Cycles containing an odd number of negative edges are considered to be unbalanced, guaranteeing total balance therefore only in networks containing no such cycles. This strict condition makes it quite unlikely for a signed network to be totally balanced. The literature on signed networks suggests many different formulae to measure balance. These measures are useful for detecting total balance and imbalance, but for intermediate cases their performance is not clear and has not been systematically studied.

\subsection*{\textbf{Our contribution in Chapter \ref{ch:1}}}
The main focus of this chapter is to provide insight into measuring partial balance, as much uncertainty still exists on this. The dynamics leading to specific global structures in signed networks remain speculative even after studies with fine-grained approaches. The central thesis of this chapter is that not all measures are equally useful. We provide a numerical comparison of several measures of partial balance on a variety of undirected signed networks, both randomly generated and inferred from well-known datasets. Using theoretical results for simple classes of graphs, we suggest an axiomatic framework to evaluate these measures and shed light on the methodological details involved in using such measures.

This chapter begins by laying out the theoretical dimensions of the research in Section~\ref{1s:problem} and looks at basic definitions and terminology. In Section~\ref{1s:check} different means of checking for total balance are outlined. Section~\ref{1s:measure} discusses some approaches to measuring partial balance in Eq. \eqref{1eq1.5} -- \eqref{1eq5.1}, categorised into three families of measures \ref{1ss:familyc} -- \ref{1ss:familyf} and summarised in Table~\ref{1tab1}. Numerical results on synthetic data are provided in Figures \ref{1fig1} -- \ref{1fig2} in Section~\ref{1s:basic}. Section~\ref{1s:special} is concerned with analytical results on synthetic data in closed-form formulae in Table~\ref{1tab2} and visually represented in Figures \ref{1fig3} -- \ref{1fig5}. Axioms and desirable properties are suggested in Section~\ref{1s:axiom} to evaluate the measures systematically. Section~\ref{1s:recom} concerns recommendations for choosing a measure of balance. Numerical results on real signed networks are presented in Section~\ref{1s:real}. Finally, Section~\ref{1s:conclu} summarises the chapter. 

\section{Problem statement and notation} \label{1s:problem}
Throughout this chapter, the terms signed graph and signed network will be used interchangeably to refer to a graph with positive and negative edges. We use the term cycle only as a shorthand for referring to simple cycles of the graph. While several definitions of the concept of balance have been suggested, this chapter will only use the definition for undirected signed graphs unless explicitly stated.

We consider an undirected signed network $G = (V,E,\sigma)$ where $V$ and $E$ are the sets of vertices and edges, and $\sigma$ is the sign function 
$\sigma: E\rightarrow\{-1,+1\}$. The set of nodes is denoted by $V$, with $|V| = n$. The set of edges is represented by $E$ including $m^-$ negative edges and $m^+$ positive edges adding up to a total of $m=m^+ + m^-$ edges. We denote the graph density by $\rho= 2m/(n(n-1))$. The symmetric \emph{signed adjacency matrix} and the \emph{unsigned adjacency matrix} are denoted by \textbf{A} and ${|\textbf{A}|}$ respectively. Their entries are defined in \eqref{1eq1} and \eqref{1eq1.1}. 
\begin{equation}\label{1eq1}
a{_u}{_v} =
\left\{
\begin{array}{ll}
\sigma_{(u,v)} & \mbox{if } {(u,v)}\in E \\
0 & \mbox{if } {(u,v)}\notin E
\end{array}
\right.
\end{equation}

\begin{equation}\label{1eq1.1}
|a{_u}{_v}| =
\left\{
\begin{array}{ll}
1 & \mbox{if } {(u,v)}\in E \\
0 & \mbox{if } {(u,v)}\notin E
\end{array}
\right.
\end{equation}

The \emph{positive degree} and \emph{negative degree} of node $i$ are denoted by $d^+ _{i}$ and $d^- _{i}$ representing the number of positive and negative edges incident on node $i$ respectively. They are calculated based on $d^+ _{i}=(\sum_j |a_{ij}| + \sum_j a_{ij})/2$ and $d^- _{i}=(\sum_j |a_{ij}| - \sum_j a_{ij})/2$. The \emph{degree} of node $i$ is represented by $d_i$ and equals the number of edges incident on node $i$. It is calculated based on $d_{i}= d^+ _{i} + d^- _{i}= \sum_j |a_{ij}|$. 

A \emph{walk} of length $k$ in $G$ is a sequence of nodes $v_0,v_1,...,v_{k-1},v_k$ such that for each $i=1,2,...,k$ there is an edge from $v_{i-1}$ to $v_i$. If $v_0=v_k$, the sequence is a \emph{closed walk} of length $k$. If all the nodes in a closed walk are distinct except the endpoints, it is a \emph{cycle} (simple cycle) of length $k$. The \emph{sign of a cycle} is the product of the signs of its edges. A cycle is \emph{balanced} if its sign is positive and is \emph{unbalanced} otherwise. The total number of balanced cycles (closed walks) of length $k$ is denoted by $O_k ^+$ ($Q_k ^+$). Similarly, $O_k ^-$ ($Q_k ^-$) denotes the total number of unbalanced cycles (closed walks) of length $k$. The total number of cycles (closed walks) of length $k$ is represented by $O_k=O_k ^+ + O_k ^-$ ($Q_k=Q_k ^+ + Q_k ^-$).

We use $G_r=(V,E,\sigma_r)$ to denote a reshuffled graph in which the sign function $\sigma_r$ is a random mapping of $E$ to $\{-1,+1\}$ that preserves the number of negative edges. The reshuffling process preserves the underlying graph structure. 

\section{Checking for balance}\label{1s:check}
It is essential to have an algorithmic means of checking for balance. We recall several known methods here. The characterisation of \emph{bi-polarity} (also called bipartitionability), that a signed graph is balanced if and only if its vertex set can be partitioned into two subsets such that each negative edge joins vertices belonging to different subsets \cite{harary_notion_1953}, leads to an algorithm of complexity $\mathcal{O}(m)$ \cite{harary_simple_1980} similar to the usual algorithm for determining whether a graph is bipartite. 
An alternate algebraic criterion is that the eigenvalues of the signed and unsigned adjacency matrices are equal if and only if the signed network is balanced \cite{acharya_spectral_1980} which results in an algorithm of complexity $\mathcal{O}(n^2)$ to check for balance. For our purposes the following additional method of detecting balance is also important. We define the \emph{switching function} $g(X)$ operating over a set of vertices $X\subseteq V$ as follows.
\begin{equation} \label{1eq1.2}
\sigma^{g(X)} _{(u,v)}=
\left\{
\begin{array}{ll}
\sigma_{(u,v)} & \mbox{if } {u,v}\in X  \ \text{or} \ {u,v}\notin X  \\
-\sigma_{(u,v)} & \mbox{if } (u \in X \ \text{and} \ v\notin X) \ \text{or} \ (u \notin X \ \text{and} \ v \in X)
\end{array}
\right.
\end{equation}
As the sign of cycles remains the same when $g$ is applied, any balanced graph can switch to an all-positive signature \cite{hansen_labelling_1978}. Accordingly, a balance detection algorithm of complexity $\mathcal{O}(n^2)$ 
can be developed by constructing a switching rule on a spanning tree and a root vertex, as suggested in \cite{hansen_labelling_1978}.
Finally, another method of checking for balance in connected signed networks makes use of the signed Laplacian matrix defined by $\textbf{L}=\textbf{D}-\textbf{A}$ where $\textbf{D}_{ii} = \sum_j |a_{ij}|$ is the diagonal matrix of degrees. The signed Laplacian matrix, $\textbf{L}$, is positive-semidefinite i.e. all of its eigenvalues are nonnegative \cite{zaslavsky1983signed,zaslavsky_matrices_2010}. The smallest eigenvalue of $\textbf{L}$ equals 0 if and only if the graph is balanced \cite[Section 8A]{zaslavsky1983signed}. This leads to an $\mathcal{O}(n^2)$ balance checking algorithm. 

\section{Measures of partial balance} \label{1s:measure}
Several ways of measuring the extent to which a graph is balanced have been introduced by researchers. We discuss three families of measures here and summarise them in Table~\ref{1tab1}.

\subsection{Measures based on cycles}\label{1ss:familyc}

The simplest of such measures is the \textit{degree of balance} suggested by Cartwright and Harary \cite{cartwright_structural_1956}, which is the fraction of balanced cycles:
\begin{equation}\label{1eq1.5}
D(G)= \frac {\sum \limits_{k=3}^n O_k ^+ } {\sum \limits_{k=3}^n O_k}
\end{equation}

There are other cycle-based measures closely related to $D(G)$. The \emph{relative $k$-balance}, denoted by $D_k(G)$ and formulated in Eq.\ \eqref{1eq1.6} is a cycle-based measure where the sums defining the numerator and denominator of $D(G)$ are restricted to a single term of fixed index $k$ \cite{harary_structural_1957,harary1977graphing}. The special case $k=3$ is called the \emph{triangle index}, denoted by $T(G)$.

\begin{equation}\label{1eq1.6}
D_k(G)= \frac {O_k ^+} {O_k}
\end{equation}

Giscard et al.\ have recently introduced efficient algorithms for counting simple cycles \cite{giscard2016general} making it possible to use various measures related to $D_k(G)$ to evaluate balance in signed networks \cite{Giscard2016}. 

A generalisation is \emph{weighted degree of balance}, obtained by weighting cycles based on length as in Eq.\ \eqref{1eq1.7}, in which $f(k)$ is a monotonically decreasing nonnegative function of the length of the cycle. 

\begin{equation}\label{1eq1.7}
C(G)=\frac{\sum \limits_{k=3}^n f(k) O_k ^+ }{\sum \limits_{k=3}^n f(k) O_k}
\end{equation}

The selection of an appropriate weighting function is briefly discussed by Norman and Roberts \cite{norman_derivation_1972}, suggesting functions such as $1/k,1/k^2,1/2^k$, but no objective criterion for choosing such a weighting function is known. We consider two weighting functions $1/k$ and $1/k!$ for evaluating $C(G)$ in this chapter. Given the typical distribution of cycles of different lengths, $f(k)=1/k$ makes $C(G)$ mostly dominated by longer cycles that are more frequent while $f(k)=1/k!$ makes $C(G)$ mostly determined by shorter cycles.

Although fast algorithms are developed for counting and listing cycles of undirected graphs \cite{birmele2013optimal, giscard2016general}, the number of cycles grows 
exponentially with the size of a typical real-world network. 
To tackle the computational complexity, Terzi and Winkler \cite{terzi_spectral_2011} used $D_3(G)$ in their study and made use of the equivalence between triangles and closed walks of length $3$. The triangle index can be calculated efficiently by the formula in \eqref{1eq1.8} where $\Tr(\textbf{A})$ denotes the trace\footnote{The trace of a matrix is the sum of its diagonal entries.} of ${\textbf{A}}$. 
\begin{align}\label{1eq1.8}
T(G)= D_3(G) = \frac{O_3 ^+}{O_3} = \frac{\Tr({\textbf{A}}^3)+\Tr({|\textbf{A}|}^3)}{2 \Tr({|\textbf{A}|}^3)} 
\end{align}

The \emph{relative signed clustering coefficient} is suggested as a measure of balance by Kunegis \cite{kunegis_applications_2014}, taking insight from the classic clustering coefficient. After normalisation, this measure is equal to the triangle index. Having access to an easy-to-compute formula \cite{terzi_spectral_2011} for $T(G)$ obviates the need for a clustering-based calculation which requires iterating over all triads\footnote{groups of three nodes} in the graph.

Bonacich argues that dissonance and tension are unclear in cycles of length greater than three \cite{bonacich_introduction_2012}, justifying the use of the triangle index to analyse structural balance. However, the neglected interactions may represent potential tension and dissonance, though not as strong as that represented by unbalanced triads%, still determinant of network structure
. One may consider a smaller weight for longer cycles, thereby reducing their impact rather than totally disregarding them. Note that $C(G)$ is a generalisation of both $D(G)$ and $D_3(G)$. 

In all the cycle-based measures, we consider a value of $1$ for the case of division by zero. This allows the measures $D(G)$ and $C(G)$ ($D_{k}(G)$) to provide a value for acyclic graphs (graphs with no $k$-cycle).

\subsection{Measures related to eigenvalues}\label{1ss:familye}

Beside checking cycles, there are computationally easier approaches to evaluating structural balance such as the walk-based approach. The \emph{walk-based measure of balance} is suggested by Pelino and Maimone \cite{pelino2012towards} with more weight placed on shorter closed walks than the longer ones. Let $\Tr(e^\textbf{A})$ and $\Tr(e^{|\textbf{A}|})$ denote the trace of the matrix exponential\footnote{The matrix exponential is a matrix function similar to the ordinary exponential function.} for $\textbf{A}$ and $|\textbf{A}|$ respectively. In Eq.\ \eqref{1eq2}, closed walks are weighted by a function with a relatively fast rate of decay compared to functions suggested in \cite{norman_derivation_1972}. The weighted ratio of balanced to total closed walks is formulated in Eq.\ \eqref{1eq2}. 
\begin{equation}\label{1eq2}
W(G)= \frac {K(G)+1}{2}, \quad K(G)=\frac{\sum \limits_{k}\frac{Q_k ^+ - Q_k ^-}{k!}}{\sum \limits_{k} \frac{Q_k ^+ + Q_k ^-}{k!}}=\frac{\Tr(e^\textbf{A})}{\Tr(e^{|\textbf{A}|})}  
\end{equation}
Regarding the calculation of $\Tr(e^\textbf{A})$, one may use the standard fact that $\textbf{A}$ is a symmetric matrix for undirected graphs. It follows that $\Tr(e^\textbf{A})=\sum _{i} e^{\lambda_i}$ in which $\lambda_i$ ranges over eigenvalues of $\textbf{A}$. The idea of a walk-based measure was then used by Estrada and Benzi \cite{estrada_walk-based_2014}. They have tested their measure on five signed networks resulting in values inclined towards imbalance which were in conflict with some previous observations \cite{facchetti_computing_2011,kunegis_applications_2014}. The walk-based measure of balance suggested in \cite{estrada_walk-based_2014} have been scrutinised in the subsequent studies \cite{Giscard2016,singh2017measuring}. Giscard et al. discuss how using closed-walks in which the edges might be repeated results in mixing the contribution of various cycle lengths and leads to values that are difficult to interpret \cite{Giscard2016}. Singh et al. criticise the walk-based measure from another perspective and explains how the inverse factorial weighting distorts the measure towards showing imbalance \cite{singh2017measuring}.

The idea of another eigenvalue-based measure comes from spectral graph theory \cite{kunegis_spectral_2010}. The smallest eigenvalue of the signed Laplacian matrix (defined in Section \ref{1s:check}) provides a measure of balance for connected graphs called \textit{algebraic conflict} \cite{kunegis_spectral_2010}. Algebraic conflict, denoted by $\lambda(G)$, equals zero if and only if the graph is balanced. Positive-semidefiniteness of $\textbf{L}$ results in $\lambda(G)$ representing the amount of imbalance in a signed network. Algebraic conflict is used in \cite{kunegis_applications_2014} to compare the level of balance in online signed networks of different sizes. Moreover, Pelino and Maimone analysed signed network dynamics based on $\lambda(G)$ \cite{pelino2012towards}. Bounds for $\lambda(G)$ are investigated by \cite{Hou2004} leading to recent applicable results in \cite{Belardo2014,Belardo2016}. Belardo and Zhou prove that $\lambda(G)$ for a fixed $n$ is maximised by the complete all-negative graph of order $n$ \cite{Belardo2016}. Belardo shows that $\lambda(G)$ is bounded by $\lambda_\text{max}(G)=\overline{d}_{\text{max}} -1$ in which $\overline{d}_{\text{max}}$ represents the maximum average degree of endpoints over graph edges \cite{Belardo2014}. We use this upper bound to normalise algebraic conflict. \textit{Normalised algebraic conflict}, denoted by $A(G)$, is expressed in Eq.\ \eqref{1eq3}. 
\begin{equation}\label{1eq3}
A(G)=1- \frac {\lambda(G)}{\overline{d}_{\text{max}} -1} , \quad \overline{d}_{\text{max}}=\max_{(u,v)\in E} (d_u+d_v)/2
\end{equation}

\subsection{Measures based on frustration}\label{1ss:familyf}

A quite different measure is the \emph{frustration index} \cite{abelson_symbolic_1958,harary_measurement_1959,zaslavsky_balanced_1987} that is also referred to as the \emph{line index for balance} \cite{harary_measurement_1959}. A set $E^*$ of edges is called \textit{minimum deletion set} if deleting all edges in $E^*$ results in a balanced graph and deleting edges from no smaller set leads to a balanced graph. The frustration index equals the cardinality of a minimum deletion set as in Eq.\ \eqref{1eq5.0}. 

\begin{equation}\label{1eq5.0}
L(G)= {|E^*|}
\end{equation}

Each edge in $E^*$ lies on an unbalanced cycle and every unbalanced cycle of the network contains an odd number of edges in $E^*$. Iacono et al. showed that $L(G)$ equals the minimum number of unbalanced fundamental cycles induced over all spanning trees\footnote{A minimal set of cycles which may be formed from any spanning tree of a given graph, through choosing the cycles formed by combining a path from the tree with a single edge from outside the tree.} of the graph \cite{iacono_determining_2010}. The graph resulted from deleting all edges in $E^*$ is called a \textit{balanced transformation} of a signed graph. 

Similarly, in a setting where each vertex is given a black or white colour, if the endpoints of positive (negative) edges have different colours (same colour), they are ``frustrated". The frustration index is therefore the smallest number of frustrated edges over all possible 2-colourings of the nodes.

$L(G)$ is hard to compute as the special case with all edges being negative is equivalent to the MAXCUT problem \cite{gary1979computers}, which is known to be NP-hard. There are upper bounds for the frustration index such as $L(G)\leq m^-$ which states the obvious result of removing all negative edges. 

Facchetti, Iacono, and Altafini have used computational methods related to Ising spin glass models to estimate the frustration index in relatively large online social networks \cite{facchetti_computing_2011}. Using an estimation of the frustration index obtained by a heuristic algorithm, they concluded that the online signed networks are extremely close to total balance; an observation that contradicts some other research studies like \cite{estrada_walk-based_2014}. 

The number of frustrated edges in special Erd\H{o}s-R\'{e}nyi graphs, $G(n,p)$, is analysed by El Maftouhi, Manoussakis and Megalakaki \cite{el_maftouhi_balance_2012}. It follows a binomial distribution with parameters $n(n-1)/2$ and $p/2$ in which $p$ represents the sum of equal probabilities for positive and negative edges in Erd\H{o}s-R\'{e}nyi graph $G(n,p)$. Therefore, the expected number of frustrated edges is $n(n-1)p/4$. They also prove that such a network is almost always not balanced when $p \geq (\log2)/n$. It is straightforward to prove that frustration index is equal to the minimum number of negative edges over all switching functions \cite{zaslavsky_matrices_2010}. Petersdorf \cite{petersdorf_einige_1966} proves that the frustration index is bounded by $\lfloor \left(n-1 \right)^2/4 \rfloor$.

Bounds for the largest number of frustrated edges for a graph with $n$ nodes and $m$ edges are provided in \cite{akiyama_balancing_1981}. It follows that ${L(G)} \leq {m}/{2}$; an upper bound that is not necessarily tight.

Another upper bound for the frustration index is reported in \cite{iacono_determining_2010} referred to as the worst-case upper bound on the consistency deficit. However, the frustration index values in complete graphs with all negative edges shows that the upper bound is incorrect.

In order to compare with the other indices which take values in the unit interval and give the value $1$ for balanced graphs, we suggest \emph{normalised frustration index}, denoted by $F(G)$ and formulated in Eq.\ \eqref{1eq5.1}. 

\begin{equation}\label{1eq5.1}
F(G)=1- \frac {L(G)}{m/2}
\end{equation}

Using a different upper bound for normalising the frustration index, we discuss another frustration-based measure in Subsection \ref{1ss:normal} and formulate it in Eq.\ \eqref{1eq5.8}.

\subsection{Other methods of evaluating balance}

Balance can also be analysed by blockmodeling\footnote{Blockmodeling is a method for dividing network vertices into particular sets called blocks.} based on iteratively calculating Pearson moment correlations\footnote{The Pearson moment correlation is a measure of correlation which quantifies the strength and the direction of relationship between two variables.} from the columns of $\textbf{A}$ \cite{doreian_generalized_2005}. Blockmodeling reveals increasingly homogeneous sets of vertices. 

Doreian and Mrvar discuss this approach in partitioning signed networks \cite{doreian_partitioning_2009}. Applying the method to Correlates of War data on positive and negative international relationships, they refute the hypothesis that signed networks gradually move towards balance using blockmodeling alongside some variations of $D(G)$ and $L(G)$ \cite{patrick_doreian_structural_2015}. 

Moreover, there are probabilistic methods that compare the expected number of balanced and unbalanced triangles in the signed network and its reshuffled version \cite{leskovec_signed_2010,yap_why_2015, Szell_multi, Szell_measure}. As long as these measures are used to evaluate balance, the result will not be different to what $T(G)$ provides alongside a basic statistical testing of its value against reshuffled networks.

Some researchers suggest that studying the structural dynamics of signed networks is more important than measuring balance \cite{cai2015particle,ma_memetic_2015}. This approach is usually associated with considering an energy function to be minimised by local graph operations decreasing the energy. However, the energy function is somehow a measure of network imbalance which requires a proper definition and investigation of axiomatic properties. Seven measures of partial balance investigated in this chapter are outlined in Table~\ref{1tab1}.

\begin{table}[ht]
	\centering
	\caption{Measures of partial balance summarised}
	\label{1tab1}
	\begin{tabularx}{\textwidth}{ll}
 \hline
		Measure & \multicolumn{1}{c}{Name, Reference, and Description}                                                 \\ \hline
		$D(G)$    & \textit{Degree of balance} \cite{cartwright_structural_1956, harary_measurement_1959}                      \\
		& A cycle-based measure representing the ratio of balanced cycles                          \\
		$C(G)$    & \textit{Weighted degree of balance} \cite{norman_derivation_1972}                                 \\
		& An extension of $D(G)$ using cycles weighted by a function of length  \\
		$D_k(G)$& \textit{Relative $k$-balance} \cite{harary_structural_1957, harary1977graphing} \\
		& A variant of $C(G)$ placing a non-zero weight only on cycles of length $k$  \\
		$T(G)$    &  \textit{Triangle index} \cite{terzi_spectral_2011, kunegis_applications_2014}                             \\
		& A triangle-based measure representing the ratio of balanced triangles           \\
		$W(G)$    & \textit{Walk-based measure of balance} \cite{pelino2012towards, estrada_walk-based_2014}          \\
		& A simplified extension of $D(G)$ replacing cycles by closed walks                        \\
		$A(G)$    & \textit{Normalised algebraic conflict} \cite{kunegis_spectral_2010, kunegis_applications_2014}      \\
		& A normalised measure using least eigenvalue of the Laplacian matrix          \\
		$F(G)$    & \textit{Normalised frustration index} \cite{harary_measurement_1959, facchetti_computing_2011}                    \\
		& Normalised minimum number of edges whose removal results in balance                                \\ \hline 
	\end{tabularx}
\end{table}

\subsection*{\textbf{Outline of the rest of the chapter}}
We started by discussing balance in signed networks in Sections \ref{1s:problem} and \ref{1s:check} and reviewed different measures in Section~\ref{1s:measure}. We will provide some observations on synthetic data in Figures \ref{1fig1} -- \ref{1fig5} in Sections \ref{1s:basic} and \ref{1s:special} to demonstrate the values of different measures. The reader who is not particularly interested in the analysis of measures using synthetic data may directly go to Section~\ref{1s:axiom} in which we introduce axioms and desirable properties for measures of partial balance. In Section~\ref{1s:recom}, we provide some recommendations on choosing a measure and discuss how using unjustified measures has led to conflicting observations in the literature. The numerical results on real signed networks are presented in Section~\ref{1s:real}.

\section{Numerical results on synthetic data} \label{1s:basic}
In this section, we start with a brief discussion on the relationship between negative edges and imbalance in networks. According to the definition of structural balance, all-positive signed graphs (merely containing positive edges) are totally balanced. Intuitively, one may expect that all-negative signed graphs are very unbalanced. Perhaps another intuition derived by assuming symmetry is that increasing the number of negative edges in a network reduces partial balance proportionally. We analyse partial balance in randomly generated graphs to evaluate these intuitions. Our motivation for analysing balance in such graphs is to gain an understanding of the behaviour of the measures and their connections with signed graph parameters like $m^-,n$, and $\rho$. %We see that the underlying structure is not important for this analysis. So, we use randomly generated graphs.

\subsection{Erd\H{o}s-R\'{e}nyi random network with various $m^-$}
\label{1ss:erdos}
\begin{figure}
	
	\subfloat[The mean values of seven measures of partial balance]{\includegraphics[height=2.4in]{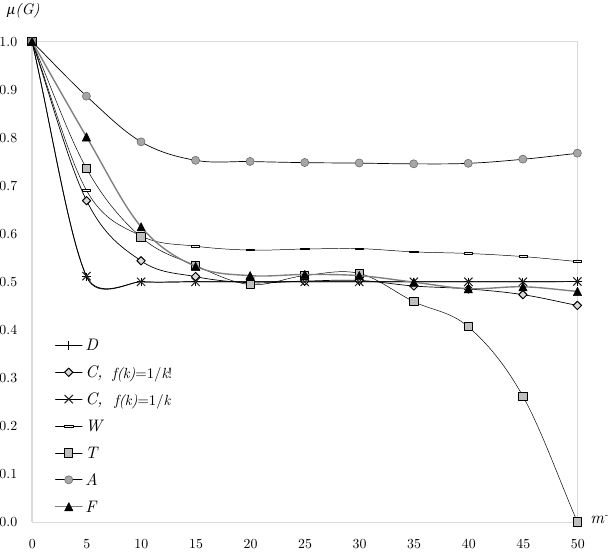}%
		\label{1fig1_first_case}}
	\hfil
	\subfloat[The mean values of relative $k$-balance $D_k$]{\includegraphics[height=2.4in]{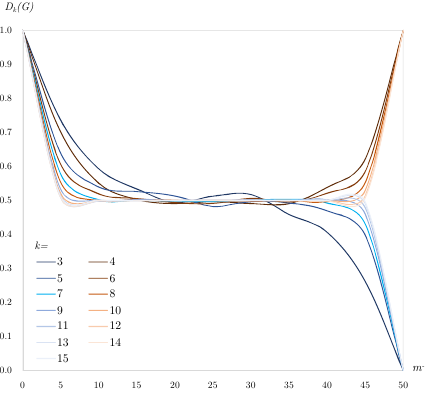}%
		\label{1fig1_second_case}}
	
	~
	
	\subfloat[The standard deviation of $D$ and $C$]{\includegraphics[height=2.5in]{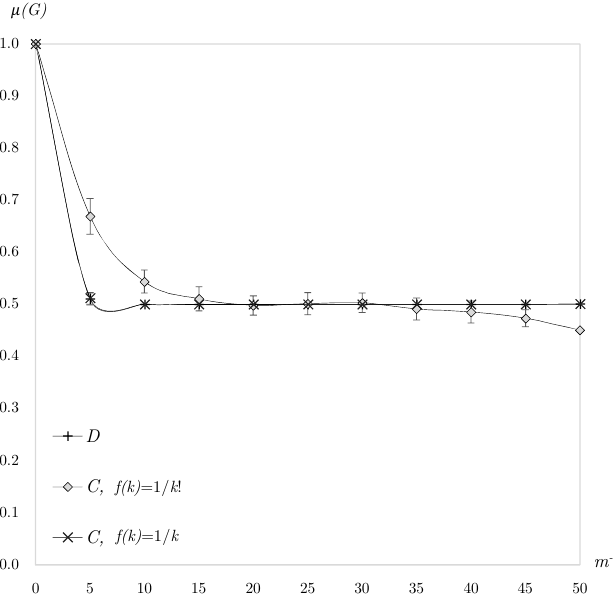}%
		\label{1fig1_third_case}}
	\hfil
	\subfloat[The standard deviation of $W, T, A$ and $F$]{\includegraphics[height=2.5in]{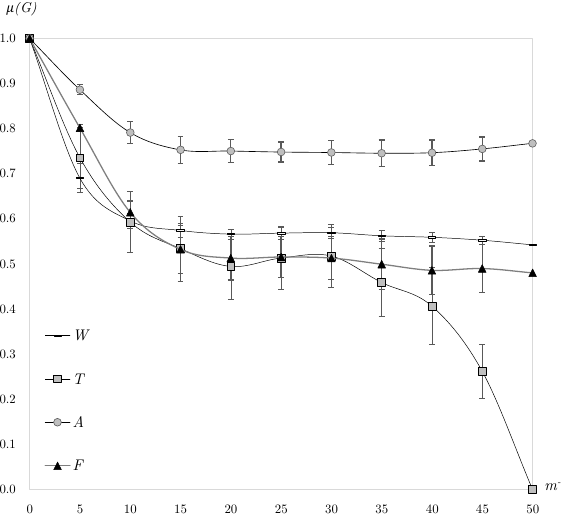}%
		\label{1fig1_fourth_case}}
	
	\caption{Partial balance measured by different methods in Erd\H{o}s-R\'{e}nyi network, $G(n,M)$, with various number of negative edges}
	\label{1fig1}
\end{figure}

We calculate measures of partial balance, denoted by $\mu(G)$, for an Erd\H{o}s-R\'{e}nyi random network, $G(n,M)$, with 15 nodes, 50 edges, and a various number of negative edges. Figure~\ref{1fig1} demonstrates the partial balance measured by different methods. For each data point, we report the average of 50 runs, each assigning negative edges at random to the fixed underlying graph. The subfigures (c) and (d) of Figure~\ref{1fig1} show the mean along with $\pm 1$ standard deviation. 

Measures $D(G)$ and $C(G)$ with $f(k)=1/k$ are observed to tend to $0.5$ where $m^- > 5$, not differentiating partial balance in graphs with a non-trivial number of negative edges. Given the typical distribution of cycles of different lengths, we expect $D(G)$ and $C(G)$ with $f(k)=1/k$ to be mostly determined by longer cycles that are much more frequent. For this particular graph, cycles with a length of 10 and above account for more than $96 \%$ of the total cycles in the graph. Such long cycles tend to be balanced roughly half the time for almost all values of $m^-$ (for all the values within the range of $5 \leq m^- \leq 45$ in the network considered here). The perfect overlap of data points for $D(G)$ and $C(G)$ with $f(k)=1/k$ in Figure~\ref{1fig1} shows that using a linear rate of decay does not make a difference. One may think that if we use $D_k(G)$ which does not mix cycles of different lengths, it may circumvent the issues. However, subfigure (b) of Figure~\ref{1fig1} demonstrating values of $D_k(G)$ for different cycle lengths shows the opposite. It shows not only does $D_k(G)$ not resolve the problems of lack of sensitivity and clustering around $0.5$, but it behaves unexpectedly with substantially different values based on the parity of $k$ when $m^- > 35$. $C(G)$ weighted by $f(k)=1/k!$ and mostly determined by shorter cycles, decreases slower than $D(G)$ and then provides values close to $0.5$ for $m^-\geq10$. $W(G)$ drops below $0.6$ for $m^-=10$ and then clusters around $0.55$ for $m^->10$. $T(G)$ is the measure with a wide range of values symmetric to $m^-$. The single most striking observation to emerge is that $A(G)$ seems to have a completely different range of values, which we discuss further in Subsection \ref{1ss:normal}. A steady linear decrease is observed from $F(G)$ for $m^-\leq10$. 

\subsection{4-regular random networks of different orders}\label{1ss:regular}
\begin{figure}
	\centering
	\includegraphics[width=0.65\textwidth]{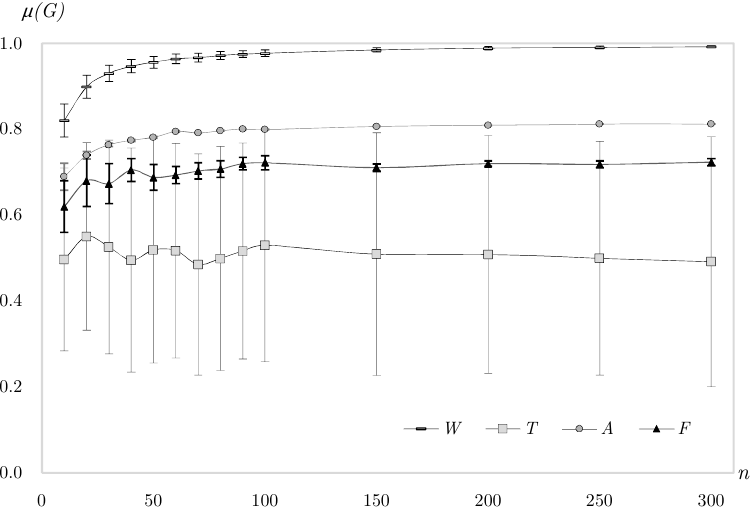}
	\caption{Partial balance measured by different methods in 50\% negative 4-regular graphs of different orders $n$ and decreasing densities $4/(n-1)$}
	\label{1fig2}
\end{figure}

To investigate the impacts of graph order (number of nodes) and density on balance, we computed the measures for randomly generated 4-regular graphs with 50 percent negative edges. Intuitively we expect values to have low variation and no trends for similarly structured graphs of different orders. Figure~\ref{1fig2} demonstrates the analysis in a setting where the degree of all the nodes remains constant, but the density ($4/(n-1)$) is decreasing in larger graphs. For each data point the average and standard deviation of 100 runs are reported. In each run, negative weights are randomly assigned to half of the edges in a fixed underlying 4-regular graph of order $n$.

According to Figure~\ref{1fig2}, the four measures differ not only in the range of values, but also in their sensitivity to the graph order and density. First, $W(G) \rightarrow 1$ when $n \rightarrow \infty$ for larger graphs although the graphs are structurally similar, which goes against intuition. Clustered around $0.5$ is $T(G)$ which features a substantial standard deviation for 4-regular random graphs. Values of $A(G)$ are around $0.8$ and do not seem to change substantially when $n$ increases. $F(G)$ provides stationary values around $0.7$ when $n$ increases. While $\lambda(G)$ and $L(G)$ depend on the graph order and size, the relative constancy of $A(G)$ and $F(G)$ values suggest the normalised measures $A(G)$ and $F(G)$ are largely independent of the graph size and order, as our intuition expects. We further discuss the normalisation of $A(G)$ and $F(G)$ in Subsection \ref{1ss:normal}.

\section{Analytical results on synthetic data} \label{1s:special}
In this section, we analyse the capability of measuring partial balance in some families of specially structured graphs. Closed-form formulae for the measures in specially structured graphs are provided in Table~\ref{1tab2}. We will describe two families of complete signed graphs in \ref{1ss:kna} and \ref{1ss:knc}. %

\subsection{Minimally unbalanced complete graphs with a single negative edge}\label{1ss:kna}
The first family includes complete graphs with a single negative edge, denoted by ${K}_n^a$. Such graphs are only one edge away from a state of total balance. It is straight-forward to provide closed-form formulae for $\mu({K}_n^a)$ as expressed in Eq.\ \eqref{1eq15} -- \eqref{1eq18} in Appendix \ref{1s:calc}.

\begin{table}[ht]
	\centering
	\caption{Balance in minimally and maximally unbalanced graphs ${K}_n^a$ (\ref{1ss:kna}) and ${K}_n^c$ (\ref{1ss:knc})}
	\label{1tab2}
	\begin{tabular}{lll}
		\hline
		$\mu(G)$ & ${K}_n^a$                                                                              & ${K}_n^c$                                                                  \\ \hline
		$D(G)$      & $\sim 1- {2}/{n}$ & $\sim \frac{1}{2} + (-1)^n e^{-2}$     \\
		$C(G), f(k)=1/k!$      & $\sim 1- {1}/{n}$ &  $\sim \frac{1}{2} - \frac{3n\log n}{2^n}$\\
		$D_k(G)$    & $1-{2k}/{n(n-1)}$ & $0 , 1$                                       \\
		$W(G)$      & $\sim 1-{2}/{n}$  & $\sim \frac{1+e^{2-2n}}{2}$                 \\
		$A(G)$      & $\sim 1-{4}/{n^2}$& $0$                                        \\
		$F(G)$      & $1-{4}/{n(n-1)}$  & $\frac{1}{n},\frac{1}{n-1}$                \\ \hline
	\end{tabular}
\end{table}

\begin{figure}
	\centering
	\includegraphics[width=0.9\textwidth]{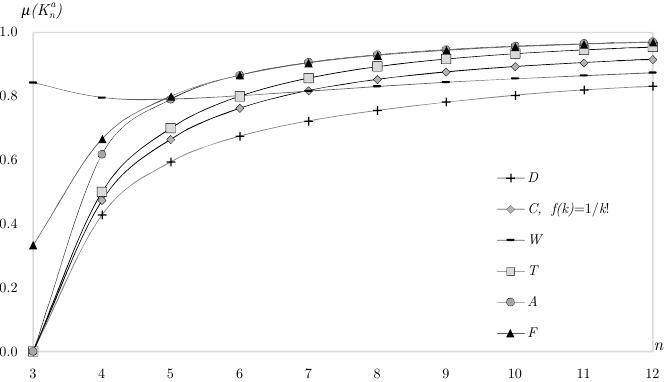}
	\includegraphics[width=0.9\textwidth]{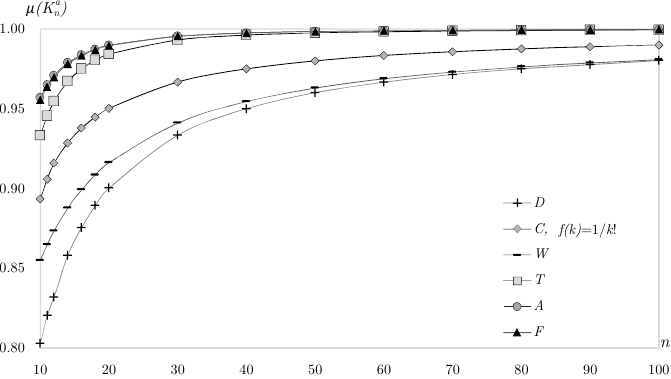}
	\caption{Partial balance measured by different methods for ${K}_n^a$ (\ref{1ss:kna})}
	\label{1fig3}
\end{figure}
In ${K}_n^a$, intuitively we expect $\mu({K}_n^a)$ to increase with $n$ and $\mu({K}_n^a) \rightarrow 1$ as $n \rightarrow \infty$. We also expect the measure to detect the imbalance in ${K}_3^a$ (a triangle with one negative edge). Figure~\ref{1fig3} demonstrates the behaviour of different indices for complete graphs with one negative edge. $W({K}_n^a)$ gives unreasonably large values for $n < 5$. Except for $W({K}_n^a)$, the measures are co-monotone\footnote{Excluding $W({K}_n^a)$, we observe a consistent order among the values of the other five measures within the given range of $n$.} over the given range of $n$.

\subsection{Maximally unbalanced complete graphs with all-negative edges}\label{1ss:knc}
The second family of specially structured graphs to analyse includes all-negative complete graphs denoted by ${K}_n^c$. The indices are calculated in Eq.\ \eqref{1eq21} -- \eqref{1eq25} in Appendix \ref{1s:calc}. 

\begin{figure}
	\centering
	\includegraphics[width=0.9\textwidth]{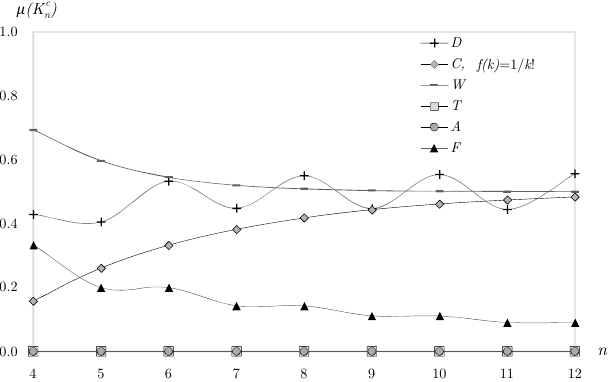}
	\includegraphics[width=0.9\textwidth]{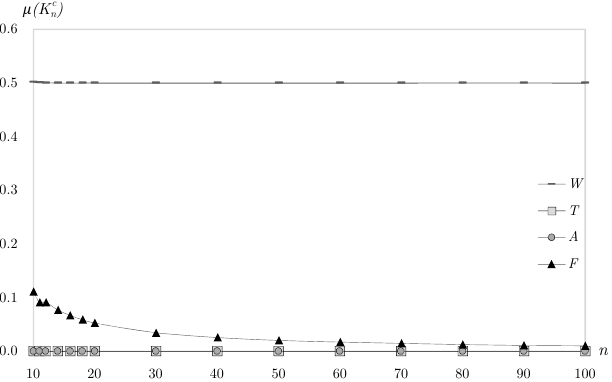}
	\caption{Partial balance measured by different methods for ${K}_n^c$ (\ref{1ss:knc})}
	\label{1fig5}
\end{figure}

Intuitively, we expect a measure of partial balance to represent the lack of balance in ${K}_n^c$ by providing a value close to $0$. Figure~\ref{1fig5} illustrates $D({K}_n^c)$ oscillating around $0.5$ and $W({K}_n^c),C({K}_n^c) \rightarrow 0.5$ as $n$ increases. We explain the oscillation of $D({K}_n^c)$ in Appendix \ref{1s:calc}. Clearly, measures $D(G)$, $C(G)$ and, $W(G)$ provide values for ${K}_n^c$ that go against our intuition. Figure~\ref{1fig5} shows that $F({K}_n^c) \rightarrow 0$ as $n \rightarrow \infty$ as expected based on Table~\ref{1tab2}.

\subsection{Normalisation of the measures}\label{1ss:normal}
It is worth mentioning that measures of partial balance may lead to different maximally unbalanced complete graphs. Based on $\lambda(G)$ and $L(G)$, ${K}_n^c$ are maximally unbalanced graphs \cite{Belardo2016, petersdorf_einige_1966} (also see Subsections \ref{1ss:familye} and \ref{2ss:bounds}), while it is merely one family among the maximally unbalanced graphs according to $T(G)$. Estrada and Benzi have found complete graphs comprised of one cycle of $n$ positive edges with the remaining pairs of nodes connected by negative edges to be a family of maximally unbalanced graphs based on $W(G)$ \cite{estrada_walk-based_2014}, while this argument is not supported by any other measures. It is difficult to find the structure of maximally unbalanced graphs under the cycle-based measures $D(G)$ and $C(G)$ partly because the signs of cycles in a graph are not independent. This is a major obstacle in finding a suitable way to normalise cycle-based measures.

A simple comparison of $L({K}_n^c)$ (calculations provided in Eq.\ \eqref{1eq24} in Appendix \ref{1s:calc}) and the proposed upper bound $m/2=(n^2-n)/4$ reveals substantial gaps. These gaps equal $n/4$ for even $n$ and $(n-1)/4$ for odd $n$. This supports the previous discussions on looseness of $m/2$ as an upper bound for frustration index. As ${K}_n^c$ is maximally unbalanced under $L(G)$, $\lfloor m/2-(n-1)/4 \rfloor$ can be used as a tight upper bound for normalising the frustration index. This allows a modified version of normalised frustration index, denoted by $F^\prime (G)$ and defined in Eq.\ \eqref{1eq5.8}, to take the value zero for ${K}_n^c$.
\begin{equation}
\label{1eq5.8}
F^\prime (G)=1-L(G)/ \lfloor m/2-(n-1)/4 \rfloor
\end{equation}

Similarly, the upper bound, $\lambda_\text{max}(G)$, used to normalise algebraic conflict, is not tight for many graphs. For instance, in the Erd\H{o}s-R\'{e}nyi graph, $G(n,M)$, studied in Section~\ref{1s:basic} with $m=m^-$, the existence of an edge with $\overline{d}_{\text{max}}=9$ makes $\lambda_\text{max}(G)=8$, while $\lambda(G)=1.98$. 

The two observations mentioned above suggest that tighter upper bounds can be used for normalisation. However, the statistical analysis we use in Section~\ref{1s:real} to evaluate balance in real networks is independent of the normalisation method, so we do not pursue this question further now.

\subsection{Expected values of the cycle-based measures}\label{1ss:expected}

Relative $k$-balance, $D_k (G)$, is proved by El Maftouhi, Manoussakis and Megalakaki \cite{el_maftouhi_balance_2012} to tend to $0.5$ for Erd\H{o}s-R\'{e}nyi graphs, $G(n,p)$, such that the probability of an edge being negative is equal to $0.5$. Moreover, Giscard et al. discuss the probability distribution of  $1 - D_k(G)$. Their discussion is based on a model in which the sign of any edge is negative with a fixed probability \cite[Section 4.2]{Giscard2016}. We use the same model to present some simple observations that appear not to have been noticed by previous authors advocating for the use of cycle-based measures. We are going to take a different approach from that of Giscard et al. and merely calculate the expected values of cycle-based measures in general, rather than the full distribution under additional assumptions. Note that for an arbitrary graph, ${O_k ^+}/ {O_k}$ gives the probability that a randomly chosen $k$-cycle is balanced and is denoted by $B_{(k,q)}$.
%\begin{theorem}
	\label{1thm:random cycle}
	Let $G$ be a graph and consider the sign function obtained by independently choosing each edge to be negative with probability $q$, and positive otherwise. Then, the expected value of $D_k(G)$,
	\begin{equation}\label{1eq1.9}
	E(D_k(G))  = (1+{(1-2q)}^k)/2 . %\\
	\end{equation}
%\end{theorem}
\begin{proof}
	Note that a cycle is balanced if and only if it has an even number of negative edges. Thus $$E\left(\frac {O_k ^+} {O_k}\right) = \sum \limits_{i \: \text{even}} {\binom{k}{i}} q^i (1-q)^{k-i}$$ (compare with \cite[Eq. 4.1]{Giscard2016}). This simplifies to the stated formula (details of calculations are given in Appendix \ref{1s:calc}).
\end{proof}

Note that the expected values are independent of the graph structure and obtaining them does not require making any assumptions on the signs of cycles being independent random variables. As the signs of the edges are independent random variables, the expected value of $B_{(k,q)}$ can be obtained by summing on all cases having an even number of negative signs in the $k$-cycle. 

Based on \eqref{1eq1.9}, $E(D_k(G)) = 1$ when $q = 0$ and $E(D_k(G)) = 0.5$ when $q = 0.5$ supporting our intuitive expectations. However, when $q = 1$, $E(D_k(G))$ takes extremal values based on the parity of $k$ which is a major problem as previously observed in the subfigure (b) of Figure~\ref{1fig1}. It is clear to see that the parity of $k$ makes a substantial difference to $D_k(G)$ when a considerable proportion of edges are negative.

%\begin{theorem}
	\label{1thm:randomallcycle}
	Let $G$ be a graph and consider the sign function obtained by independently choosing each edge to be negative with probability $q$, and positive otherwise. Then
	\begin{equation}
	E(D(G))  = \frac{1}{2} \frac { \sum \limits_{k=3}^n (1+{(1-2q)}^k)(O_k) } {\sum \limits_{k=3}^n O_k} \label{1eq1.10}
	\end{equation}
%\end{theorem}

\begin{proof}
	The random variable ${O_k ^+}$ can be written as ${O_k ^+}= B_{(k,q)}\cdot{O_k}$. Taking expected value from the two sides gives $E({O_k ^+})= {O_k}\cdot E(B_{(k,q)})$ as $O_k$ is a constant for a fixed $k$. This completes the proof using the result from Eq.\ \eqref{1eq1.9}.
\end{proof}

Note that the exponential decay of the factor $(1-2q)^k$ reduces the contribution for large $k$, and small values of $k$ will dominate for many graphs. For example, if $q=0.2$ the expression for $E(D(G))$ simplifies to 
$$
\frac{1}{2} + \frac{1}{2} \frac{ \sum \limits_{k=3}^n{0.6}^k O_k}{\sum \limits_{k=3}^n O_k}.
$$
For many graphs encountered in practice, $O_k$ will initially grow with $k$ (exponentially, but at a rate less than $1/0.6$) and then decrease, so the tail contribution will be small. Larger values of $q$ only make this effect more pronounced. Thus we expect that $E(D(G))$ will often be very close to $0.5$ in signed graphs with a reasonably large fraction of negative edges (we have already seen such a phenomenon in Subsection~\ref{1ss:erdos}). A similar conclusion can be made for $C(G)$. This casts doubt on the usefulness of the measures that mix cycles of different lengths whether weighted or not. 

While we have also observed many problems involving values of cycle-based measures on synthetic data in other parts of Sections \ref{1s:basic} and \ref{1s:special}, we will continue evaluating their axiomatic properties in Section~\ref{1s:axiom} and then summarise the methodological findings in Section~\ref{1s:recom}.

\section{Axiomatic framework of evaluation} \label{1s:axiom}
The results in Section~\ref{1s:basic} and Section~\ref{1s:special} indicate that the choice of measure substantially affects the values of partial balance. Besides that, the lack of a standard measure calls for a framework of comparing different methods. Two different sets of axioms are suggested in \cite{norman_derivation_1972}, which characterise the measure $C(G)$ inside a smaller family (up to the choice of $f(k)$). Moreover, the theory of structural balance itself is axiomatised in \cite{schwartz_friend_2010}. However, to our knowledge, axioms for general measures of balance have never been developed. Here we provide the first set of axioms and desirable properties for measures of partial balance, in order to shed light on their characteristics and performance.

\subsection{Axioms for measures of partial balance}
We define a measure of partial balance to be a function $\mu$ taking each signed graph to an element of $[0,1]$. Worthy of mention is that some of these measures were originally defined as a measure of imbalance (algebraic conflict, frustration index and the original walk-based measure) calibrated at $0$ for completely balanced structures, so that some normalisation was required, and perhaps our normalisation choices can be improved on (see Subsection \ref{1ss:normal}). As the choice of $m/2$ as the upper bound for normalising the line index of balance was somewhat arbitrary, another normalised version of frustration index is defined in Eq.\ \eqref{1eq6}. 
\begin{equation}
\label{1eq6}	
X(G)=1-L(G)/{m^-}
\end{equation}

Before listing the axioms, we justify the need for an axiomatic evaluation of balance measures. As an attempt to understand the need for axiomatising measures of balance, we introduce two unsophisticated and trivial measures that come to mind for measuring balance. The fraction of positive edges, denoted by $Y(G)$, is defined in Eq.\ \eqref{1eq7} on the basis that all-positive signed graphs are balanced. Moreover, a binary measure of balance, denoted by $Z(G)$, is defined in Eq.\ \eqref{1eq8}. While $Y(G)$ and $Z(G)$ appear to be irrelevant, there is currently no reason not to use such measures.
\begin{equation}
\label{1eq7}	
Y(G)=m^{+}/m \quad 
\end{equation}
\begin{equation}
\label{1eq8}
Z(G) =
\left\{
\begin{array}{ll}
1 & \mbox{if } $G \:$ \mbox{is totally balanced} \\
0 & \mbox{if } $G \:$ \mbox{is not balanced}
\end{array} \right.
\end{equation}

We consider the following notation for referring to basic operations on signed graphs:

\noindent
$G^{g(X)}$ denotes signed graph $G$ switched by $g(X)$ (switched graph).\\
$G\oplus H$ denotes the disjoint union of two signed graphs $G$ and $H$ (disjoint union).\\
$G\ominus e$ denotes $G$ with $e$ deleted (removing an edge).\\
$G\ominus E^*$ denotes $G$ after removing the edges in a minimum deletion set (balanced transformation).\\
$G\oplus C^+_3$ denotes the disjoint union of graphs $G$ and a positive 3-cycle (adding a balanced 3-cycle).\\
$G\oplus C^-_3$ denotes the disjoint union of graphs $G$ and a negative 3-cycle (adding an unbalanced 3-cycle).\\
$e\in E^*$ denotes an edge in a minimum deletion set.\\
$G \ominus E^* \oplus e$ denotes a balanced transformation of a graph with an edge $e$ added to it.\\

We list the following axioms:
\begin{description}
	\item[A1] $0 \leq \mu(G) \leq 1$. 
	\item[A2] $\mu(G) = 1$ if and only if $G$ is balanced. 
	\item[A3] If $\mu(G) \leq \mu(H)$, then $\mu(G) \leq \mu(G\oplus H) \leq \mu(H)$.	 
	\item[A4] $\mu(G^{g(X)}) = \mu(G)$.
\end{description}

The justifications for such axioms are connected to very basic concepts in balance theory. We consider A1 essential in order to make meaningful comparisons between measures. Introducing the notion of partial balance, we argue that total balance, being the extreme case of partial balance, should be denoted by an extremal value as in A2. In A3, the argument is that the overall balance of two disjoint graphs is bounded between their individual balances. This also covers the basic requirement that the disjoint union of two copies of graph $G$ must have the same value of partial balance as $G$. Switching nodes should not change balance \cite{zaslavsky_matrices_2010} as in A4.

Table \ref{1tab3} shows how some measures fail on particular axioms. The results provide important insights into how some of the measures are not suitable for measuring partial balance. A more detailed discussion on the proof ideas and counterexamples related to Table~\ref{1tab3} is provided in Appendix \ref{1s:counter}.
\begin{table}[hbtp]
	\centering
	\caption{Different measures satisfying or failing axioms}
	\label{1tab3}
		\begin{tabular}{llllllllll}
 \hline
			   & $D(G)$ & $C(G)$ & $W(G)$ & $D_k(G)$ & $A(G)$ & $F(G)$ & $X(G)$ & $Y(G)$ & $Z(G)$ \\ \hline
			A1 & \cmark & \cmark & \cmark & \cmark &  \cmark & \cmark & \cmark & \cmark & \cmark \\
			A2 & \cmark & \cmark & \cmark & \xmark &  \xmark & \cmark & \cmark & \xmark & \cmark \\
			A3 & \cmark & \cmark & \cmark & \cmark &  \xmark & \cmark & \cmark & \cmark & \cmark \\
			A4 & \cmark & \cmark & \cmark & \cmark &  \cmark & \cmark & \xmark & \xmark & \cmark \\
			 \hline
		\end{tabular}
\end{table}
\subsection{Some other desirable properties}

We also consider four desirable properties that formalise our expectations of a measure of partial balance. We do not consider the following as axioms in that they are based on adding or removing 3-cycles and edges which may bias the comparison in favour of cycle-based and frustration-based measures.

Positive and negative 3-cycles are very commonly used to explain the theory of structural balance which makes B1 and B2 obvious requirements. Removing an edge which belongs to a minimum deletion set, should not decrease balance as in B3. Finally, if a balanced transformation of graph $G$ becomes unbalanced by adding an edge, the addition of such an edge to the graph $G$ should not increase balance as in B4.

\begin{description}
		\item[B1]  If $\mu(G) \neq 1$, then $\mu(G\oplus C^+_3) > \mu(G)$. 
		\item[B2]  If $\mu(G) \neq 0$, then $\mu(G\oplus C^-_3) < \mu(G)$
		\item[B3]  If $e\in E^*$, then $\mu(G \ominus e) \geq \mu(G)$.
		\item[B4]  If $\mu(G)\neq 0$ and $\mu(G \ominus E^* \oplus e)\neq 1$, then $\mu(G \oplus e) \leq \mu(G)$.
\end{description}

Table \ref{1tab3.5} shows how some measures fail on particular desirable properties. It is worth mentioning that the evaluation in Tables \ref{1tab3}--\ref{1tab3.5} is somewhat independent of parametrisation: for each strictly increasing function $h$ such that $h(0)=0$ and $h(1)=1$, the results in Tables \ref{1tab3}--\ref{1tab3.5} hold for $h(\mu(G))$. Proof ideas and counterexamples related to Table~\ref{1tab3.5} is provided in Appendix \ref{1s:counter}.

\begin{table}[hbtp]
	\centering
	\caption{Different measures satisfying or failing desirable properties}
	\label{1tab3.5}
	\begin{tabular}{llllllllll}
		\hline
		   & $D(G)$ & $C(G)$ & $W(G)$ & $D_k(G)$ & $A(G)$  & $F(G)$ & $X(G)$ & $Y(G)$ & $Z(G)$ \\ \hline
		B1 & \cmark & \cmark & \cmark & \xmark &  \cmark & \cmark & \xmark & \xmark & \xmark \\
		B2 & \cmark & \cmark & \xmark & \xmark &  \xmark & \xmark & \xmark & \xmark & \cmark \\
		B3 & \xmark & \xmark & \xmark & \xmark &  \xmark & \cmark & \cmark & \xmark & \xmark \\
		B4 & \xmark & \xmark & \xmark & \xmark &  \xmark & \cmark & \cmark & \xmark & \cmark \\ 
		\hline
	\end{tabular}
\end{table}

Another desirable property, which we have not formulated as a formal requirement owing to its vagueness, is that the measure takes on a wide range of values. For example, $D(G)$ and $C(G)$ tend rapidly to $0.5$ as $n$ increases which makes their interpretation and possibly comparison with other measures difficult. A possible way to formalise it would be expecting $\mu(G)$ to give $0$ and $1$ on each complete graph of order at least $3$, for some assignment of signs of edges. This condition would be satisfied by $T(G)$ and $A(G)$, as well as $F^\prime(G)$. However, $D(G),C(G)$ and $W(G)$ would not satisfy this condition due to the existence of balanced cycles and closed walks in complete signed graphs of general orders.
Moreover, the very small standard deviation of $D(G)$, $C(G)$, and $W(G)$ makes statistical testing against the balance of reshuffled networks complicated. The measures $D(G)$, $C(G)$, and $W(G)$ also have shown some unexpected behaviours on various types of graphs discussed in Section~\ref{1s:basic} and Section~\ref{1s:special}.

\section{Discussion on methodological findings} \label{1s:recom}

Taken together, the findings in Sections \ref{1s:basic} -- \ref{1s:axiom} give strong reason not to use cycle-based measures $D(G)$ and $C(G)$, regardless of the weights. The major issues with cycle-based measures $D(G)$ and $C(G)$ include the very small variance in randomly generated and reshuffled graphs, lack of sensitivity and clustering of values around 0.5 for graphs with a non-trivial number of negative edges. Recall the numerical analysis of synthetic data in Section~\ref{1s:basic}, analytical results on the expected values of cycle-based measures in Subsection \ref{1ss:expected}, and the numerical values which are difficult to interpret like the oscillation of $D(G)$ and values of $C(G)$ for ${K}_n^c$ graphs in Table~\ref{1tab2} and Figure~\ref{1fig5}.

The relative $k$-balance which is ultimately from the same family of measures, seems to resolve some, but not all the problems discussed above. However, it fails on several axioms and desirable properties. It is easy to compute $D_{3}(G)$ based on closed walks of length 3 \cite{terzi_spectral_2011} and there are recent methods resolving the computational burden of computing $D_{k}(G)$ for general $k$ \cite{Giscard2016,giscard2016general}. However, $D_{k>3}(G)$ cannot be used for cyclic graphs that do not have $k$-cycles. Besides, for networks with a large proportion of negative edges, the parity of $k$ substantially distorts the values of $D_{k}(G)$. Accepting all these shortcomings, one may use $D_{k}(G)$ when cycles of a particular length have a meaningful interpretation in the context of study.

Walk-based measures like $W(G)$ require a more systematic way of weighting to correct for the double-counting of closed walks with repeated edges. The shortcomings of $W(G)$ involving the weighting method and contribution of non-simple cycles are also discussed in \cite{Giscard2016,singh2017measuring}. Recall that $W(G) \rightarrow 1$ in 4-regular graphs when we increase $n$ as in our discussion in Subsection \ref{1ss:regular}. Besides, $W({K}_n^c) \rightarrow 0.5$ as $n$ increases as discussed in Subsection \ref{1ss:knc}. The commonly observed clustering of values near 0.5 may also present problems. Moreover, the model behind $W(G)$ is strange as signs of closed walks do not represent balance or imbalance. For these reasons we do not recommend $W(G)$ for future use.

The major weakness of the normalised algebraic conflict, $A(G)$, seems to be its incapability of evaluating the overall balance in graphs that have more than one connected component. Note that some of the failures observed for $A(G)$ on axioms and desirable properties stem from its dependence on $\lambda(G)$ the smallest eigenvalue of the signed Laplacian matrix. $\lambda(G)$ might be determined by a component of the graph disconnected from other components and in turn not capturing the overall balance of the graph as a whole. For analysing graphs with just one cyclic connected component, one may use $A(G)$ while disregarding the acyclic components. However, if a graph has more than one cyclic connected component, using $A(G)$ or $\lambda(G)$ is similar to disregarding all but the most balanced connected component in the graph.

The three trivial measures, namely $X(G)$, $Y(G)$ and $Z(G)$, fail on various basic axioms and desirable properties in Tables \ref{1tab3} and \ref{1tab3.5}, and also show a lack of sensitivity to the graph, making them inappropriate to be used as measures of balance.

Satisfying almost all the axioms and desirable properties, $F(G)$ seems to measure something different from what is obtained using all cycles or all $k$-cycles, and be worth pursuing in future. Note that $L(G)$ equals the minimum number of unbalanced fundamental cycles \cite{iacono_determining_2010}; suggesting a connection between the frustration and unbalanced cycles yet to be explored further. We recommend using $F(G)$ for all graphs as long as their size allows computing $L(G)$ (to be further discussed in Chapters \ref{ch:3} -- \ref{ch:4}). The optimisation models discussed in Chapters \ref{ch:2} -- \ref{ch:3} are shown to be capable of computing the frustration index in graphs with up to thousands of nodes and edges. For larger graphs, exact computation of $L(G)$ would be time consuming and it can be approximated using a nonzero optimality gap tolerance with the optimisation models in Chapters \ref{ch:2} -- \ref{ch:3}. Alternatively, $A(G)$ and $D_k(G)$ seem to be the other options. Depending on the type of the graph, $k$-cycles might not necessarily capture global structural properties. For instance, this would make $D_3(G)$ an improper choice for some specific graphs like sparse 4-regular graphs (as in Subsection \ref{1ss:regular}), square grids, and sparse graphs with a small number of 3-cycles. Similarly, $A(G)$ is not suitable for graphs that have more than one connected component (including many sparse graphs). 

\subsection*{\textbf{Notes on previous work}}

In the literature, balance theory is widely used on directed signed graphs. It seems that this approach is questionable in two ways. First, it neglects the fact that many edges in signed digraphs are not reciprocated. Bearing that in mind, investigating balance theory in signed digraphs deals with conflict avoidance when one actor in such a relationship may not necessarily be aware of good will or ill will on the part of other actors. This would make studying balance in directed networks analogous to studying how people avoid potential conflict resulting from potentially unknown ties. Secondly, balance theory does not make use of the directionality of ties and the concepts of sending and receiving positive and negative links. 

Leskovec, Huttenlocher and Kleinberg compare the reliability of predictions made by competing theories of social structure: balance theory and status theory (a theory that explicitly includes direction and gives quite different predictions) \cite{leskovec_signed_2010}. The consistency of these theories with observations is investigated through large signed directed networks such as Epinions, Slashdot, and Wikipedia. The results suggest that status theory predicts accurately most of the time while predictions made by balance theory are incorrect half of the time. This supports the inefficacy of balance theory for structural analysis of signed digraphs. For another comparison of the theories on signed networks, one may refer to a study of 8 theories to explain signed tie formation between students \cite{yap_why_2015}.

In a parallel line of research on network structural analysis, researchers differentiate between classical balance theory and structural balance specifically in the way that the latter is directional \cite{bonacich_introduction_2012}. They consider another setting for defining balance where absence of ties implies negative relationships. This assumption makes the theory limited to complete signed digraphs. Accordingly, 64 possible structural configurations emerge for three nodes. These configurations can be reduced to 16 classes of triads, referred to as 16 MAN triad census, based on the number of Mutual, Asymmetric, and Null relationships they contain. There are only 2 out of 16 classes that are considered balanced. New definitions are suggested by researchers in order to make balance theory work in a directional context. According to Prell \cite{prell_social_2012}, there is a second, a third, and a fourth definition of permissible triads allowing for 3, 7, and 9 classes of all 16 MAN triads. However, there have been many instances of findings in conflict with expectations \cite{prell_social_2012}.

Apart from directionality, the interpretation of balance measures is very important. Numerous studies have compared balance measures with their extremal values and found that signed networks are far from balanced, for example \cite{estrada_walk-based_2014}. However, with such a strict criterion, we must be careful not to look for properties that are almost impossible to satisfy. A much more systematic approach is to compare values of partial balance in the signed graphs in question to the corresponding values for reshuffled graphs \cite{Szell_multi, Szell_measure} as we have done in Section~\ref{1s:real}. 

So far we formalised the notion of partial balance and compared various measures of balance based on their values in different graphs where the underlying structure was not important. We also evaluated the measures based on their axiomatic properties and ruled out the measures that we could not justify. In the next section, we focus on exploring real signed graphs based on the justified methods.

\section{Results on real signed networks} \label{1s:real}

\begin{figure}
	\subfloat[Highland tribes network (G1), a signed network of 16 tribes of the Eastern Central Highlands of New Guinea \cite{read_cultures_1954}]{\includegraphics[width=2.5in]{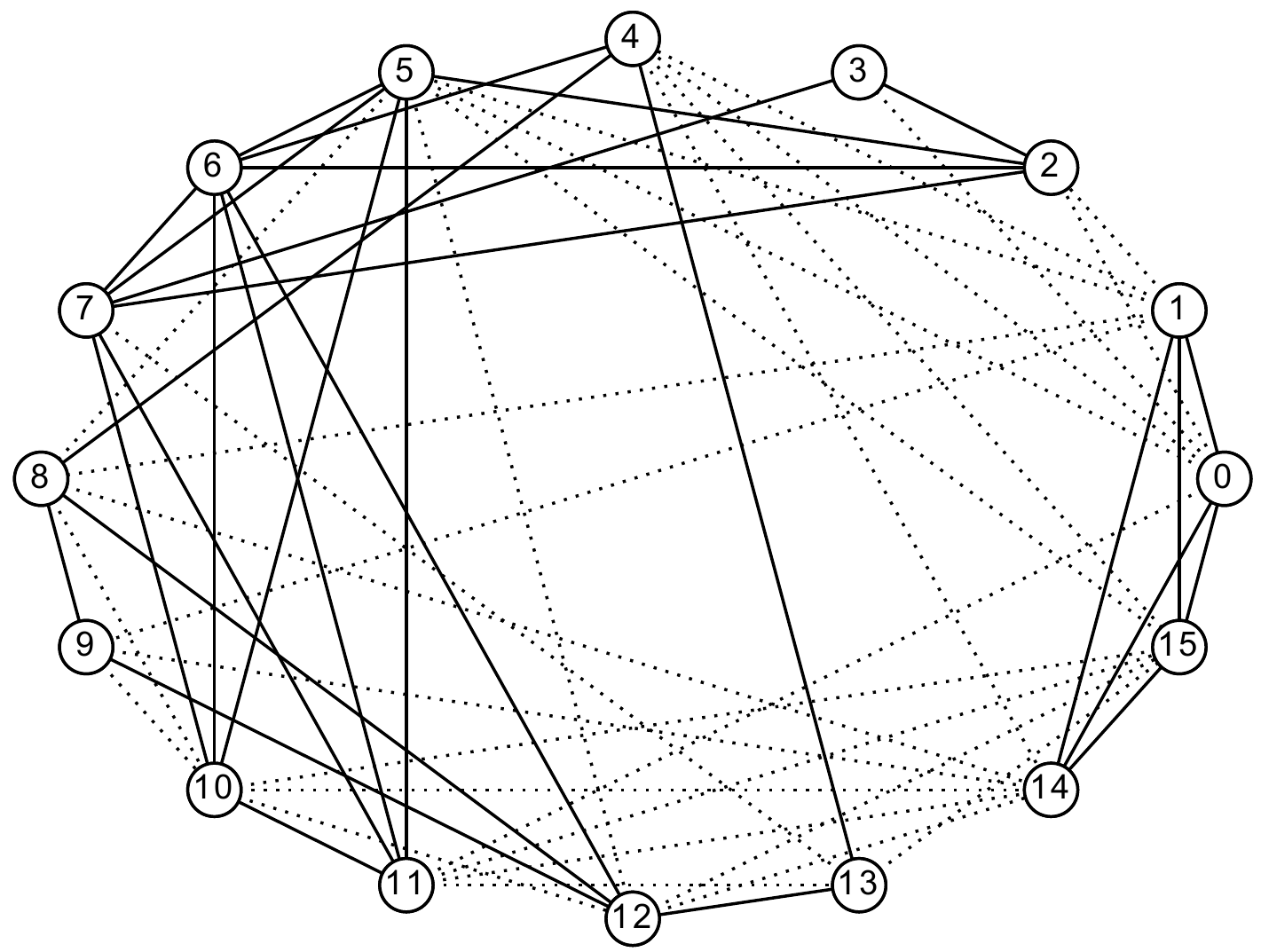}%
		\label{1fig6_first_case}}
	\hfil
	\subfloat[Monastery interactions network (G2) of 18 New England novitiates inferred from the integration of all positive and negative relationships \cite{sampson_novitiate_1968}]{\includegraphics[width=2.5in]{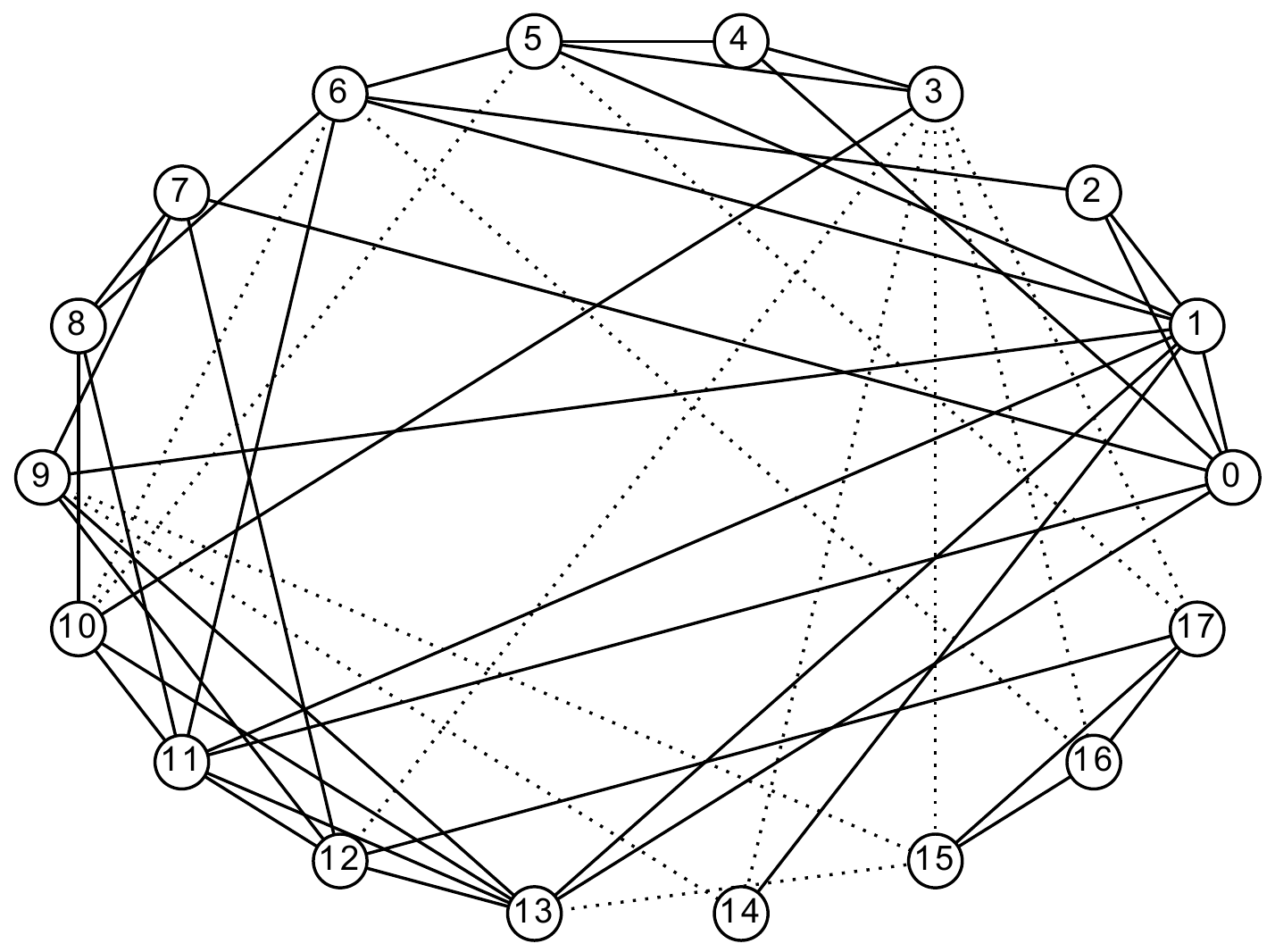}%
		\label{1fig6_second_case}}
	~
	
	\subfloat[Fraternity preferences network (G3) of 17 boys living in a pseudo-dormitory inferred from ranking data of the last week in \cite{newcomb_acquaintance_1961}]{\includegraphics[width=2.5in]{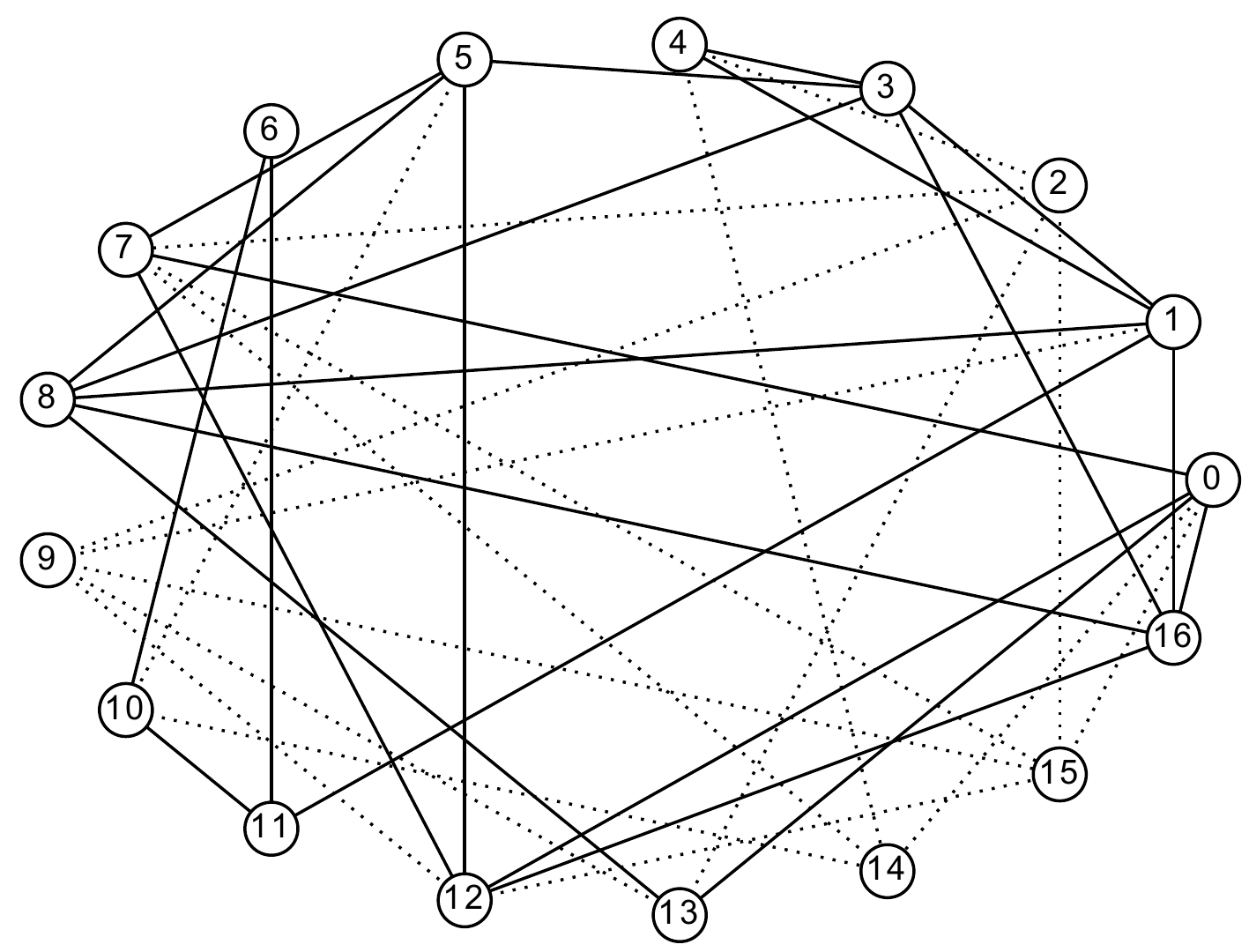}%
		\label{1fig6_third_case}}
	\hfil
	\subfloat[College preferences network (G4) of 17 girls at an Eastern college inferred from ranking data of house B in \cite{lemann_group_1952}]{\includegraphics[width=2.5in]{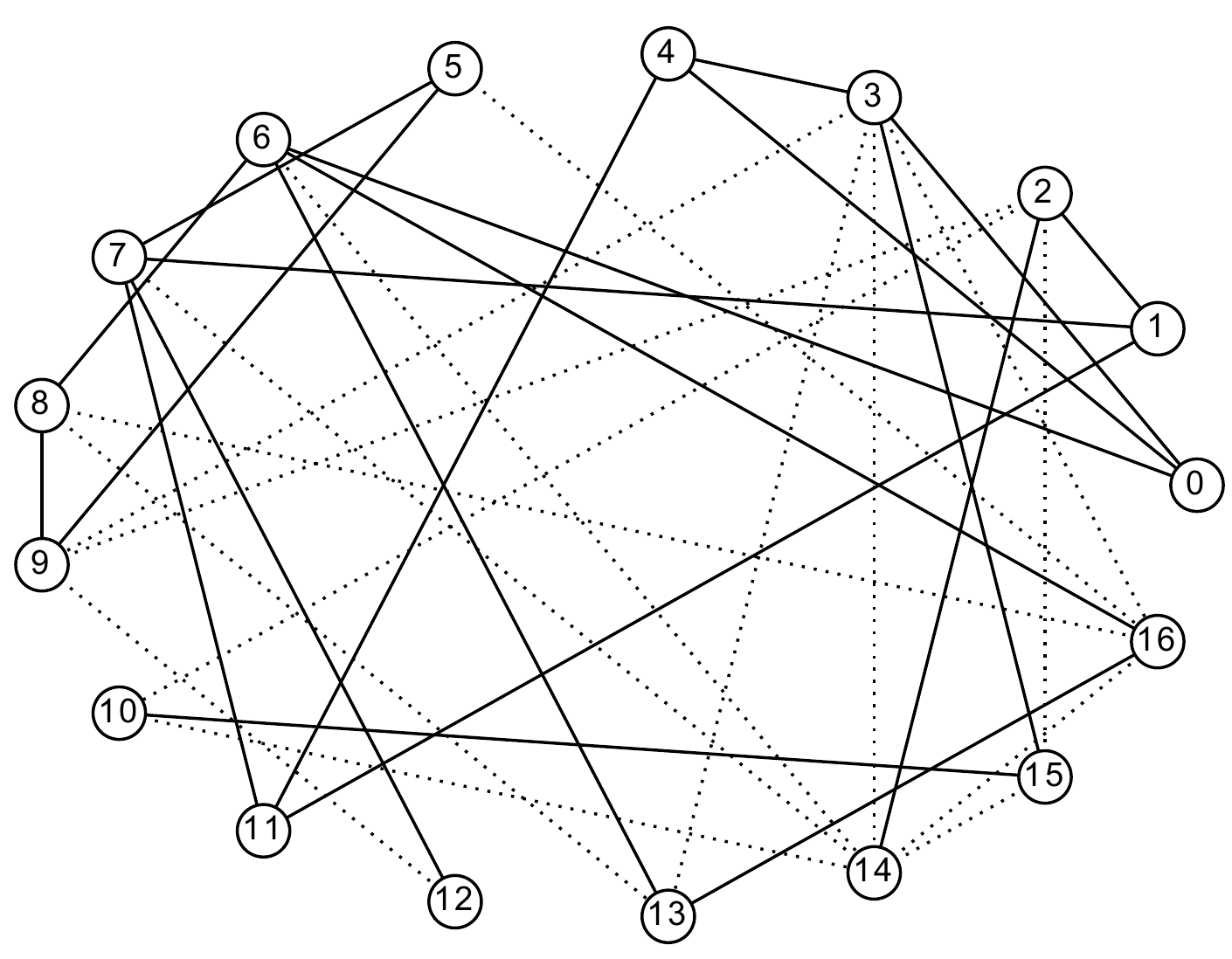}%
		\label{1fig6_fourth_case}}
	\caption{Four small signed networks visualised where dotted lines represent negative edges and solid lines represent positive edges}
	\label{1fig6}
\end{figure}

In this section, we analyse partial balance for a range of signed networks inferred from datasets of positive and negative interactions and preferences. Read's dataset for New Guinean highland tribes \cite{read_cultures_1954} is demonstrated as a signed graph (G1) in Figure~\ref{1fig6}(a), where dotted lines represent negative edges and solid lines represent positive edges. The fourth time window of Sampson's dataset for monastery interactions \cite{sampson_novitiate_1968} (G2) is drawn in Figure~\ref{1fig6}(b). We also consider datasets of students' choice and rejection (G3 and G4) \cite{newcomb_acquaintance_1961,lemann_group_1952} as demonstrated in Figure~\ref{1fig6}(c) and Figure~\ref{1fig6}(d). The last three are converted to undirected signed graphs by considering mutually agreed relations. A further explanation on the details of inferring signed graphs from the choice and rejection data is provided in Appendix \ref{1s:infer}. 

\begin{figure}
	\centering
	\subfloat[The bill co-sponsorship network (G5) of senators \cite{neal_backbone_2014}]{\includegraphics[width=3.5in]{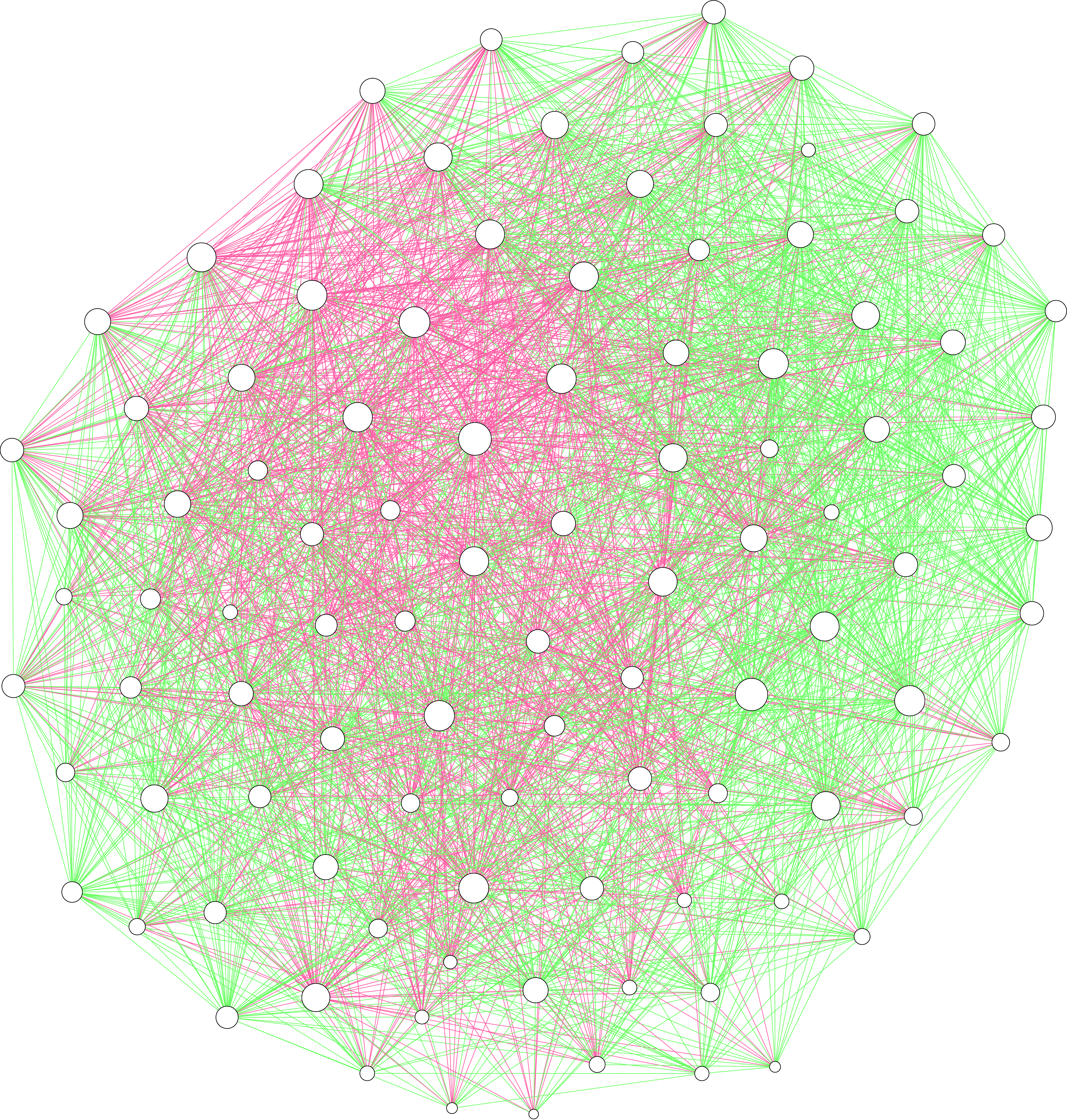}%
		\label{1fig_first_case}}
	
	~
	
	\subfloat[The gene regulatory network (G6) of \textit{Saccharomyces cerevisiae} \cite{Costanzo2001yeast}]{\includegraphics[width=2.5in]{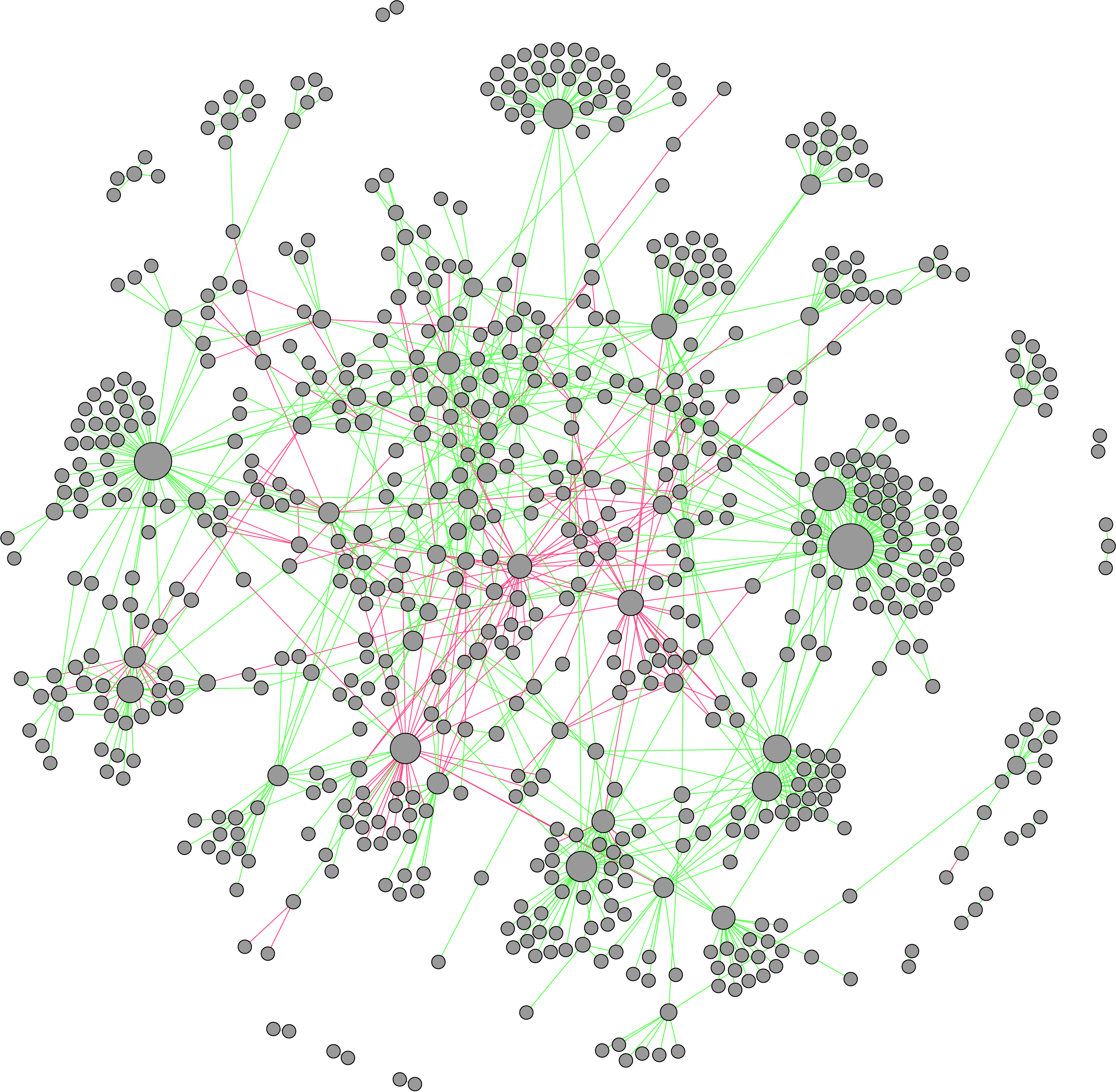}%
		\label{1fig_second_case}}
	\hfil
	\subfloat[The gene regulatory network (G7) of the \textit{Escherichia coli} \cite{salgado2006ecoli}]{\includegraphics[width=2.5in]{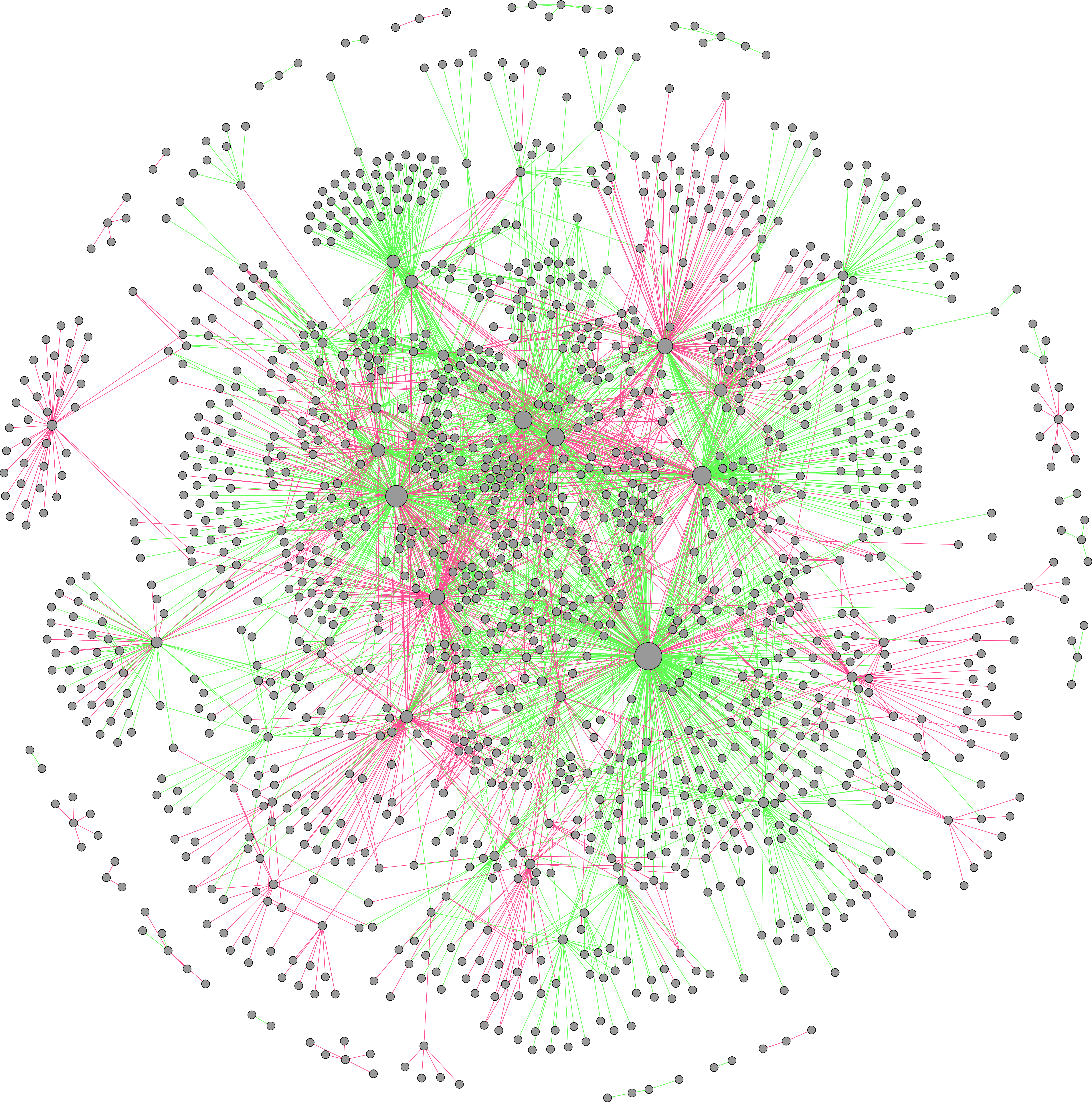}%
		\label{1fig_third_case}}
	
	\caption{Three larger signed datasets illustrated as signed graphs in which red lines represent negative edges and green lines represent positive edges}

	\label{1fig6.5}
\end{figure}

A larger signed network (G5) is inferred by \cite{neal_backbone_2014} through implementing a stochastic degree sequence model on Fowler's data on Senate bill co-sponsorship \cite{fowler_legislative_2006}. Besides the signed social network datasets, large scale biological networks can be analysed as signed graphs. There are relatively large signed biological networks analysed by \cite{dasgupta_algorithmic_2007} and \cite{iacono_determining_2010} from a balance viewpoint under a different terminology where \textit{monotonocity} is the equivalent for balance. The two gene regulatory networks we consider are related to two organisms: a eukaryote (the yeast \textit{Saccharomyces cerevisiae}) and a bacterium (\textit{Escherichia coli}). Graphs G6 and G7 represent the gene regulatory networks of \textit{Saccharomyces cerevisiae} \cite{Costanzo2001yeast} and \textit{Escherichia coli} \cite{salgado2006ecoli} respectively. Note that the densities of these networks are much smaller than the other networks introduced above. In gene regulatory networks, nodes represent genes. Positive and negative edges represent \textit{activating connections} and \textit{inhibiting connections} respectively. Figure~\ref{1fig6.5} shows the bill co-sponsorship network as well as biological signed networks. The colour of edges correspond to the signs on the edges (green for $+1$ and red for $-1$). For more details on the biological datasets, one may refer to \cite{iacono_determining_2010}.

As Figure~\ref{1fig6.5} shows, graphs G6 and G7 have more than one connected component. Besides the giant component, there are a number of small components that we discard in order to use $A(G)$ and $\lambda(G)$. Note that this procedure does not change $T(G)$ and $L(G)$ as the small components are all acyclic. The values of $(n, m, m^-)$ for giant components of G6 and G7 are $(664, 1064, 220)$ and $(1376, 3150, 1302)$ respectively.

The results are shown in Table~\ref{1tab4}. Although neither of the networks is completely balanced, the small values of $L(G)$ suggest that removal of relatively few edges makes the networks completely balanced. Table~\ref{1tab4} also provides a comparison of partial balance between different datasets of similar sizes. In this regard, it is essential to know that the choice of measure can make a substantial difference. For instance among G1--G4, under $T(G)$, G1 and G3 are respectively the most and the least partially balanced networks. However, if we choose $A(G)$ as the measure, G1 and G3 would be the least and the most partially balanced networks respectively. This confirms our previous discussions on how choosing a different measure can substantially change the results and helps to clarify some of the conflicting observations in the literature \cite{facchetti_computing_2011,kunegis_applications_2014} and \cite{estrada_walk-based_2014}, as previously discussed in Section~\ref{1s:recom}.

\begin{table}[ht]
	\centering
	\caption{Partial balance computer for signed graphs (G1--7) and reshuffled graphs}
	\label{1tab4}
	\begin{tabularx}{1\textwidth}{p{3.1cm}p{0.6cm}llllll}
		\hline
		Graph:$(n, m, m^-)$                    & $\rho$                &                         & $T$   & $A$   & $F$   & $\lambda$ & $L$    \\ \hline
		\multirow{4}{*}{G1: (16, 58, 29)}       & \multirow{4}{*}{0.483} & $\mu(G)$                & 0.87  & 0.88  & 0.76  & 1.04      & 7      \\
		&                        & $\text{mean}(\mu(G_r))$ & 0.50  & 0.76  & 0.49  & 2.08      & 14.65  \\
		&                        & $\text{SD}(\mu(G_r))$   & 0.06  & 0.02  & 0.05  & 0.20      & 1.38   \\
		&                        & Z-score                 & 6.04  & 5.13  & 5.54  & $-5.13$     & $-5.54$  \\ \hline
		\multirow{4}{*}{G2: (18, 49, 12)}       & \multirow{4}{*}{0.320} & $\mu(G)$                & 0.86  & 0.88  & 0.80  & 0.75      & 5      \\
		&                        & $\text{mean}(\mu(G_r))$ & 0.55  & 0.79  & 0.60  & 1.36      & 9.71   \\
		&                        & $\text{SD}(\mu(G_r))$   & 0.09  & 0.03  & 0.05  & 0.18      & 1.17   \\
		&                        & Z-score                 & 3.34  & 3.37  & 4.03  & $-3.37$     & $-4.03$  \\ \hline
		\multirow{4}{*}{G3:(17, 40, 17)}       & \multirow{4}{*}{0.294} & $\mu(G)$                & 0.78  & 0.90  & 0.80  & 0.50      & 4      \\
		&                        & $\text{mean}(\mu(G_r))$ & 0.49  & 0.82  & 0.62  & 0.89      & 7.53   \\
		&                        & $\text{SD}(\mu(G_r))$   & 0.11  & 0.06  & 0.06  & 0.30      & 1.24   \\
		&                        & Z-score                 & 2.64  & 1.32  & 2.85  & $-1.32$     & $-2.85$  \\ \hline
		\multirow{4}{*}{G4: (17, 36, 16)}       & \multirow{4}{*}{0.265} & $\mu(G)$                & 0.79  & 0.88  & 0.67  & 0.71      & 6      \\
		&                        & $\text{mean}(\mu(G_r))$ & 0.49  & 0.87  & 0.64  & 0.79      & 6.48   \\
		&                        & $\text{SD}(\mu(G_r))$   & 0.14  & 0.03  & 0.06  & 0.17      & 1.08   \\
		&                        & Z-score                 & 2.16  & 0.50  & 0.45  & $-0.50$     & $-0.45$  \\ \hline
		\multirow{4}{*}{G5: (100, 2461, 1047)}  & \multirow{4}{*}{0.497} & $\mu(G)$                & 0.86  & 0.87  & 0.73  & 8.92      & 331    \\
		&                        & $\text{mean}(\mu(G_r))$ & 0.50  & 0.75  & 0.22  & 17.46     & 965.6  \\
		&                        & $\text{SD}(\mu(G_r))$   & 0.00  & 0.00  & 0.01  & 0.02      & 9.08   \\
		&                        & Z-score                 & 118.5 & 387.8 & 69.89 & $-387.8$    & $-69.89$ \\ \hline
		\multirow{4}{*}{G6:(690, 1080, 220)}   & \multirow{4}{*}{0.005} & $\mu(G)$                & 0.54  & 1.00  & 0.92  & 0.02      & 41     \\
		&                        & $\text{mean}(\mu(G_r))$ & 0.58  & 1.00  & 0.77  & 0.02      & 124.3  \\
		&                        & $\text{SD}(\mu(G_r))$   & 0.07  & 0.00  & 0.01  & 0.00      & 4.97   \\
		&                        & Z-score                 & $-0.48$ & 8.61  & 16.75 & $-8.61$     & $-16.75$ \\ \hline
		\multirow{4}{*}{G7:(1461, 3215, 1336)} & \multirow{4}{*}{0.003} & $\mu(G)$                & 0.50  & 1.00  & 0.77  & 0.06      & 371    \\
		&                        & $\text{mean}(\mu(G_r))$ & 0.50  & 1.00  & 0.59  & 0.06      & 653.4  \\
		&                        & $\text{SD}(\mu(G_r))$   & 0.02  & 0.00  & 0.00  & 0.00      & 7.71   \\
		&                        & Z-score                 & $-0.33$ & 3.11  & 36.64 & $-3.11$     & $-36.64$ \\ \hline
				
	\end{tabularx}
\end{table}

In Table~\ref{1tab4}, the mean and standard deviation of measures for the reshuffled graphs $(G_r)$, denoted by $\text{mean}(\mu(G_r))$ and $\text{SD}(\mu(G_r))$, are also provided for comparison. We implement a very basic statistical analysis as in \cite{Szell_multi, Szell_measure} using $\text{mean}(\mu(G_r))$ and $\text{SD}(\mu(G_r))$ of 500 reshuffled graphs. Reshuffling the signs on the edges 500 times, we obtain two parameters of balance distribution for the fixed underlying structure. For measures of balance, Z-scores are calculated based on Eq.\ \eqref{1eq26}.
\begin{equation}
\label{1eq26}
Z=\frac{\mu(G)-\text{mean}(\mu(G_r))}{\text{SD}(\mu(G_r))}
\end{equation}

The Z-score shows how far the balance is with regards to balance distribution of the underlying structure. Positive values of Z-score for $T(G)$, $A(G)$, and $F(G)$ can be interpreted as existence of more partial balance than the average random level of balance. 

It is worth pointing out that the statistical analysis we have implemented is independent of the normalisation method used in $A(G)$ and $F(G)$. The two right columns of \ref{1tab4} provide $\lambda(G)$ and $L(G)$ alongside their associated Z-scores. 

The Z-scores show that as measured by the frustration index and algebraic conflict, signed networks G1--G7 exhibit a level of partial (but not total) balance beyond what is expected by chance. Based on these two measures, the level of partial balance is high for graphs G1, G2, G5, G6, and G7 while the numerical results for G3 and G4 do not allow a conclusive interpretation. It indicates that most of the real signed networks investigated are relatively consistent with the theory of structural balance. However, the Z-scores obtained based on the triangle index for G6--G7 show totally different results. Note that G6 and G7 are relatively sparse graphs which only have 70 and 1052 triangles. This may explain the difference between Z-scores of $T(G)$ and that of other measures. The numerical results using the algebraic conflict and frustration index support previous observations of real-world networks' closeness to balance \cite{facchetti_computing_2011,kunegis_applications_2014}.

\section{Conclusion of Chapter \ref{ch:1}} \label{1s:conclu}

In this chapter, we started by discussing balance in signed networks in Sections~\ref{1s:problem} and \ref{1s:check} and introduced the notion of partial balance. We discussed different ways to measure partial balance in Section~\ref{1s:measure} and provided some observations on synthetic data in Sections~\ref{1s:basic} and \ref{1s:special}. After gaining an understanding of the behaviour of different measures, basic axioms and desirable properties were used in Section~\ref{1s:axiom} to rule out the measures that cannot be justified.

We have discussed various methodologies and how they have led to conflicting observations in the literature in Section~\ref{1s:recom}. Taking axiomatic properties of the measures into account, using the common cycle-based measures denoted by $D(G)$ and $C(G)$ and the walk-based measure $W(G)$ is not recommended. $D_k(G)$ and $A(G)$ may introduce some problems, but overall using them seems to be more appropriate compared to $D(G)$, $C(G)$ and $W(G)$. The observations on synthetic data taken together with the axiomatic properties, recommend $F(G)$ as the best overall measure of partial balance. However, considering the difficulty of computing the exact value of $L(G)$ for very large graphs (to be discussed in Chapters \ref{ch:2} -- \ref{ch:3}), one may approximate it using a nonzero optimality gap tolerance with exact optimisation-based computational models. Alternatively, $A(G)$ and $D_k(G)$ seem to be the other options accepting their potential shortcomings.

Using the three measures $F(G)$, $T(G)$, and $A(G)$, each representing a family of measures, we compared balance in real signed graphs and analogous reshuffled graphs having the same structure in Section~\ref{1s:real}. Table~\ref{1tab4} provides this comparison showing that different results are obtained under different measures. 

Returning to the questions posed at the beginning of this chapter, it is now possible to state that under the frustration index and algebraic conflict many signed networks exhibit a level of partial (but not total) balance beyond that expected by chance. However, the numerical results in Table~\ref{1tab4} show that the level of balance observed using the triangle index can be totally different. One of the more significant findings to emerge from this chapter is that methods suggested for measuring balance may have different context and may require some justification before being interpreted based on their values. This chapter confirms that some measures of partial balance cannot be taken as a reliable static measure to be used for analysing network dynamics. 

One gap in this chapter is that we avoid using structural balance theory for analysing directed networks, making directed signed networks like Epinions, Slashdot, and Wikipedia Elections \cite{leskovec_signed_2010,estrada_walk-based_2014,Giscard2016} datasets untested by our approach. However, see our discussion in Section~\ref{1s:recom}. Although a numerical part of this chapter is based on signed networks with less than a few thousand nodes, the analytical findings that were not restricted to a particular size suggest the inefficacy of some methods for analysing larger networks as well.

From a practical viewpoint, international relations is a crucial area to implement signed network structural analysis. Having an efficient measure of partial balance in hand, international relations can be investigated in terms of evaluation of partial balance over time for networks of states (to be discussed in Chapter \ref{ch:4}). 

%\section*{Acknowledgement}
%We are grateful for the valuable comments from the anonymous reviewers that have improved this chapter.

%\appendix
\section{Appendix}
\subsection{Details of calculations} \label{1s:calc}
In order to simplify the sum $E(D_k(G)) = \sum \limits_{i \: \text{even}} {\binom{k}{i}} q^i (1-q)^{k-i}$, one may add the two following equations and divide the result by 2:
\begin{equation}
\begin{split}
	\sum \limits_{i} {\binom{k}{i}} q^i (1-q)^{k-i} = (q+(1-q))^k \\
	\sum \limits_{i} {\binom{k}{i}} (-q)^i (1-q)^{k-i} = (-q+(1-q))^k
\end{split}
\end{equation}

In ${K}_n^a$, a $k$-cycle is specified by choosing $k$ vertices in some order, then correcting for the overcounting by dividing by $2$ (the possible directions) and $k$ (the number of starting points, namely the length of the cycle). If the unique negative edge is required to belong to the cycle, by orienting this in a fixed way we need choose only $k-2$ further elements in order, and no overcounting occurs. The numbers of negative cycles and total cycles are as follows.

\begin{equation}
\sum_{k=3}^n O_k^- 
=\sum_{k=3}^n\frac{(n-2)!}{(n-k)!}
,\quad
\sum_{k=3}^n O_k 
=\sum_{k=3}^n \frac{n!}{2k(n-k)!}
\end{equation}

Asymptotic approximations for these sums can be obtained by introducing the exponential generating function. For example, letting 
$$
a_n = \sum_{1\leq k\leq n} \frac{n!}{(n-k)!k}
$$
we have
$$
\sum_{n\geq 0} \frac{a_n}{n!}x^n = \sum_{n,k} \sum_{k\leq n} \frac{n!}{(n-k)!k} x^n = \sum_{k\geq 1} \frac{1}{k} \sum_{n\geq k} \frac{1}{(n-k)!} x^n 
$$
$$ 
=\sum_{k\geq 1} \frac{x^k}{k} \sum_{m\geq 0} \frac{1}{m!}x^m = e^x \log\left(\frac{1}{1-x}\right).
$$
Similarly we obtain
$$
\sum_{n\geq 0} \sum_{k\geq 0} \frac{1}{(n-k)!} x^n = \frac{e^x}{1-x}.
$$
Standard singularity analysis methods \cite{flajolet2009analytic} show the denominator of the expression for $D(K^a_n)$ to be asymptotic to $(n-1)!e/2$ while the number of negative cycles is asymptotic to $(n-2)!e$. Similarly the weighted sum defining $C(K^a_n)$, where we choose $f(k) = 1/k!$, can be expressed using the ordinary generating function, which for the denominator turns out to be 
$$\frac{1}{1-x} \log\left( \frac{1-2x}{1-x}\right).$$ Again, singularity analysis techniques yield an approximation $2^n/n$. The numerator is easier, and asymptotic to $2^n/(n^2-n)$. This yields the result.

The unsigned adjacency matrix $|\textbf{A}|$ of the complete graph has the form $\textbf{E} - \textbf{I}$ where $\textbf{E}$ is the matrix of all $1$'s. The latter matrix has rank 1 and nonzero eigenvalue $n$. Thus $|\textbf{A}| _{({K}_n^a)}$ has eigenvalues $n-1$ (with multiplicity 1) and $-1$ (with multiplicity $n-1$). The matrix $\textbf{A} _{({K}_n^a)}$ has a similar form and we can guess eigenvectors of the form $(-1,1,0, \dots, 0)$ and $(a,a,1,1, \dots , 1)$. Then $a$ satisfies a quadratic $2a^2 + (n-3)a - (n-2) = 0$. Solving for $a$ and the corresponding eigenvalues, we obtain eigenvalues $(n-4 \pm \sqrt{(n-2)(n+6)})/2, 1, -1$ (with multiplicity $n-3$)).

This yields 
$$K{({K}_n^a)} = \frac{(n-3)e^{-1} + e + e^{\frac{n-4 - \sqrt{(n-2)(n+6)}}{2}} + e^{\frac{n-4 + \sqrt{(n-2)(n+6)}}{2}}}{(n-1)e^{-1} + e^{n-1}}$$ which results in $W({K}_n^a) \sim \frac{1+e^{-4/n}}{2}$. 

Furthermore, since every node of $K_n$ has degree $n-1$, the eigenvalues of $\textbf{L}:=(n-1)\textbf{I}-\textbf{A}$ are precisely of the form $n-1 - \lambda$ where $\lambda$ is an eigenvalue of $\textbf{A}$.

Measures of partial balance for ${K}_n^a$ can therefore be expressed by the formulae \eqref{1eq15} -- \eqref{1eq18}:

\begin{equation}\label{1eq15}
D({K}_n^a)=1- \frac{\sum_{k=3}^n \frac{(n-2)!}{(n-k)!}}{\sum_{k=3}^n \frac{n!}{2k(n-k)!}} \sim 1 - \frac{2}{n}
\end{equation}
\begin{equation}\label{1eq15.5}
C({K}_n^a)=1- \frac{\sum_{k=3}^n \frac{(n-2)!}{(n-k)!k!}}{\sum_{k=3}^n \frac{n!}{2k(n-k)!k!}} \sim 1 - \frac{1}{n}
\end{equation}
\begin{equation}\label{1eq17}
D_k({K}_n^a)=1-\frac{\frac{(n-2)!}{(n-k)!}}{\frac{n!}{2k(n-k)!}}=1-\frac{2k}{n(n-1)} \sim 1 - \frac{2k}{n^2}
\end{equation}
\begin{equation}\label{1eq16}
W({K}_n^a) \sim \frac{1+e^{-4/n}}{2} \sim 1-\frac{2}{n}
\end{equation}
\begin{equation}\label{1eq17.2}
\lambda({K}_n^a)= n-1 - (n-4 + \sqrt{(n-2)(n+6)})/2 = (n+2 - \sqrt{(n-2)(n+6)})/2
\end{equation}
\begin{equation}\label{1eq17.4}
A({K}_n^a)= 1- \frac{n+2 - \sqrt{(n-2)(n+6)}}{2n-4} \sim 1-\frac{4}{n^2}
\end{equation}
\begin{equation}\label{1eq18}
F({K}_n^a)=1-\frac{2}{n(n-1)/2}=1-\frac{4}{n(n-1)} \sim 1 - \frac{4}{n^2}
\end{equation}

In ${K}_n^c$, all cycles of odd length are unbalanced and all cycles of even length are balanced. Therefore:
\begin{equation}
\sum_{k=3}^n O_k^+ 
= \sum_{\textnormal{even}}^n \frac{n!}{2k(n-k)!}
\end{equation}

It follows that $D_k({K}_n^c)$ equals 0 for odd $k$ and 1 for even $k$. Based on maximality of $\lambda(G)$ in ${K}_n^c$, $A({K}_n^c)=0$.

Using the above generating function techniques we obtain that the numerator of $D(K^c_n)$ is asymptotic to $(n-1)!(e+(-1)^ne^{-1})/4$. The denominator we know from above is asymptotic to $(n-1)!e/2$. This yields $D(K^c_n) \sim 1/2 + (-1)^n e^{-2}$. Note that $e^{-2} \approx 0.135$ and this explains the oscillation in Figure~\ref{1fig5}. Similarly we obtain results for $C(K^c_n)$.

$|\textbf{A}|_{({K}_n^c)}$ has eigenvalues $n-1$ (with multiplicity 1) and $-1$ (with multiplicity $n-1$). The matrix $\textbf{A}_{({K}_n^c)}$ has a similar form and the corresponding eigenvalues would be $1-n$ (with multiplicity 1) and $1$ (with multiplicity $n-1$). This yields
$K{({K}_n^c)} = \frac{(n-1)e^{1} + e^{1-n}}{(n-1)e^{-1} + e^{n-1}}$ which results in $W({K}_n^c) \sim \frac{1+e^{2-2n}}{2}$.

Moreover, a closed-form formula for $L({K}_n^c)$ can be expressed based on a maximum cut which gives a function of $n$ equal to an upper bound of the frustration index under a different name in \cite{abelson_symbolic_1958}. Measures of partial balance for ${K}_n^c$ can be expressed via the closed-form formulae as stated in Eq.\ \eqref{1eq21} -- \eqref{1eq25}:

\begin{equation}\label{1eq21}
D({K}_n^c)= \frac{\sum_{\textnormal{$k$ even}}^n \frac{n!}{2k(n-k)!}}{\sum_{k=3}^n \frac{n!}{2k(n-k)!}} \sim \frac{1}{2} + (-1)^n e^{-2}
\end{equation}
\begin{equation}\label{1eq22}
C({K}_n^c)= \frac{\sum_{\textnormal{$k$ even}}^n \frac{n!}{2k(n-k)!k!}}{\sum_{k=3}^n \frac{n!}{2k(n-k)!k!}} \sim \frac{1}{2} - \frac{3n\log n}{2^n}
\end{equation}
\begin{equation} \label{1eq22.5}
D_k({K}_n^c)=
\left\{
\begin{array}{ll}
1 & \mbox{if } k \ \text{is even} \\
0 & \mbox{if } k \ \text{is odd}
\end{array}
\right.
\end{equation}
\begin{equation}\label{1eq23}
W({K}_n^c) \sim \frac{1+e^{2-2n}}{2}
\end{equation}
\begin{equation}\label{1eq23.3}
\lambda({K}_n^c)=\lambda_\text{max}=\overline{d}_{\text{max}}-1=n-2
\end{equation}
\begin{equation} \label{1eq24}
L({K}_n^c)=
\left\{
\begin{array}{ll}
({n^2-2n})/{4} & \mbox{if } n  \ \text{is even} \\
({n^2-2n+1})/{4} & \mbox{if } n  \ \text{is odd}
\end{array}
\right.
\end{equation}
\begin{equation} \label{1eq25}
F({K}_n^c)=
\left\{
\begin{array}{ll}
1-\frac{n(n-2)/4}{n(n-1)/4}=\frac{1}{n-1} & \mbox{if } n  \ \text{is even} \\
1-\frac{(n-1)(n-1)/4}{n(n-1)/4}=\frac{1}{n} & \mbox{if } n  \ \text{is odd}
\end{array}
\right.
\end{equation}

Our calculations for $L({K}_n^c)$ show that the upper bound suggested for the frustration index in \cite{iacono_determining_2010} is incorrect.

\subsection{Counterexamples and proof ideas for the axioms and desirable properties} \label{1s:counter}

Axioms: 

Axiom 1 holds in all the measures introduced due to the systematic normalisation implemented.

$D_k(G)$, $A(G)$, and $Y(G)$ do not satisfy Axiom 2. All $k$-cycles being balanced, $D_k(G)$ fails to detect the imbalance in graphs with unbalanced cycles of different lengths. $A(G \oplus C^+)=1$ for unbalanced graphs which makes $A(G)$ fail Axiom 2. $Y(G)$ fails on detecting balance in completely bi-polar signed graphs that are indeed balanced.

As long as $\mu(G\oplus H)$ can be written in the form of $(a+c)/(b+d)$ where $\mu(G)=a/b$ and $\mu(H)=c/d$, $\mu$ satisfies Axiom 3. So all the measures considered satisfy Axiom 3, except for $A(G)$. In case of $\lambda(G) < \lambda(H)$ and $\lambda_\text{max}(G) < \lambda_\text{max}(H)$, $A(G \oplus H)= 1 - \frac{\lambda(G)}{\lambda_\text{max}(H)} > A(H)$ which shows that $A(G)$ fails Axiom 3.

The sign of cycles (closed walks), the Laplacian eigenvalues \cite{Belardo2016}, and the frustration index \cite{zaslavsky_matrices_2010} will not change by applying the switching function introduced in Eq.\ \eqref{1eq1.2}. Therefore, Axiom 4 holds for all the measures discussed except for $X(G)$ and $Y(G)$ because they depend on $m^-$, which changes in switching.

\noindent
Desirable properties:

Clearly in B1, $C^+_3$ contributes positively to $D(G)$ and $C(G)$, whereas for $D_k(G)$ it depends on $k$ which makes it fail B1. As $W(C^+_3)$ equals 1, $\Tr(e^{\textbf{A}})/\Tr(e^{|\textbf{A}|})$ would be added by equal terms in both the numerator and denominator leading to $W(G)$ satisfying B1. $A(G)$ satisfies B1 because $A(G\oplus C^+_3)=1$. As $m$ increases by 3, $F(G)$ satisfies B1. The dependency of $X(G)$ and $Y(G)$ on $m^-$ and incapability of the binary measure, $Z(G)$, in providing values between 0 and 1 make them fail B1.

$C^-_3$ adds only to the denominators of $D(G)$ and $C(G)$, whereas for $D_k(G)$ it depends on $k$ which makes it fail B2. Following the addition of a negative 3-cycle, $W(G)$ is observed to increase resulting in its failure in B2 (for example, take $G=K_5$ with single negative edge, and $C^-_3$ having a single negative edge). As $A(G)\neq0$ and $A(C^-_3)=0$, the value of $A(G \oplus C^-_3)$ does not change when a negative 3-cycle is added. Therefore, it fails B2.	
Moreover, $F(G)$ fails B2 whenever $L(G) \geq m/3$ as observed in a family of graphs in Subsection \ref{1ss:knc}. However, $F^\prime (G)$ introduced in Eq.\ \eqref{1eq5.8} which only differs in normalisation, satisfies this desirable property. The measures $X(G)$ and $Y(G)$ fail B2, but the binary measure, $Z(G)$, satisfies it.
	
All the cycle-based measures, namely $D(G),C(G)$, and $D_3(G)$ fail B3 (for example, take $G=K_4$ with two symmetrically located negative edges). $W(G)$ is also observed to fail B3 (for instance, take $G$ as the disjoint union of a 3-cycle and a 5-cycle each having 1 negative edge). It is known that $\lambda(G \ominus e) \leq \lambda(G)$ \cite{Belardo2016}. However in some cases where $\lambda_\text{max}(G \ominus e) < \lambda_\text{max} (G)$ counterexamples are found showing $A(G)$ fails on B3 (consider a graph with $n=8, m^+=10, m^-=3, \min{|E^*|}=3$ in which $\lambda_\text{max} (G)=6$ and $\lambda_\text{max}(G \ominus e)=3$). $Y(G)$ and $Z(G)$ fail B3. Moreover, $F(G)$ satisfies B3 because $L(G \ominus e) = L(G) -1$.

The cycle-based measures and $W(G)$ do not satisfy B4. For $D_3(G)$, we tested a graph with $n=7, m=15, |E^*|=3$ and we observed $D_3(G \oplus e) > D_3(G)$. According to Belardo and Zhou, $\lambda(G \oplus e) \geq \lambda(G)$ \cite{Belardo2016}. However in some cases where $\lambda_\text{max}(G \oplus e) > \lambda_\text{max} (G)$ counterexamples are found showing $A(G)$ fails on B4. counterexamples showing $D(G),C(G),W(G)$, and $A(G)$ fail B4, are similar to that of B3. Moreover, $F(G)$ satisfies B4 as do $X(G)$ and $Z(G)$, while $Y(G)$ fails B4 when $e$ is positive.

\subsection{Inferring undirected signed graphs} \label{1s:infer}
Sampson collected different sociometric rankings from a group of 18 monks at different times \cite{sampson_novitiate_1968}. The data provided includes rankings on like, dislike, esteem, disesteem, positive influence, negative influence, praise, and blame. We have considered all positive and negative rankings. Then only the reciprocated relations with similar signs are considered to infer an undirected signed edge between two monks (see \cite{doreian_partitioning_2009} and how the authors inferred a directed signed graph in their Table 5 by summing the influence, esteem and respect relations).

Newcomb reported rankings made by 17 men living in a pseudo-dormitory \cite{newcomb_acquaintance_1961}. We used the ranking data of the last week which includes complete ranks from 1 to 17 gathered from each man. As the gathered data is related to complete ranking, we considered ranks 1-5 as one-directional positive relations and 12-17 as one-directional negative relations. Then only the reciprocated relations with similar signs are considered to infer an undirected signed edge between two men (see \cite{doreian_partitioning_2009} and how the authors converted the top three and bottom three ranks to a directed signed edges in their Fig. 4.).

Lemann and Solomon collected ranking data based on multiple criteria from female students living in off-campus dormitories \cite{lemann_group_1952}. We used the data for house B which is resulted by integrating top and bottom three rankings for multiple criteria. As the gathered data itself is related to top and bottom rankings, we considered all the ranks as one-directional signed relations. Then only the reciprocated relations with similar signs are considered to infer an undirected signed edge between two women (see \cite{doreian_multiple_2008} and how the author inferred a directed signed graph in their Fig. 5 from the data for house B.).

% references section

%Instructions	%1. remove bbl file
%2. set bibliographystyle to {imaiai} and run twice to get the reference list in the correct format
%3. set bibliographystyle to {plain} and run twice to get the correct style for in-text reference

%\bibliographystyle{imaiai} 
%\bibliographystyle{plainnat}
%\bibliographystyle{abbrv}
%\bibliography{refs}

%\end{document}
 
\cleardoublepage

\chapter{Computing the Line Index of Balance Using Integer Programming Optimisation}\label{ch:2}

\maketitle

%\cleardoublepage
\section*{Abstract}
	An important measure of signed graphs is the line index of balance which has applications in many fields. However, this graph-theoretic measure was underused for decades because of the inherent complexity in its computation which is closely related to solving NP-hard graph optimisation problems like MAXCUT.
	We develop new quadratic and linear programming models to compute the line index of balance exactly. Using the Gurobi integer programming optimisation solver, we evaluate the line index of balance on real-world and synthetic datasets. The synthetic data involves Erd\H{o}s-R\'{e}nyi graphs, Barab\'{a}si-Albert graphs, and specially structured random graphs. We also use well-known datasets from the sociology literature, such as signed graphs inferred from students' choice and rejection, as well as datasets from the biology literature including gene regulatory networks. 
	The results show that exact values of the line index of balance in relatively large signed graphs can be efficiently computed using our suggested optimisation models. We find that most real-world social networks and some biological networks have a small line index of balance which indicates that they are close to balanced.
	
	%\textbf{Keywords:} 
%	Integer programming,
%	Optimisation,
%	Frustration index,
%	Branch and bound,
%	Signed graphs,
%	Structural balance theory
	
	%\textbf{Mathematics Subject Classification (MSC 2010):}
	%05C22 15C22 90C09 90C11 90C90 90C35

\cleardoublepage
\section{Introduction to Chapter \ref{ch:2}} \label{2s:intro}

Graphs with positive and negative edges are referred to as \textit{signed graphs} \cite{zaslavsky_mathematical_2012} which are very useful in modelling the dual nature of interactions in various contexts. Graph-theoretic conditions \cite{harary_notion_1953,cartwright_structural_1956} of the structural balance theory \cite{heider_social_1944,harary_notion_1953} define the notion of \textit{balance} in signed graphs. If the vertex set of a signed graph can be partitioned into $k \leq 2$ subsets such that each negative edge joins vertices belonging to different subsets, then the signed graph is balanced \cite{cartwright_structural_1956}. For graphs that are not balanced, a distance from balance (a measure of partial balance as discussed in Chapter \ref{ch:1}) can be computed.

Among various measures is the \textit{frustration index} that indicates the minimum number of edges whose removal results in balance \cite{abelson_symbolic_1958,harary_measurement_1959,zaslavsky_balanced_1987}. Originally, this number was peripherally mentioned by Abelson et al.\ \cite{abelson_symbolic_1958} and referred to as \textit{complexity}. One year later, Harary proposed the same idea much more clearly with the name \textit{line index of balance} \cite{harary_measurement_1959}. More than two decades later, Toulouse used the term \textit{frustration} to discuss the minimum energy of an Ising spin glass model \cite{toulouse_theory_1987}. Zaslavsky has made a connection between the line index of balance and spin glass concepts and introduced the name frustration index \cite{zaslavsky_balanced_1987}. We use both names, line index of balance and frustration index, interchangeably in this chapter.

\section{Literature review}\label{2s:literature}

As discussed in Chapter \ref{ch:1}, except for a normalised version of the frustration index, measures of balance used in the literature \cite{cartwright_structural_1956,norman_derivation_1972,terzi_spectral_2011,kunegis_applications_2014,estrada_walk-based_2014} do not satisfy key axiomatic properties. Using cycles \cite{cartwright_structural_1956,norman_derivation_1972}, triangles \cite{terzi_spectral_2011,kunegis_applications_2014}, Laplacian matrix eigenvalues \cite{kunegis_spectral_2010}, and closed-walks \cite{estrada_walk-based_2014} to evaluate distance from balance has led to conflicting observations \cite{leskovec_signed_2010, facchetti_computing_2011, estrada_walk-based_2014}. 

Besides applications as a measure of balance, the frustration index is a key to frequently stated problems in several fields of research (to be discussed in Chapter \ref{ch:4}). In biology, optimal decomposition of biological networks into monotone subsystems is made possible by calculating the line index of balance \cite{iacono_determining_2010}. In finance, portfolios whose underlying signed graph has negative edges and a frustration index of zero have a relatively low risk \cite{harary_signed_2002}. In physics, the line index of balance provides the minimum energy state of atomic magnets \cite{kasteleyn_dimer_1963,Sherrington, Barahona1982}. In international relations, alliance and antagonism between countries can be analysed using the line index of balance \cite{patrick_doreian_structural_2015}. In chemistry, bipartite edge frustration indicates the stability of fullerene, a carbon allotrope \cite{doslic_computing_2007}. For a discussion on applications of the frustration index, one may refer to Chapter \ref{ch:4}.

Detecting whether a graph is balanced can be solved in polynomial time \cite{hansen_labelling_1978,harary_simple_1980,zaslavsky1983signed}. However, calculating the line index of balance in general graphs is an NP-hard problem equivalent to the ground state calculation of an unstructured Ising model \cite{mezard2001bethe}. Computation of the line index of balance can be reduced from the graph maximum cut (MAXCUT) problem, in the case of all negative edges, which is known to be NP-hard \cite{huffner_separator-based_2010}.

Similar to MAXCUT for planar graphs \cite{hadlock}, the line index of balance can be computed in polynomial time for planar graphs \cite{katai1978studies}. Other special cases of related problems can be found among the works of Hartmann and collaborators who have suggested efficient algorithms for computing ground state in 3-dimensional spin glass models \cite{hartmann2015matrix} improving their previous contributions in 1-, 2-, and 3-dimensional \cite{hartmann2014exact,hartmann2011ground,hartmann2013information} spin glass models. Recently, they have used a method for solving 0/1 optimisation models to compute the ground state of 3-dimensional models containing up to $268^3$ nodes \cite{hartmann2016revisiting}. 

A review of the literature shows 5 algorithms suggested for computing the line index of balance between 1963 and 2002. The first algorithm \cite[pages 98-107]{flament1963applications} is developed specifically for complete graphs. It is a naive algorithm that requires explicit enumeration of all possible combinations of sign changes that may or may not lead to balance. With a run time exponential in the number of edges, this is clearly not practical for graphs with more than 8 nodes that require billions of cases to be checked. The second algorithm is an optimisation method suggested in \cite{hammer1977pseudo}. This method is based on solving an unconstrained binary quadratic model. We will discuss a model of this type in Subsection \ref{2ss:ubqp} and a more efficient model later in this chapter. The third computation method is an iterative algorithm suggested in \cite[algorithm 3, page 217]{hansen_labelling_1978}. The iterative algorithm is based on removing edges to eliminate negative cycles of the graph and only provides an upper bound on the line index of balance. A fourth method suggested by Harary and Kabell \cite[page 136]{harary_simple_1980} is based on extending a balance detection algorithm. This method is inefficient according to Bramsen \cite{bramsen2002further} who in turn suggests an iterative algorithm with a run time that is exponential in the number of nodes. Using Bramsen's suggested method for a graph with 40 nodes requires checking trillions of cases to compute the line index of balance which is clearly impractical. Doreian and Mrvar have recently attempted computing the line index of balance \cite{patrick_doreian_structural_2015}. However, our computations on their data show that their solutions are not optimal and thus do not give the line index of balance.

This review of literature shows that computing the line index of balance in general graphs lacks extensive and systematic investigation.

\subsection*{Our contribution in Chapter \ref{ch:2}} \label{2ss:contrib}
We provide an efficient method for computing the line index of balance in general graphs of the sizes found in many application areas. Starting with a quadratic programming model based on signed graph switching equivalents, we suggest several optimisation models. We use powerful mathematical programming solvers like Gurobi \cite{gurobi} to solve the optimisation models. 

This chapter begins with the preliminaries in Section~\ref{2s:prelim}. Three mathematical programming models are developed in Section~\ref{2s:qmodel}. The results on synthetic data are provided in Section~\ref{2s:results}. Numerical results on real social and biological networks are provided in Section~\ref{2s:real} including graphs with up to 3215 edges. Section~\ref{2s:conclu} summarises the key highlights of the chapter. 

\section{Preliminaries} \label{2s:prelim}

We recall some standard definitions.

\subsection{Basic notation} \label{2s:problem}
We consider undirected signed networks $G = (V,E,\sigma)$. The ordered set of nodes is denoted by $V$, with $|V| = n$. The set $E$ of edges is partitioned into the set of positive edges $E^+$ and the set of negative edges $E^-$ with $|E^-|=m^-$, $|E^+|=m^+$, and $|E|=m=m^- + m^+$. For clarity, we sometimes use $m^-_{(G)}$ to refer to the number of negative edges in $G$. The sign function, denoted by $\sigma$, is a mapping of edges to signs $\sigma: E\rightarrow\{-1,+1\}$. We represent the $m$ undirected edges in $G$ as ordered pairs of vertices $E = \{e_1, e_2, ..., e_m\} \subseteq \{ (i,j) \mid i,j \in V , i<j \}$, where a single edge $e_k$ between nodes $i$ and $j$, $i<j$, is denoted by $e_k=(i,j) , i<j$. We denote the graph density by $\rho= 2m/(n(n-1))$. The entries of the adjacency matrix $\textbf{A}=(a_{ij})$ are defined in Eq.\ \eqref{2eq1}. 
\begin{equation}\label{2eq1}
a{_i}{_j} =
\left\{
\begin{array}{ll}
\sigma_{(i,j)} & \mbox{if } (i,j) \in E \\
\sigma_{(j,i)} & \mbox{if } (j,i) \in E \\
0 &  \text{otherwise}
\end{array}
\right.
\end{equation}

The number of positive (negative) edges incident on the node $i \in V$ is the \textit{positive degree (negative degree)} of the node and is denoted by $d^+ {(i)}$ ($d^- {(i)}$). The \textit{net degree} of a node is defined by $d^+ {(i)} -d^- {(i)}$. The \textit{degree} of node $i$ is represented by $d {(i)} = d^+ {(i)} +d^- {(i)}$ and equals the total number of edges incident on node $i$.

A \emph{walk} of length $k$ in $G$ is a sequence of nodes $v_0,v_1,...,v_{k-1},v_k$ such that for each $i=1,2,...,k$ there is an edge between $v_{i-1}$ and $v_i$. If $v_0=v_k$, the sequence is a \emph{closed walk} of length $k$. If the nodes in a closed walk are distinct except for the endpoints, it is a \emph{cycle} of length $k$. The \emph{sign} of a cycle is the product of the signs of its edges. A balanced graph is one with no negative cycles \cite{cartwright_structural_1956}.

\subsection{Node colouring and frustration count}

For each signed graph $G=(V, E, \sigma)$, we can partition $V$ into two sets, denoted by $X \subseteq V$ and $\bar X=V \setminus X$. We think of $X$ as specifying a colouring of the nodes, where each node $i \in X$ is coloured black, and each node $i \in \bar X$ is coloured white. 

We let $x_i$ denote the colour of node $i \in V$ under $X$, where $x_i=1$ if $i \in X$ and $x_i=0$ otherwise. We say that an edge $(i,j) \in E$ is {\em frustrated} under $X$ if either edge $(i,j)$ is a positive edge (\ $(i,j) \in E^+$) but nodes $i$ and $j$ have different colours ($x_i \ne x_j$), or edge $(i,j)$ is a negative edge (\ $(i,j) \in E^-$) but nodes $i$ and $j$ share the same colour ($x_i = x_j$). We define the {\em frustration count} $f_G(X)$ as the number of frustrated edges in $G$ under $X$: $$f_G(X) = \sum_{(i,j) \in E} f_{ij}(X)$$
where for $(i,j) \in E$:
\begin{equation} \label{2eq1.1}
f_{ij}(X)=
\begin{cases}
0, & \text{if}\ x_i = x_j \text{ and } (i,j) \in E^+ \\
1, & \text{if}\ x_i = x_j \text{ and } (i,j) \in E^- \\
0, & \text{if}\ x_i \ne x_j \text{ and } (i,j) \in E^- \\
1, & \text{if}\ x_i \ne x_j \text{ and } (i,j) \in E^+. \\
\end{cases}
\end{equation}

The frustration index $L(G)$ of a graph $G$ can be obtained by finding a subset $X^* \subseteq V$ of $G$ that minimises the frustration count $f_G(X)$, i.e., solving Eq.\ \eqref{2eq1.2}.

\begin{equation} \label{2eq1.2}
L(G)=\min_{X \subseteq V}f_G(X)\
\end{equation}

\subsection{Minimum deletion set and switching function}\label{2ss:state}

For each signed graph, there are sets of edges, called \textit{deletion sets}, whose deletion results in a balanced graph. A minimum deletion set $E^*\subseteq E$ is a deletion set with the minimum size. The frustration index $L(G)$ equals the size of a minimum deletion set: $L(G)=|E^*|$. 

We define the \emph{switching function} $g(X)$ operating over a set of vertices, called the \textit{switching set}, $X\subseteq V$ as follows in Eq.\ \eqref{2eq2}.
\begin{equation} \label{2eq2}
\sigma^{g(X)} _{(i,j)}=
\left\{
\begin{array}{rl}
\sigma_{(i,j)} & \mbox{if } {i,j}\in X \ \text{or} \ {i,j}\notin X \\
-\sigma_{(i,j)} & \mbox{if } (i \in X \ \text{and} \ j\notin X) \ \text{or} \ (i \notin X \ \text{and} \ j \in X)
\end{array}
\right.
\end{equation}
The graph resulting from applying switching function $g$ to signed graph $G$ is called $G$'s \textit{switching equivalent} and denoted by $G^g$. The switching equivalents of a graph have the same value of the frustration index, i.e. $L{(G^g)}=L(G) \ \forall \ g$ \cite{zaslavsky_matrices_2010}. It is straightforward to prove that the frustration index is equal to the minimum number of negative edges in $G^g$ over all switching functions $g$. An immediate result is that any balanced graph can switch to an equivalent graph where all the edges are positive \cite{zaslavsky_matrices_2010}. Moreover, in a switched graph with the minimum number of negative edges, called a \textit{negative minimal graph} and denoted by $G^{g^*}$, all vertices have a non-negative net degree. In other words, every vertex $ i $ in $G^{g^*}$ satisfies $d^- {(i)} \leq d^+ {(i)}$. 

\subsection{Bounds for the line index of balance}\label{2ss:bounds}
An obvious upper bound for the line index of balance is $L(G)\leq m^-$ which states the result that removing all negative edges gives a balanced graph. Recalling that acyclic signed graphs are balanced, the \textit{circuit rank} of the graph can also be considered as an upper bound for the frustration index \cite[p. 8]{flament1970equilibre}. Circuit rank, also known as the \textit{cyclomatic number} and the \textit{feedback edge set number}, is the minimum number of edges whose removal results in an acyclic graph. Moreover, the maximum number of edge-disjoint negative cycles in $G$ provides a lower bound for the frustration index \cite{zaslavsky2017}.

Petersdorf \cite{petersdorf_einige_1966} proves that among all sign functions for complete graphs with $n$ nodes, assigning negative signs to all the edges, i.e. putting $\sigma: E\rightarrow\{-1\}$, gives the maximum value of the frustration index which equals $\lfloor (n-1)^2/4 \rfloor$. %We denote these complete graphs with all-negative edges by ${K}_n^c$.
Petersdorf's proof confirms a conjecture by Abelson and Rosenberg\cite{abelson_symbolic_1958} that is also proved in \cite{tomescu_note_1973} and further discussed in \cite{akiyama_balancing_1981}.

Akiyama et al.\ provide results indicating that the frustration index of signed graphs with $n$ nodes and $m$ edges is bounded by $m/2$ \cite{akiyama_balancing_1981}. They also show that the frustration index of signed graphs with $n$ nodes is maximum in complete graphs with no positive 3-cycles and is bounded by $\lfloor (n-1)^2/4 \rfloor$ \cite[Theorem 1]{akiyama_balancing_1981}. 
Besides all-negative complete graphs, this group of graphs also contains complete graphs with nodes that can be partitioned into two classes such that all positive edges connect nodes from different classes and all negative edges connect nodes belonging to the same class \cite{tomescu_note_1973}. %We denote these graphs by ${K}_n^d$. 
Akiyama et al.\ refer to these graphs as \textit{antibalanced} \cite{akiyama_balancing_1981} which is a term coined by Harary in \cite{harary_structural_1957} and also discussed in \cite{zaslavsky_matrices_2010}. 

Iacono et al.\ suggest an upper bound for the frustration index \cite[page 227]{iacono_determining_2010} referred to as the worst-case
upper bound on the consistency deficit. However, values of the frustration index in all-negative complete graphs show that this upper bound is incorrect (take a complete graph with 9 nodes and 36 negative edges which has a frustration index of 16 while the bound suggested in \cite{iacono_determining_2010} gives a value of 15).

%It is easy to see that ${K}_n^c$ and ${K}_n^d$ are switching equivalents which leads to $L({K}_n^d)=L({K}_n^c)$. 

\section{Mathematical programming models} \label{2s:qmodel}

In this section, we formulate three mathematical programming models in Eq.\ \eqref{2eq4}, \eqref{2eq7}, and \eqref{2eq8} to calculate the frustration index by optimising an objective function formed using integer variables.

\subsection{A quadratically constrained quadratic programming model}

We formulate a mathematical programming model in Eq.\ \eqref{2eq4} to maximise $Z_1$, the sum of entries of $\textbf{A}^g$, the adjacency matrix of the graph switched by $g$, over different switching functions. Bearing in mind that the frustration index is the number of negative edges in a negative minimal graph, $L(G)=m^{-}{(G^{g^*})}$, then maximising $Z_1$ will effectively calculate the line index of balance. We use decision variables, $y_{i} \in \{-1,1\}$ to define node colours. Then $X=\{i \mid y_{i}=1\}$ gives the black-coloured nodes (alternatively nodes in the switching set). The restriction $y_{i} \in \{-1,1\}$ for the variables is formulated by $n$ quadratic constraints $y^2_{i}=1$. Note that the switching set $X=\{i \mid y_{i}=1\}$ creates a negative minimal graph with the adjacency matrix entries given by $a_{ij} y_{i} y_{j}$. The model can be represented as Eq.\ \eqref{2eq4} in the form of a continuous Quadratically Constrained Quadratic Programming (QCQP) model with $n$ decision variables and $n$ constraints.

\begin{equation}\label{2eq4}
\begin{split}
\max_{y_i} Z_1 &= \sum\limits_{i \in V} \sum\limits_{j \in V} a_{ij} y_{i} y_{j} \\
\text{s.t.} \quad
y^2_{i}&=1 \quad \forall i \in V
\end{split}
\end{equation}

Maximising $\sum_{i \in V} \sum _{j \in V} a_{ij} y_{i} y_{j}$ is equivalent to computing $m^{-}{(G^{g^*})}=|\{(i,j) \in E: a_{ij} y_{i} y_{j}=-1\}|$. Note that choosing $y_i$ to maximise $\sum_{i \in V} \sum _{j \in V} a_{ij} y_{i} y_{j}$ is equivalent to choosing $g$ to minimise $m^-(G^g)$. The optimal value of the objective function, $Z_1^*$, is equal to the sum of entries in the adjacency matrix of a negative minimal graph which can be represented by $Z_1^*= 2m^{+}{(G^{g^*})} - 2m^{-}{(G^{g^*})} = 2m - 4L(G)$. Therefore, the graph frustration index can be calculated by $L(G)= (2m - Z_1^*)/4$.

While the model expressed in Eq.\ \eqref{2eq4} is quite similar to the non-linear energy function minimisation model used in \cite{facchetti2012exploring, facchetti_computing_2011, esmailian_mesoscopic_2014, ma_memetic_2015} and the Hamiltonian of Ising models with $\pm 1$ interactions \cite{Sherrington}, the feasible region in model \eqref{2eq4} is neither convex nor a second order cone. Therefore, the QCQP model in Eq.\ \eqref{2eq4} only serves as an easy-to-understand optimisation model describing the connection between colouring nodes (alternatively selecting nodes to switch) and computing the line index of balance.

\subsection{An unconstrained binary quadratic programming model}\label{2ss:ubqp}

The optimisation model \eqref{2eq4} can be converted into an Unconstrained Binary Quadratic Programming (UBQP) model \eqref{2eq7} by changing the decision variables into binary variables $y_{i}=2x_{i}-1$ where $x_{i} \in \{0,1\}$. Note that the binary variables, $x_{i}$, that take value $1$ define the black-coloured nodes $X=\{i \mid x_{i}=1\}$ (alternatively, nodes in the switching set). The optimal solution represents a colouring set $X^* \subseteq V$ that minimises the resulting frustration count. 

Furthermore, by substituting $y_{i}=2x_{i}-1$ into the objective function in Eq.\ \eqref{2eq4} we get \eqref{2eq5}. The terms in the objective function can be modified as shown in Eq.\ \eqref{2eq5}--\eqref{2eq6} in order to have an objective function whose optimal value, $Z^*_2$, equals $L(G)$.
\begin{equation}\label{2eq5}
\begin{split}
Z_1 &= \sum\limits_{i \in V} \sum\limits_{j \in V} (4a_{ij} x_{i} x_{j} - 2 x_{i} a_{ij} - 2 x_{j} a_{ij} + a_{ij}) \\
&= \sum\limits_{i \in V} \sum\limits_{j \in V} (4 a_{ij} x_{i} x_{j} - 4 x_{i} a_{ij}) + (2m - 4m^{-}_{(G)})
\end{split}
\end{equation}

\begin{equation}\label{2eq6}
Z_2 = (2m - Z_1)/4
\end{equation}

Note that the binary quadratic model in Eq.\ \eqref{2eq7} has $n$ decision variables and no constraints.

\begin{equation}\label{2eq7}
\begin{split}
\min_{x_i} Z_2 &= \sum\limits_{i \in V} \sum\limits_{j \in V} (a_{ij}x_{i} - a_{ij}x_{i}x_{j})  + m^{-}_{(G)}\\
\text{s.t.} \quad x_{i} &\in \{0,1\} \ i \in V
\end{split}
\end{equation}

The optimal value of the objective function in Eq.\ \eqref{2eq7} represents the frustration index directly as shown in Eq.\ \eqref{2eq7.5}.

\begin{equation}\label{2eq7.5}
Z_2^* = (2m - Z^*_1)/4 = (2m - (2m - 4L(G)))/4 = L(G)
\end{equation}

The objective function in Eq.\ \eqref{2eq7} can be interpreted as initially starting with $m^{-}_{(G)}$ and then adding 1 for each positive frustrated edge (positive edge with different endpoint colours) and -1 for each negative edge that is not frustrated (negative edge with different endpoint colours). This adds up to the total number of frustrated edges.

\subsection{The 0/1 linear model}

The linearised version of \eqref{2eq7} is formulated in Eq.\ \eqref{2eq8}. The objective function of \eqref{2eq7} is first modified as shown in Eq.\ \eqref{2eq7.6} and then its non-linear term $x_{i} x_{j}$ is replaced by $|E|$ additional binary variables $x_{ij}$. The new decision variables $x_{ij}$ are defined for each edge $(i,j) \in E$ and take value 1 whenever $x_{i}=x_{j}=1$ and $0$ otherwise.
Note that $\sum_{j \in V}a_{ij}$ is a constant that equals the net degree of node $i$.

\begin{equation}\label{2eq7.6}
\begin{split}
Z_2 &= \sum\limits_{i \in V} \sum\limits_{j \in V} a_{ij}x_{i} - \sum\limits_{i \in V} \sum\limits_{j \in V}a_{ij}x_{i}x_{j}  + m^{-}_{(G)}\\
&= \sum\limits_{i \in V} x_{i} \sum\limits_{j \in V}a_{ij} - \sum\limits_{i \in V} \sum\limits_{j \in V, j>i} 2a_{ij}x_{i}x_{j}  + m^{-}_{(G)}%\\
%&=\sum\limits_{i \in V} x_{i} d_{i} - \sum\limits_{i \in V} \sum\limits_{j \in V, j>i} 2a_{ij}x_{i}x_{j}  + m^{-}_{(G)}
\end{split}
\end{equation}

The dependencies between the $x_{ij}$ and $x_{i},x_{j}$ values are taken into account by considering a constraint for each new variable. Therefore, the 0/1 linear model has $n+m$ variables and $m$ constraints, as shown in Eq.\ \eqref{2eq8}.

\begin{equation}\label{2eq8}
\begin{split}
\min_{x_i, x_{ij}} Z_2 &= \sum\limits_{i \in V} x_{i} \sum\limits_{j \in V}a_{ij} - \sum\limits_{(i,j) \in E} 2a_{ij}x_{ij}  + m^{-}_{(G)}\\
\text{s.t.} \quad
x_{ij} &\leq (x_{i}+x_{j})/2 \quad \forall (i,j) \in E^+ \\
x_{ij} &\geq x_{i}+x_{j}-1 \quad \forall (i,j) \in E^- \\
x_{i} &\in \{0,1\} \ i \in V\\
x_{ij} &\in \{0,1\} \ (i,j) \in E
\end{split}
\end{equation}

\subsection{Additional constraints for the 0/1 linear model}\label{2ss:additional}

The structural properties of the model allow us to restrict the model by adding additional valid inequalities. Valid inequalities are utilised by our solver, Gurobi, as additional non-core constraints that are kept aside from the core constraints of the model. Upon violation by a solution, valid inequalities are efficiently pulled in to the model. Pulled-in valid inequalities cut away a part of the feasible space and restrict the model. Additional restrictions imposed on the model can often speed up the solver algorithm if they are valid and useful \cite{Klotzpractical}. Properties of the optimal solution can be used to determine these additional constraints. Two properties we use are the connection between the degree of a node and its optimal colour and the fact that in every cycle there is always an even number of edges that change sign when applying the switching function $g$.

An obvious structural property is that the colouring of the graph leading to the minimum frustration count gives node $i$ a colour that makes at most half of the incident edges frustrated. This can be proved by contradiction. 

Assume an optimal colouring (which minimises the frustration count) gives a node a colour that makes more than than half of the connected edges frustrated. It follows that, changing the colour decreases the number of frustrated edges incident on it, which is in contradiction with the minimality of frustration count under the optimal colouring.

A \textit{node degree constraint} can be added to the model for each node restricting all variables associated with the incident edges. This structural property can be formulated as inequality \eqref{2eq10} which is valid for the optimal solution of the problem. As $x_i$ represents the colour of a node, ${(1-a_{ij})}/{2} + a_{ij}(x_i +x_j - 2x_{ij})$ takes value $1$ if the edge $(i,j) \in E$ is frustrated and takes value $0$ otherwise.
\begin{equation}\label{2eq10}
\sum\limits_{j: (i,j)\in E \text{ or } (j,i)\in E} {(1-a_{ij})}/{2} + a_{ij}(x_i +x_j - 2x_{ij}) \leq d {(i)}/2 \quad \forall i \in V
\end{equation}

Another structural property we observe is related to the edges making a cycle. According to the definition of the switching function \eqref{2eq2}, switching one node negates all edges (changes the signs on all edges) incident on that node. Because there are two edges incident on each node in a cycle, in every cycle there is always an even number of edges that change sign when switching function $g$ is applied to signed graph $G$.

As listing all cycles of a graph is computationally intensive, this structural property can be applied to cycles of a limited length. For instance, we may apply this structural property to the edge variables making triangles in the graph. 
This structural property can be formulated as valid inequalities in Eq.\ \eqref{2eq11} in which $T=\{(i,j,k)\in V^3 \mid (i,j),(i,k),(j,k) \in E \}$ contains ordered 3-tuples of nodes whose edges form a triangle. Note that $(x_{i} + x_{j} -2x_{ij})$ equals $1$ if edge $(i,j) \in E$ is negated and equals $0$ otherwise. The expression in Eq.\ \eqref{2eq11} denotes the total number of negated edges in the triangle formed by three edges $(i,j),(i,k),(j,k)$.

\begin{equation}\label{2eq11}
\begin{split}
&x_{i} + x_{j} -2x_{ij}+x_{i} + x_{k} -2x_{ik}+x_{j} + x_{k} -2x_{jk}\\ 
&= 0 \text{ or } 2 \quad \forall (i,j,k) \in T
\end{split} 
\end{equation}

Eq.\ \eqref{2eq11} can be linearised to Eq.\ \eqref{2eq12} as follows. \textit{Triangle constraints} can be applied to the model as four constraints per triangle, restricting three edge variables and three node variables per triangle.

\begin{equation}\label{2eq12}
\begin{split}
x_{i}+x_{jk} &\geq x_{ij} + x_{ik} 										  \quad \forall (i,j,k) \in T\\
x_{j}+x_{ik} &\geq x_{ij} + x_{jk}  											\quad \forall (i,j,k) \in T\\
x_{k}+x_{ij} &\geq x_{ik} + x_{jk}  											\quad \forall (i,j,k) \in T\\
1 + x_{ij} + x_{ik} + x_{jk} &\geq x_{i} + x_{j} + x_{k} \quad \forall (i,j,k) \in T
\end{split} 
\end{equation}

%More restrictions can be imposed using the minimum of the two upper bounds discussed in \ref{2ss:bounds}. This is implemented as a constraint in Eq.\ \eqref{2eq13}. 

%\begin{equation}\label{eq13}
%Z_2 \leq \min\{m/2 ,m^{-}\}
%\end{equation}

In order to speed up the model in Eq.\ \eqref{2eq8}, we consider fixing a node colour to increase the root node objective function in the solver's branch and bound process.
We conjecture the best node variable to fix is the one associated with the highest unsigned node degree (to be further discussed in Subsection \ref{3ss:branch}). This constraint is formulated in Eq.\ \eqref{2eq15} which our experiments show speeds up the branch and bound algorithm by increasing the lower bound.

\begin{equation}\label{2eq15}
x_{k} = 1 \quad k= \text{arg} \max_{i \in V} d(i)
\end{equation}

The complete formulation of the 0/1 linear model with further restrictions on the feasible space includes the objective function and core constraints in Eq.\ \eqref{2eq8} and valid inequalities in Eq.\ \eqref{2eq10}, Eq.\ \eqref{2eq12}, and Eq.\ \eqref{2eq15}. The model has $n+m$ binary variables, $m$ core constraints, and $n+4|T|+1$ additional constraints. 

Table \ref{2tab1} provides a comparison of the three optimisation models based on their variables, constraints, and objective functions. In the next sections, we mainly focus on the 0/1 linear model solved in conjunction with the valid inequalities (additional constraints).

\begin{table}
	\centering
	\caption{Comparison of the three optimisation models}
	\label{2tab1}
	\begin{tabular}{p{2.5cm}p{2.5cm}p{2.5cm}l} 
		\hline\noalign{\smallskip}		& QCQP \eqref{2eq4}  & UBQP \eqref{2eq7}   & 0/1 linear model \eqref{2eq8}       \\ 
		\hline
		Variables       & $n$                & $n$       & $n+m$          \\
		Constraints     & $n$                & $0$       & $m$            \\
		Variable type   & continuous         & binary    & binary         \\
		Constraint type & quadratic       & -         & linear            \\
		Objective       & quadratic          & quadratic & linear         \\ \noalign{\smallskip}\hline\noalign{\smallskip}
	\end{tabular}
\end{table}

\section{Numerical results in random graphs} \label{2s:results}
In this section, the frustration index of various random networks is computed by solving the 0/1 linear model \eqref{2eq8} coupled with the additional constraints. The \textit{NetworkX} package in Python is used for generating different types of random graphs. We use Gurobi version 7 on a desktop computer with an Intel Core i5 4670 @ 3.40 GHz and 8.00 GB of RAM running 64-bit Microsoft Windows 7. The models were created using Gurobi's Python interface. All four processor cores available were used by Gurobi.

To verify our software implementation, we manually counted the number of frustrated edges given by our software's proposed node colouring for a number of test problems, and confirmed that this matched the frustration count reported by our software. These tests showed that our models and implementations were performing as expected.

\subsection{Performance of the 0/1 linear model on random graphs} \label{2ss:perform}

In this subsection we discuss the time performance of Gurobi's branch and bound algorithm for solving the 0/1 linear model. In order to evaluate the performance of the 0/1 linear model \eqref{2eq8} coupled with the additional constraints, we generate 10 decent-sized Erd\H{o}s-R\'{e}nyi random graphs, $G(n,M)$, \cite{bollobas2001random} as test cases with various densities and percentages of negative edges. Results are provided in Table~\ref{2tab2} in which B\&B nodes stands for the number of branch and bound nodes (in the search tree of the branch and bound algorithm) explored by the solver.

\begin{table}
	\centering
	\caption{Performance measures of Gurobi solving the 0/1 linear model in Eq.\ \eqref{2eq8} for the random networks}
	\label{2tab2}
	\begin{tabular}{lllllllll}
		\hline\noalign{\smallskip}
		TestCase	&	$n$	&	$m$	&	$m^-$	&	$\rho$	&	$\frac{m^-}{m}$	&	$L(G)$	&	B\&B nodes	&	time(s)\\
		\hline
		1	&	65	&	570	&	395	&	0.27	&	0.69	&	189	&	5133	&	65.4\\
		2	&	68	&	500	&	410	&	0.22	&	0.82	&	162	&	4105	&	27.3\\
		3	&	80	&	550	&	330	&	0.17	&	0.60	&	170	&	11652	&	153.3\\
		4	&	50	&	520	&	385	&	0.42	&	0.74	&	185	&	901		&	22.4\\
		5	&	53	&	560	&	240	&	0.41	&	0.43	&	193	&	292		&	13.5\\
		6	&	50	&	510	&	335	&	0.42	&	0.66	&	178	&	573		&	13.8\\
		7	&	59	&	590	&	590	&	0.34	&	1.00	&	213	&	1831	&	46.0\\
		8	&	56	&	600	&	110	&	0.39	&	0.18	&	110	&	0		&	0.4\\
		9	&	71	&	500	&	190	&	0.20	&	0.38	&	155	&	6305	&	77.7\\
		10	&	80	&	550	&	450	&	0.17	&	0.82	&	173	&	12384	&	138.0\\
		\noalign{\smallskip}\hline\noalign{\smallskip}
	\end{tabular}
\end{table}

The results in Table~\ref{2tab2} show that random test cases based on Erd\H{o}s-R\'{e}nyi graphs, $G(n,M)$, with 500-600 edges can be solved to optimality in a reasonable time. The branching process for these test cases explores various numbers of nodes ranging between 0 and 12384. These numbers also depend on the number of threads and the heuristics that the solver uses automatically. %Solving the test cases for a second time often leads to a different number of branch and bound nodes.

\subsection{Impact of negative edges on the frustration index}

\begin{figure}
	\centering
	\includegraphics[width=0.8\textwidth]{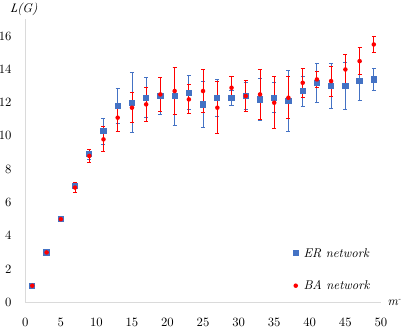}
	\caption{The frustration index in Erd\H{o}s-R\'{e}nyi (ER) networks, $G(n,M)$, with 15 nodes and 50 edges and Barab\'{a}si-Albert (BA) networks with 15 nodes and 50 edges and various number of negative edges}
	\label{2fig1}
\end{figure}

In this subsection we use both Erd\H{o}s-R\'{e}nyi graphs, $G(n,M)$, and Barab\'{a}si-Albert graphs \cite{bollobas2001random} as synthetic data for computing the line index of balance. In this analysis, we use the same randomly generated graphs with different numbers of negative edges (assigned by a uniform random distribution) as test cases over 50 runs per experiment setting. Figure~\ref{2fig1} shows the average and standard deviation of the line index of balance in these random signed networks with $n=15,m=50$. It is worth mentioning that we have observed similar results in other types of random graphs including small world, scale-free, and random regular graphs \cite{bollobas2001random}.

Figure \ref{2fig1} shows similar increases in the line index of balance in the two graph classes as $m^-$ increases. It can be observed that the maximum frustration index is still smaller than $m/3$ for both graphs. This shows a gap between the values of the line index of balance in random graphs and the theoretical upper bound of $m/2$. It is important to know whether this gap is proportional to graph size and density.

\subsection{Impact of graph size and density on the frustration index}

\begin{figure}
	\centering
	\includegraphics[width=1\textwidth]{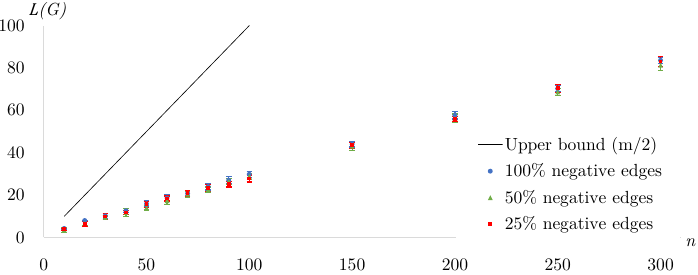}
	\caption{The frustration index in random 4-regular networks of different orders $n$ and decreasing densities}
	\label{2fig2}
\end{figure}

In order to investigate the impact of graph size and density, 4-regular random graphs with a constant fraction of randomly assigned negative edges are analysed averaging over 50 runs per experiment setting. The frustration index is computed for 4-regular random graphs with 25\%, 50\%, and 100\% negative edges and compared with the upper bound $m/2$. Figure~\ref{2fig2} demonstrates the average and standard deviation of the frustration index where the degree of all nodes remains constant, but the density of the 4-regular graphs, $\rho=4/n-1$, decreases as $n$ and $m$ increase.

An observation to derive from Figure~\ref{2fig2} is the similar frustration index values obtained for networks of the same sizes, even if they have different percentages of negative edges. It can be concluded that starting with an all-positive graph (which has a frustration index of $0$), making the first quarter of graph edges negative increases the frustration index much more than making further edges negative. Future research is required to get a better understanding of how the frustration index and minimum deletion sets change when the number of negative edges is increased (on a fixed underlying structure). Another observation is that the gap between the frustration index values and the theoretical upper bound increases with increasing $n$. 

\section{Numerical results in real signed networks} \label{2s:real}
In this section, the frustration index is computed for nine real networks by solving the 0/1 linear model \eqref{2eq8} coupled with the additional constraints using Gurobi version 7 on a desktop computer with an Intel Core i5 4670 @ 3.40 GHz (released in 2013) and 8.00 GB of RAM running 64-bit Microsoft Windows 7.

We use well studied signed social network datasets representing communities with positive and negative interactions and preferences including Read's dataset for New Guinean highland tribes \cite{read_cultures_1954} and Sampson's dataset for monastery interactions \cite{sampson_novitiate_1968} which we denote respectively by G1 and G2. We also use graphs inferred from datasets of students' choice and rejection, denoted by G3 and G4 \cite{newcomb_acquaintance_1961,lemann_group_1952}. A further explanation on the details of inferring signed graphs from choice and rejection data can be found in Subsection \ref{1s:infer}. Moreover, a larger signed network, denoted by G5, is inferred by \cite{neal_backbone_2014} through implementing a stochastic degree sequence model on Fowler's data on Senate bill co-sponsorship \cite{fowler_legislative_2006}.

As well as the signed social network datasets, large scale biological networks can be analysed as signed graphs. We use the four signed biological networks analysed by \cite{dasgupta_algorithmic_2007} and \cite{iacono_determining_2010}. Graph G6 represents the gene regulatory network of \textit{Saccharomyces cerevisiae} \cite{Costanzo2001yeast} and graph G7 is related to the gene regulatory network of \textit{Escherichia coli} \cite{salgado2006ecoli}. The Epidermal growth factor receptor pathway \cite{oda2005} is represented as graph G8. Graph G9 represents the molecular interaction map of a macrophage \cite{oda2004molecular}. For more details on the four biological datasets, one may refer to \cite{iacono_determining_2010}. The data for real networks used in this chapter is publicly available on the \urllink{https://figshare.com/articles/Signed_networks_from_sociology_and_political_science_biology_international_relations_finance_and_computational_chemistry/5700832}{Figshare} research data repository \cite{Aref2017data}.

We use $G_r=(V,E,\sigma_r)$ to denote a reshuffled graph in which the sign function $\sigma_r$ is a random mapping of $E$ to $\{-1,+1\}$ that preserves the number of negative edges. The reshuffling process preserves the underlying graph structure. The numerical results on the frustration index of the nine signed graphs and reshuffled versions of these graphs are shown in Table~\ref{2tab3} where, for each graph $G$, the average and standard deviation of the line index of balance in 500 reshuffled graphs, denoted by $L(G_r)$ and $\text{SD}$, are also provided for comparison. 

\begin{table}
	\centering
	\caption{The frustration index in various signed networks}
	\label{2tab3}
	\begin{tabular}{p{1.5cm}p{1cm}p{1cm}p{1cm}p{1cm}p{2.5cm}l}
		\hline\noalign{\smallskip}
		Graph & $n$ & $m$ & $m^-$ & $L(G)$ & $L(G_r) \pm \text{SD}$ & Z score \\ 
		\hline
		G1    & 16  & 58  & 29    & 7    & $14.65 \pm 1.38$ &  -5.54 \\
		G2    & 18  & 49  & 12    & 5    & $9.71  \pm 1.17$ &  -4.03 \\
		G3    & 17  & 40  & 17    & 4    & $ 7.53 \pm 1.24$ &  -2.85 \\
		G4    & 17  & 36  & 16    & 6    & $ 6.48 \pm 1.08$ &  -0.45 \\
		G5    & 100 & 2461& 1047  & 331  & $ 965.6\pm 9.08$& -69.89 \\ 
		G6    & 690 & 1080& 220   &  41  & $ 124.3\pm 4.97$& -16.75 \\ 
		G7    & 1461& 3215& 1336  & 371  & $ 653.4\pm 7.71$& -36.64 \\ 	
		G8    & 329 & 779 & 264   & 193  & $ 148.96\pm 5.33$&   8.26 \\ 
		G9    & 678 & 1425&  478  & 332  & $ 255.65\pm 8.51$&  8.98 \\ 	
		\noalign{\smallskip}\hline\noalign{\smallskip}	
	\end{tabular}
\end{table}

\begin{figure}
	\subfloat[Highland tribes network (G1), a signed network of 16 tribes of the Eastern Central Highlands of New Guinea \cite{read_cultures_1954}. Minimum deletion set comprises 7 negative edges.]{\includegraphics[width=2.5in]{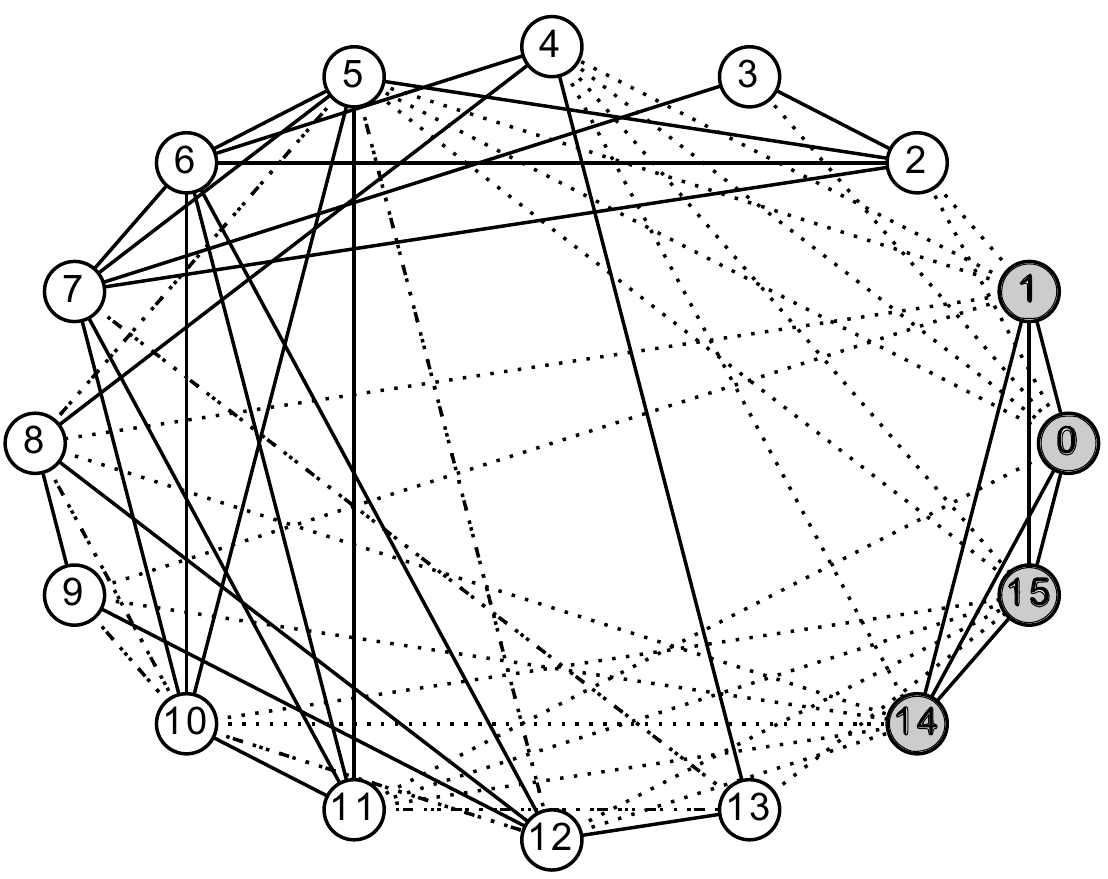}}%
	\hfil
	\subfloat[Monastery interactions network (G2) of 18 New England novitiates inferred from the integration of all positive and negative relationships \cite{sampson_novitiate_1968}. Minimum deletion set comprises 2 positive and 3 negative edges.]{\includegraphics[width=2.5in]{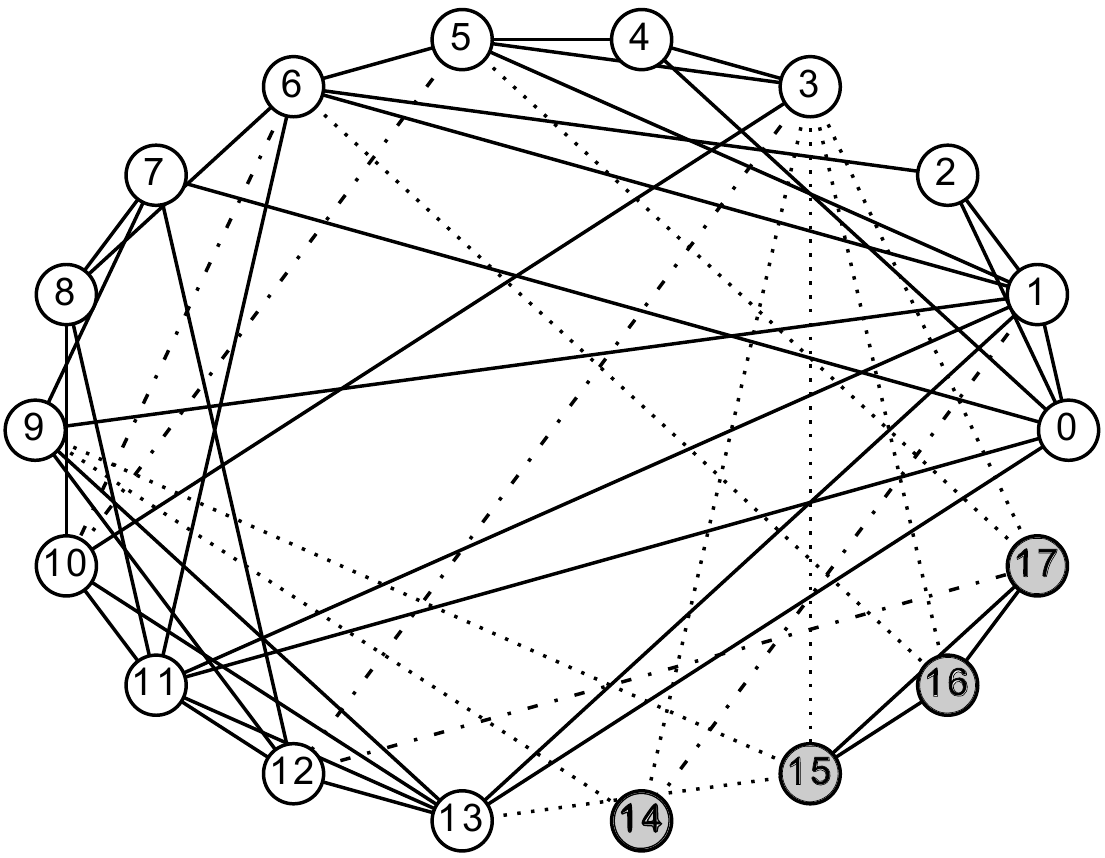}}%
	\quad
	\subfloat[Fraternity preferences network (G3) of 17 boys living in a pseudo-dormitory inferred from ranking data of the last week in \cite{newcomb_acquaintance_1961}. Minimum deletion set comprises 4 negative edges.]{\includegraphics[width=2.5in]{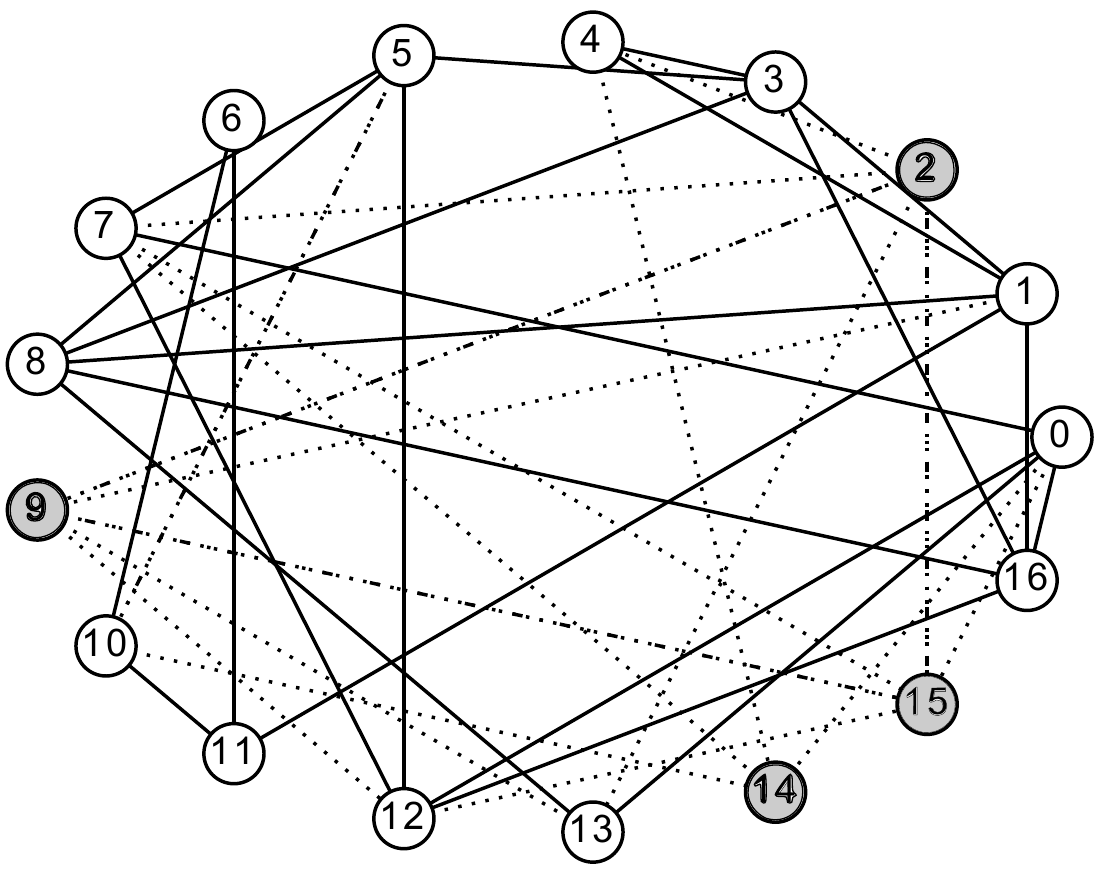}}%
	\hfil
	\subfloat[College preferences network (G4) of 17 girls at an Eastern college inferred from ranking data of house B in \cite{lemann_group_1952}. Minimum deletion set comprises 3 positive and 3 negative edges.]{\includegraphics[width=2.5in]{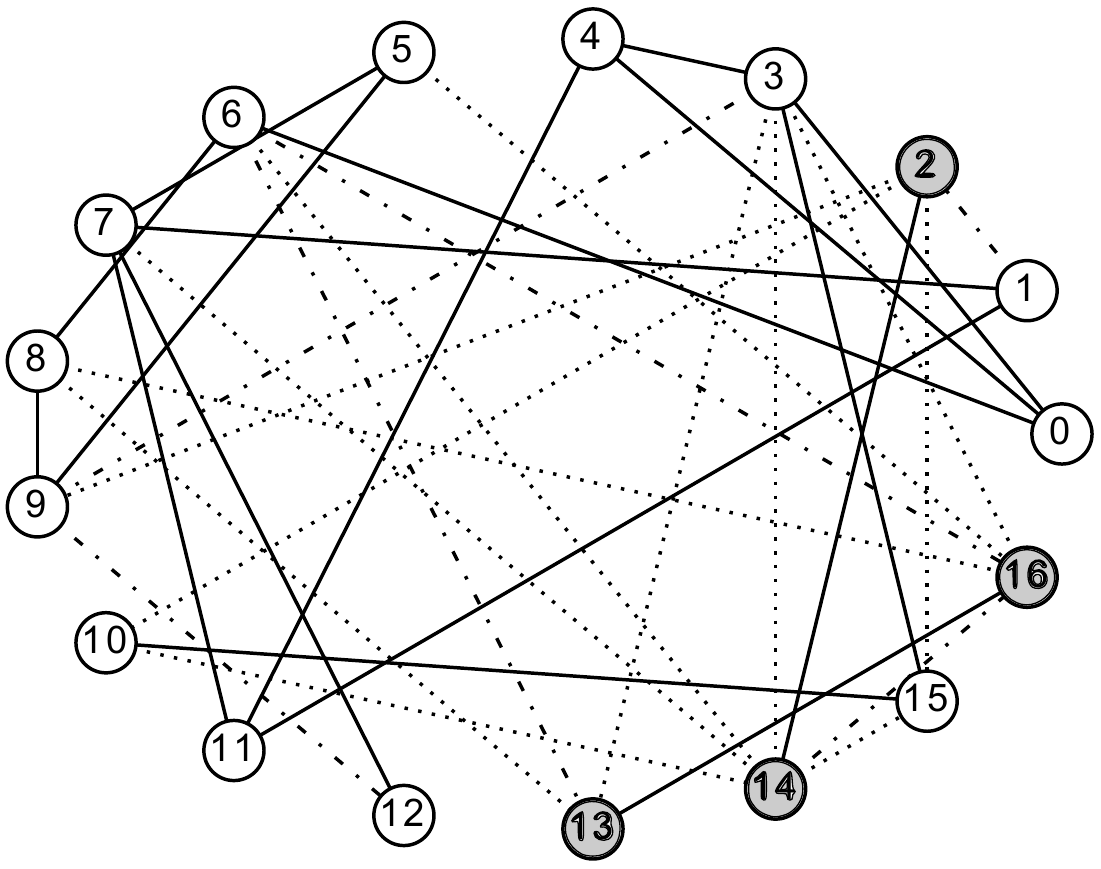}}%
	\caption{The frustrated edges represented by dotdash lines for four small signed networks inferred from the sociology datasets}
	\label{2fig3}
\end{figure}

Although the signed networks are not balanced, the relatively small values of $L(G)$ suggest a low level of frustration in some of the networks. Figure \ref{2fig3} shows how the small signed networks G1 -- G4 can be made balanced by negating (or removing) the edges from a minimum deletion set. Dotted lines represent negative edges, solid lines represent positive edges, and frustrated edges are indicated by dotdash lines regardless of their original signs. The node colourings leading to the minimum frustration counts are also shown in Figure~\ref{2fig3}. Note that it is pure coincidence that there are an equal number of nodes coloured black for each graph G1 -- G4 in Figure~\ref{2fig3}. Visualisation of graphs G1 -- G4 without node colours and minimum deletion sets can be found in Figure \ref{1fig6}.

In order to be more precise in evaluating the relative levels of frustration in G1 -- G9, we have implemented a very basic statistical analysis using Z scores, where $Z={(L(G)-L(G_r))}/{\text{SD}}$. The Z scores, provided in the right column of Table~\ref{2tab3}, show how far the frustration index is from the values obtained through random allocation of signs to the fixed underlying structure (unsigned graph). 
Negative values of the Z score can be interpreted as a lower level of frustration than the value resulting from a random allocation of signs. G1, G2, G5, G6, and G7 exhibit a level of frustration lower than what is expected by chance, while the opposite is observed for G8 and G9. The numerical results for G3 and G4 do not allow a conclusive interpretation. Refer to Sections \ref{4s:d1}--\ref{4s:d2} for an in-depth interpretation of the results on these real networks.

\section{Conclusion of Chapter \ref{ch:2}} \label{2s:conclu}

This chapter focuses on frustration index as a measure of balance in signed networks and the findings may well have a bearing on the applications of the line index of balance in the other disciplines (to be discussed in Chapter \ref{ch:4}). This chapter has suggested a novel method for computing a measure of structural balance that can be used for analysing dynamics of signed networks. It contributes additional evidence that suggests signed social networks and biological gene regulatory networks exhibit a relatively low level of frustration (compared to the expectation when allocating signs at random). In Chapter \ref{ch:3}, we continue the same line of research with more focus on operations research aspects of computing the frustration index.

\textit{Point index of balance} is a similarly defined measure of balance based on removing a minimum number of nodes to achieve balance \cite{harary_measurement_1959}. The computations of this measure may also be considered as a niche point to be explored using exact and heuristic computational methods \cite{gutin2009fixed}.
%\begin{acknowledgement}
%	The authors are grateful for the extremely valuable comments of the anonymous reviewers that have prevented incorrect attributions in the literature review section and helped improve the discussions in this chapter.
%\end{acknowledgement} 

%\bibliographystyle{abbrv}
%\bibliography{Ch1/refs}

%\end{document}
 
\cleardoublepage

\chapter{Efficient Computation of the Frustration Index in Signed Networks}\label{ch:3}

\maketitle

%\cleardoublepage
\section*{Abstract}
Computing the frustration index of a signed graph is a key step toward solving problems in many fields including social networks, physics, chemistry, and biology. The frustration index determines the distance of a network from a state of total structural balance. Although the definition of the frustration index goes back to 1960, its exact algorithmic computation, which is closely related to classic NP-hard graph problems, has only become a focus in recent years. 
We develop three new binary linear programming models to compute the frustration index exactly and efficiently as the solution to a global optimisation problem. Solving the models with prioritised branching and valid inequalities in Gurobi, we can compute the frustration index of real signed networks with over 15000 edges in less than a minute on inexpensive hardware. 
We provide extensive performance analysis for both random and real signed networks and show that our models outperform all existing approaches by large factors.
Based on solve time, algorithm output, and effective branching factor we highlight the superiority of our models to both exact and heuristic methods in the literature.
%\footnote{The current authors have published a book chapter under the title ``Computing the Line Index of Balance Using Integer Programming Optimisation''\cite{aref2017computing}. This paper is a continuation of the same line of research with a focus on developing new models to reduce solve time by large factors and facilitate processing larger instances.}

%\textbf{Keywords:} 
%0-1 programming,
%Frustration index,
%Branch and bound,
%Signed graphs,
%Balance theory\\

\cleardoublepage
\section{Introduction to Chapter \ref{ch:3}} \label{3s:intro}
Local ties between entities lead to global structures in networks. Ties can be formed as a result of interactions and individual preferences of the entities in the network. The dual nature of interactions in various contexts means the ties may form in two opposite types, namely positive ties and negative ties. In a social context, this is interpreted as friendship versus enmity or trust versus distrust between people. The term \textit{signed network} embodies a multitude of concepts involving relationships characterisable by ties with plus and minus signs. \textit{Signed graphs} are used to model such networks where edges have positive and negative signs. \textit{Structural balance} in signed graphs is a macro-scale structural property that has become a focus in network science.

Structural balance theory was the first attempt to understand the sources of tensions and conflicts in groups of people with signed ties \cite{heider_social_1944}. According to balance theory, some structural configurations of people with signed ties lead to social tension and therefore are not balanced. Using graph-theoretic concepts, Cartwright and Harary identified cycles of the graph %(closed-walks with distinct nodes)
as the origins of tension, in particular cycles containing an odd number of negative edges \cite{cartwright_structural_1956}. By definition, signed graphs in which no such cycles are present satisfy the property of structural balance. The vertex set of a \textit{balanced} signed network can be partitioned into $k \leq 2$ subsets such that each negative edge joins vertices belonging to different subsets \cite{cartwright_structural_1956}. For graphs that are not totally balanced, a distance from total balance (a measure of partial balance) can be computed. Among various measures is the \textit{frustration index} that indicates the minimum number of edges whose removal (or equivalently, negation) results in balance \cite{abelson_symbolic_1958,harary_measurement_1959,zaslavsky_balanced_1987}. In what follows, we discuss previous works related to the frustration index (also called the \textit{line index of balance} \cite{harary_measurement_1959}). We use both names, line index of balance and frustration index, interchangeably in this chapter.
\subsection{Motivation}

In the past few decades, different measures of balance \cite{cartwright_structural_1956,norman_derivation_1972,terzi_spectral_2011,kunegis_applications_2014,estrada_walk-based_2014} have been suggested and deployed to analyse balance in real-world signed networks resulting in conflicting observations \cite{leskovec_signed_2010, facchetti_computing_2011, estrada_walk-based_2014}. Measures based on cycles \cite{cartwright_structural_1956,norman_derivation_1972}, triangles \cite{terzi_spectral_2011,kunegis_applications_2014}, and closed-walks \cite{estrada_walk-based_2014} are not generally consistent and do not satisfy key axiomatic properties (as discussed in Chapter \ref{ch:1}). Among all the measures, a normalised version of the frustration index is shown to satisfy many basic axioms. This measure provides a clear understanding of the transition to balance in terms of the number of edges to be modified to reduce the tension, as opposed to graph cycles that were first suggested as origins of tension in unbalanced networks \cite{cartwright_structural_1956}.

The frustration index is a key to frequently stated problems in many different fields of research \cite{iacono_determining_2010,harary_signed_2002,kasteleyn_dimer_1963,patrick_doreian_structural_2015,doslic_computing_2007}. In biological networks, optimal decomposition of a network into monotone subsystems %---which is essential for understanding Drosophila segment polarity---
is made possible by computing the signed graph frustration index \cite{iacono_determining_2010}. In finance, performance of a portfolio is related to the balance of its underlying signed graphs \cite{harary_signed_2002}. In physics, the frustration index provides the minimum energy state of magnetic materials \cite{kasteleyn_dimer_1963}. In international relations, signed clustering of countries in a region can be investigated using the frustration index \cite{patrick_doreian_structural_2015}. In chemistry, bipartite edge frustration has applications to the stability of fullerene, a carbon allotrope \cite{doslic_computing_2007}. For a discussion on applications of the frustration index, one may refer to Chapter \ref{ch:4}.

\subsection{Complexity}
Computing the frustration index is related to the well-known unsigned graph optimisation problem EDGE-BIPARTIZATION, which requires minimisation of the number of edges whose deletion makes the graph bipartite \cite{huffner_separator-based_2010}. Given an instance of the latter problem, by declaring each edge to be negative we convert it to the problem of computing the frustration index. Since EDGE-BIPARTIZATION is known to be NP-hard \cite{yannakakis1981edge}, so is computing the frustration index. %In the converse direction there is a reduction of the frustration index problem to EDGE-BIPARTIZATION which increases the number of edges by a factor of at most $2$ \cite{huffner_separator-based_2010}. %If the reduction preserves planarity, the frustration index can be computed in polynomial time for such planar graphs \cite{grotschel1981weakly}, which is equivalent to the ground state calculation of a two-dimensional spin glass model with no periodic boundary conditions and no magnetic field \cite{DeSimone1995,hartmann2016revisiting}. 

The classic graph optimisation problem MAXCUT is also a special case of the frustration index problem, as can be seen by assigning all edges to be negative (an edge is frustrated if and only if it does not cross the cut). Similar to MAXCUT for planar graphs \cite{hadlock}, the frustration index can be computed in polynomial time for planar graphs \cite{katai1978studies}.

\subsection{Approximation}
In general graphs, the frustration index is even NP-hard to approximate within any constant factor (assuming Khot's Unique Games Conjecture \cite{khot2002power}) \cite{huffner_separator-based_2010}. That is, for each $C>0$, the problem of finding an approximation to the frustration index that is guaranteed to be within a factor of $C$ is believed to be NP-hard.

%The computational complexity of the problem might have played a role in the lack of systematic investigation of computing the frustration index while there are many studies on approximating it.
%Efficient approximations have long existed for MAXCUT including Goemans and Williamson's approximation featuring an approximation guarantee of 0.878 \cite{goemans_improved_1995}. The semidefinite programming algorithm of Goemans and Williamson leads to the same approximation guarantee for the frustration index \cite{thagard1998coherence, dasgupta_algorithmic_2007}. %%SA This is probably wrong because we cannot approximate it to a constant factor
The frustration index can be approximated to a factor of $\mathcal{O}(\sqrt{\log n})$ \cite{agarwal2005log} or $\mathcal{O}(k \log k)$ \cite{avidor2007multi} where $n$ is the number of vertices and $k$ is the frustration index. Coleman et al.\ provide a review on the performance of several approximation algorithms of the frustration index \cite{coleman2008local}.

\subsection{Heuristics and local optimisation}
Doreian and Mrvar have reported numerical values as the line index of balance and suggest that determining this index is in general a polynomial-time hard problem \cite{patrick_doreian_structural_2015}. However, their algorithm does not provide optimal solutions and the results are not equal to the line index of balance (as discussed in Chapter \ref{ch:2}).
Data-reduction schemes \cite{huffner_separator-based_2010} and ground state search heuristics \cite{iacono_determining_2010} are used to obtain bounds for the frustration index. 
%Iacono et al.\ showed that the frustration index equals the minimum number of fundamental negative cycles induced over all spanning trees of the graph \cite{iacono_determining_2010}. Originally discussed in the biology context, the terminology used in \cite{iacono_determining_2010} is different where \textit{monotonocity} and \textit{consistency deficit} are the equivalents for balance and frustration.
Facchetti, Iacono, and Altafini suggested a non-linear energy function minimisation model for finding the frustration index \cite{facchetti_computing_2011}. Their model was solved using various techniques \cite{iacono_determining_2010, esmailian_mesoscopic_2014, ma_memetic_2015, ma_decomposition-based_2017}. Using the ground state search heuristic algorithms \cite{iacono_determining_2010}, the frustration index is estimated in biological networks with $n \approx 1.5\times10^3$ \cite{iacono_determining_2010} and social networks with $n \approx 10^5$ \cite{facchetti_computing_2011, facchetti2012exploring}. 

\subsection{Exact computation}
Using a parametrised algorithmics approach, H\"{u}ffner, Betzler, and Niedermeier show that the frustration index (under a different name) is \textit{fixed parameter tractable} and can be computed in $\mathcal{O}(2^k m^2)$ \cite{huffner_separator-based_2010} where $m$ is the number of edges and $k$ is the fixed parameter (the frustration index). %The values of $k$ we have observed in signed graphs inferred from the literature make this approach impractical (see Subsection~\ref{3s:real} for numerical results on real networks).
In Chapter \ref{ch:2}, we suggested binary (quadratic and linear) programming models as methods for computing the frustration index capable of processing graphs with $m \approx 10^3$ edges.

%Their suggested heuristic is reported to solve networks with up to $n \leq 10^5$ within $99\%$ of optimality. However, not only their main theorem (Theorem 1 in \cite{esmailian_mesoscopic_2014}) is incorrect, but Mendon{\c{c}}a et al.\ has also cast doubt on their main conclusion regarding the role of negative ties in signed graphs \cite{mendoncca2015relevance}.

\subsection*{Our contribution in Chapter \ref{ch:3}} \label{3ss:contrib}
The principal focus of this chapter is to provide further insight into computing the frustration index by developing efficient computational methods outperforming previous methods by large factors. We systematically investigate several formulations for exact computation of the frustration index and compare them based on solve time as well as other performance measures.

The advantage of formulating the problem as an optimisation model is not only exploring the details involved in a fundamental NP-hard problem, but also making use of powerful mathematical programming solvers like Gurobi \cite{gurobi} to solve the NP-hard problem exactly and efficiently. We provide numerical results on a variety of undirected signed networks, both randomly generated and inferred from well-known datasets (including real signed networks with over 15000 edges).

In Chapter \ref{ch:2}, we investigated computing the frustration index in smaller scales using quadratic and linear optimisation models showing a considerable overlap in the objectives with the current chapter. In this Chapter, we provide three new binary linear formulations which not only outperform the models in Chapter \ref{ch:2} by large factors, but also facilitate a more direct and intuitive interpretation. We discuss more efficient speed-up techniques that require substantially fewer additional constraints compared to the models in Chapter \ref{ch:2}. This allows Gurobi's branch and bound algorithm to start with a better root node solution and explore considerably fewer nodes %(0.001-0.006)
leading to a faster solve time. Moreover, our new models handle larger instances that were not solvable by the models in Chapter \ref{ch:2}. %In this study, we provide a more straight forward interpretation of the optimisation models. %obviating the need to discuss some convoluted signed graph concepts. 
We provide in-depth performance analysis using extensive numerical results showing the solve times of our worst-performing model to be $2 - 9$ times faster than the best-performing model in Chapter \ref{ch:2}.

This chapter begins by laying out the theoretical dimensions of the research in Section~\ref{3s:prelim}. Linear programming models are formulated in Section~\ref{3s:model}. Section~\ref{3s:speed} provides different techniques to improve the formulations and reduce solve time. The numerical results on the models' performance are presented in Section~\ref{3s:results}. Section~\ref{3s:evaluate} provides comparison against the literature using both random and real networks. Recent developments on a closely related problem are discussed in Section~\ref{3s:related} followed by two extensions to the models in Section~\ref{3s:future}. Section~\ref{3s:conclu} presents the findings of the research and sums up the research highlights. 

\section{Preliminaries} \label{3s:prelim}

We recall some standard definitions.

\subsection{Basic notation} \label{3s:problem}
We consider undirected signed networks $G = (V,E,\sigma)$. The ordered set of nodes is denoted by $V=\{1,2,\dots,n\}$, with $|V| = n$. The set $E$ of edges is partitioned into the set of positive edges $E^+$ and the set of negative edges $E^-$ with $|E^-|=m^-$, $|E^+|=m^+$, and $|E|=m=m^- + m^+$. The sign function is denoted by $\sigma: E\rightarrow\{-1,+1\}$.

We represent the $m$ undirected edges in $G$ as ordered pairs of vertices $E = \{e_1, e_2, ..., e_m\} \subseteq \{ (i,j) \mid i,j \in V , i<j \}$, where a single edge $e_k$ between nodes $i$ and $j$, $i<j$, is denoted by $e_k=(i,j) , i<j$.  We denote the graph density by $\rho= 2m/(n(n-1))$. The entries $a{_i}{_j}$ of the \emph{signed adjacency matrix}, \textbf{A}, are defined in \eqref{3eq1}. 

\begin{equation}\label{3eq1}
a{_i}{_j} =
\left\{
\begin{array}{ll}
\sigma_{(i,j)} & \mbox{if } (i,j) \in E \\
\sigma_{(j,i)} & \mbox{if } (j,i) \in E \\
0 &  \text{otherwise}
\end{array}
\right.
\end{equation}
%The number of edges incident on the node $i \in V$ represents the degree of node $i$ and is denoted by $d {(i)}$. A \emph{walk} of length $k$ in $G$ is a sequence of nodes $v_0,v_1,...,v_{k-1},v_k$ such that for each $i=1,2,...,k$ there is an edge from $v_{i-1}$ to $v_i$. If $v_0=v_k$, the sequence is a \emph{closed walk} of length $k$. If the nodes in a closed walk are distinct except for the endpoints, it is a directed cycle (for simplicity \emph{cycle}) of length $k$. The \emph{sign} of a cycle is the product of the signs of its edges. A cycle with negative sign is unbalanced. A balanced cycle is one with positive sign. A balanced graph is one with no negative cycles.

The number of edges incident on the node $i \in V$ represents the degree of node $i$ and is denoted by $d {(i)}$. A directed cycle (for simplicity \emph{cycle}) of length $k$ in $G$ is a sequence of nodes $v_0,v_1,...,v_{k-1},v_k$ such that for each $i=1,2,...,k$ there is an edge from $v_{i-1}$ to $v_i$ and the nodes in the sequence except for $v_0=v_k$ are distinct. The \emph{sign} of a cycle is the product of the signs of its edges. A cycle with negative sign is unbalanced. A balanced cycle is one with positive sign. A balanced graph is one with no negative cycles.

\subsection{Node colouring}\label{3ss:nodecolouring}

\textit{Satisfied} and \textit{frustrated} edges are defined based on colourings of the nodes. Colouring each node with black or white, a frustrated (satisfied) edge $(i,j)$ is either a positive (negative) edge with different colours on the endpoints $i,j$ or a negative (positive) edge with the same colours on the endpoints $i,j$.

Subfigure \ref{3fig1a} illustrates an example signed graph in which positive and negative edges are represented by solid lines and dotted lines respectively. Subfigures \ref{3fig1b} and \ref{3fig1c} illustrate node colourings and their impacts on the frustrated edges that are represented by thick lines.

\subsection{Max (2,2)-CSP formulation and theoretical results}
In this subsection, we formulate the problem of computing the frustration index as a constraint satisfaction problem in \eqref{3eq12} and provide theoretical results on the fastest known algorithms. Computation of the frustration index can be formulated as a Maximum 2-Constraint Satisfaction Problem with 2 states per variable (Max (2,2)-CSP) with $n$ variables and $m$ constraints.

The signed graph, $G(V,E,\sigma)$, is the input constraint graph. We consider a score for each edge $(i,j)$ depending on its sign $\sigma_{ij}$ and the assignment of binary values to its endpoints. In the formulation provided in \eqref{3eq12}, the dyadic score function $S_{(i,j)}:\{0,1\}^2 \rightarrow \{0,1\}$ determines the satisfaction of edge $(i,j)$ accordingly (score $1$ for satisfied and score $0$ for frustrated). The output of solving this problem is the colouring function $\phi: V \rightarrow \{0,1\}$ which maximises the total number of satisfied edges as score function $S(\phi)$. 

\begin{equation} \label{3eq12}
	\begin{split}
		\max_\phi S(\phi)=\sum_{(i,j) \in E} S_{(\phi(i), \phi(j))} \\
		%\max_\phi S(\phi)=\sum_{(u,v) \in E^-} S^-_{(\phi(u),\phi(v))} + \sum_{(u,v) \in E^+} S^+_{(\phi(u),\phi(v))} \\ 
		S_{(i,j)}=\{ ((0,0),(1+\sigma_{ij})/2),\\
		((0,1),(1-\sigma_{ij})/2),\\
		((1,0),(1-\sigma_{ij})/2),\\
		((1,1),(1+\sigma_{ij})/2)\} 
		%S^-_{(u,v)}=\{ ((0,0),0), ((0,1),1),((1,0),1),((1,1),0)\} \\
		%S^+_{(u,v)}=\{ ((0,0),1), ((0,1),0),((1,0),0),((1,1),1)\}
	\end{split}
\end{equation}

Denoting the maximum score function value by $S^*(\phi)$, the frustration index can be calculated as the number of edges that are not satisfied $m-S^*(\phi)$.

According to worst-case analyses, the fastest known algorithm \cite{koivisto2006optimal} with respect to $n$ solves Max (2,2)-CSP in $\mathcal{O}(nm2^{n\omega/3})$, where $\omega$ is the matrix multiplication exponent. Since $\omega < 2.373$ \cite{LeGall2014}, the running time of the algorithm from \cite{koivisto2006optimal} is $\mathcal{O}(1.7303^n)$. It improves on the previous fastest algorithm \cite{WILLIAMS2005} only in the polynomial factor of the running time. With respect to $n$, the algorithm in \cite{koivisto2006optimal} is the fastest known algorithm for MAXCUT, and therefore for computing the frustration index as well. Both algorithms \cite{koivisto2006optimal,WILLIAMS2005} use exponential space and it is open whether MAXCUT can be solved in $\mathcal{O}(c^n)$ for some $c<2$ when only polynomial space is allowed.

With respect to the number $m$ of edges, the Max (2,2)-CSP formulation in \eqref{3eq12} enables the use of algorithms from \cite{Gaspers2017} and \cite{SCOTT2007}. The first algorithm uses $2^{(9m/50 + \mathcal{O}(m))}$ time and polynomial space \cite{Gaspers2017}, while the second algorithm uses $2^{(13m/75 + \mathcal{O}(m))}$ time and exponential space. With respect to the number of edges, these two algorithms \cite{Gaspers2017,SCOTT2007} are also the fastest algorithms known for MAXCUT, and therefore for computing the frustration index.

Having provided the theoretical results, the rest of this chapter focuses on the practical aspects of computing the frustration index using mathematical programming formulations different from the formulation expressed in \eqref{3eq12}.

\subsection{Frustration count}\label{3ss:frustrationcount}

In this subsection, we provide definitions that are central to the rest of the chapter.

\textbf{Definition.}
%For any signed graph $G=(V, E, \sigma)$, we can partition $V$ into two sets, denoted $X \subseteq V$ and $\overline{X}=V \setminus X$. We call $X$ the colouring set and we think of this partitioning as specifying a colouring of the nodes, where each node $i \in X$ is coloured black, and $i \in \overline{X}$ is coloured white. We let $x_i$ denote the colour of node $i \in V$ under $X$, where $x_i=1$ if $i \in X$ and $x_i=0$ otherwise. 
Let $X \subseteq V$ be a subset of vertices. This defines a partition $(X, V \setminus X)$ of $V$. We call $X$ a colouring set. 

\textbf{Definition.}
Let binary variable $x_i$ denote the colour of node $i \in V$ under colouring set $X$. We consider $x_i=1$ if $i \in X$ (black node) and $x_i=0$ if $i \in V \setminus X$ (white node). %We say that each node $i \in X$ is black and each node $i \in V \setminus X$ is white. 

%We say that an edge $(i,j)$ is frustrated under $X$ if either edge $(i,j)$ is a positive edge (i.e.\ $(i,j) \in E^+$) but nodes $i$ and $j$ have different colours ($x_i \ne x_j$), or edge $(i,j)$ is a negative edge (i.e.\ $(i,j) \in E^-$) but nodes $i$ and $j$ share the same colour ($x_i = x_j$). 

\textbf{Definition.}
We define the {\em frustration count} of signed graph $G$ under colouring $X$ as $f_G(X) := \sum_{(i,j) \in E} f_{ij}(X)$
where $f_{ij}(X)$ is the frustration state of edge $(i,j)$, given by
\begin{equation} \label{3eq2}
f_{ij}(X)=
\begin{cases}
0, & \text{if}\ x_i = x_j \text{ and } (i,j) \in E^+ \\
1, & \text{if}\ x_i = x_j \text{ and } (i,j) \in E^- \\
0, & \text{if}\ x_i \ne x_j \text{ and } (i,j) \in E^- \\
1, & \text{if}\ x_i \ne x_j \text{ and } (i,j) \in E^+ \\
\end{cases}
\end{equation}

The optimisation problem is finding a colouring set $X^* \subseteq V$ of $G$ that minimises the frustration count $f_G(X)$, i.e., solving Eq.\ \eqref{3eq3}. The globally optimal solution to this problem gives the frustration index $L(G)$ of signed graph $G$.
\begin{equation} \label{3eq3}
L(G) = \min_{X \subseteq V}f_G(X)\
\end{equation}

\begin{figure}[ht]
	\subfloat[An example graph with four nodes, two positive edges, and three negative edges]{\includegraphics[height=1.2in]{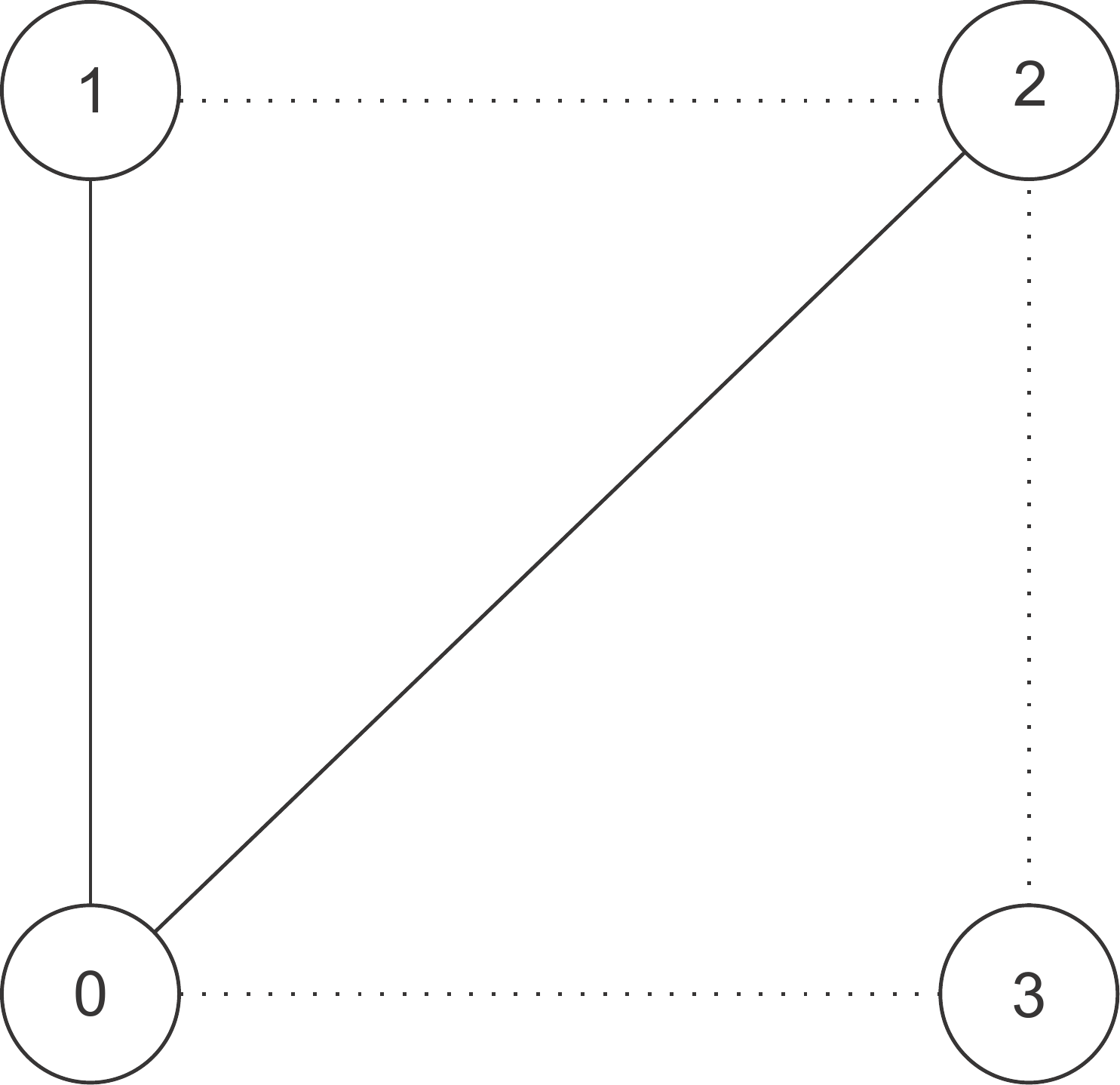}%
		\label{3fig1a}}
	\hfil
	\subfloat[An arbitrary node colouring resulting in two frustrated edges (0,2), (2,3)]{\includegraphics[height=1.2in]{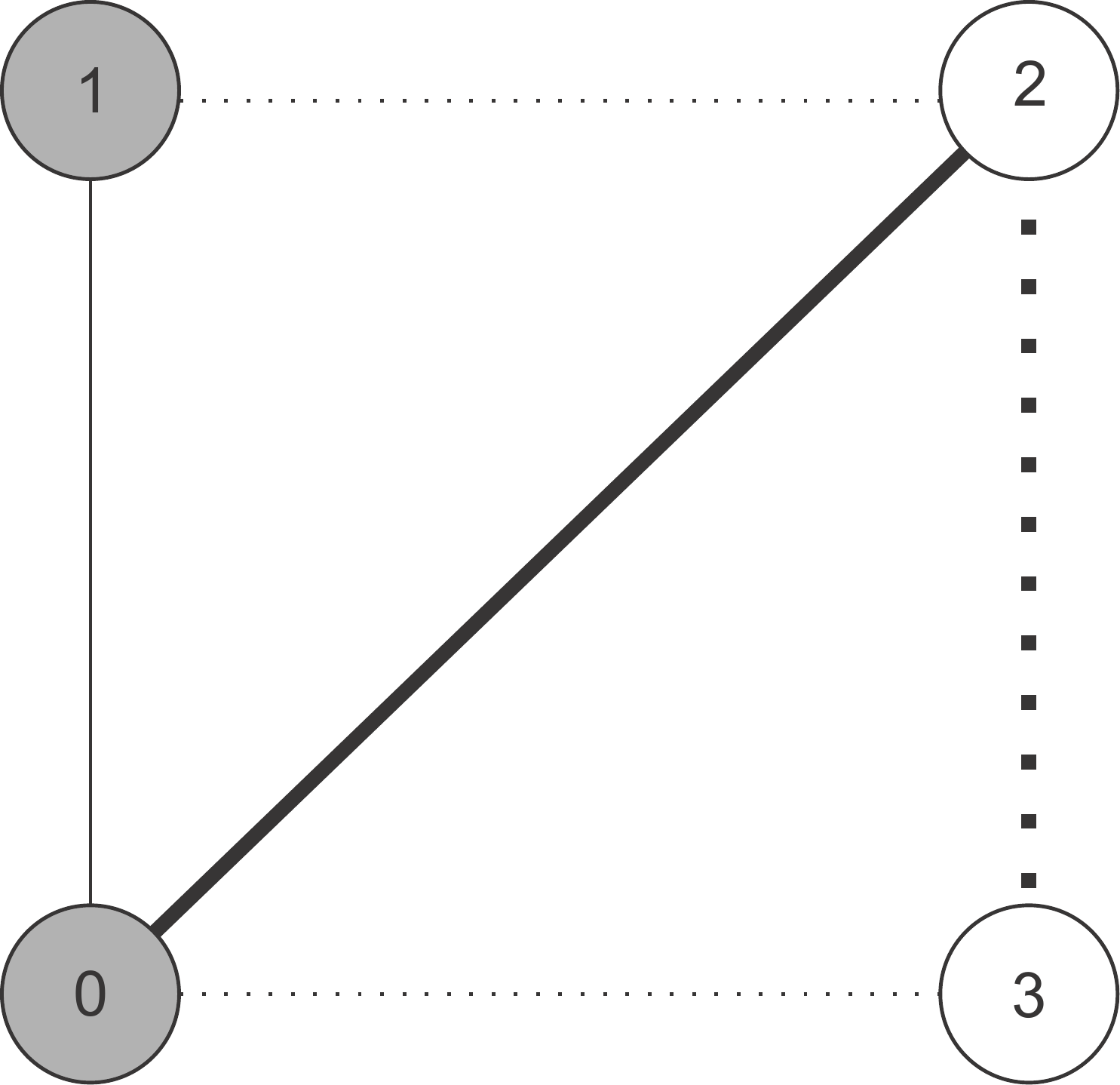}%
		\label{3fig1b}} 
	\hfil
	\subfloat[Another node colouring resulting in one frustrated edge (1,2)]{\includegraphics[height=1.2in]{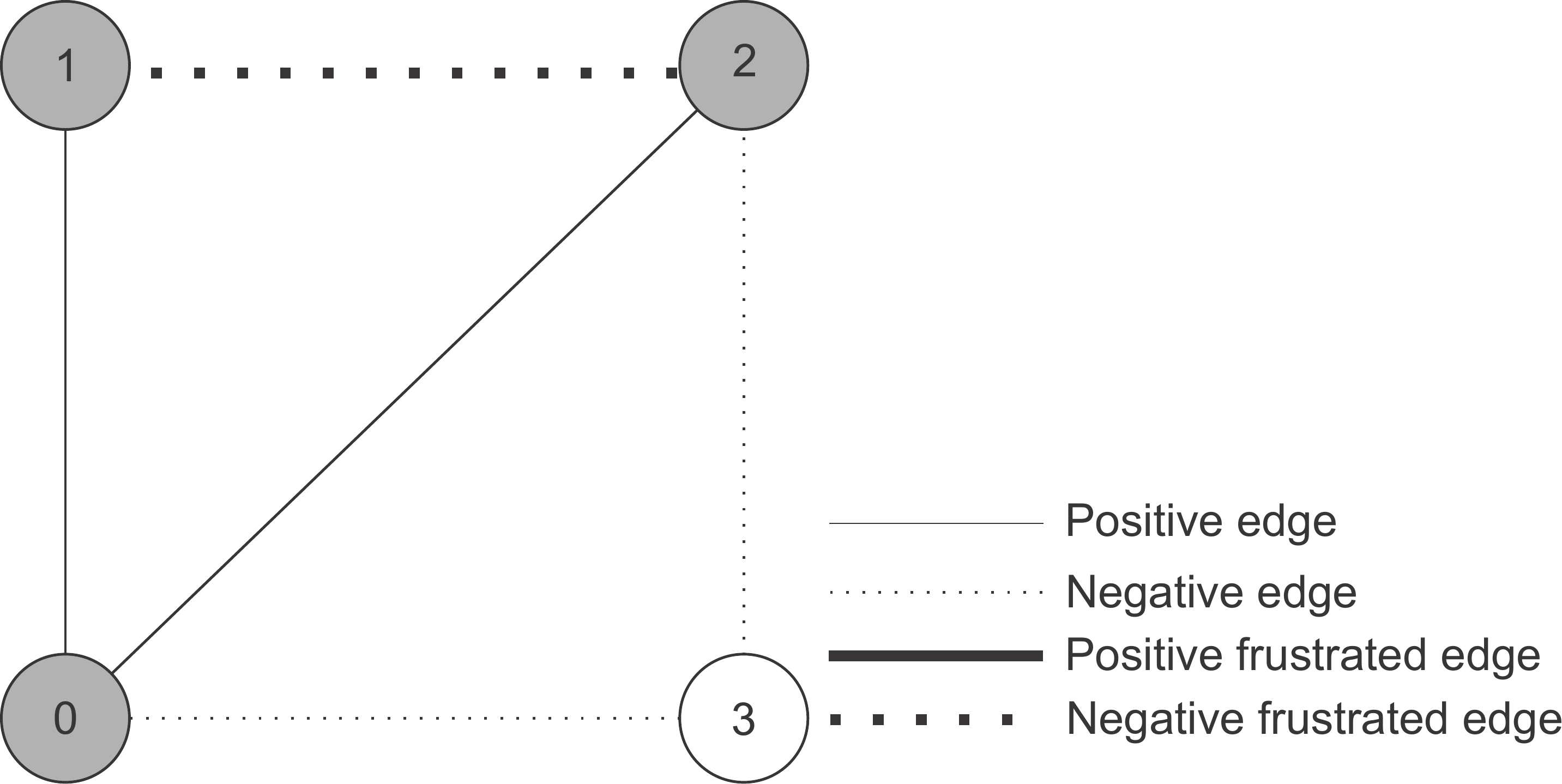}%
		\label{3fig1c}}
	\caption{Node colourings and the respective frustrated edges for an example signed graph}% The figure is produced using Adobe Illustrator.}
	\label{3fig1}
\end{figure}

It follows that $f_G(X)$ gives an upper bound on $L(G)$ for any $X \subseteq V$. Note that the colouring in Subfigure \ref{3fig1b} does not minimise $f_G(X)$, while in Subfigure \ref{3fig1c} the frustration count is minimum.

\section{Binary linear programming formulations} \label{3s:model}
In this section, we introduce three 0/1 linear programming models in Eq.\ \eqref{3eq3.5} -- \eqref{3eq5} to minimise the frustration count as the objective function. There are various ways to form the frustration count using variables defined over graph nodes and edges which lead to various mathematical programming models that we discuss in this section. 

\subsection{The AND model}
We start with an objective function to minimise the frustration count. Note that the frustration of a positive edge $(i,j)$ can be represented by $f_{ij}=x_i +x_j - 2(x_i \text{AND} x_j),$ $\forall (i,j) \in E^+$ using the two binary variables $x_i , x_j\in\{0,1\}$ for the endpoint colours. For a negative edge, we have $f_{ij}= 1- (x_i +x_j - 2(x_i \text{AND} x_j)),$ $\forall (i,j) \in E^-$.  

The term $x_i \text{AND} x_j$ can be replaced by binary variables $x_{ij}=x_i \text{AND} x_j$ for each edge $(i,j)$ that take value 1 whenever $x_{i}=x_{j}=1$ (both endpoints are coloured black) and 0 otherwise. This gives our first 0/1 linear model in Eq.\ \eqref{3eq3.5} that calculates the frustration index in the minimisation objective function.
\begin{equation}\label{3eq3.5}
\begin{split}
\min_{x_i: i \in V, x_{ij}: (i,j) \in E} Z &= \sum\limits_{(i,j) \in E^+} x_{i} + x_{j} - 2x_{ij}  + \sum\limits_{(i,j) \in E^-} 1 - (x_{i} + x_{j} - 2x_{ij})\\
\text{s.t.} \quad
x_{ij} &\leq x_{i} \quad \forall (i,j) \in E^+ \\
x_{ij} &\leq x_{j} \quad \forall (i,j) \in E^+ \\
x_{ij} &\geq x_{i}+x_{j}-1 \quad \forall (i,j) \in E^- \\
x_{i} &\in \{0,1\} \quad  \forall i \in V \\
x_{ij} &\in \{0,1\} \quad \forall (i,j) \in E 
\end{split}
\end{equation}

The optimal solution represents a subset $X^* \subseteq V$ of $G$ that minimises the frustration count. The optimal value of the objective function in Eq.\ \eqref{3eq3.5} is denoted by $Z^*$ which represents the frustration index.

The dependencies between the $x_{ij}$ and $x_{i},x_{j}$ values are taken into account using standard AND constraints. The AND model has $n+m$ variables and $2m^+ + m^-$ constraints. Note that $x_{ij}$ variables are dependent variables because of the constraints. Therefore, we may drop the integrality constraint of the $x_{ij}$ variables and consider them as continuous variables in the unit interval, $x_{ij} \in [0,1]$. The next subsection discusses an alternative binary linear model for calculating the frustration index.

\subsection{The XOR model}
Minimising the frustration count can be directly formulated as a binary linear model. The XOR model is designed to directly count the frustrated edges using binary variables $f_{ij}\in\{0,1\}, \forall (i,j) \in E$. As before, we use $x_i\in\{0,1\}, \forall i \in V$ to denote the colour of node $i$. This model is formulated by observing that the frustration state of a positive edge $(i,j) \in E^+$ is given by $f_{ij}(X)=x_i \text{XOR} x_j$. Similarly for $(i,j) \in E^-$, we have $f_{ij}(X)=1- x_i \text{XOR} x_j$. Therefore, the minimum frustration count under all node colourings is obtained by solving \eqref{3eq4}.

The dependencies between the $f_{ij}$ and $x_{i},x_{j}$ values are taken into account using two standard XOR constraints per edge. Therefore, the XOR model has $n+m$ variables and $2m$ constraints. Note that $f_{ij}$ variables are dependent variables because of the constraints and the positive coefficients in the minimisation objective function. Therefore, we may specify $f_{ij}$ variables as continuous variables in the unit interval, $f_{ij} \in [0,1]$. 
\begin{equation}\label{3eq4}
\begin{split}
\min_{x_i: i \in V, f_{ij}: (i,j) \in E} Z &= \sum\limits_{(i,j) \in E}  f_{ij}  \\
\text{s.t.} \quad
f_{ij} &\geq x_{i}-x_{j} \quad \forall (i,j) \in E^+ \\
f_{ij} &\geq x_{j}-x_{i} \quad \forall (i,j) \in E^+ \\
f_{ij}  &\ge  x_{i} + x_{j} -1 \quad \forall (i,j) \in E^- \\
f_{ij}  &\ge  1-x_{i} - x_{j}  \quad \forall (i,j) \in E^- \\
x_{i} &\in \{0,1\} \quad  \forall i \in V \\
f_{ij} &\in \{0,1\} \quad \forall (i,j) \in E 
\end{split}
\end{equation}
A third linear formulation of the problem is provided in the next subsection.

\subsection{The ABS model}
In this subsection, we propose the ABS model, a binary linear model in which two edge variables are used to represent the frustration state of an edge. 

We start by observing that for a node colouring, $|x_i - x_j|=1$ for a positive frustrated edge and $|x_i - x_j|=0$ for a positive satisfied edge $(i,j) \in E^+$. Similarly, $1-|x_i - x_j|=|x_i + x_j -1|$ gives the frustration state of a negative edge $(i,j) \in E^-$.

To model the absolute value function, we introduce additional binary variables $e_{ij}, h_{ij}\in\{0,1\},$ $\forall (i,j) \in E$. We observe that for a positive edge if $x_i - x_j =  e_{ij} - h_{ij}$ then $|x_i - x_j| =  e_{ij} + h_{ij}$. Similarly, for a negative edge if $x_i + x_j -1 =  e_{ij} - h_{ij}$ then $|x_i + x_j -1| = e_{ij} + h_{ij}$. This allows us to formulate the linear model in Eq.\ \eqref{3eq5}.

The objective function, being the total number of frustrated edges, sums the aforementioned absolute value terms to compute the frustration count in Eq.\ \eqref{3eq5}. The conditions observed for positive and negative edges are expressed as linear constraints in Eq.\ \eqref{3eq5}. Therefore, the ABS model has $n+2m$ variables and $m$ constraints.
\begin{equation}\label{3eq5}
\begin{split}
\min_{x_i: i \in V, e_{ij},h_{ij}: (i,j) \in E} Z &=\sum\limits_{(i,j) \in E}  e_{ij} + h_{ij} \\
\text{s.t.} \quad 
x_{i} - x_{j} &=  e_{ij} - h_{ij} \quad \forall (i,j) \in E^+ \\
x_{i} + x_{j} -1 &= e_{ij} - h_{ij} \quad \forall (i,j) \in E^- \\
x_{i} &\in \{0,1\} \quad  \forall i \in V \\
e_{ij} &\in \{0,1\} \quad \forall (i,j) \in E \\
h_{ij} &\in \{0,1\} \quad \forall (i,j) \in E 
\end{split} 
\end{equation}

\subsection{Comparison of the models} \label{3ss:compare}
In this subsection we compare the three models introduced above and two of the models suggested in Chapter \ref{ch:2}, based on the number and type of constraints. Table~\ref{3tab1} summarises the comparison.

\begin{table}[ht]
	\centering
	\caption{Comparison of optimisation models developed for computing the frustration index}
	\label{3tab1}
	\begin{tabular}{lp{1.8cm}p{2cm}p{1.7cm}p{1.8cm}p{1.4cm}} \hline
		& UBQP model \eqref{2eq7} &  0/1 linear model \eqref{2eq8}  & \multirow{2}{*}{AND \eqref{3eq3.5}} & \multirow{2}{*}{XOR \eqref{3eq4}} & \multirow{2}{*}{ABS \eqref{3eq5} } \\ \hline
		Variables                    & $n$   & $n+m$    & $n+m$             & $n+m$             & $n+2m$          \\
		Constraints                  & $0$    & $m^+ + m^-$   & $2m^+ + m^-$               & $2m^+ + 2m^-$              & $m^+ + m^-$             \\
		Constraint type     & -       & linear  & linear            & linear            & linear      \\
		Objective            & quadratic & linear  & linear            & linear            & linear         \\ \hline
	\end{tabular}
\end{table}

Eq.\ \eqref{3eq6} shows that the three linear models (AND, XOR, and ABS) are mathematically equivalent. Note that not only does the number of constraints scale linearly with graph size, each constraint involves at most 4 variables. Thus the worst-case space usage for solving these models is $\mathcal{O}(n^2)$.

\begin{equation}\label{3eq6}
f_{ij} = e_{ij} + h_{ij} = {(1-a_{ij})}/{2} + a_{ij}(x_i +x_j - 2x_{ij}) 
\end{equation}

The three linear models perform differently in terms of solve time and the number of branch and bound (B\&B) nodes required to solve a given instance. 

Solving large-scale binary programming models is not easy \cite{Bilitzky2005} and therefore there is a limit to the size of the largest graph whose frustration index can be computed in a given time. In the next section, we discuss some techniques for improving the performance of Gurobi in solving the binary linear models.

\section{Speed-up techniques}\label{3s:speed}

In this section we discuss techniques to speed up the branch and bound algorithm for solving the binary linear models described in the previous section. %To improve the branch and bound, one may use certain conditions based on the structure of the binary programming problem \cite{Bilitzky2005}.
Two techniques often deployed in solving Integer Programming (IP) models are valid inequalities and branching priority, which we discuss briefly. %They both are well-known techniques in solving integer linear programming models using the branch and bound algorithm.

The key feature of valid inequalities is that they are satisfied by any integer solution to some original formulation. Furthermore, we hope to find valid inequalities that strengthen the formulation by reducing the feasible region of the linear programming relaxation. A "lazy constraint" is a constraint that is given to the solver, but the solver does not add it to the model unless it is violated by a solution \cite{Klotzpractical}.
%We implement some valid inequalities as constraints that are kept aside from the original constraints of the model and pulled into the model only if they are violated by a solution \cite{Klotzpractical}. %Such implementation of additional restrictions is referred to as lazy constraints. 
Implementing the valid inequalities as lazy constraints restricts the model by cutting away a part of the feasible space. If these restrictions are valid and useful, they speed up the solver algorithm \cite{Klotzpractical}.

%Branching is the main process in the branch and bound algorithm to progress in searching the feasible space for integer solutions.

The branch and bound algorithm can be provided with a list of prioritised variables for branching which may speed up the solver if the prioritised list is more effective in making integer values.

We report the improvement of speed-up techniques at the end of each subsection. The solve time improvement evaluation is based on 100 Erd\H{o}s-R\'{e}nyi graphs, $G(n,p)$, with uniformly random parameters from the ranges $40 \leq n \leq 50$, $0 \leq p \leq 1$, and $0 \leq m^-/m \leq 1$.

\subsection{Pre-processing data reduction} \label{3ss:pre}

Standard graph pre-processing can be used to reduce graph size and order without changing the frustration index. This may reduce solve time in graphs containing nodes of degree $0$ and $1$ (also called isolated and pendant vertices respectively) and nodes whose removal increases the number of connected components (also called articulation points). We implement some of the data-reduction schemes in \cite{huffner_separator-based_2010}. H\"{u}ffner et al.\ suggest different ways to reduce nodes and edges of the graph that are separated by a small set of vertices called a \textit{separator} \cite{huffner_separator-based_2010}.

We have tested iterative reduction of isolated and pendant vertices as well as decomposing graphs by cutting them into smaller subgraphs using articulation points. These operations are referred to as data reduction using separators of size 0 and 1 in \cite{huffner_separator-based_2010}. Our experiments show that reducing isolated and pendant vertices does not considerably affect the solve time. Moreover, the scarcity of articulation points in many graphs in which isolated and pendant vertices have been removed, makes decomposition based on articulation points not particularly useful. However, H\"{u}ffner et al.\ suggest their data-reduction schemes using separators of size up to 3 to be very effective on reducing solve time in their experiments \cite{huffner_separator-based_2010}.

\subsection{Branching priority and fixing a colour} \label{3ss:branch}

%In binary programming models, the root node solution is integral if the constraint matrix is unimodular and the right hand side vector is integral. However, 

%Most practical integer optimisation problems, including the problem under investigation, do not have an integral root node solution \cite{Bilitzky2005} and therefore require an algorithm like branch and bound for finding integral solutions.

%In order to speed up the algorithms, we consider adding an additional constraint to increase the root node objective function value.

We relax the integrality constraints and observe in the Linear Programming relaxation (LP relaxation) of all three models that there always exists a fractional solution of $x_i=0.5, \forall i \in V$ which gives an optimal objective function value of 0. We can increase the root node objective by fixing one node variable, $x_k=1$, which breaks the symmetry that exists and allows changing all node colours to give an equivalent solution. %This idea is similar to fixing the \textit{ghost spin} in the ground state calculation of a spin glass model \cite{DeSimone1995}. 

Fixing the colour of node $k$ by imposing $x_k=1$ leads to the maximum amount of increase to the LP relaxation optimal solution when the optimal values of other node variables do not change, i.e., $x_i=0.5, \forall i \in V \setminus \{k\}$. In this case, all edges incident on node $k$ contribute $0.5$ to the LP relaxation objective function resulting in a root node objective value of $d {(k)}/2$. This observation shows that the best node variable to be fixed is the one associated with the highest degree which allows for a potential increase of $\max_{i \in V} d {(i)}/2$ in the LP relaxation optimal solution. We formulate this as a constraint in Eq.\ \eqref{3eq6.5}.
\begin{equation}\label{3eq6.5}
x_{k} = 1 \quad k= \text{arg} \max_{i \in V} d {(i)}
\end{equation}

In our experiments, we always observed an improvement in the root node objective value when Eq.\ \eqref{3eq6.5} was added, which shows it is useful. We provide more detailed results on the root node objective values for several instances in Section~\ref{3s:evaluate}.

Based on the same idea, we may modify the branch and bound algorithm so that it branches first on the node with the highest degree. This modification is implemented by specifying a branching priority for the node variables in which variable $x_i$ has a priority given by its degree $d {(i)}$.

Our experiments on random graphs show that fixing a colour and using prioritised branching lead to 60\%, 88\%, and 72\% 
%65\%, 70\%, and 43\% 
%43\%, 63\%, and 44\% 
reduction in the average solve time of AND, XOR, and ABS models respectively.

\subsection{Unbalanced triangle valid inequalities} \label{3ss:unbalanced}
%The structural properties of the problem allow us to restrict the model by adding valid inequalities as additional constraints \cite{Bilitzky2005}. Structural properties of signed graphs can be used to determine valid inequalities. 
%%SA zaslavsky_balance_2010 is not the right reference. We can prove this:
%We know that changing signs of the frustrated edges leads to balance. We also know that all cycles are positive in a balanced graph. When we change the sign of an even number of edges in a cycle, the sign of cycle does not change. Therefore, in order to make all negative cycle positive, an odd number of edges on each of them must change sign.
We consider adding one inequality for each negative cycle of length 3 (unbalanced triangle) in the graph. Every negative cycle of the graph contains an odd number of frustrated edges. This means that any colouring of the nodes in an unbalanced triangle must produce at least one frustrated edge. Recalling that under colouring $X$, the variable $f_{ij}$ is 1 if edge $(i,j)$ is frustrated (and 0 otherwise), then for any node triple $(i,j,k)$ defining an unbalanced triangle in $G$, we have the inequality \eqref{3eq7} which is valid for all feasible solutions of the problem.
\begin{equation}\label{3eq7}
f_{ij} + f_{ik} + f_{jk} 
\geq 1 \quad \forall (i,j,k) \in T^-
\end{equation}

in Eq.\ \eqref{3eq7}, $T^-=\{(i,j,k)\in V^3 \mid a{_i}{_j} a{_i}{_k} a{_j}{_k} = -1 \}$ denotes the set of node triples that define an unbalanced triangle. The expression in inequality \eqref{3eq7} denotes the sum of frustration states for the three edges $(i,j),(i,k),(j,k)$ making an unbalanced triangle. Note that in order to implement the unbalanced triangle valid inequality \eqref{3eq7}, $f_{ij}$ must be represented using the decision variables in the particular model. Eq.\ \eqref{3eq6} shows how $f_{ij}$ can be defined in the AND and ABS models.
%Note that, $ \min_{f_{ij}: (i,j) \in E} \sum f_{ij}$ subject to \eqref{eq16} with $m$ binary variables $f_{ij} &\in \{0,1\}$ and $|T^-|$ constraints can be considered as another formulation of the problem.

The fractional solution $x_i=0.5, \forall i \in V \setminus \{k\}$ violates the valid inequality in \eqref{3eq7}. Therefore, the valid inequality in \eqref{3eq7} is useful and adding it to the model leads to an increase in the root node objective. From a solve time perspective, our experiments on random graphs show that implementing this speed-up technique leads to 38\%, 24\%, and 12\% 
%65\% and 87\% and <5\%
%68\%, 93\%, and 8\% 
reduction in the average solve time of AND, XOR, and ABS models respectively. 

\subsection{Overall improvement made by the speed-up techniques}\label{3ss:overall}
%In Section~\ref{s:speed}, we discussed the solve time improvement made by the individual implementation of the speed-up techniques on the binary linear models in Section~\ref{s:speed}. %The solve time improvement percentages reported are based on 100 random graphs with $n=30,m=300,m^-=150$.
The total solve time reduction observed when both speed-up techniques (\ref{3ss:branch} -- \ref{3ss:unbalanced}) are implemented is 67\% for the AND model, 90\% for the XOR model, and 78\% for the ABS model.

Table \ref{3tab1.5} shows the solve time improvements made by implementing the speed-up techniques individually and collectively on 100 Erd\H{o}s-R\'{e}nyi graphs, $G(n,p)$, with uniformly random parameters $40 \leq n \leq 50$, $0 \leq p \leq 1$, and $0 \leq m^-/m \leq 1$.

\begin{table}[ht]
	\centering
	\caption{Usefulness of the speed-up techniques based on 100 Erd\H{o}s-R\'{e}nyi graphs $G(n,p)$}
	\label{3tab1.5}
	\begin{tabular}{llllllll}
		\hline
		\multirow{2}{*}{}          & \multicolumn{3}{l}{Average solve time (s)} &  & \multicolumn{3}{l}{Time improvement (\%)} \\ \cline{2-4} \cline{6-8} 
		& AND          & XOR          & ABS          &  & AND            & XOR            & ABS           \\ \hline
		Without speed-up           & 14.80        & 41.60        & 19.71        &  & -              & -              & -             \\
		With branching priority    & 5.90         & 4.91         & 5.50         &  & 60\%           & 88\%           & 72\%          \\
		With triangle inequalities & 9.21         & 31.72        & 17.26        &  & 38\%           & 24\%           & 12\%          \\
		With both speed-ups        & 4.93         & 4.08         & 4.42         &  & 67\%           & 90\%           & 78\%          \\ \hline
	\end{tabular}
\end{table}

\section{Computational performance} \label{3s:results}
In this section, various random instances are solved by our optimisation models using 64-bit Gurobi version 7.5.2 on a desktop computer with an Intel Core i5 7600 @ 3.50 GHz (released in 2017) and 8.00 GB of RAM running 64-bit Microsoft Windows 10. We use the \textit{NetworkX} package in Python for generating different types of random graphs. The models were created using Gurobi's Python environment in 64-bit Anaconda3 5.0.1 Jupyter.

\subsection{Comparison of the models' performance} \label{3ss:perform}
In this subsection, we discuss the time performance of Gurobi's algorithms for solving the extended binary linear models.

\begin{table}[ht]
	\centering
	\caption{Solve time comparison of the three models based on test cases of 10 Barab\'{a}si-Albert graphs}
	\label{3tab2}
	\begin{tabular}{llllllll}
		\hline
		\multirow{2}{*}{$n$} & \multirow{2}{*}{$m$} &  \multirow{2}{*}{$\rho$} & \multirow{2}{*}{$\frac{m^-}{m}$} & \multicolumn{1}{l}{\multirow{2}{*}{$Z^*$}} & \multicolumn{3}{c}{Solve time (s) mean  $\pm$  SD}                          \\ \cline{6-8} 
		&                        &                          &                          & \multicolumn{1}{r}{}                       & \multicolumn{1}{c}{AND \eqref{3eq3.5}} & \multicolumn{1}{c}{XOR \eqref{3eq4}} & \multicolumn{1}{c}{ABS \eqref{3eq5}} \\ \hline
		60                   & 539                  & 0.3                     & 0.3                      & 157.4                                      & 1.13  $\pm$ 0.48        & 1.59 $\pm$ 0.3          & 0.84 $\pm$ 0.1          \\
		&                      &                         & 0.5                      & 185.0                                      & 1.48 $\pm$ 0.56         & 2.95 $\pm$ 0.28         & 1.1 $\pm$ 0.19          \\
		&                      &                         & 0.7                      & 172.9                                      & 1.07 $\pm$ 0.41         & 2.55 $\pm$ 0.8          & 0.84 $\pm$ 0.16         \\
		&                      &                         & 1                        & 55.0                                       & 0.04 $\pm$ 0.01         & 0.04 $\pm$ 0.01         & 0.06 $\pm$ 0.02         \\
		& 884                  & 0.5                     & 0.3                      & 262.4                                      & 1.4 $\pm$ 0.16          & 0.45 $\pm$ 0.08         & 0.41 $\pm$ 0.04         \\
		&                      &                         & 0.5                      & 325.8                                      & 37.41 $\pm$ 11.53       & 27.09 $\pm$ 27.09       & 25.15 $\pm$ 8.46        \\
		&                      &                         & 0.7                      & 329.4                                      & 36.73 $\pm$ 8.28        & 39.8 $\pm$ 7.82         & 30.44 $\pm$ 5.73        \\
		&                      &                         & 1                        & 272.4                                      & 1 $\pm$ 0.17            & 0.77 $\pm$ 0.26         & 6.12 $\pm$ 4.61         \\
		70                   & 741                  & 0.3                     & 0.3                      & 217.0                                      & 4.07 $\pm$ 1.67         & 4.55 $\pm$ 0.77         & 1.52 $\pm$ 0.34         \\
		&                      &                         & 0.5                      & 260.6                                      & 4.56 $\pm$ 0.89         & 12.28 $\pm$ 1.72        & 2.84 $\pm$ 0.46         \\
		&                      &                         & 0.7                      & 248.0                                      & 2.94 $\pm$ 0.37         & 9.72 $\pm$ 2.32         & 1.87 $\pm$ 0.26         \\
		&                      &                         & 1                        & 78.0                                       & 0.07 $\pm$ 0            & 0.05 $\pm$ 0.01         & 0.1 $\pm$ 0.03          \\
		& 1209                 & 0.5                     & 0.3                      & 361.7                                      & 3.27 $\pm$ 0.34         & 0.76 $\pm$ 0.09         & 0.96 $\pm$ 0.1          \\
		&                      &                         & 0.5                      & 460.4                                      & 471.18 $\pm$ 77.27      & 322.99 $\pm$ 112.29     & 324.72 $\pm$ 131.86     \\
		&                      &                         & 0.7                      & 457.7                                      & 308.05 $\pm$ 130.31     & 369.14 $\pm$ 208.88     & 251.21 $\pm$ 96.75      \\
		&                      &                         & 1                        & 382.2                                      & 4.07 $\pm$ 1.08         & 2.93 $\pm$ 1.31         & 20.67 $\pm$ 14.28       \\ \hline
	\end{tabular}
\end{table}

In order to compare the performance of the three linear models, we consider 12 test cases each containing 10 Barab\'{a}si-Albert random graphs with various combinations of density and proportion of negative edges.
The results in Table~\ref{3tab2} show that the three models have similar performance in terms of solve time. Comparing values of the same column, it can be seen that graphs with a higher density (more edge variables) have a longer solve time. For graphs of a given order and density, we observe the shortest solve times for $m^-/m \in \{0.3, 1\}$ in most cases which are also associated with the two smallest averages of $Z^*$. 

\subsection{Convergence of the models with and without the speed-up techniques}

\begin{figure}
	\centering
	\subfloat[The AND model, ER graph]{\includegraphics[height=2.3in]{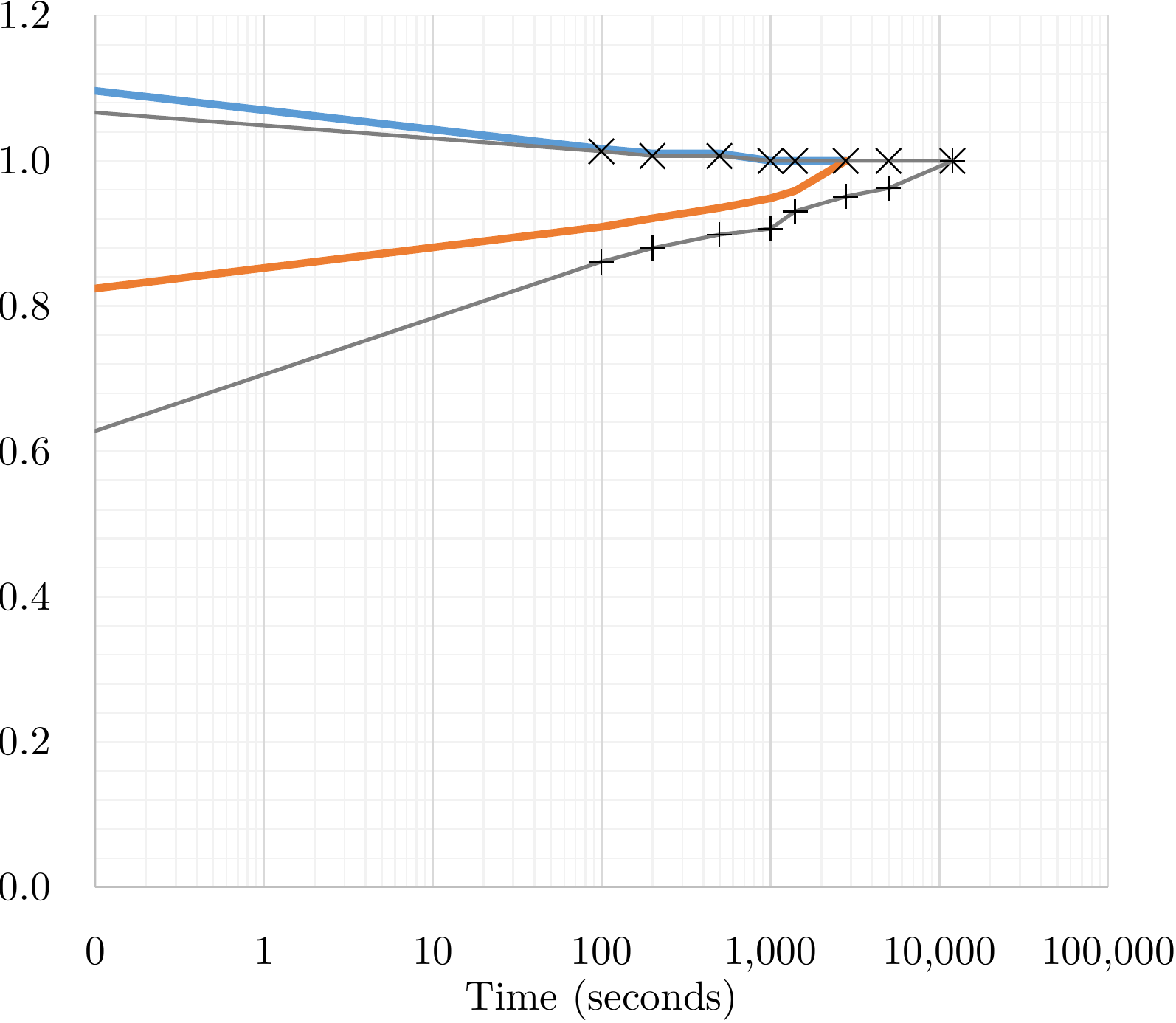}%
		\label{3fig2a}}
	%\hfil
	\subfloat[The AND model, BA graph]{\includegraphics[height=2.3in]{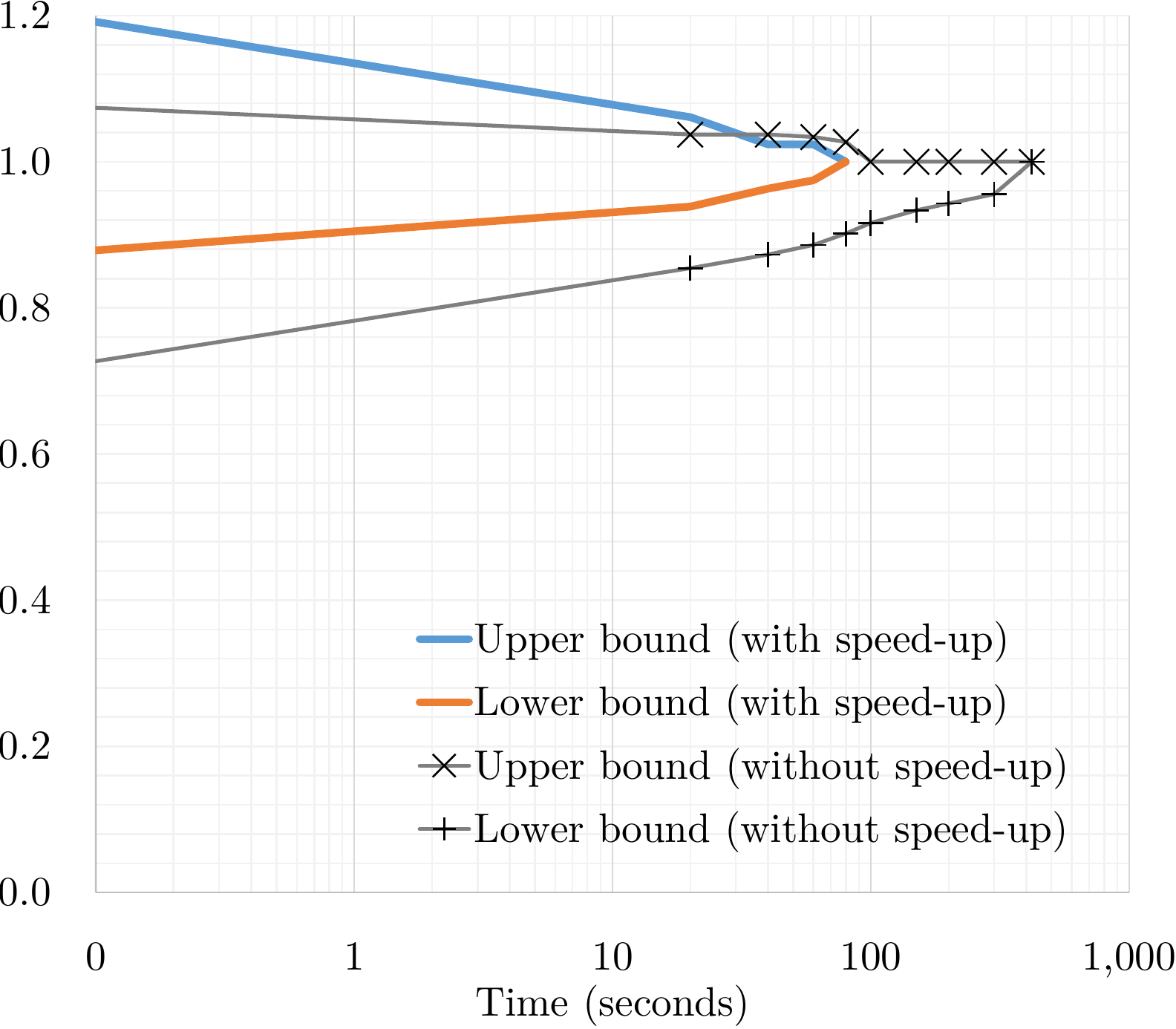}%
		\label{3fig2d}}
	\hfil
	\subfloat[The XOR model, ER graph]{\includegraphics[height=2.3in]{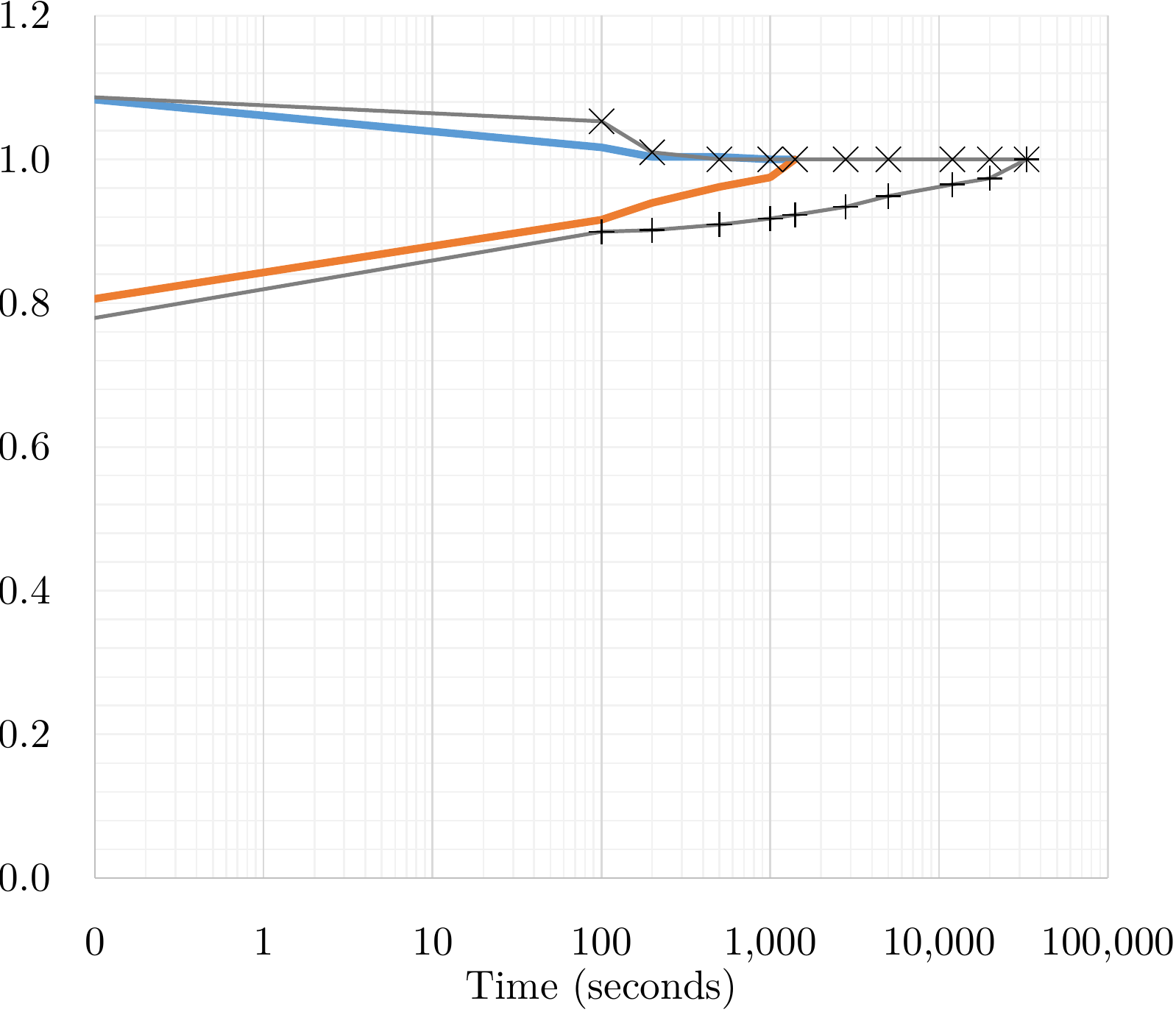}%
		\label{3fig2b}}				 
	%\hfil
	\subfloat[The XOR model, BA graph]{\includegraphics[height=2.3in]{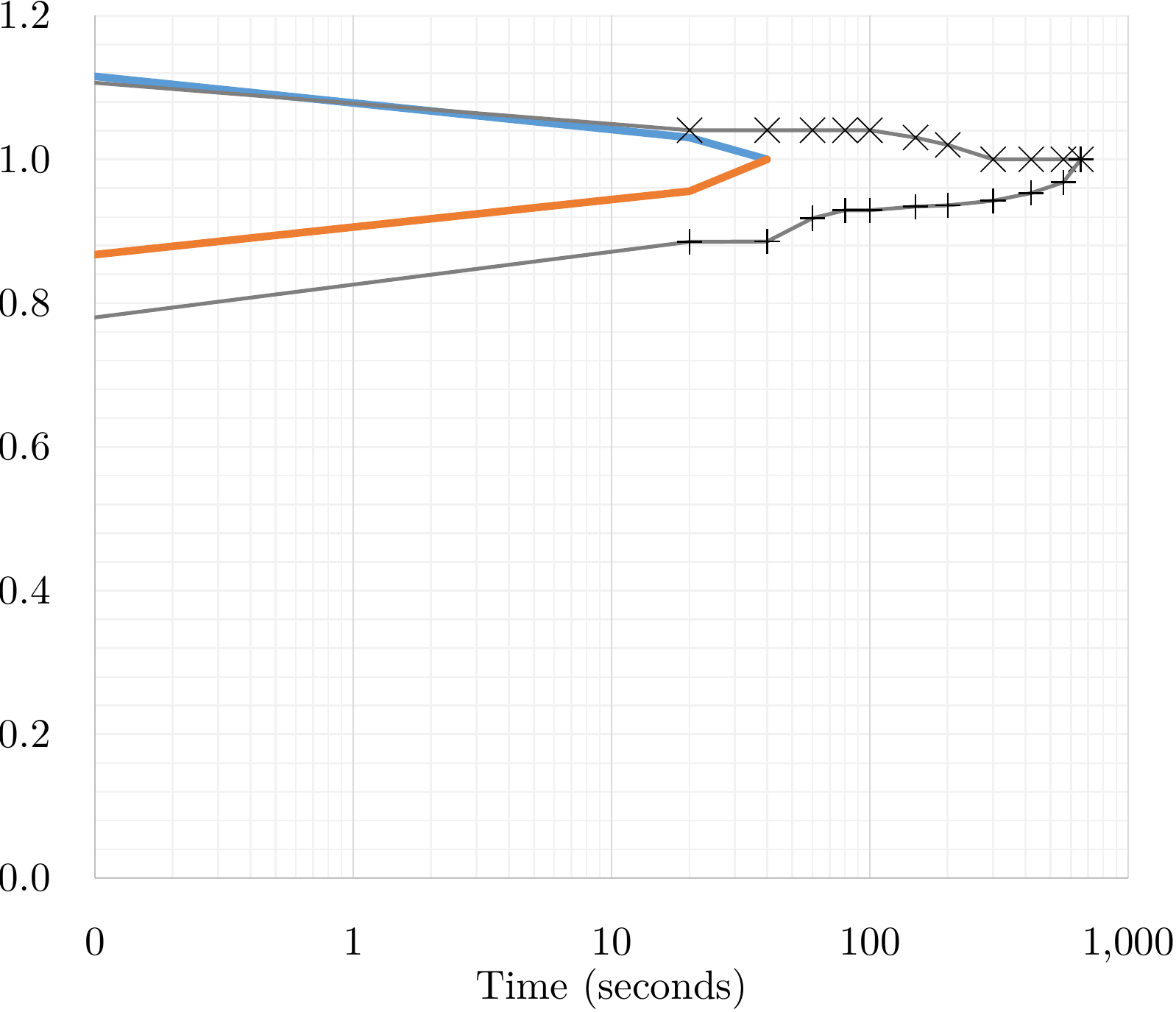}%
		\label{3fig2e}}
	\hfil
	\subfloat[The ABS model, ER graph]{\includegraphics[height=2.3in]{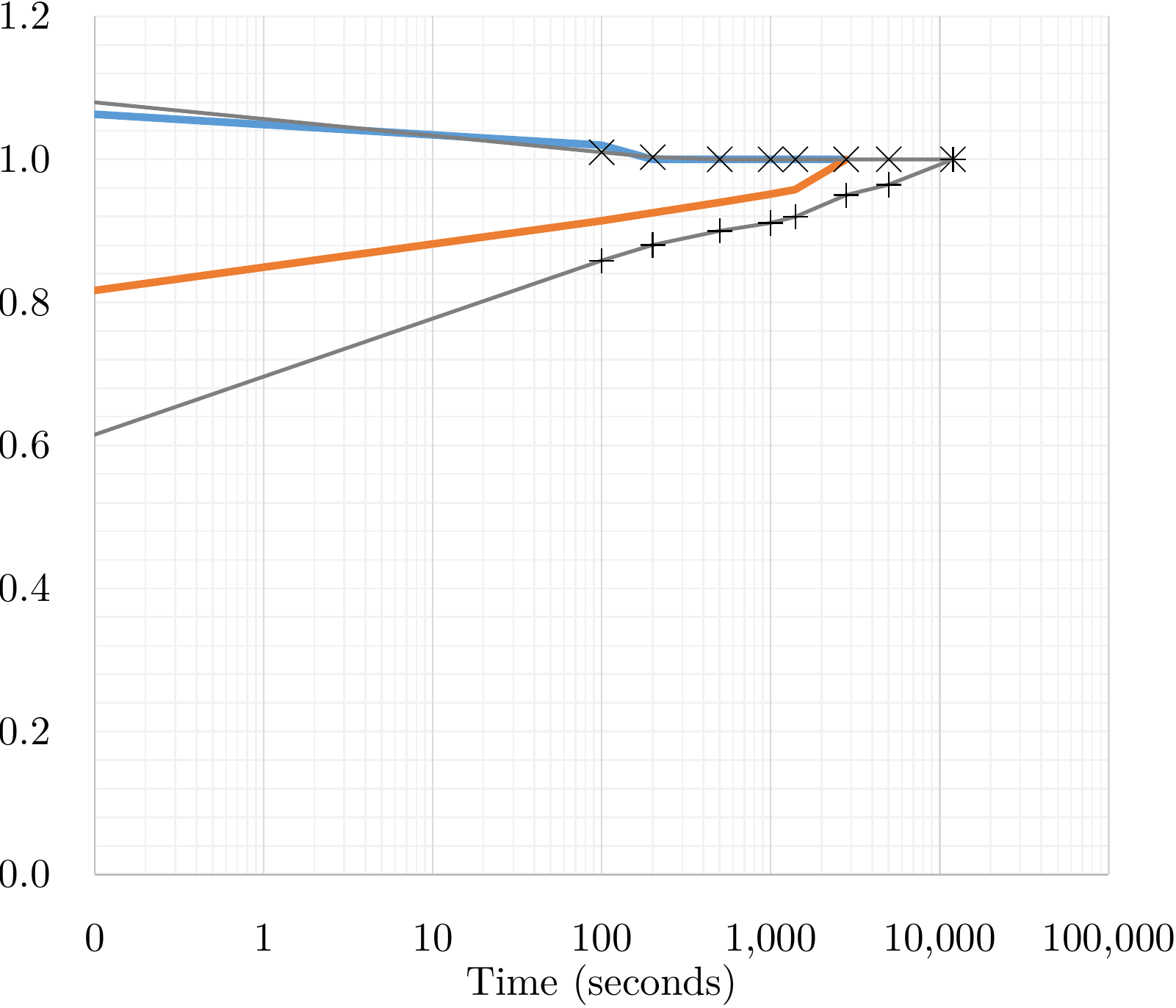}%
		\label{3fig2c}}
	%\hfil
		\subfloat[The ABS model, BA graph]{\includegraphics[height=2.3in]{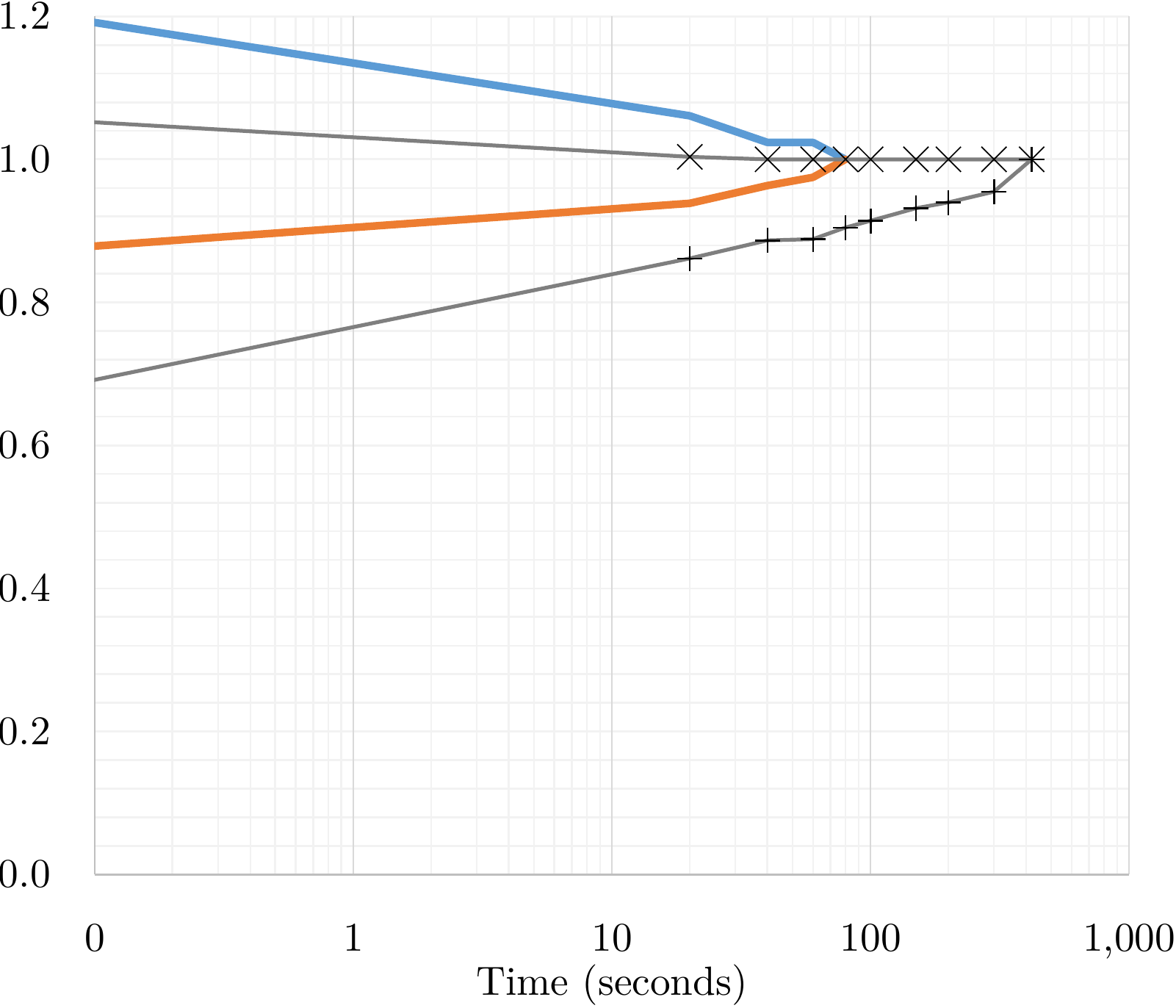}%
		\label{3fig2f}}
	%\hfil
	\caption{Solve time and normalised bounds with and without the speed-up techniques for random graphs with $n=100, m=900, m^-=600$ on a logarithmic scale (vertical axes show normalised upper and lower bounds.)}
	\label{3fig2}
\end{figure}

We investigate the algorithm convergence by running the three models with and without the speed-up techniques for an Erd\H{o}s-R\'{e}nyi (ER) graph, $G(n,M)$, and a Barab\'{a}si-Albert (BA) graph with $n=100, m=900, m^-=600$ and plotting the upper and lower bounds over time. Figure~\ref{3fig2} shows the upper and lower bounds on a log scale where the vertical axes represent upper and lower bounds normalised by dividing by the optimal solution.

For the randomly generated Erd\H{o}s-R\'{e}nyi graph, $G(n,M)$, in Subfigures \ref{3fig2a}, \ref{3fig2b}, and \ref{3fig2c}, the solve times of all three models without the speed-up techniques are over 12000 seconds (and in one case 33000 seconds). These solve times are reduced to less than 2800 seconds (and in one case 1400 seconds) when the speed-up techniques are implemented. 

Subfigures \ref{3fig2d}, \ref{3fig2e}, and \ref{3fig2f} show a considerable solve time improvement for the randomly generated Barab\'{a}si-Albert graph. It takes 420 seconds (80 seconds) for the AND model and the ABS model to find the optimal solution without (with) the speed up techniques. The XOR model without (with) the speed up techniques reaches optimality in 655 seconds (40 seconds).
%We provide more extensive comparison of the models equipped with the two speed-up techniques in the next subsection.

\subsection{Largest instances solvable in 10 hours}

It might be also interesting to know the size of the largest graph whose frustration index can be computed in a reasonable time using an extended binary linear model. Two important factors must be taken into consideration while answering this question: network properties and processing capacities.

As it is expected from our degree-based prioritised branching in \ref{3ss:branch}, network properties such as degree heterogeneity could have an impact on the solve time. %The results in \ref{3fig2} show Barab\'{a}si-Albert graphs have a shorter solve time compared to Erd\H{o}s-R\'{e}nyi graphs of the same size, order, and proportion of negative edges. 
Moreover, the numerical results in Chapter \ref{ch:2} suggest that reaching optimality in real signed networks takes a considerably shorter time compared to randomly generated signed networks of comparable size and order, confirming the observations of \cite{dasgupta_algorithmic_2007, huffner_separator-based_2010}.

Equally relevant to the size of the largest instance solvable in a given time, are the processing capacities of the computer that runs the optimisation models. Gurobi's algorithms make efficient use of multiple processing cores for exploring the feasible space in parallel\cite{gurobi}. Besides, exploring a large binary tree may require a considerable amount of memory which might be a determining factor in solve time of some instances due to memory limits.

Given a maximum solve time of 10 hours on the current hardware configuration (Intel Core i5 7600 @ 3.50 GHz and 8.00 GB of RAM), random instances with up to 2000 edges were observed to be solvable to global optimality. Regarding real signed graphs which have regularities favouring Gurobi's solver performance, graphs with up to 30000 edges are solvable (to global optimality) within 10 hours.

If we use more advanced processing capacities (32 Intel Xeon CPU E5-2698 v3 @ 2.30 GHz processors and 32 GB of RAM), real signed graphs with up to 100000 edges are solvable (to global optimality) within 10 hours (to be discussed in Chapter \ref{ch:4}). 

We have observed in most of our numerical experiments that the branch and bound algorithm finds the globally optimal solution in a fraction of the total solve time, but it takes more time and computations to ensure the optimality. To give an example, Subfigures \ref{3fig2a}, \ref{3fig2b}, and \ref{3fig2c} show that a considerable proportion of the solve time, ranging in 30\% - 90\%, is used for ensuring optimality after finding the globally optimal solution. One may consider using a non-zero mixed integer programming gap to find solutions within a guaranteed proximity of the optimal solution even if the instance has more than 100000 edges.

\section{Evaluating performance against the literature}\label{3s:evaluate}
In this section, we use both random and real networks to evaluate not only the solve time, but also the output of our models against other methods in the literature.

\subsection{Solve time in random graphs}
In this subsection, we compare the solve time of our algorithm against other algorithms suggested for computing the frustration index. Besides the models in Chapter \ref{ch:2}, our review of the literature finds only two methods claiming exact computation of the frustration index \cite{brusco_k-balance_2010, huffner_separator-based_2010}. Brusco and Steinley suggested a branch and bound algorithm for minimising the overall frustration (under a different name) for a predefined number of colours \cite{brusco_k-balance_2010}. H\"{u}ffner, Betzler, and Niedermeier have suggested a data-reduction schemes and an iterative compression algorithm for computing the frustration index \cite{huffner_separator-based_2010}.

Brusco and Steinley have reported running times for very small graphs with only up to $n=21$ vertices. While, their exact algorithm fails to solve graphs as large as $n=30$ in a reasonable time \cite{brusco_k-balance_2010}, our binary linear models solve such instances in split seconds.

H\"{u}ffner, Betzler, and Niedermeier have generated random graphs of order $n$ with low densities $(\rho \leq 0.04)$ %by specifying $n$, degree distribution, clustering coefficient, and the percentage of negative edges
to test their algorithm \cite{huffner_separator-based_2010}. The largest of such random graphs solvable by their algorithm in 20 hours has $500$ nodes. They also reported that only 3 out of 5 random graphs with $n \in\{100,200,300,400,500\}$ can be solved by their method in 20 hours. Our three binary linear models solve all such instances in less than 100 seconds.

\subsection{Solve time and algorithm output in real networks} \label{3s:real}
In this section we use signed network datasets from biology and international relations. The frustration index of biological networks has been a subject of interest to measure the network distance to \textit{monotonicity} \cite{dasgupta_algorithmic_2007,iacono_determining_2010}. In international relations, the frustration index is used to measure distance to balance for a network of countries \cite{patrick_doreian_structural_2015}. In this section, the frustration index is computed in real biological and international relations networks by solving the three binary linear models coupled with the two speed-up techniques \ref{3ss:branch} -- \ref{3ss:unbalanced}.

We use \textit{effective branching factor} as a performance measure. If the solver explores $b$ branch and bound nodes to find the optimal solution of a model with $v$ variables, the effective branching factor is $\sqrt[v]b$. The most effective branching is obtained when the solver only explores 1 branch and bound node to reach optimality. The effective branching factor for such a case would take value 1 which represents the strength of the mathematical formulation.
% using the Gurobi 7.0.2 Python interface and a desktop computer with an Intel Corei5 4670 @ 3.40 GHz and 8.00 GB of RAM running 64-bit Microsoft Windows 7.
\subsubsection{Biological datasets}
We use the four signed biological networks that were previously analysed by \cite{dasgupta_algorithmic_2007} and \cite{iacono_determining_2010}. The epidermal growth factor receptor (EGFR) pathway \cite{oda2005} is a signed network with 779 edges. The molecular interaction map of a macrophage (macro.) \cite{oda2004molecular} is another well studied signed network containing 1425 edges. We also investigate two gene regulatory networks, related to two organisms: a eukaryote, the \textit{yeast Saccharomyces cerevisiae} (yeast), \cite{Costanzo2001yeast} and a bacterium, \textit{Escherichia coli} (E.coli) \cite{salgado2006ecoli}. The yeast and E.coli networks have 1080 and 3215 edges respectively. The datasets for real networks used in this chapter are publicly available on the \urllink{https://figshare.com/articles/Signed_networks_from_sociology_and_political_science_biology_international_relations_finance_and_computational_chemistry/5700832}{Figshare} research data sharing website \cite{Aref2017data}. For more details on the four biological datasets, one may refer to \cite{iacono_determining_2010}.

We use root node objective, Number of B\&B nodes, effective branching factor, and solve time as performance measures. The performance of three binary linear models can be compared based on these measures in Table~\ref{3tab4}. Values for the running the models without the speed-ups are provided between parentheses.

\begin{table}[ht]
	\centering
	\caption{Performance measures for the three binary linear models with (and without) the speed-up techniques}
	\label{3tab4}
	\begin{tabular}{llllll}
		\hline
		\multicolumn{2}{l}{\begin{tabular}[c]{@{}l@{}}Graph\\ $n, m$\end{tabular}} & \begin{tabular}[c]{@{}l@{}}EGFR\\ 329, 779\end{tabular} & \begin{tabular}[c]{@{}l@{}}Macro.\\ 678, 1425\end{tabular} & \begin{tabular}[c]{@{}l@{}}Yeast\\ 690, 1080\end{tabular} & \begin{tabular}[c]{@{}l@{}}E.coli\\ 1461, 3215\end{tabular} \\ \hline
		\multirow{6}{*}{\begin{tabular}[c]{@{}l@{}}Root\\ node\\ objective\end{tabular}}        & \multirow{2}{*}{AND} & 28.5            & 67                 & 11.5              & 130.5               \\
		&                      & (13)            & (53)               & (0)               & (4)                 \\
		& \multirow{2}{*}{XOR} & 28.5            & 67                 & 11.5              & 130.5               \\
		&                      & (13)            & (53)               & (0)               & (4)                 \\
		& \multirow{2}{*}{ABS} & 28.5            & 67                 & 11.5              & 130.5               \\
		&                      & (13)            & (53)               & (0)               & (4)                 \\ \hline
		\multirow{6}{*}{\begin{tabular}[c]{@{}l@{}}Number\\ of B\&B\\ nodes\end{tabular}}       & \multirow{2}{*}{AND} & 3               & 1                  & 1                 & 31                  \\
		&                      & (91)            & (199)              & (7)               & (279)               \\
		& \multirow{2}{*}{XOR} & 1               & 1                  & 1                 & 3                   \\
		&                      & (25)            & (1)                & (1)               & (19)                \\
		& \multirow{2}{*}{ABS} & 1               & 1                  & 3                 & 36                  \\
		&                      & (47)            & (456)              & (7)               & (357)               \\ \hline
		\multirow{6}{*}{\begin{tabular}[c]{@{}l@{}}Effective\\ branching\\ factor\end{tabular}} & \multirow{2}{*}{AND} & 1.0010          & 1                  & 1                 & 1.0007              \\
		&                      & (1.0041)        & (1.0025)           & (1.0011)          & (1.0012)            \\
		& \multirow{2}{*}{XOR} & 1               & 1                  & 1                 & 1.0002              \\
		&                      & (1.0029)        & (1)                & (1)               & (1.0006)            \\
		& \multirow{2}{*}{ABS} & 1               & 1                  & 1.0004            & 1.0006              \\
		&                      & (1.0027)        & (1.0022)           & (1.0008)          & (1.0010)            \\ \hline
	\end{tabular}
\end{table}

DasGupta et al.\ have suggested approximation algorithms \cite{dasgupta_algorithmic_2007} that are later tested on the four biological networks in \cite{huffner_separator-based_2010}. Their approximation method provides $196 \leq L(G)_{\text{EGFR}} \leq 219$ which our exact model proves to be incorrect. The bounds obtained by implementing their approximation are not incorrect for the other three networks, but they have very large gaps between lower and upper bounds.

H\"{u}ffner, Betzler, and Niedermeier have previously investigated frustration in the four biological networks suggesting a data-reduction schemes and (an attempt at) an exact algorithm \cite{huffner_separator-based_2010}. Their suggested data-reduction schemes can take more than 5 hours for yeast, more than 15 hours for EGFR, and more than 1 day for macrophage if the parameters are not perfectly tuned. Besides the solve time issue, their algorithm provides $L(G)_{\text{EGFR}}=210, L(G)_{\text{macrophage}}=374$, both of which are incorrect. They report their algorithm failed to terminate for E.coli \cite{huffner_separator-based_2010}.

Iacono et al.\ have also investigated frustration in the four networks \cite{iacono_determining_2010}. Their heuristic algorithm provides upper and lower bounds for EGFR, macrophage, yeast, and E.coli with 96.37\%, 90.96\%, 100\%, and 98.38\% ratio of lower to upper bound respectively. The comparison of our outputs against those reported in the literature is provided in Table~\ref{3tab5}.

Iacono et al.\ also suggest an upper bound for the frustration index \cite[page 227]{iacono_determining_2010}. However, some values of the frustration index in complete graphs show that the upper bound is incorrect (as discussed in Subsection \ref{2ss:bounds}). For a more detailed discussion on bounds for the frustration index, one may refer to Subsection \ref{2ss:bounds} and \cite{martin2017frustration}.

\begin{table}[ht]
	\centering
	\caption{Our algorithm output against the best results reported in the literature}
	\label{3tab5}
	\begin{tabular}{p{1.6cm}p{1.7cm}p{1.4cm}p{1.6cm}p{1cm}p{0.7cm}p{0.7cm}p{0.7cm}}
		\hline
		
		Author Reference     & DasGupta \quad et al.\ \cite{dasgupta_algorithmic_2007} & H\"{u}ffner \quad et al.\ \cite{huffner_separator-based_2010} & Iacono \quad et al.\ \cite{iacono_determining_2010} & Eq.\ \eqref{2eq8} &AND \eqref{3eq3.5} & XOR \eqref{3eq4} &ABS \eqref{3eq5}    \\ \hline
		
		EGFR & {[}196, 219{]}*                                                           & 210*                                                          & {[}186, 193{]}                                                     & 193  & 193 & 193& 193   \\
		Macro. & {[}218,383{]}                                                            & 374*                                                          & {[}302, 332{]}                                                     & 332  & 332  & 332  & 332    \\
		Yeast & {[}0, 43{]}                                                              & 41                                                           & 41                                                                & 41   & 41 & 41 & 41    \\
		E.coli & {[}0, 385{]}                                                             & $\dagger$                                                & {[}365, 371{]}                                                   & 371    & 371  & 371 & 371   \\ \hline
		\multicolumn{5}{l}{* incorrect results}&&&\\  \multicolumn{5}{l}{$\dagger$ the algorithm does not converge} &&&\\ 
	\end{tabular}
\end{table}
\FloatBarrier

We also compare our solve times to the best results reported for heuristics and approximation algorithms in the literature. 

H\"{u}ffner et al.\ have provided solve time results for their suggested algorithm \cite{huffner_separator-based_2010} (if parameters are perfectly tuned for each instance) as well as the algorithm suggested by DasGupta et al.\ \cite{dasgupta_algorithmic_2007}. Iacono et al.\ have only mentioned that their heuristic requires a fairly limited amount of time (a few minutes on an ordinary PC \cite{iacono_determining_2010}) that we conservatively interpret as 60 seconds. Table~\ref{3tab5.5} sums up the solve time comparison of our suggested models against the literature in which the values for running our models without the speed up techniques are provided between parentheses. As the hardware configuration is not reported in \cite{dasgupta_algorithmic_2007, iacono_determining_2010}, we conservatively evaluate the order-of-magnitude improvements in solve time with respect to the differences in computing power in different years.

\begin{table}[ht]
	\centering
	\caption{Algorithm solve time in seconds with (and without) the speed ups against the results reported in the literature}
	\label{3tab5.5}
	\begin{tabular}{llllllll}
		\hline
		Year &2010&2010&2010&2018&2018&2018&2018\\ 
		Reference     & \cite{dasgupta_algorithmic_2007} & \cite{huffner_separator-based_2010} &  \cite{iacono_determining_2010} & Eq.\ \eqref{2eq8} &AND \eqref{3eq3.5} & XOR \eqref{3eq4} &ABS \eqref{3eq5}    \\ \hline

		EGFR & 420                                                                   & 6480                                                      & \textgreater60                                             & 0.68 & 0.27 (0.82) & 0.21 (0.67)& 0.23 (0.66)\\
		Macro. & 2640                                                                  & 60                                                        & \textgreater60                                                & 1.85 & 0.34 (1.24) & 0.26 (1.37) & 0.49 (1.30) \\
		Yeast & 4620                                                                  & 60                                                        & \textgreater60                                                & 0.33 & 0.18 (0.45) & 0.11 (0.28) & 0.15 (0.39) \\
		E.coli & $\ddagger$                                                           & $\dagger$                                                & \textgreater60                                               &  18.14&  0.99 (1.91) & 1.97 (4.73)  & 0.74 (1.86) \\ \hline   \multicolumn{5}{l}{$\dagger$ the algorithm does not converge} &&&\\  \multicolumn{5}{l}{$\ddagger$ not reported}
	\end{tabular}
\end{table}

According to Moore's law \cite{moore1965}, the exponential increase in transistor density on integrated circuits leads to computer power doubling almost every two years. Moore's prediction has been remarkably accurate from 1965 to 2013, while the actual rate of increase in %the maximum number of transistors to fit on an integrated circuit chip
computer power has slowed down since 2013 \cite{mack2015multiple}.

Moore's law ballpark figures allow us to compare computations executed on different hardware in different years. We conservatively estimate a factor of 16 times for the improvements in computer power between 2010 and 2018 to be attributable to hardware improvements.
%Based on Moore's law, we expect 16 times faster computation in 2018 compared to 2010. 
The solve times of the slowest (fastest) model among AND, XOR, and, ABS in Table~\ref{3tab5.5} shows a factor of improvement ranging between $30 - 333$ ($81 - 545$) compared to the fastest solve time in 2010 \cite{dasgupta_algorithmic_2007, huffner_separator-based_2010, iacono_determining_2010}. This shows our solve time improvements are not merely resulted from hardware differences.

While data-reduction schemes \cite{huffner_separator-based_2010} can take up to 1 day for these datasets and heuristic algorithms \cite{iacono_determining_2010} only provide bounds with up to 9\% gap from optimality, our three binary linear models equipped with the speed-up techniques (\ref{3ss:branch} -- \ref{3ss:unbalanced}) solve the four instances to optimality in a few seconds. 

\subsubsection{International relations datasets}

We also compute the frustration index for two datasets of international relations networks. In international relation networks, countries and their relations are represented by nodes and edges of signed graphs. We use the Correlates of War (CoW) \cite{correlatesofwar2004} dataset which has 51 instances of networks with up to 1247 edges \cite{patrick_doreian_structural_2015} and the United Nations General Assembly (UNGA) \cite{macon2012community} dataset which has 62 instances with up to 15531 edges when converted into signed networks as discussed in \cite{figueiredo2014maximum}. %Figueiredo and Frota provide detailed explanation on the process of creating signed networks from the UNGA data \cite{figueiredo2014maximum}. %The Correlates of War dataset contains 51 time windows of a temporal network representing signed international relations among countries starting with 1946-1949 time window and ending with 1996-1999 time window \cite{correlatesofwar2004}. Both size and order change in each time window.
%The first time window of the temporal network has 362 edges while the last time window contains 1247 edges.

The CoW signed network dataset is constructed by Doreian and Mrvar \cite{patrick_doreian_structural_2015} based on international relations data for the 1946-1999 period. In their analysis, some numerical results provided on the CoW dataset are referred to as line index \cite{patrick_doreian_structural_2015}. However, the values of $L(G)$ we have obtained using our optimisation models prove that values reported in \cite{patrick_doreian_structural_2015} for the 51 time frames of the network are never the smallest number of edges whose removal results in balance. Doreian and Mrvar have not reported any solve time, but have suggested that determining their line index is in general a polynomial-time hard problem \cite{patrick_doreian_structural_2015}. The solve times of our models for each instance of the CoW dataset is $\leq 0.1$ seconds.

\begin{figure}[ht]
	\includegraphics[height=1.78in]{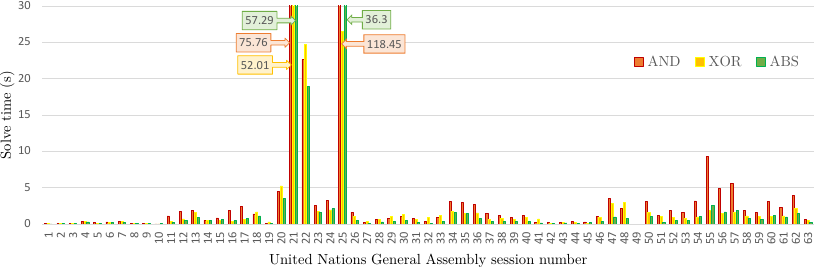}%
	\caption{Solve times of AND, XOR, and ABS models tested on the UNGA instances}
	\label{3fig2.5}
\end{figure}

We also tested our three models on the UNGA instances. The UNGA dataset is based on voting on the UN resolutions. In this dataset, instances refer to annual UNGA sessions between 1946 and 2008. Figure~\ref{3fig2.5} shows the solve times instances of this dataset. 

Figure~\ref{3fig2.5} shows that most UNGA instances can be solved in less that 5 seconds using any of the three models. The XOR and the ABS models solve all UNGA instances in less than a minute, while solving the AND model for instance 21 and instance 25 takes about 75 and 118 seconds respectively. These two harder instances have the highest values of the frustration index ($L(G)=616$ and $L(G)=611$ respectively) in the UNGA dataset. 

\section{Related works}\label{3s:related}
Despite the lack of exact computational methods for the frustration index, a closely related and more general problem in signed graphs has been investigated comprehensively. According to Davis's definition of \textit{generalised balance}, a signed network is \textit{weakly balanced} ($k$-balanced) if and only if its vertex set can be partitioned into $k$ subsets such that each negative edge joins vertices belonging to different subsets \cite{Davis}. The problem of finding the minimum number of frustrated edges for general $k$ (an arbitrary number of subsets) is referred to as the \textit{Correlation Clustering} problem \cite{bansal2004correlation}.

For every fixed $k$, there is a polynomial-time approximation scheme for the correlation clustering problem \cite{Giotis}. For arbitrary $k$, exact \cite{brusco_k-balance_2010, figueiredo2013mixed} and heuristic methods \cite{drummond2013efficient,levorato2015ils,levorato2017evaluating} are developed based on a mixed integer programming model \cite{demaine2006correlation}. Denoting the order of a graph by $n$, exact algorithms fail for $n>21$ \cite{brusco_k-balance_2010} and $n>40$ \cite{figueiredo2013mixed}, while greedy algorithms \cite{drummond2013efficient} and local search heuristics \cite{levorato2015ils} are capable of providing good solutions for $n \approx 10^3$ and $n \approx 10^4$ respectively.

After extending the non-linear energy minimisation model suggested by Facchetti et al.\ \cite{facchetti_computing_2011} to generalised balance, Ma et al.\ has provided good solutions for the correlation clustering problem in networks with $n \approx 10^5$ using various heuristics \cite{ma_memetic_2015, ma_decomposition-based_2017}. Esmailian et al.\ have also extended the work of Facchetti et al.\ \cite{facchetti_computing_2011} focusing on the role of negative ties in signed graph clustering \cite{esmailian_mesoscopic_2014, esmailian2015community}.

\section{Extensions to the models} \label{3s:future}
In this section we formulate two extensions to the 2-colour minimum frustration count optimisation problem.

\subsection{Weighted minimum frustration count optimisation problem}
We extend the 2-colour minimum frustration count optimisation problem for a graph with weights $w_{ij} \in [-1,1]$ instead of the signs $a_{ij} \in \{-1,1\}$ on the edges. We call such a graph a \textit{weighted signed graph}.

Taking insights from \eqref{3eq6}, the frustration of edge $(i,j) \in E$ with weight $w_{ij}$ can be represented by $f_{ij}={(1-w_{ij})}/{2} + w_{ij}(x_i +x_j - 2x_{ij}) $ using the binary variables $x_i , x_j, x_{ij}$ of the AND model \eqref{3eq3.5}. Note that, the frustration of an edge in a weighted signed graph is a continuous variable in the unit interval $f_{ij} \in [0,1]$.

Note that, $a_{ij}x_{ij} \leq (3a_{ij}-1)(x_i + x_j)/4 + {(1-a_{ij})}/{2}$ embodies all constraints for edge $(i,j)$ in the AND model regardless of the edge sign. Similarly, the constraints of the AND model can be represented using weights $w_{ij}$.  The weighted minimum frustration count optimisation problem can be formulated as a binary linear programming model in Eq.\ \eqref{3eq9}.
\begin{equation}\label{3eq9}
\begin{split}
\min_{x_i: i \in V, x_{ij}: (i,j) \in E} Z &= \sum\limits_{(i,j) \in E} {(1-w_{ij})}/{2} + w_{ij}(x_i +x_j - 2x_{ij})\\
\text{s.t.} \quad
w_{ij}x_{ij} &\leq (3w_{ij}-1)(x_i + x_j)/4 + {(1-w_{ij})}/{2} \quad \forall (i,j) \in E\\
x_{i} &\in \{0,1\} \quad  \forall i \in V \\
x_{ij} &\in \{0,1\} \quad \forall (i,j) \in E 
\end{split}
\end{equation}

We have generated random weighted signed graphs to test the model in Eq.\ \eqref{3eq9}. Our preliminary results show that the weighted version of the problem \eqref{3eq9} is solved faster than the original models for signed graphs.

\subsection{Multi-colour minimum frustration count optimisation problem}\label{3ss:multi}

\begin{figure}[ht]
	\subfloat[An example graph with $n=4,$ $m^-=4,$ $ m^+=1$]{\includegraphics[height=1in]{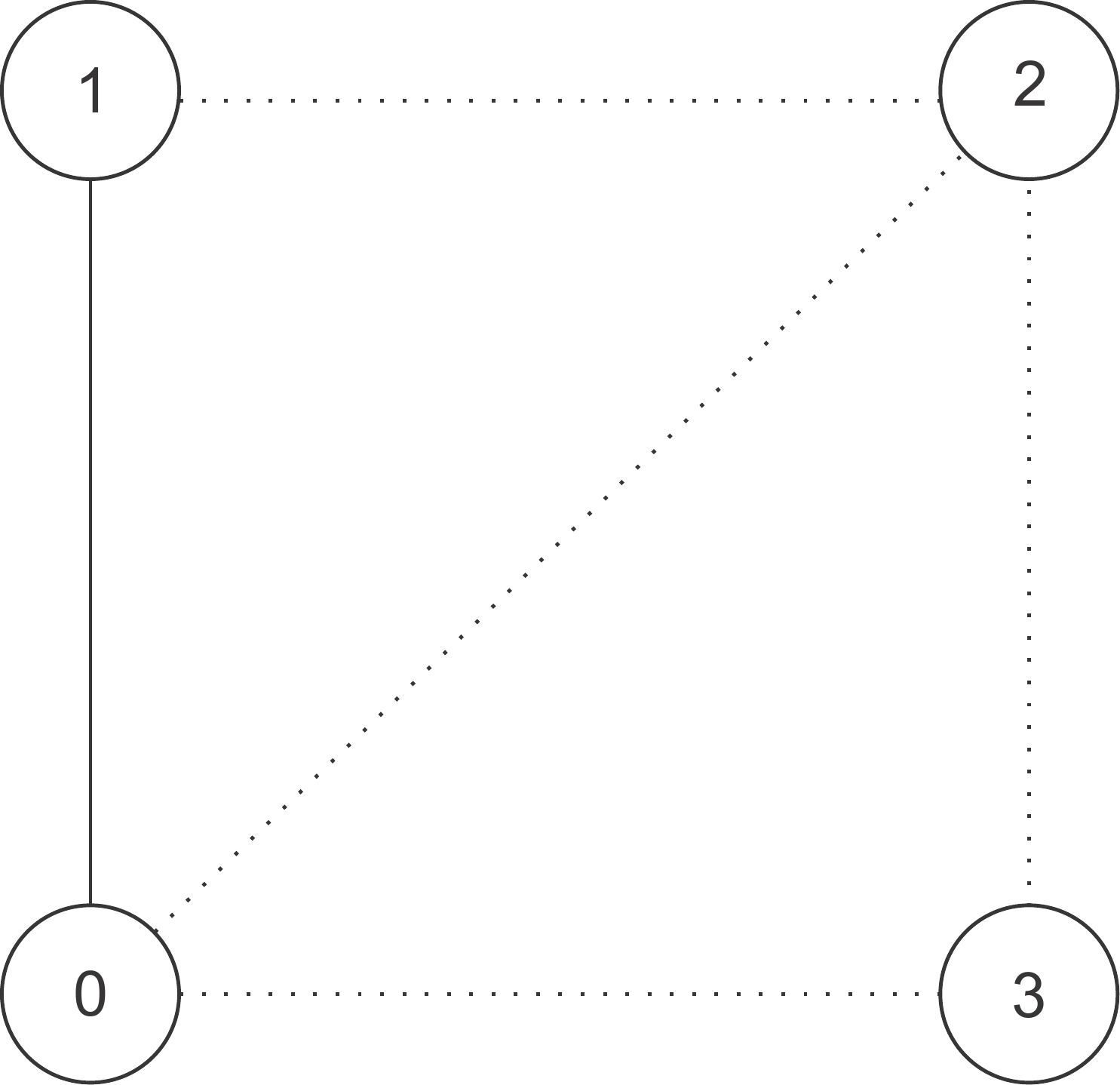}%
		\label{3fig3a}} 
	\hfil
	\subfloat[One colour resulting in four frustrated edges]{\includegraphics[height=1in]{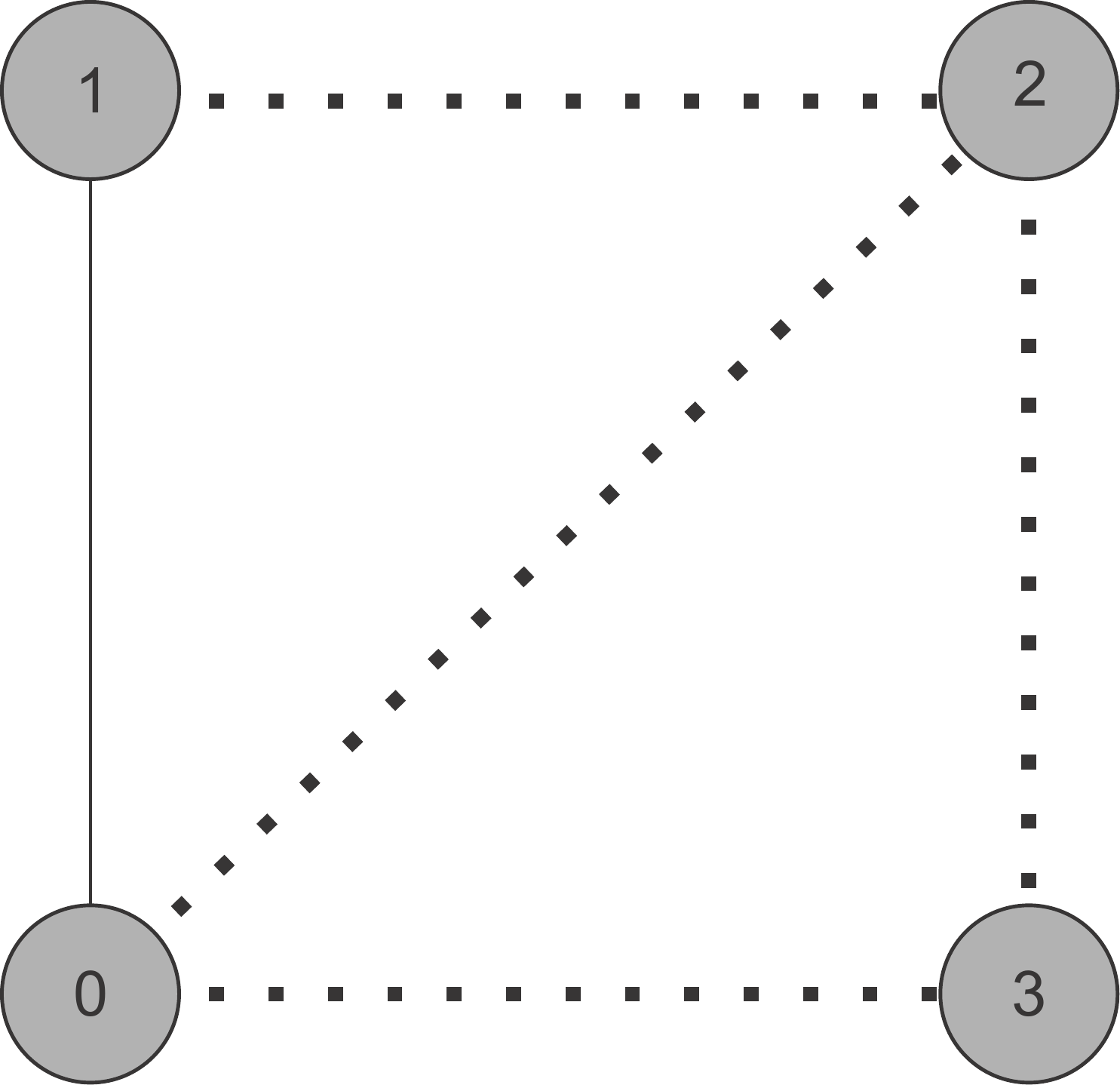}%
		\label{3fig3b}}
	\hfil
	\subfloat[Two colours resulting in one frustrated edge]{\includegraphics[height=1in]{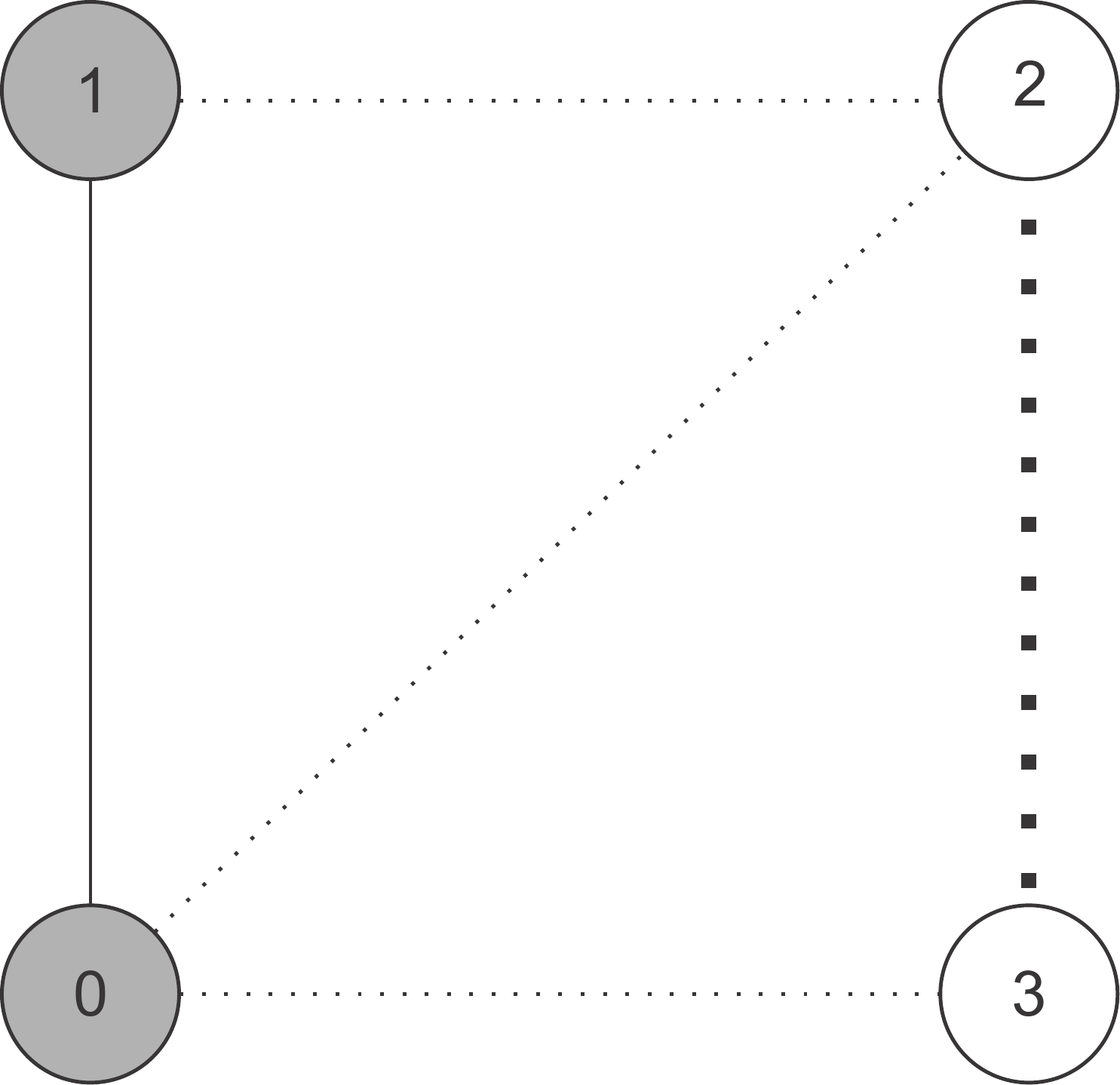}%
		\label{3fig3c}}
	\hfil
	\subfloat[Three colours resulting in no frustrated edge]{\includegraphics[height=1in]{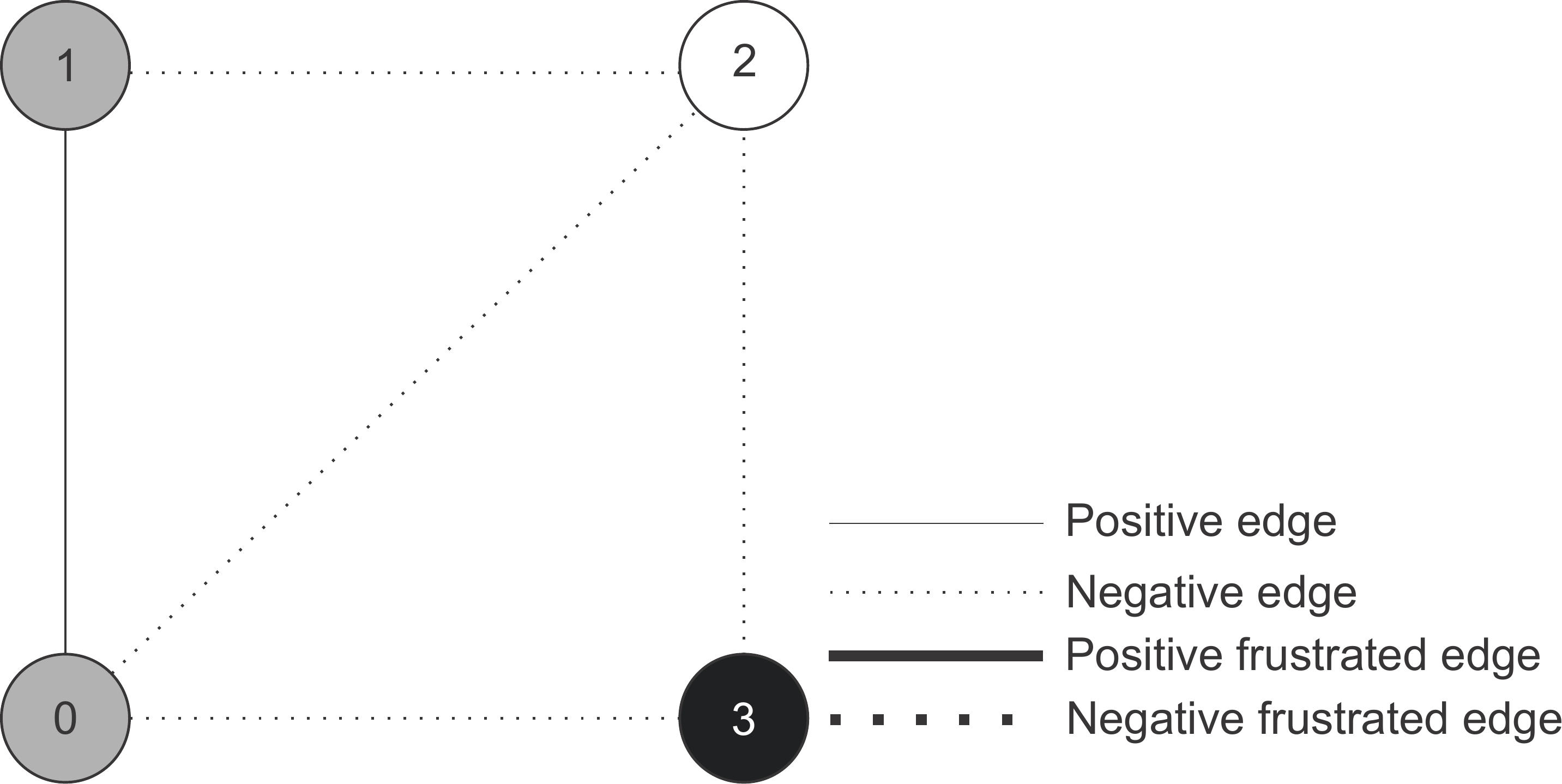}%
		\label{3fig3d}}
	\caption{The frustrated edges represented by dashed lines for the multi-colour minimum frustration count optimisation problem.}% The figure is produced using Adobe Illustrator.}
	\label{3fig3}
\end{figure}

We formulate another extension to the 2-colour minimum frustration count optimisation problem by allowing more than 2 colours to be used. As previously mentioned in Section~\ref{3s:related}, a signed network is $k$-balanced if and only if its vertex set can be partitioned into $k$ subsets (for some fixed $k\geq 2$) such that each negative edge joins vertices belonging to different subsets \cite{Davis}. Figure~\ref{3fig3} demonstrates an example graph and the frustrated edges for various numbers of colours. Subfigure~\ref{3fig3d} shows that the graph is weakly balanced. 

The harder problem of finding the minimum number of frustrated edges where $k$ is not specified in advance (an arbitrary number of node colours) is referred to as the Correlation Clustering problem. As mentioned in Section~\ref{3s:related}, another integer linear programming formulation for the correlation clustering problem is suggested by \cite{demaine2006correlation} which is widely used in the literature \cite{figueiredo2013mixed, drummond2013efficient, levorato2015ils}.

In the multi-colour minimum frustration count optimisation problem, each node may be given one of a set of colours $C=\{1, 2, 3,..., k:=|C|\}$. Assume $c_i \in C$ is the colour of node $i$. We consider that a positive edge $(i,j) \in E^+$ is frustrated (indicated by $f_{ij}=1$) if its endpoints $i$ and $j$ are coloured differently, i.e.,\ $c_i \ne c_j$; otherwise it is not frustrated (indicated by $f_{ij}=0$). A negative edge $(i,j) \in E^-$ is frustrated (indicated by $f_{ij}=1$) if $c_i = c_j$; otherwise it is not frustrated (indicated by $f_{ij}=0$). %This gives rise to the optimisation problem in Eq.\ \eqref{3eq10}.
%\begin{equation}\label{3eq10}
%\min_{(c_1, c_2, ..., c_{|V|})  \in C^{|V|}} \sum_{(i,j) \in E}  f_{ij}
%\end{equation}
Using binary variables $x_{ic}=1$ if node $i \in V$ has colour $c\in C$ (and  $x_{ic}=0$ otherwise), we formulate this as the following integer programming model in Eq.\ \eqref{3eq11}.
\begin{equation} \label{3eq11}
\begin{split}
\min \sum_{(i,j) \in E}   f_{ij}  \\
\text{s.t.} \quad \sum_{c \in C} x_{ic} &= 1 \quad \forall i \in V \\
f_{ij}  &\ge  x_{ic} - x_{jc} \quad \forall (i,j) \in E^+,  ~\forall c \in C \\
f_{ij}  &\ge  x_{ic} + x_{jc} -1 \quad \forall (i,j) \in E^-,  ~\forall c \in C \\
x_{ic} &\in \{0,1\} \quad  \forall i \in V,  ~\forall c \in C \\
f_{ij} &\in \{0,1\} \quad \forall (i,j) \in E 
\end{split}
\end{equation}

If we have just two colours, then we use  $x_{i} \in \{0,1\}$ to denote the colour of node $i$. This gives the XOR model expressed in Eq.\ \eqref{3eq4}. Solving the problem in Eq.\ \eqref{3eq11} provides us with the minimum number of frustrated edges in the $k$-colour setting. This number determines how many edges should be removed to make the network $k$-balanced. For a more general formulation of partitioning graph vertices into $k$ sets, one may refer to \cite{ales2016polyhedral} where numerical results for graphs with up to $20$ nodes are provided. 

\section{Conclusion of Chapter \ref{ch:3}} \label{3s:conclu}

In this chapter, we provided an efficient method for computing a standard measure in signed graphs which has many applications in different disciplines. This chapter suggested efficient mathematical programming models and speed-up techniques for computing the frustration index in graphs with up to 15000 edges on inexpensive hardware.

We developed three new binary optimisation models which outperform previous methods by large factors. We also suggested prioritised branching and valid inequalities which make the binary linear optimisation models several times (see Table \ref{3tab5.5}) faster than the models in Chapter \ref{ch:2} and capable of processing larger instances.

Extensive numerical results on random and real networks were provided to evaluate computational performance and underline the superiority of our models in both solve time and algorithm output. We also provided two extensions to the model for future investigation.

\cleardoublepage

\chapter{Balance and Frustration in Signed Networks}\label{ch:4}

\maketitle

%\cleardoublepage
\section*{Abstract}
	{The frustration index is a key measure for analysing signed networks, which has been underused due to its computational complexity. We use an exact optimisation-based method to analyse frustration as a global structural property of signed networks coming from diverse application areas. In the classic friend-enemy interpretation of balance theory, a by-product of computing the frustration index is the partitioning of nodes into two internally solidary but mutually hostile groups.
	The main purpose of this chapter is to present general methodology for answering questions related to partial balance in signed networks, and apply it to a range of representative examples that are now analysable because of advances in computational methods.
	We provide exact numerical results on social and biological signed networks, networks of formal alliances and antagonisms between countries, and financial portfolio networks. Molecular graphs of carbon and Ising models are also considered. We point out several mistakes in the signed networks literature caused by inaccurate computation, implementation errors or inappropriate measures.}
	
%	\textbf{Keywords:} 
%	{Frustration index,
%	Line index of balance,
%	Signed graph,
%	Integer programming,
%	Optimisation,
%	Balance theory}

%Mathematics Subject Classification (MSC 2010): 05C22, 90C35, 90C90, 90C09, 90C10

\cleardoublepage
\section{Introduction to Chapter \ref{ch:4}} \label{4s:intro}

The theory of structural balance introduced by Heider \cite{heider_social_1944} is an essential tool in the context of social relations for understanding the impact of local interactions on the global structure of signed networks. Following Heider, Cartwright and Harary identified cycles containing an odd number of negative edges \cite{cartwright_structural_1956} as a source of tension that may influence the structure of signed networks in particular ways. Signed networks in which no such cycles are present satisfy the property of structural balance, which is considered as a state with minimum tension \cite{cartwright_structural_1956}. For graphs that are not balanced, a distance from balance (a measure of partial balance) can be computed (as discussed in Chapter \ref{ch:1}). 

Among various measures \cite{cartwright_structural_1956,norman_derivation_1972,terzi_spectral_2011,facchetti_computing_2011,estrada_walk-based_2014} is the \textit{frustration index} that indicates the minimum number of edges whose removal (or equivalently, negation) results in balance \cite{abelson_symbolic_1958,harary_measurement_1959,zaslavsky_balanced_1987}. The clear definition of the frustration index allows for an intuitive interpretation of its values as the minimum number of edges that keep the network away from a state of total balance (an edge-based distance from balance). In this chapter, we focus on applications of the frustration index, also known as \textit{the line index of balance} \cite{harary_measurement_1959}, in different contexts beyond the structural balance of signed social networks. 

Satisfying essential axiomatic properties as a measure of partial balance (as discussed in Chapter \ref{ch:1}), the frustration index is a key to frequently stated problems in many different fields of research \cite{iacono_determining_2010,kasteleyn_dimer_1963,patrick_doreian_structural_2015,doslic_computing_2007}. In biological networks, optimal decomposition of a network into monotone subsystems %-- which is essential for understanding Drosophila segment polarity -- 
is made possible by calculating the frustration index of the underlying signed graph \cite{iacono_determining_2010}. In physics, the frustration index provides the ground state of atomic magnet models \cite{kasteleyn_dimer_1963,hartmann2011ground}. In international relations, the dynamics of alliances and enmities between countries can be investigated using the frustration index \cite{patrick_doreian_structural_2015}. Frustration index can also be used as an indicator of network bi-polarisation in practical examples involving financial portfolios. For instance, some low-risk portfolios are shown to have an underlying balanced signed graph containing negative edges \cite{harary_signed_2002}. In chemistry, bipartite edge frustration can be used as a stability indicator of carbon allotropes known as fullerenes \cite{doslic2005bipartivity, doslic_computing_2007}.

\section{Computing the frustration index} \label{4s:computing}

From a computational viewpoint, computing the frustration index of a signed graph is an NP-hard problem equivalent to the ground state calculation of an Ising model without special structure \cite{Sherrington, Barahona1982,mezard2001bethe}. Computation of the frustration index also reduces from classic unsigned graph optimisation problems (EDGE-BIPARTIZATION and MAXCUT) which are known to be NP-hard \cite{huffner_separator-based_2010}.

The frustration index can be computed in polynomial time for planar graphs \cite{katai1978studies}. In general graphs; however, the frustration index is believed to be NP-hard to approximate within any constant factor \cite{huffner_separator-based_2010}. %(assuming Khot's Unique Games Conjecture \cite{khot2002power}).
There has been a lack of systematic investigations for computing the exact frustration index of large-scale networks (as discussed in Chapters \ref{ch:2} -- \ref{ch:3}). In small graphs with fewer than 40 nodes, exact computational methods \cite{flament1963applications,hammer1977pseudo,bramsen2002further,brusco_k-balance_2010} are used to obtain the frustration index. Some recent studies focus on approximating \cite{dasgupta_algorithmic_2007,agarwal2005log,avidor2007multi,coleman2008local} the frustration index. In Chapters \ref{ch:2} -- \ref{ch:3}, we discussed methods for exact computation of the frustration index in large signed graphs with at least thousands of edges.
%Regarding the approximation methods, the semidefinite programming algorithm of Goemans and Williamson \cite{goemans_improved_1995} leads to an approximation guarantee of 0.878 for the frustration index \cite{thagard1998coherence, dasgupta_algorithmic_2007}. The frustration index can be approximated to a factor of $O(\sqrt{\log n})$ \cite{agarwal2005log} or $O(k \log k)$ \cite{avidor2007multi} where $n$ is the number of vertices and $k$ is the frustration index. Coleman et al.\ provides a review on the performance of several approximation algorithms of the frustration index \cite{coleman2008local}. Using a parametrised algorithmics approach, H\"{u}ffner, Betzler, and Niedermeier show that the frustration index (under a different name) is \textit{fixed parameter tractable} and can be computed in $O(2^k m^2)$ \cite{huffner_separator-based_2010} where $m$ is the number of edges and $k$ is the fixed parameter (the frustration index). The values of $k$ observed in signed graphs inferred from the literature makes this approach impractical \cite{aref2016exact}.

%Despite the lack of exact computational methods for the frustration index, a closely-related and more general problem in signed graphs has been investigated comprehensively.

A closely related and more general problem (that is beyond our discussions in this chapter) is finding the minimum number of edges whose removal results in a weakly balanced signed graph (as in Davis's definition of \textit{weak balance} \cite{Davis}). This problem is referred to as the \textit{Correlation Clustering} problem \cite{figueiredo2013mixed, ma_memetic_2015} which is investigated more comprehensively in the literature \cite{Giotis,brusco_k-balance_2010, figueiredo2013mixed,drummond2013efficient,levorato2015ils,levorato2017evaluating}. A comparison of mathematical programming models for computing the frustration index and correlation clustering can be found in Subsection \ref{3ss:multi}.

% According to Davis's definition of \textit{generalized balance}, a signed network is \textit{weakly balanced} ($k$-balanced) iff its vertex set can be partitioned into $k$ subsets such that each negative edge joins vertices belonging to different subsets \cite{Davis}.  For every fixed $k$, there is a polynomial time approximation scheme for the correlation clustering problem \cite{Giotis}. For arbitrary $k$, exact \cite{brusco_k-balance_2010, figueiredo2013mixed} and heuristic methods \cite{drummond2013efficient,levorato2015ils} are developed based on a mixed integer programming model \cite{demaine2006correlation}. Denoting the order of a graph by $n$, exact algorithms fail for $n>21$ \cite{brusco_k-balance_2010} and $n>40$ \cite{figueiredo2013mixed}, while greedy algorithms \cite{drummond2013efficient} and local search heuristics \cite{levorato2015ils} are capable of providing good solutions for $n \approx 10^3$ and $n \approx 10^4$ respectively.

%Efficient data reduction schemes \cite{huffner_separator-based_2010} and ground state search heuristics \cite{iacono_determining_2010} are suggested that provide bounds for the frustration index. Iacono et al.\ showed that the frustration index equals the minimum number of fundamental negative cycles induced over all spanning trees of the graph \cite{iacono_determining_2010}. Originally discussed in the biology context, the terminology used in \cite{iacono_determining_2010} is different where \textit{monotonocity} and \textit{inconsistency} are the equivalences for balance and frustration. 
Facchetti, Iacono, and Altafini suggested a non-linear energy function minimisation model for finding the frustration index \cite{facchetti_computing_2011}. Their model was used as the basis of various non-exact optimisation techniques \cite{iacono_determining_2010, esmailian_mesoscopic_2014, ma_memetic_2015, ma_decomposition-based_2017, Wang2016}. Using heuristic algorithms \cite{iacono_determining_2010}, estimations of the frustration index have been provided for biological networks up to $1.5\times10^3$ nodes \cite{iacono_determining_2010} and social networks with up to $10^5$ nodes \cite{facchetti_computing_2011, facchetti2012exploring}. 
Doreian and Mrvar \cite{patrick_doreian_structural_2015} have provided some upper bounds on the frustration index of signed international relation networks \cite{correlatesofwar2004}. We use their dataset in Section~\ref{4s:temporal} and analyse it using the exact values of the frustration index.
%Their claims must be evaluated while maintaining a certain level of healthy scepticism.

%After extending the non-linear energy minimization model to weak structural balance, Ma et al.\ provided good solutions for the correlation clustering problem in networks up to $n \leq 10^5$  using various heuristics \cite{ma_memetic_2015, ma_decomposition-based_2017}. Esmailian et al.\ have also extended the work of Facchetti, Iacono, and Altafini, focusing on the role of negative ties in signed graph clutering \cite{esmailian_mesoscopic_2014, esmailian2015community}.

%Their suggested heuristic is reported to solve networks with up to $n \leq 10^5$ within $99\%$ of optimality. However, not only their main theorem (Theorem 1 in \cite{esmailian_mesoscopic_2014}) is incorrect, but Mendon{\c{c}}a et al.\ has also cast doubt on their main conclusion regarding the role of negative ties in signed graphs \cite{mendoncca2015relevance}.

%An analysis of the literature shows that there are only a few studies on exact methods for computing the frustration index in signed graphs of non-trivial size and order \cite{huffner_separator-based_2010, aref2016exact}. Among the these methods, the integer linear programming models suggested by Aref et al.\ \cite{aref2016exact, aref2017computing} are the only computational methods capable of processing the signed networks that we aim to analyse in this chapter.

In this chapter, we use an exact optimisation model (the XOR model in Eq.\ \eqref{3eq4}) to compute the frustration index of large-scale signed networks exactly and efficiently.

\subsection*{\textbf{Our contribution in Chapter \ref{ch:4}}} \label{4ss:contrib}

We focus on the frustration index of signed networks, a standard measure of balance mostly estimated or approximated for decades due to the inherent combinatorial complexity. We follow a line of research begun in Chapter \ref{ch:1} (which compared various measures of partial balance and suggested that the frustration index should be more widely used), continued in Chapter \ref{ch:2} (which explained how integer linear optimisation models can be used to compute the frustration index) and Chapter \ref{ch:3} (which substantially improves the efficiency of such computations using algorithmic refinements and powerful mathematical programming solvers).

The purpose of this chapter is to present a single general methodology for studying signed networks and to demonstrate its relevance to applications. We consider a variety of signed networks arising from several disciplines. These networks differ substantially in size and the computational results require different interpretations. The current implementation of our algorithms can efficiently provide exact results on networks with up to 100000 edges. A by-product of exactly computing the frustration index is an optimal partitioning of nodes into two groups where the number of intra-group negative edges and inter-group positive edges is minimised.

%We use a recently developed optimisation model as the efficient method for analysing various signed networks.
%The advantage of formulating the problem as an optimisation model is not only exploring the details involved in a fundamental NP-hard problem, but also making use of powerful mathematical programming solvers like Gurobi \cite{gurobi} to solve the NP-hard problem exactly and efficiently. 
%We provide a numerical comparison of the frustration index on a variety of undirected signed networks, both randomly generated and inferred from well-known datasets.

This chapter begins by laying out the theoretical dimensions of the research in Section~\ref{4s:prelim}. The computational method is briefly discussed in Section~\ref{4s:model} followed by a discussion on its efficiency. %Section \ref{s:efficient} provides different techniques to speed up the computation and evaluates their effectiveness. 
%The results on synthetic data are provided in Section which also contains a closed-form formula for the frustration index in specially structured graphs.
Numerical results on signed networks of six disciplines are provided in Sections \ref{4s:d1} -- \ref{4s:d56}. Section~\ref{4s:conclu} provides a short conclusion.% Details of the computational method are briefly discussed in the appendix \ref{app:a}.%--\ref{app:c}. 
Along the way we point out several mistakes in the signed networks literature caused by inappropriate measures and inaccurate computation.

\section{Preliminaries} \label{4s:prelim}

We recall some standard definitions.

\subsection{Notation} 
We consider undirected signed networks $G = (V,E,\sigma)$. The ordered set of nodes is denoted by $V=\{1, 2, \dots, n\}$, with $|V| = n$. The set $E$ of edges can be partitioned into the set of positive edges $E^+$ and the set of negative edges $E^-$ with $|E|=m$, $|E^-|=m^-$, and $|E^+|=m^+$ where $m=m^- + m^+$. The sign function is denoted by $\sigma: E\rightarrow\{-1,+1\}$.

We represent the $m$ undirected edges in $G$ as ordered pairs of vertices $E = \{e_1, e_2, \dots , e_m\} \subseteq \{ (i,j) \mid i,j \in V , i<j \}$, where a single edge between nodes $i$ and $j$, $i<j$, is denoted by $(i,j) , i<j$. We denote the graph density by $\rho= 2m/(n(n-1))$.

% The expression $u\sim v$ denotes the adjacency of two nodes, regardless of sign.
The entries of the symmetric adjacency matrix $\textbf{A}$ are defined in Eq.\ \eqref{4eq1}. %We denote by $|\textbf{A}|$ the entrywise absolute value of $\textbf{A}$, which we call the \emph{unsigned adjacency matrix}.
\begin{equation}\label{4eq1}
a{_i}{_j} =
\left\{
\begin{array}{ll}
\sigma_{(i,j)} & \mbox{if } (i,j) \in E \\
\sigma_{(j,i)} & \mbox{if } (j,i) \in E \\
0 &  \text{otherwise}
\end{array}
\right.
\end{equation}

%The number of positive (negative) edges connected to the node $i \in V$ represents positive (negative) degree of the node and is denoted by $d^+ {(i)}$ ($d^- {(i)}$). The degree of node $i$ is defined by $d {(i)} = d^+ {(i)} + d^- {(i)}$.

%The net degree of a node is calculated by $ d^+ {(i)} -d^- {(i)}$.

We use $G_r=(V,E,\sigma_r)$ to denote a reshuffled graph in which the sign function $\sigma_r$ is a random mapping of $E$ to $\{-1,+1\}$ that preserves the number of negative edges. 

A \emph{walk} of length $k$ in $G$ is a sequence of nodes $v_0,v_1,...,v_{k-1},v_k$ such that for each $i=1,2,...,k$ there is an edge from $v_{i-1}$ to $v_i$. If $v_0=v_k$, the sequence is a \emph{closed walk} of length $k$. If the nodes in a closed walk are distinct except for the endpoints, the walk is a cycle of length $k$. The \emph{sign} of a walk or cycle is the product of the signs of its edges. Cycles with positive (negative) signs are balanced (unbalanced). A balanced graph is one with no unbalanced cycles.

\subsection{Frustration count}

%\textit{Satisfied} and \textit{frustrated} edges are defined based on colourings of the nodes. Colouring the nodes with black and white, a frustrated (satisfied) edge $(i,j)$ is either a positive (negative) edge with different colours on the endpoints $i,j$ or a negative (positive) edge with the same colours on the endpoints $i,j$. 

For any signed graph $G=(V, E, \sigma)$, we can partition $V$ into two sets, denoted $X \subseteq V$ and $\bar X=V \backslash X$. We call $X$ a colouring set and we think of this partitioning as specifying a colouring of the nodes, where each node $i \in X$ is coloured black, and each node $i \in \bar X$ is coloured white. We let $x_i$ denote the colour of node $i \in V$ under $X$, where $x_i=1$ if $i \in X$ and $x_i=0$ otherwise. 

We define the {\em frustration count} $f_G(X)$ as the number of frustrated edges of $G$ under $X$. The frustration index $L(G)$ of a graph $G$ can be obtained by finding a subset $X^* \subseteq V$ of $G$ that minimises the frustration count $f_G(X)$.

\begin{figure}
	\subfloat[An example signed graph]{\includegraphics[height=1.8in]{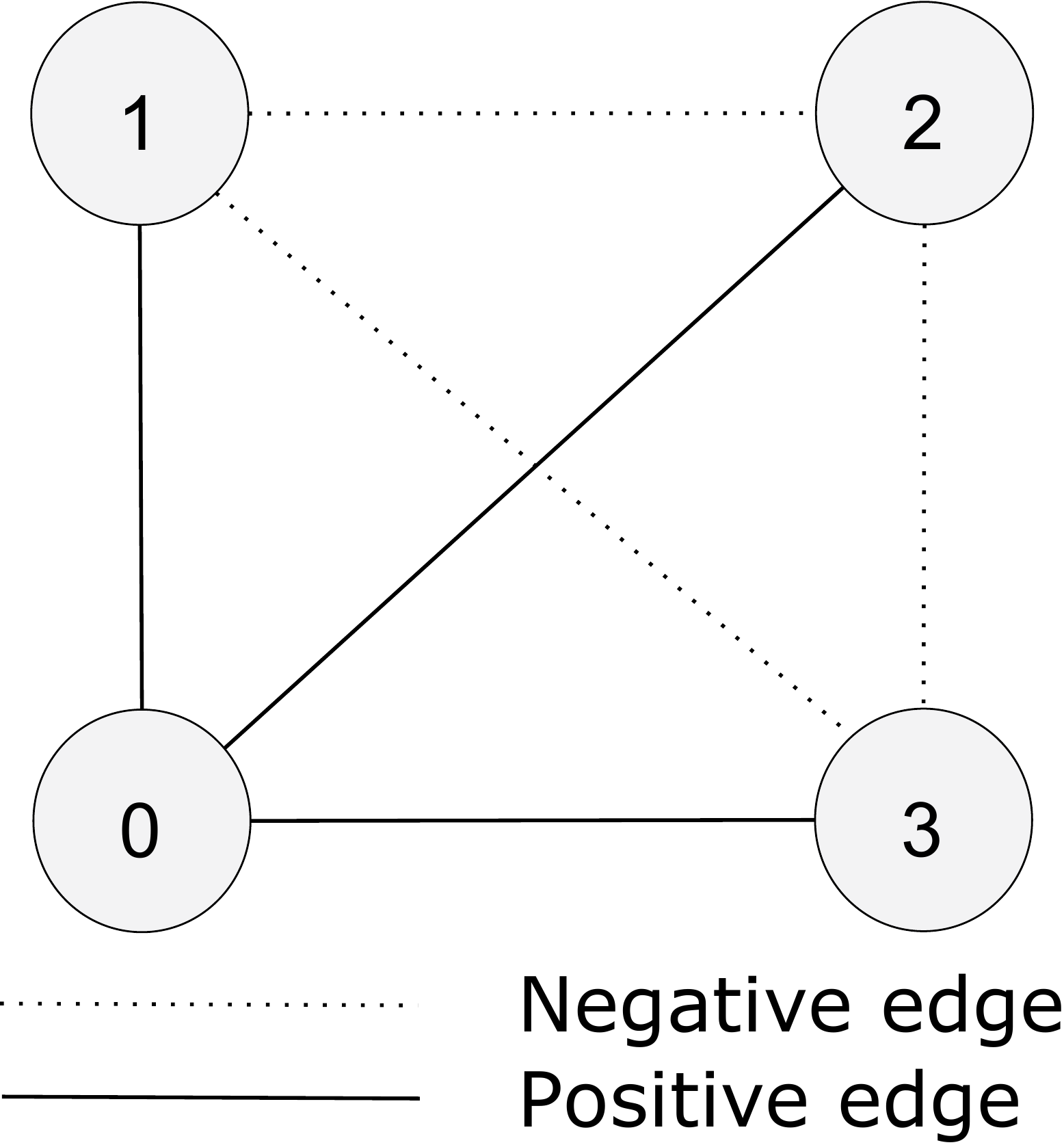}%
		\label{4fig1a}}
	\hfil
	\subfloat[Two node colourings (both optimal) and their resulting frustrated edges]{\includegraphics[height=1.8in]{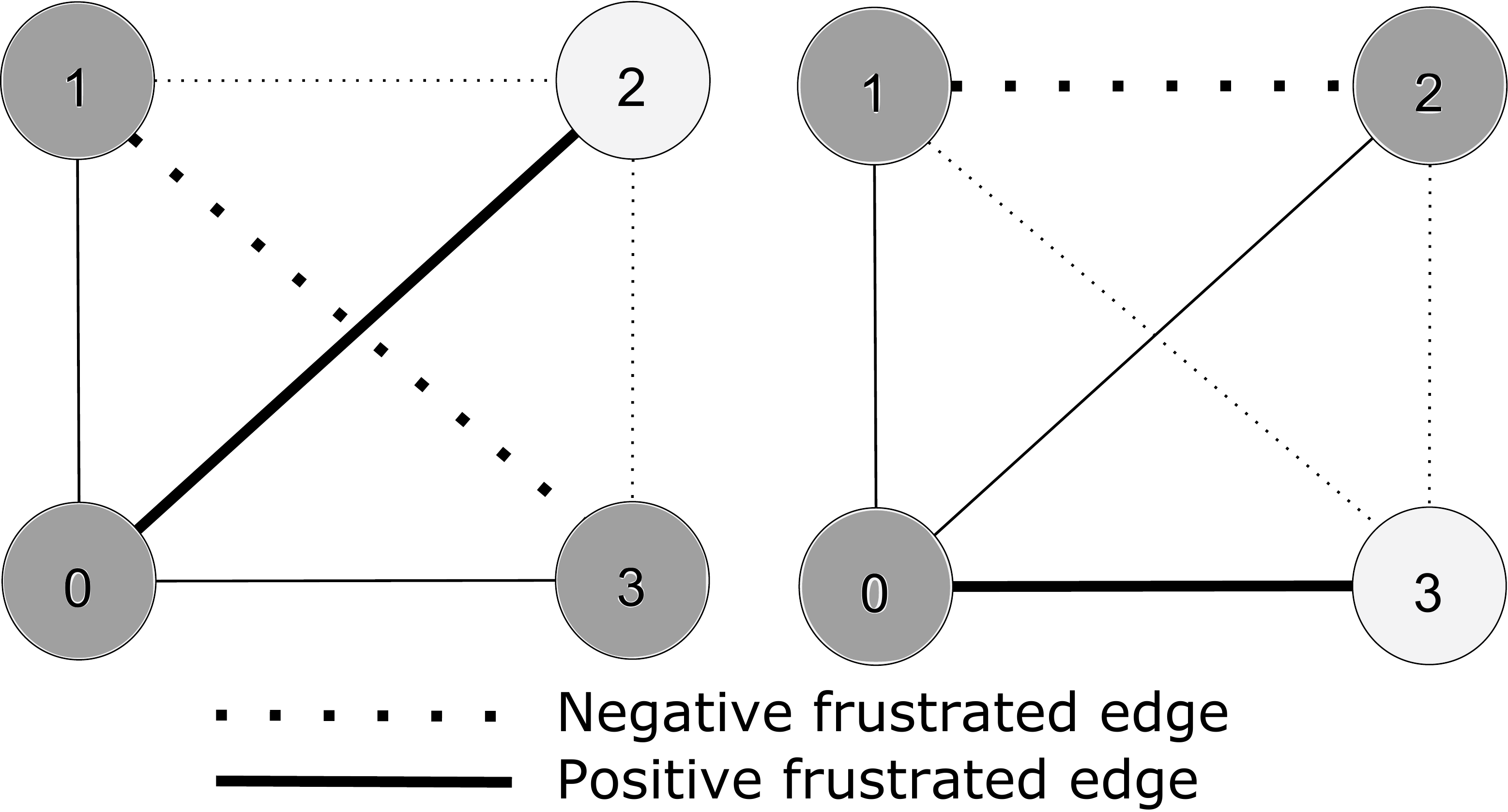}%
		\label{4fig1b}}
	\hfil
	\caption{The impact of node colouring on the frustration of edges}
	\label{4fig1}
\end{figure}

Figure \ref{4fig1} (a) demonstrates an example signed graph in which positive and negative edges are represented by solid lines and dotted lines respectively. Figure~\ref{4fig11} (b) illustrates two node colourings and the resulting frustrated edges represented by thick lines. 

\section{Methods and Materials} \label{4s:model}

In this section, we briefly discuss our methodology and datasets.

\subsection{Methodology}

In Chapters \ref{ch:2} -- \ref{ch:3}, we developed several optimisation models and tested them on synthetic and real-world datasets using ordinary desktop computers showing the efficiency of our models in computing the frustration index in comparison to other models in the literature \cite{flament1963applications,hammer1977pseudo,hansen_labelling_1978,harary_simple_1980,bramsen2002further,dasgupta_algorithmic_2007, brusco_k-balance_2010, huffner_separator-based_2010, iacono_determining_2010,patrick_doreian_structural_2015}. In this chapter, we use the XOR model in Eq.\ \eqref{3eq4} to compute the frustration index exactly and efficiently. For detailed discussions on the efficiency of the XOR model in Eq.\ \eqref{3eq4}, one may refer to Chapter \ref{ch:3}.

For comparing the level of frustration among networks of different size and order, we use the \textit{normalised frustration index}, $F(G)=1-2L(G)/m$. This standard measure of partial balance is suggested in Chapter \ref{ch:1} because it satisfies key axiomatic properties. Values of $F(G)$ are within the range of $[0,1]$ and greater values of $F(G)$ represent closeness to a state of structural balance.

Our baseline for evaluating balance comprises the average and standard deviation of the frustration index in reshuffled graphs (that have signs allocated randomly to the same underlying structure). Accordingly, we use Z score values, $Z={(L(G)-L(G_r))}/{\text{SD}}$, in order to evaluate the level of partial balance precisely.

\subsection{Materials}

We use a wide range of examples from different disciplines all being undirected signed networks. This includes four social signed networks ranging in size from 49 to 99917 edges in Section~\ref{4s:d1}, four biological signed networks with 779-3215 edges in Section~\ref{4s:d2}, one dynamic network of international relations with 51 time windows ranging in size from 362 to 1247 edges in Section~\ref{4s:d3}, six financial portfolios with 10-55 edges over 9 years in Section~\ref{4s:d4}, and molecular fullerene graphs with 270-9000 edges and Ising models with 32-79600 edges in Section~\ref{4s:d56}. The datasets used in this chapter are made publicly available on the \urllink{https://figshare.com/articles/Signed_networks_from_sociology_and_political_science_biology_international_relations_finance_and_computational_chemistry/5700832}{Figshare} research data repository \cite{Aref2017data}. We use a wide variety of datasets, rather than focusing on a specific application, in order to underline the generality of our approach.

The fundamental reason why we only use undirected signed networks is that the reliability test for predictions on directed signed networks made by balance theory shows very negative results \cite{leskovec_signed_2010}. Based on large directed signed networks such as Epinions, Slashdot, and Wikipedia, the binary predictions made by balance theory are incorrect almost half of the time \cite{leskovec_signed_2010}. This observation supports the inefficacy of balance theory for structural analysis of \emph{directed} signed graphs (as discussed in Section \ref{1s:recom}). 

%\section{Analysing real networks} \label{s:real}
The numerical results in this chapter are obtained by solving the XOR model \eqref{3eq4} coupled with three speed-up techniques (discussed in Subsections \ref{3ss:pre} -- \ref{3ss:unbalanced}) using Gurobi's Python interface \cite{gurobi}. Unless stated otherwise, the hardware used for the computational analysis is a virtual machine with 32 Intel Xeon CPU E5-2698 v3 @ 2.30 GHz processors and 32 GB of RAM running 64-bit Microsoft Windows Server 2012 R2 Standard.

\section{Social networks}\label{4s:d1}

In this section, we discuss using the frustration index to analyse social signed networks inferred from the sociology and political science datasets.

\subsection{Datasets}

We use well-studied datasets of communities with positive and negative interactions and preferences. This includes Read's dataset for New Guinean highland tribes \cite{read_cultures_1954} and the last time frame of Sampson's data on monastery interactions \cite{sampson_novitiate_1968}. %We also use graphs inferred from datasets of students' choice and rejection \cite{newcomb_acquaintance_1961,lemann_group_1952}. A further explanation on the details of inferring signed graphs from the choice and rejection data can be found in \cite{aref2015measuring}. The same datasets are used in some other studies of social networks \cite{doreian_multiple_2008,doreian_partitioning_2009,aref2015measuring}.
Our analysis also includes a signed network of US senators that is inferred in \cite{neal_backbone_2014} through implementing a stochastic degree sequence model on Fowler's Senate bill co-sponsorship data \cite{fowler_legislative_2006} for the 108th US senate.

\begin{figure} 
	\centering
	\subfloat[Optimal colouring of New Guinean tribes network with frustrated edges shown by thick lines \cite{read_cultures_1954}]{\includegraphics[height=2.6in]{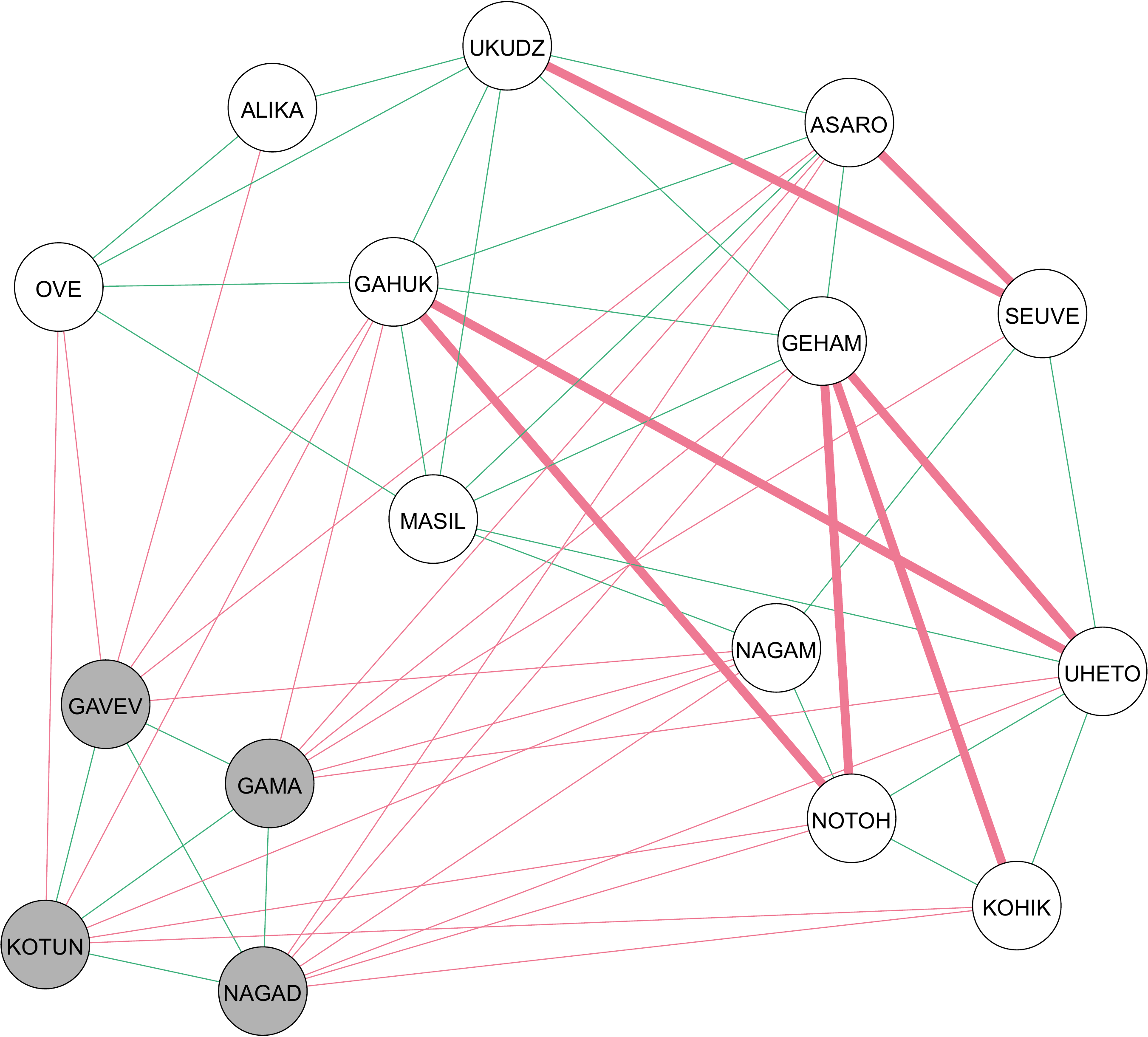}%
		\label{4fig2a}}
	\hfil
	\centering
	\subfloat[Optimal colouring of monastery interactions network with frustrated edges shown by thick lines \cite{sampson_novitiate_1968}]{\includegraphics[height=2.2in]{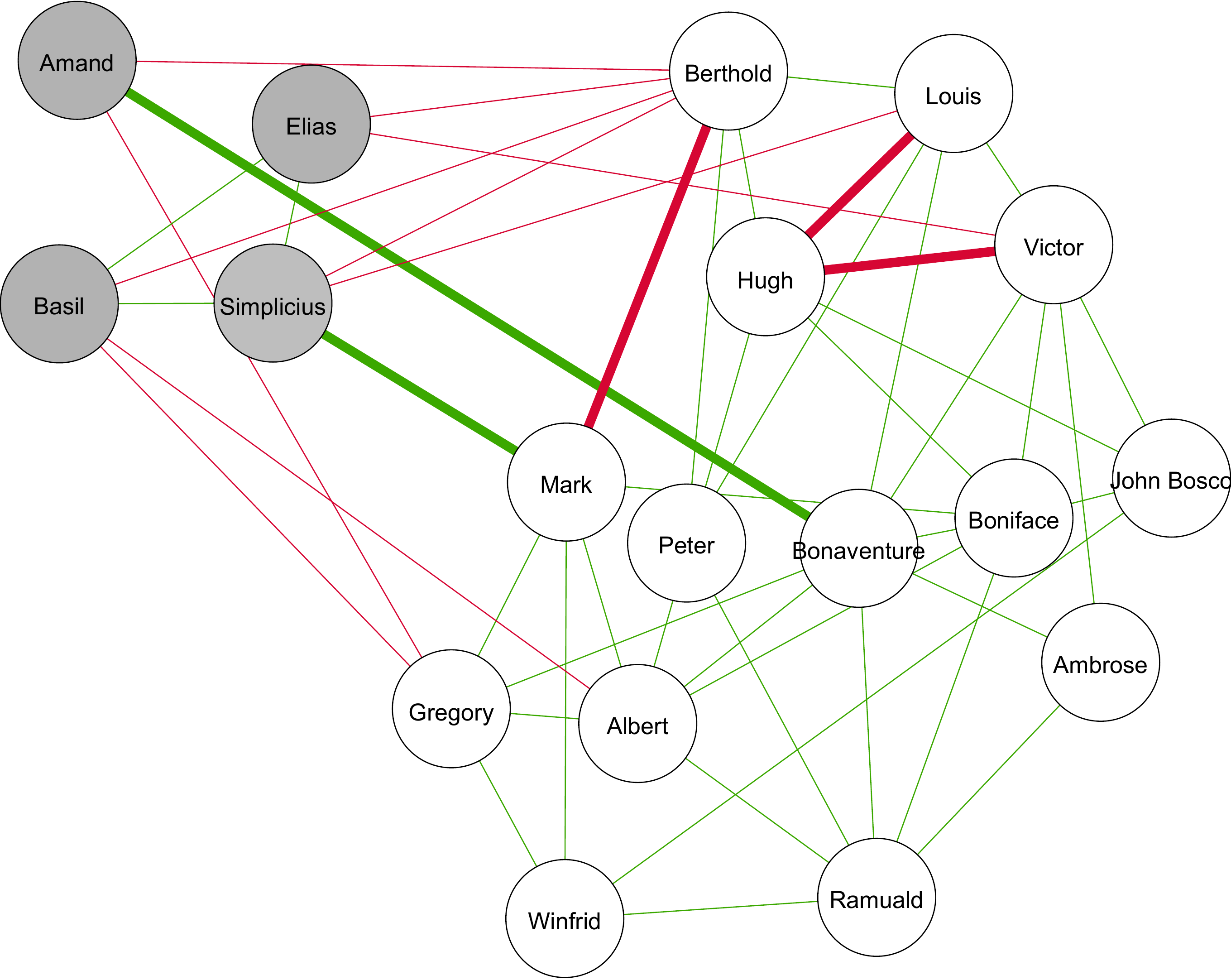}%
		\label{4fig2b}} 
	\hfil
	\centering
	\subfloat[Network of Wikipedia elections \cite{leskovec_signed_2010}]{\includegraphics[height=2.4in]{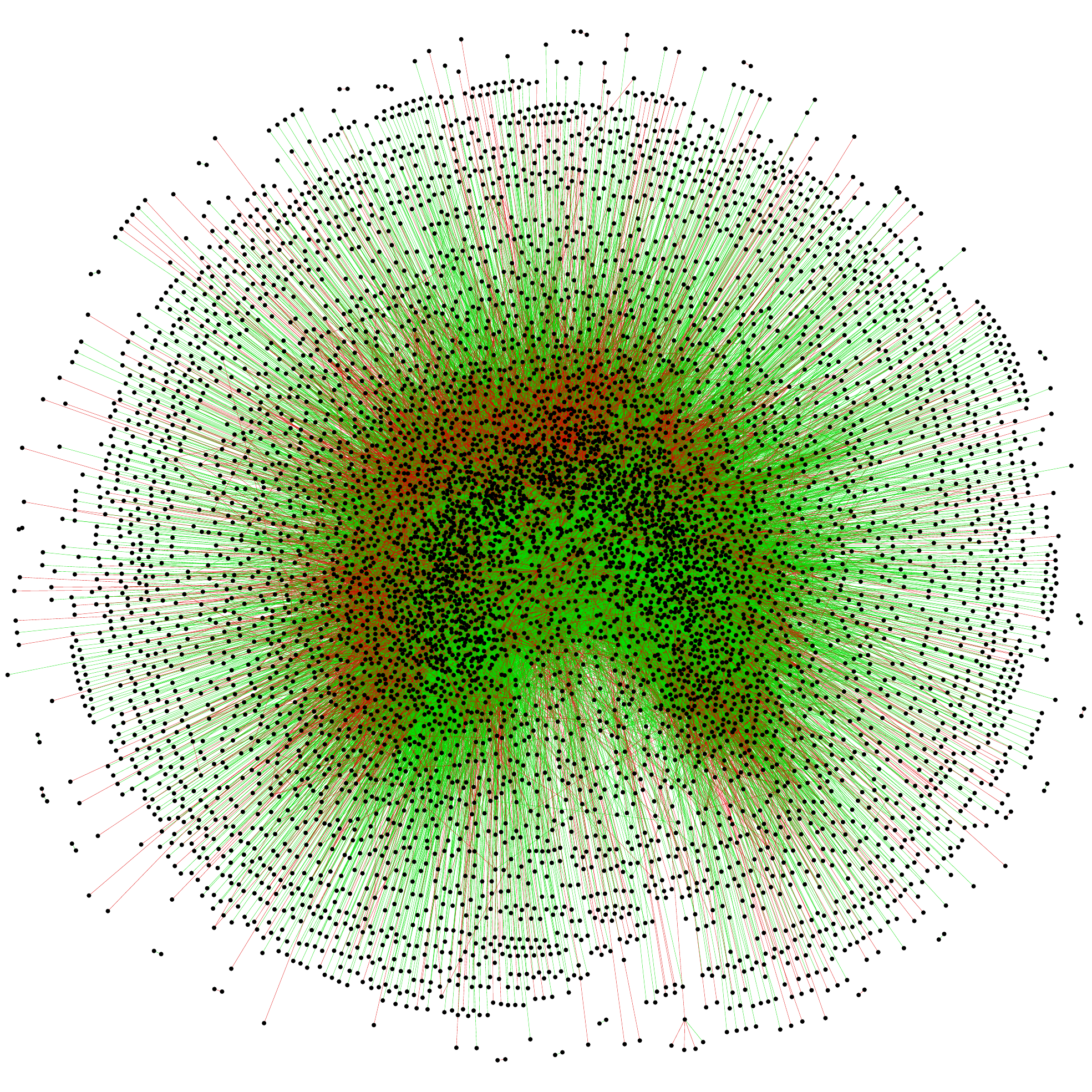}%
		\label{4fig2d}} 
	%\hfil
	%\centering
	%\subfloat[Network of the 108th US senate (nodes having mismatching party colour and optimal colour are positioned on the top and bottom.) \cite{neal_backbone_2014}]{\includegraphics[height=4in]{g5-eps-converted-to.pdf}%
	%	\label{4fig2c}} 
	%\hfil
	\caption{Three signed networks inferred from network science datasets and visualised using Gephi %(colour version online)
	}
	\label{4fig2}
\end{figure}

\begin{figure} 
\includegraphics[height=5.4in, angle =90]{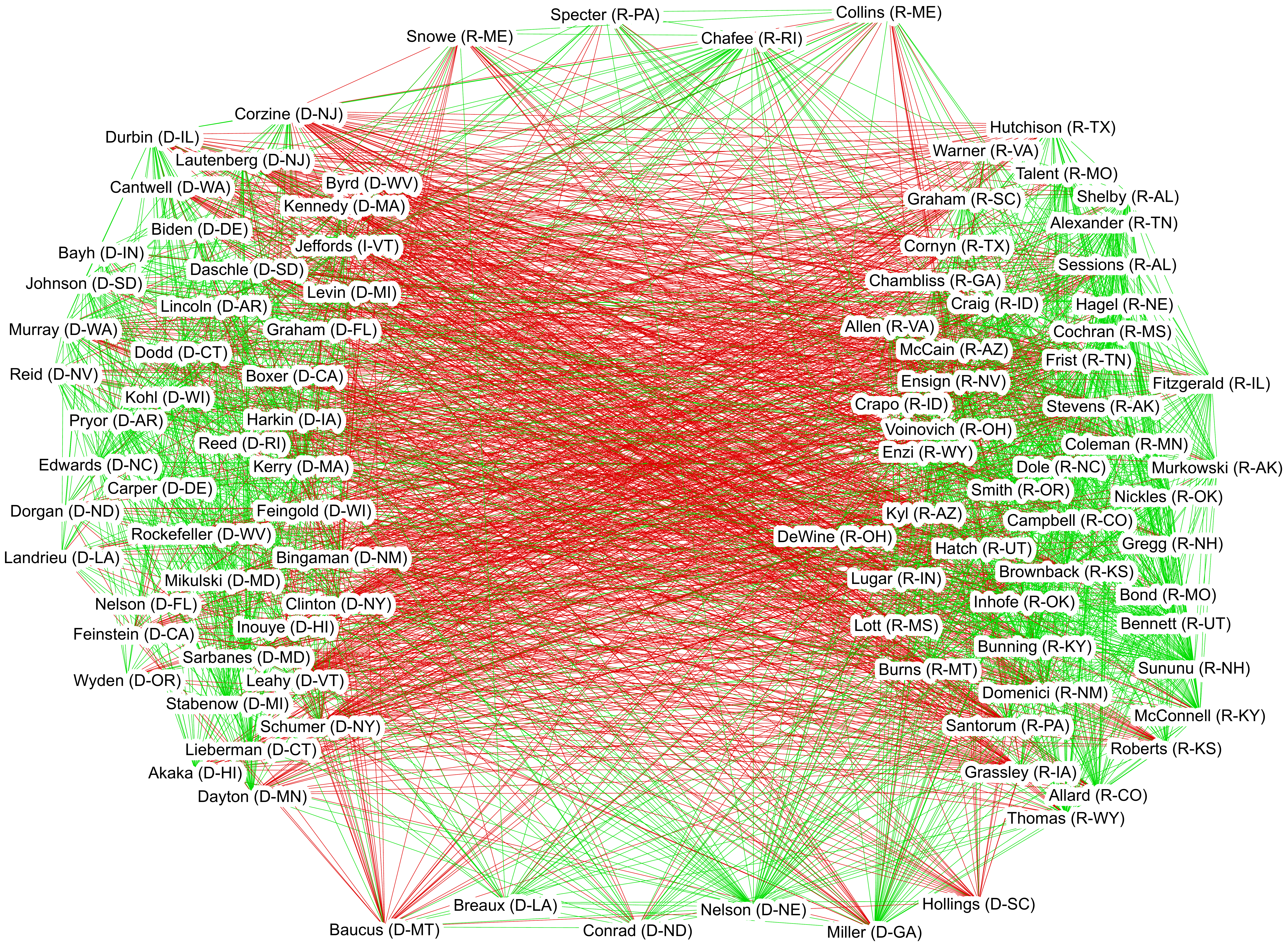}%
\caption{Network of the 108th US senate (10 nodes having mismatching party colour and optimal colour are positioned on the top and bottom of the figure.) \cite{neal_backbone_2014}}
\label{4fig2c}
\end{figure}

A larger social signed network we use is from the Wikipedia election dataset \cite{leskovec_signed_2010}. This dataset is based on all adminship elections before January 2008 in which Wikipedia users have voted for approval or disapproval of other users promotions to becoming administrators. We use an undirected version of the Wikipedia elections signed graph made publicly available in \cite{levorato2015ils}. The four social signed networks are illustrated in Figures~\ref{4fig2} -- \ref{4fig2c} where green and red edges represent positive and negative edges respectively. %There are studies on the correlation clustering problem \cite{ma_memetic_2015, ma_decomposition-based_2017} and estimating frustration index \cite{facchetti_computing_2011} using Wikipedia elections network where many inconsistent directed edges are disregarded in order to "symmetrise" the data \cite{facchetti_computing_2011}.

\subsection{Results}

Our numerical results are shown in Table~\ref{4tab1} where the average and standard deviation of the frustration index in 500 reshuffled graphs (50 reshuffled graphs for Wikipedia election network), denoted by $L(G_r)$ and $\text{SD}$, are also provided for comparison. %Reshuffling the signs on the edges 500 times, we obtain two parameters of frustration distribution for the fixed underlying structure. %The randomisation process allocates signs to edges based on random permutations while preserving the unsigned graph structure.

\begin{table}[ht]
	\centering
	\caption{The frustration index in social signed networks}
	\label{4tab1}
	\begin{tabular}{lllll}
		\hline
		Graph $(n,m,m^-)$ & $\rho$ & $L(G)$ & $L(G_r) \pm \text{SD}$ & Z score \\ \hline
		Highland tribes $(16,58,29)$  & 0.483 & $7$    & $14.65 \pm 1.38$ &  $-5.54$ \\
		Monastery interactions $(18,49,12)$  & 0.320 & $5$    & $9.71 \pm 1.17$ &  $-4.03$ \\
		US senate $(100,2461,1047)$ & 0.497 & $331$  & $ 965.6\pm 9.08$& $-69.89$ \\ 
		Wiki elections $(7112,99917,21837)$ & 0.004 & $14532$&$18936.1 \pm 45.1 ^\dagger$  & $-97.59^\dagger$ \\
	 	\hline
\multicolumn{4}{l}{$\dagger$ based on lower bounds within 15\% of the optimal solution}
	\end{tabular}
\end{table}

% Reshuffled Wiki based on upperbound with MIP Gap 0.15: $14532$ & $18959.6 \pm 46.4$ & $-28.29$

As it is expected the four social signed networks are not totally balanced. However, the relatively small values of $L(G)$ suggest low levels of frustration in these networks. 
%In such networks, macro-scale structural properties can emerge from micro-scale interactions.
In order to be more precise, we have implemented a very basic statistical analysis using Z scores $Z={(L(G)-L(G_r))}/{\text{SD}}$. These Z scores, provided in the right column of the Table~\ref{4tab1}, show how close the networks are to a state of balance. The results indicate that networks exhibit a level of frustration substantially lower than what is expected by chance.
%They demonstrate values in the standard normal distribution represented by considering observations are merely results of chance.
%Negative values of Z score can be interpreted as lower level of frustration than the random expectation. %Z score values also represent the significance when compared to the standard $(-3,3)$ interval.

%Based on the Z score values in the right column of Table~\ref{tab3}, the level of frustration is very low for the network of US senators and low for networks of tribes and monastery interactions.

%\FloatBarrier

\subsection{Optimal partitioning}

In the network of US senators, we may get insight not only from the value of the frustration index, but also the optimal node colouring leading to the minimum number of frustrated edges. As shown in node labels of Figure \ref{4fig2c}, the 108th US senate was made of 48 Democratic senators, 1 independent senator (caucusing with Democrats), and 51 Republican senators. Based on these figures, we may consider a party colour for each node $i$ in the network and position senators from different parties on the left (Democrat) and right (Republican) sides of Figure \ref{4fig2c}. The abundance of green edges on the left and right sides of Figure \ref{4fig2c} shows that most senators have positive relationship with their party senators (co-sponsor bills proposed by their party). The numerous red edges between left and right sides of Figure \ref{4fig2c} show opposition between senators from different parties (senators do not support bills put forward by the other party in most cases).

We can also compare the party colour for each node $i$ to the optimal colour ($x_i$ optimal value) that leads to the partitioning with the minimum number of frustrated edges. As expected from the bi-polar structure of US senate, the colours match in 90 out of 100 cases (considering the independent senator as a Democrat). The nodes associated with Republican (Democrat) senators who have mismatching party colour and optimal node colour are positioned on the top (bottom) of Figure \ref{4fig2c}. It can be observed from such nodes that, contrary to the other nodes, they have negative (positive) ties to their (the other) party. 

\subsection{Computations}

Regarding performance of the optimisation model, a basic optimisation formulation of the problem with no speed-up technique (such as the model formulated in Eq.\ \eqref{2eq7}) would solve the Highland tribes and Monastery interaction instances in a reasonable time on an ordinary computer (as demonstrated in Chapter \ref{ch:2}). The XOR model in Eq.\ \eqref{3eq4} solves such instances in split seconds, while for the senators network it takes a few seconds.

For Wikipedia elections network, 9.3 hours of computation is required to find the optimal solution. This considerable computation time prevents us from testing 500 reshuffled versions of the Wikipedia elections network. In order to perform the statistical analysis for the Wikipedia network in a reasonable time, we limited the number of runs to 50.

As a conservative approach, we also used the average of best lower bound obtained within 15\% of the optimal solution as $L(G_r)$ for Wikipedia elections network. This approach reduces the average computation time of each run to 4 hours. The branch and bound algorithm guarantees that the frustration index of each reshuffled graph (which remains unknown) is greater than the lower bound we use to compute the Z score for Wikipedia elections network.

\section{Biological networks}\label{4s:d2}

Some biological models are often used to describe interactions with dual nature between biological molecules in the field of systems biology. The interactions can be \textit{activation} or \textit{inhibition} and the biological molecules are enzymes, proteins or genes \cite{dasgupta_algorithmic_2007}. This explains the parallel between signed graphs and these types of biological networks. Interestingly, the concept of \textit{close-to-monotone} \cite{maayan_proximity_2008} in systems biology is analogous to being close to a state of balance. Similar to negative cycles and how they lead to unbalance, existence of \textit{negative loops} in biological networks indicates a system that does not display well-ordered behaviour \cite{maayan_proximity_2008}. 

\subsection{Datasets}

There are large-scale gene regulatory networks where nodes represent genes and positive and negative edges represent \textit{activating connections} and \textit{inhibiting connections} respectively. We use four signed biological networks previously analysed by \cite{iacono_determining_2010}. They include two gene regulatory networks, related to two organisms: a eukaryote (the yeast \textit{Saccharomyces cerevisiae}) \cite{Costanzo2001yeast} and a bacterium (\textit{Escherichia coli}) \cite{salgado2006ecoli}. Another signed network we use is based on the Epidermal Growth Factor Receptor (EGFR) pathway \cite{oda2005}. EGFR is related to the epidermal growth factor protein whose release leads to rapid cell division in the tissues where it is stored such as skin \cite{dasgupta_algorithmic_2007}. We also use a network based on the molecular interaction map of a white blood cell (macrophage) \cite{oda2004molecular}.

\begin{figure}[ht]
	\centering
	\subfloat[The gene regulatory network of the \textit{Escherichia coli} \cite{salgado2006ecoli}]{\includegraphics[height=3.6in]{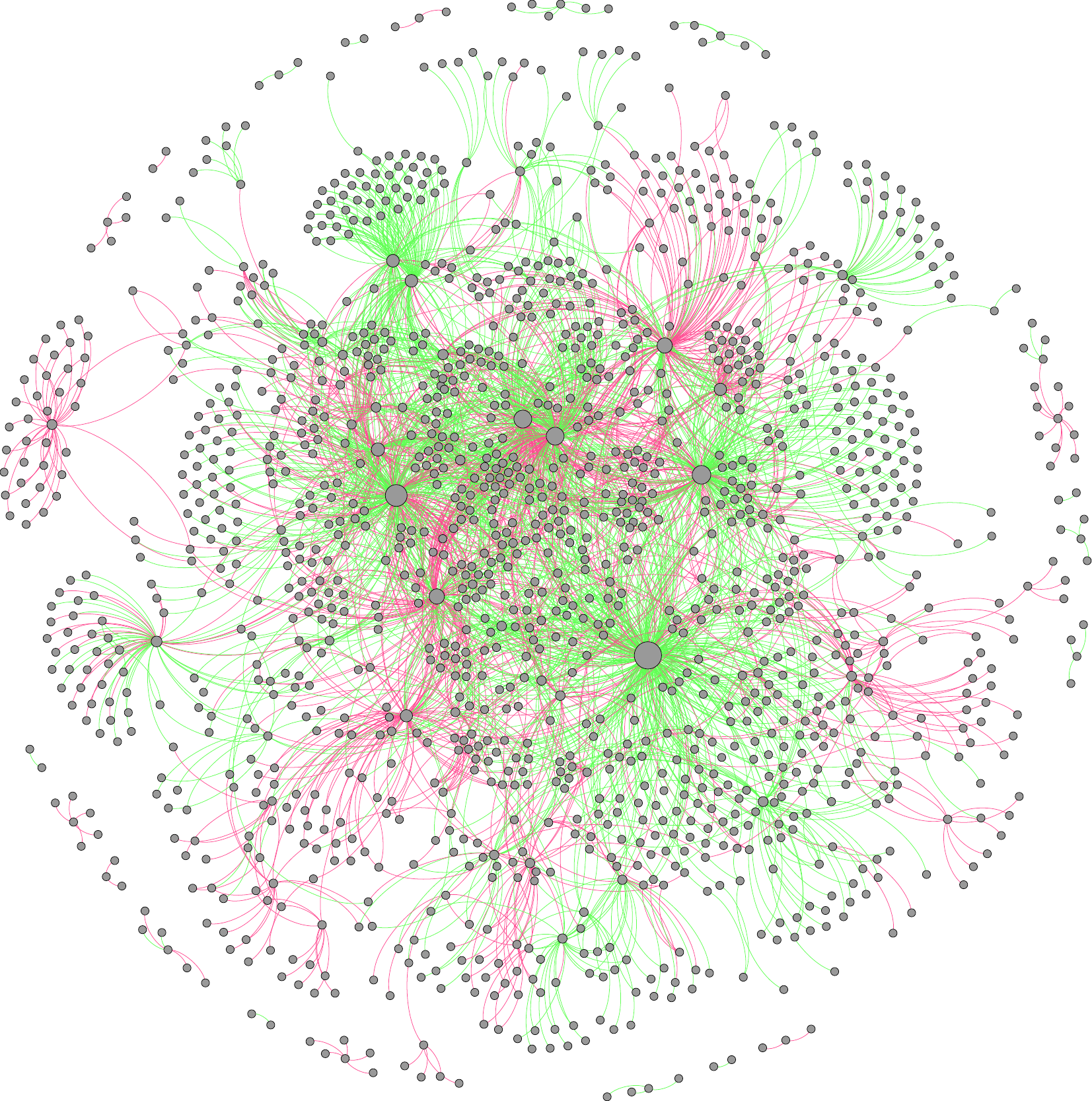}%
		\label{4fig3b}} 
	\hfil
	\centering
	\subfloat[Epidermal growth factor receptor pathway \cite{oda2005}]{\includegraphics[height=2.3in]{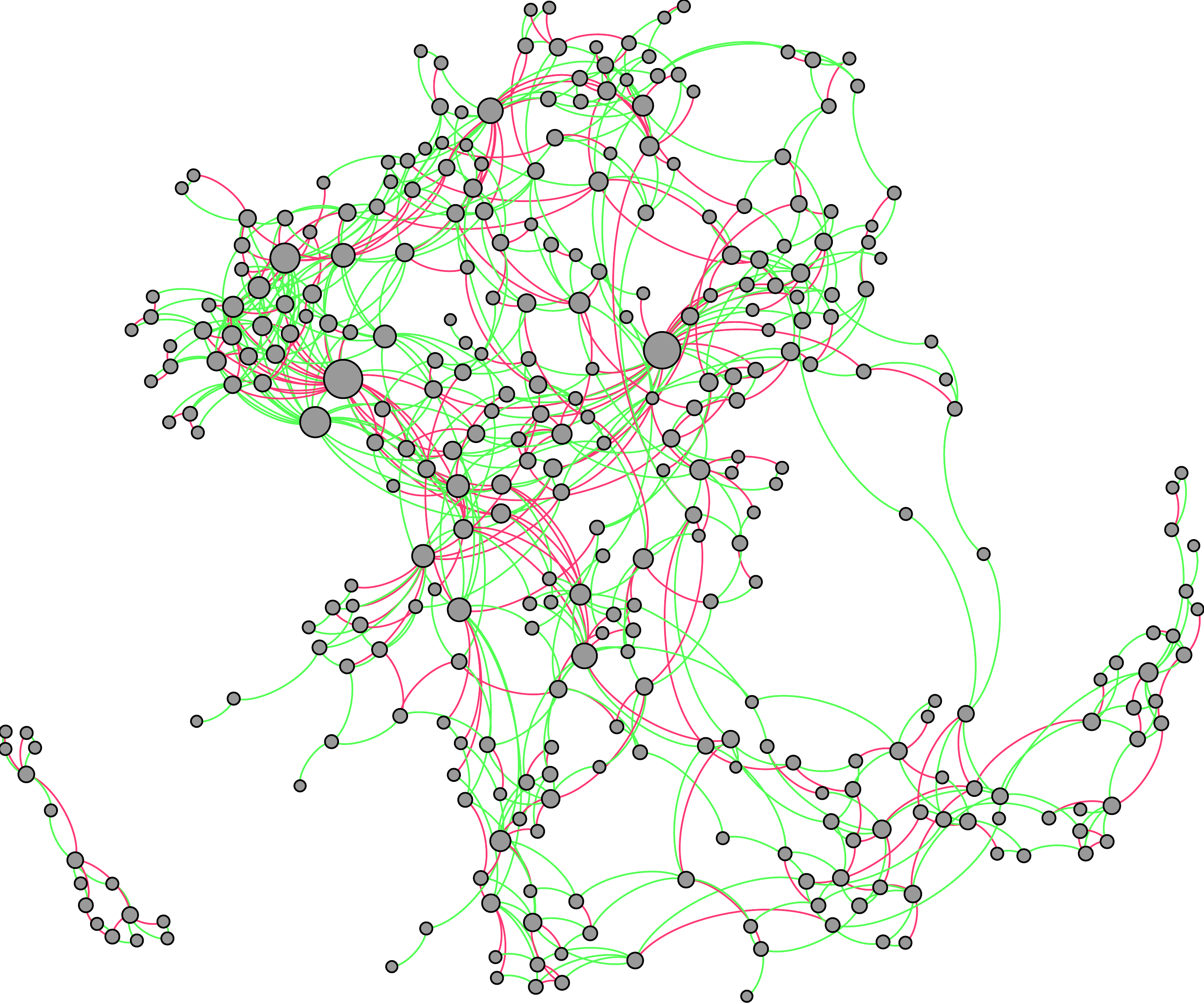}%
		\label{4fig3c}} 
	\hfil
	\centering
	\subfloat[Molecular interaction map of a macrophage \cite{oda2004molecular}]{\includegraphics[height=2.3in]{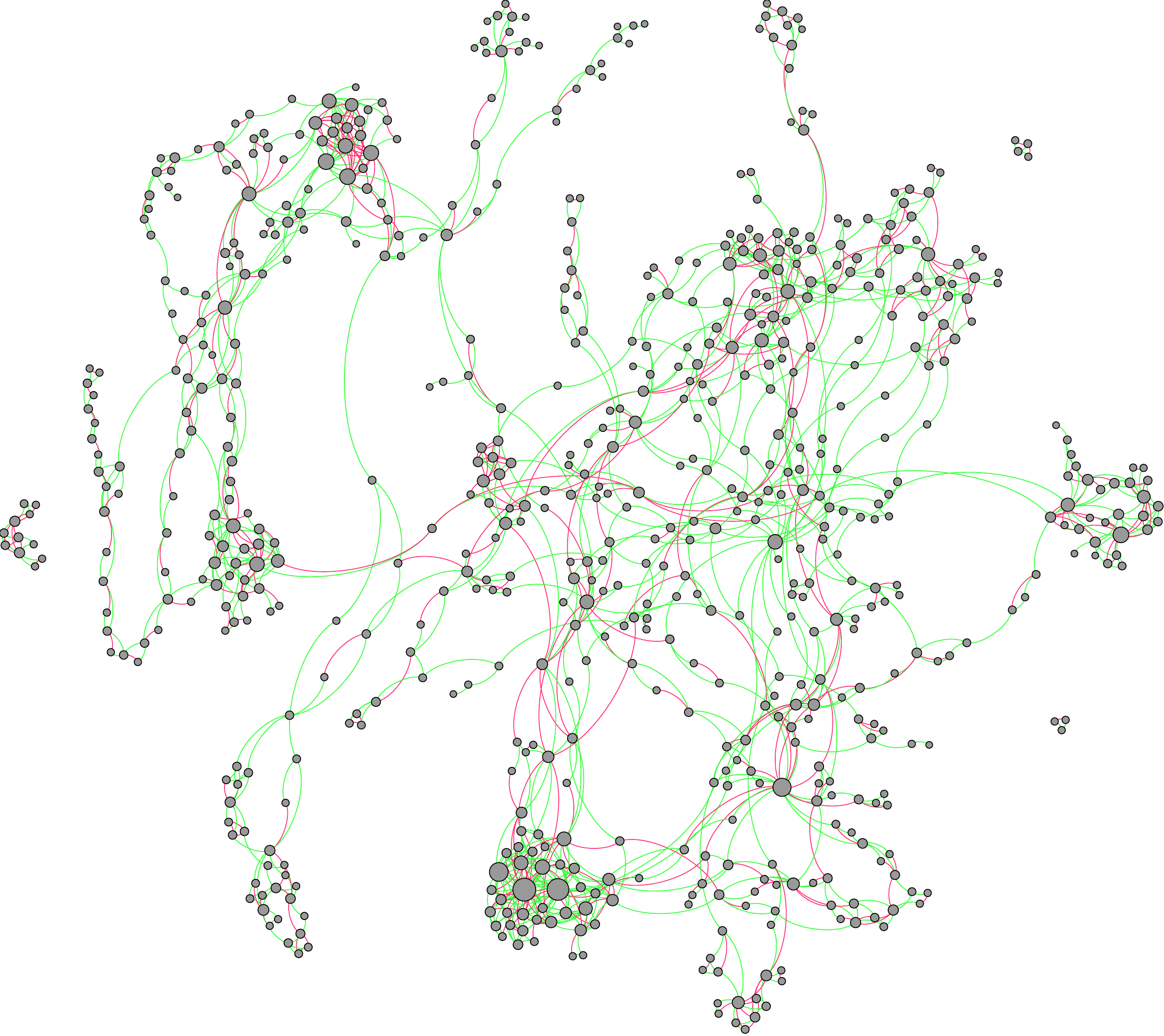}%
		\label{4fig3d}} 
	\hfil
	\caption{Three biological signed networks visualised using Gephi %(colour version online)
	}
	\label{4fig3}
\end{figure}
%\FloatBarrier
Yeast and E.coli networks are categorised as transcriptional networks while EGFR and macrophage are signalling networks \cite{iacono_determining_2010}. Figure~\ref{4fig3} shows three of these biological signed networks. The colour of edges correspond to the signs on the edges (green for activation and red for inhibition). For more details on the four biological datasets, one may refer to \cite{iacono_determining_2010}.

\subsection{Results}

Table \ref{4tab2} provides the results for the four biological networks where the average and standard deviation of the frustration index in 500 reshuffled graphs are also provided for comparison.

\begin{table}[ht]
	\centering
	\caption{The frustration index in biological networks}
	\label{4tab2}
	\begin{tabular}{lrrrrrr}
		\hline
		Graph & $n$ & $m$ & $m^-$ & $L(G)$ & $L(G_r) \pm \text{SD}$ & Z score \\ \hline
		yeast    & 690 & 1080& 220   &  41  & $ 124.3\pm 4.97$& -16.75 \\ 
		E.coli    & 1461& 3215& 1336  & 371  & $ 653.4\pm 7.71$& -36.64 \\ 	
		EGFR    & 329 & 779 & 264   & 193  & $ 148.96\pm 5.33$&   8.26 \\ 
		macrophage    & 678 & 1425&  478  & 332  & $ 255.65\pm 8.51$&  8.98 \\  \hline

	\end{tabular}
\end{table}

The results in Table~\ref{4tab2} show that the level of frustration is very low for yeast and E.coli networks. The Z score value for the yeast network is $-16.75$ which is consistent with the observation of DasGupta et al.\ (based on approximating the frustration index) that the number of edge deletions is up to 15 standard deviations away from the average for comparable random and reshuffled graphs \cite[Section 6.3]{dasgupta_algorithmic_2007}.

The Z score values in Table~\ref{4tab2} show that the transcriptional networks are close to balanced (close-to-monotone) confirming observations in systems biology \cite{maayan_proximity_2008} that in such networks the number of edges whose removal eliminates negative cycles is small compared to the reshuffled networks. This explains the stability shown by such networks in response to external stimuli \cite{maayan_proximity_2008}.

\begin{figure}
	\centering
	\subfloat[The gene regulatory network of yeast \cite{Costanzo2001yeast}]{\includegraphics[height=3.4in]{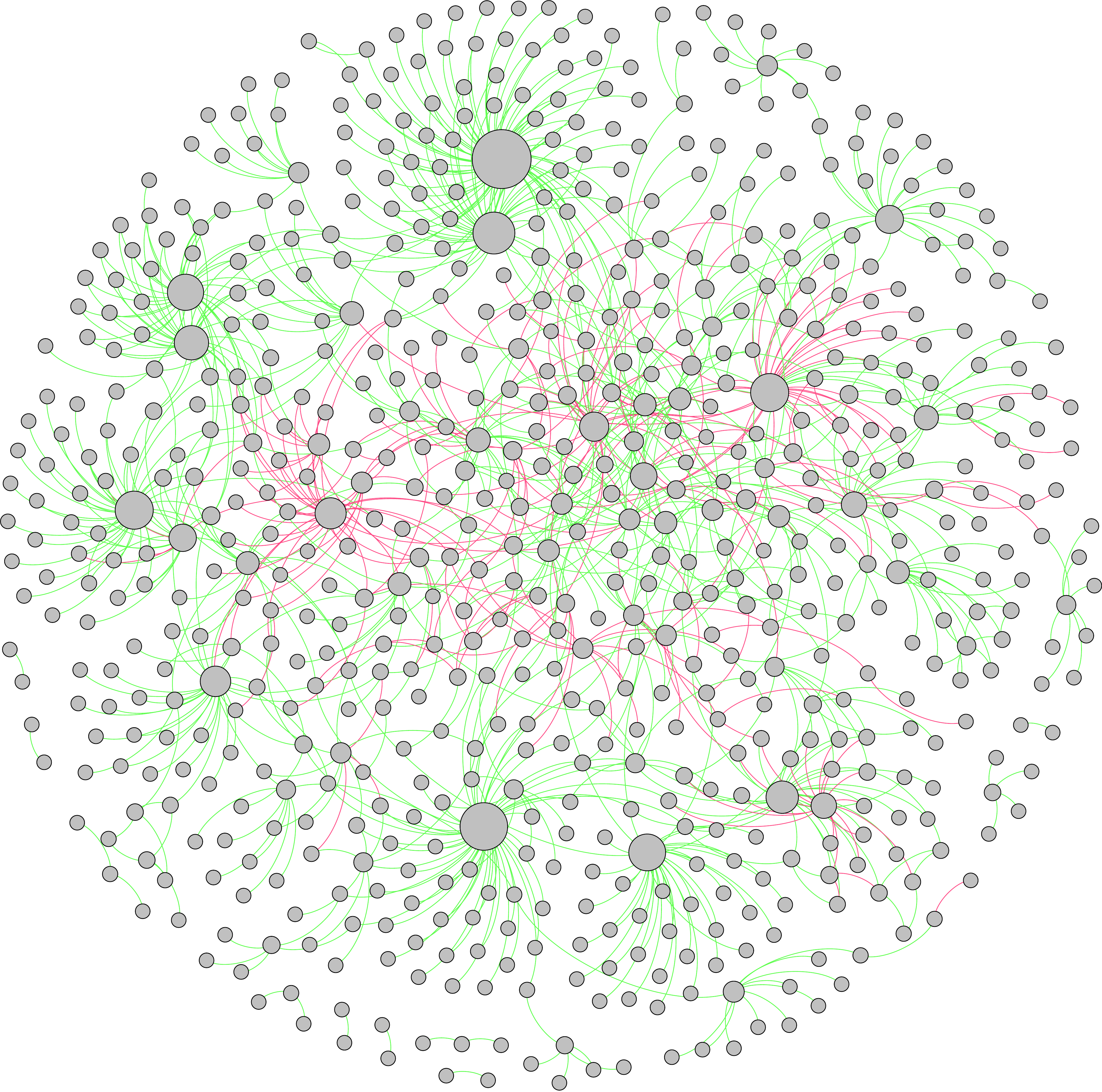}%
		\label{4fig3a}} 
	\hfil
	\centering
	\subfloat[The monotone subsystem of yeast network]{\includegraphics[height=3.4in]{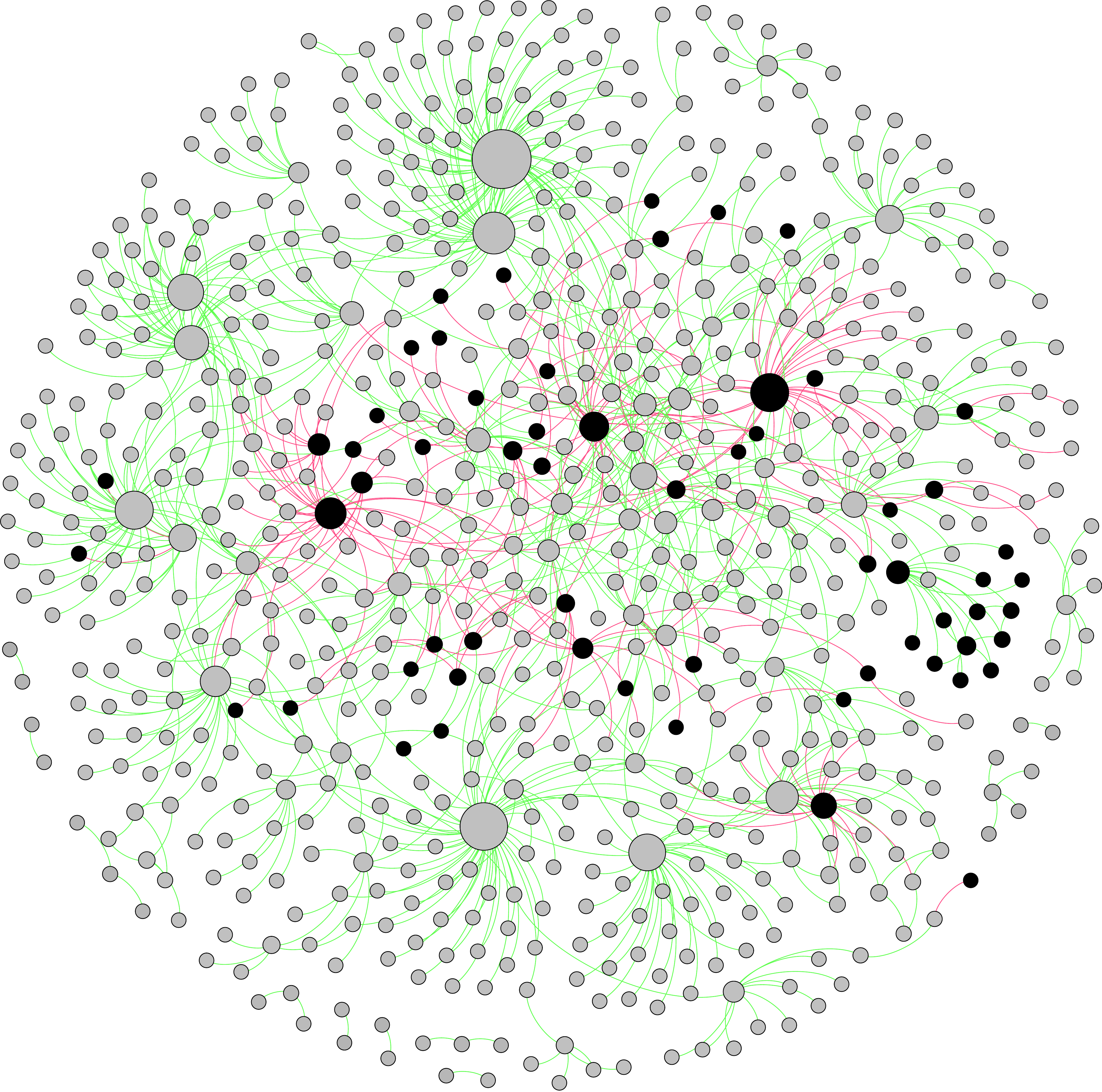}%
		\label{4fig3e}} 
	\hfil
	\caption{The gene regulatory network (a) and monotone subsystem (b) of \emph{Saccharomyces cerevisiae} obtained after removing 41 frustrated edges% and optimal node colours shown %(colour version online)
	}
	\label{4fig3.5}
\end{figure}

Removing the frustrated edges from yeast network, we obtain the monotone subsystem (induced subgraph) whose response to perturbations can be predicted from the underlying structure \cite{dasgupta_algorithmic_2007, maayan_proximity_2008}. %such as increasing the concentration on a node.
Figure \ref{4fig3.5} shows the yeast network and its corresponding monotone subsystem obtained by removing the frustrated edges. In the monotone subsystem represented in \ref{4fig3e} all walks connecting two given nodes have one specific sign. This prevents oscillation and chaotic behaviour \cite{dasgupta_algorithmic_2007, maayan_proximity_2008} and allows system biologists to predict how perturbing node A impacts on node B based on the sign of walks connecting A to B. All walks connecting nodes of the same colour (different colours) have a positive (negative) sign which leads to a monotone relationship between perturbation of one node and the impact on the other node.
%\FloatBarrier
In contrast for the two signalling networks, the level of frustration is very high, i.e., there are far more frustrated edges compared to the corresponding reshuffled networks. Networks of the EGFR protein and that of the macrophage are different from transcriptional networks in nature and our results show that they are far from balanced. This result is consistent with the discussions of Iacono et al.\ that EGFR and macrophage networks cannot be classified as close-to-monotone \cite[page 233]{iacono_determining_2010}.

Besides differences in network categories, one can see a structural difference between the transcriptional networks and signalling networks in Figures~\ref{4fig3} -- \ref{4fig3.5}. Subfigure \ref{4fig3b} and Subfigure \ref{4fig3a} show many high-degree nodes having mostly positive or mostly negative edges in the two transcriptional networks. However, such structures are not particularly common in the two signalling networks as visualised in Subfigures \ref{4fig3c} -- \ref{4fig3d}.

\subsection{Computations}

The two smallest biological networks considered here (EGFR and macrophage) are the largest networks analysed in a recent study of balancing signed networks by negating minimal edges \cite{Wang2016} in which the heuristic algorithm gives sub-optimal values of the frustration index \cite[Fig. 5]{Wang2016}. 

In Section \ref{3s:evaluate}, we compared the quality and solve time of our exact algorithm with that of recent heuristics and approximations implemented on the same datasets. While data reduction schemes \cite{huffner_separator-based_2010} may take up to 1 day for these four biological networks and heuristic algorithms \cite{iacono_determining_2010} only provide bounds with up to 9\% gap from optimality, our optimisation-based models (including the XOR model in Eq.\ \eqref{3eq4}) equipped with the speed-up techniques reach global optimality in a few seconds on an ordinary computer. %For the four biological networks, the speed-up techniques suggested by Aref et al.\ \cite{aref2016exact,aref2017computing} are useful in restricting the feasible space of optimisation models with thousands of binary variables. 

\section{International relations} \label{4s:temporal}\label{4s:d3}

International relations between countries can be analysed using signed networks models and balance theory \cite{lai_alignment_2001, doreian_partitioning_2013, Lerner2016}. In earlier studies of balance theory, Harary used signed relations between countries over different times as an example of balance theory applications in this field \cite{harary1961structural}.

\subsection{Datasets}

In this section, we analyse the frustration index in a temporal political network of international relations. Doreian and Mrvar have used the Correlates of War (CoW) datasets \cite{correlatesofwar2004} to construct a signed network with 51 sliding time windows each having a length of 4 years \cite{patrick_doreian_structural_2015}. Joint memberships in alliances, being in unions of states and sharing inter-governmental agreements are represented by positive edges. Being at war (or in conflict without military involvement) and having border disputes or sharp disagreements in ideology or policy are represented by negative edges \cite{patrick_doreian_structural_2015}.

A dynamic visualisation of the network can be viewed on the \urllink{https://youtu.be/STlNsTjYjAQ}{YouTube video sharing website} \cite{aref_youtube}. This temporal network represents more than half a century of international relations among countries in the post Second World War era starting with 1946-1949 time window and ending with 1996-1999 time window \cite{patrick_doreian_structural_2015}. One may refer to \cite[Section 3.4]{patrick_doreian_structural_2015} for a detailed explanation of using sliding time windows and other details involved in constructing the network. In the first time window of the temporal network, network parameters are $n=64$, $m=362$ and $m^-=42$. %These numbers change in each time window as a result of the changes not only in the relations, but change in the nodes and edges. 
In the last time window, these parameters are $n=151$, $m=1247$ and $m^-=147$. 

\subsection{Results}
\begin{figure}[ht]
	\centering
	\includegraphics[width=1\textwidth]{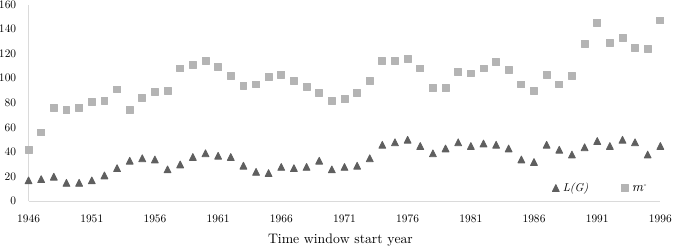}
	\caption{The number of negative edges $m^-$ and the frustration index $L(G)$ of the CoW dataset over time}
	\label{4fig4}
\end{figure}

Figure \ref{4fig4} demonstrates the number of negative edges and the frustration index in the CoW dataset. Doreian and Mrvar have attempted analysing the signed international network using the frustration index (under a different name) \cite{patrick_doreian_structural_2015} and other measures. They used a blockmodeling algorithm in \textit{Pajek} for obtaining the frustration index. However, their solutions are not optimal and thus do not give the frustration index for any of the 51 instances.
Even with reliable numerical results in hand, caution must be applied before answering whether this network has become closer to balance over the time period 1946-1999 \cite{harary_signed_2002, antal_social_2006, marvel_continuous-time_2011} or the simpler question, how close this network is to total balance (as discussed in Chapter \ref{ch:1}). 

\begin{figure}[ht]
	\centering
	\includegraphics[width=1\textwidth]{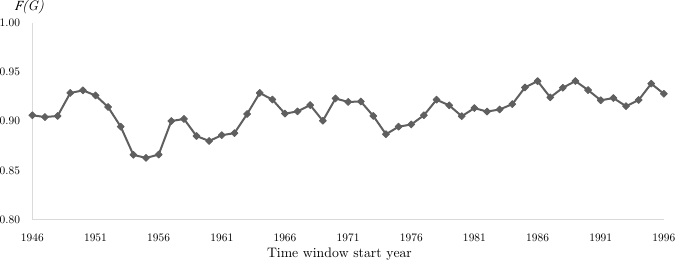}
	\caption{The normalised frustration index in the CoW dataset over time}
	\label{4fig5}
\end{figure}

Using Z scores, we observe tens of standard deviation difference between the frustration index of 51 CoW instances and the average frustration index of the corresponding reshuffled graphs. This indicates that the network has been comparatively close to a state of structural balance over the 1946-1999 period. This is contrary to the evaluation of balance by Doreian and Mrvar using their frustration index estimates \cite{patrick_doreian_structural_2015}. Bearing in mind that the size and order changes in each time window of the temporal network, we use the normalised frustration index, $F(G)=1-2L(G)/m$, in order to investigate the partial balance over time. Recall that $F(G)$ provides values in the unit interval where the value $1$ represents total balance (as discussed in Chapter \ref{ch:1}). Figure~\ref{4fig5} shows that all normalised frustration index values are greater than $0.86$. %For more information on the normalised frustration index and its interpretation, one may refer to \cite{aref2015measuring}. %We could have predicted this by the fact that $F(G) \geq 1-2m^-/m$. The largest value of $m^-/m$ in the Correlates of War dataset is $0.18$ in the 1948-1951 time window. Therefore, $F(G) \geq 0.64$ in all time windows of the Correlates of War dataset.
%\FloatBarrier

The data plotted in Figure~\ref{4fig5} can also be used to statistically test the stationarity of the normalised frustration index values. The Priestley-Subba Rao (PSR) test of non-stationarity \cite{priestley_test_1969} provides the means of a statistically rigorous hypothesis testing for stationarity of time series. We use an R implementation of the PSR test that is available in the \textit{fractal} package in the CRAN repository. The p-value of non-stationarity test for variation of $F(G)$ over time equals $0.03$ indicating that there is strong evidence to reject the null hypothesis of stationarity.

While there is no monotone trend in the values of $F(G)$, in most years the network has moved towards becoming more balanced over the 1946-1999 period. The overlap of the time period with the Cold War era may explain how the network has been close to a global state of bi-polarity with countries clustered into two antagonist sides. Doreian and Mrvar claim to have decisive evidence \cite{patrick_doreian_structural_2015} (based on frustration index estimates not showing monotonicity) against the theory \cite{harary_signed_2002, antal_social_2006, marvel_continuous-time_2011} that signed networks evolve towards becoming more balanced. Our observations based on $F(G)$ values do not reject this theory.

\subsection{Optimal partitioning}
We can investigate how the 180 countries of the CoW dataset are partitioned into two internally solidary but mutually hostile groups in this network. The optimal node colours show that 32 countries have remained in one fixed part, which we call group A, over the 1946-1999 period. Group A includes Argentina, Belgium, Bolivia, Brazil, Canada, Chile, Colombia, Costa Rica, Denmark, Dominican Republic, Ecuador, El Salvador, France, Great Britain, Guatemala, Haiti, Honduras, Iceland, Italy, Luxembourg, Mexico, Netherlands, Nicaragua, Norway, Panama, Paraguay, Peru, Portugal, Turkey, Uruguay, United States, and Venezuela. There are also 26 other countries that mostly (in over 40 time frames) belong to the same part as group A countries.
\begin{figure}
	\centering
	\includegraphics[height=5.4in, angle =90]{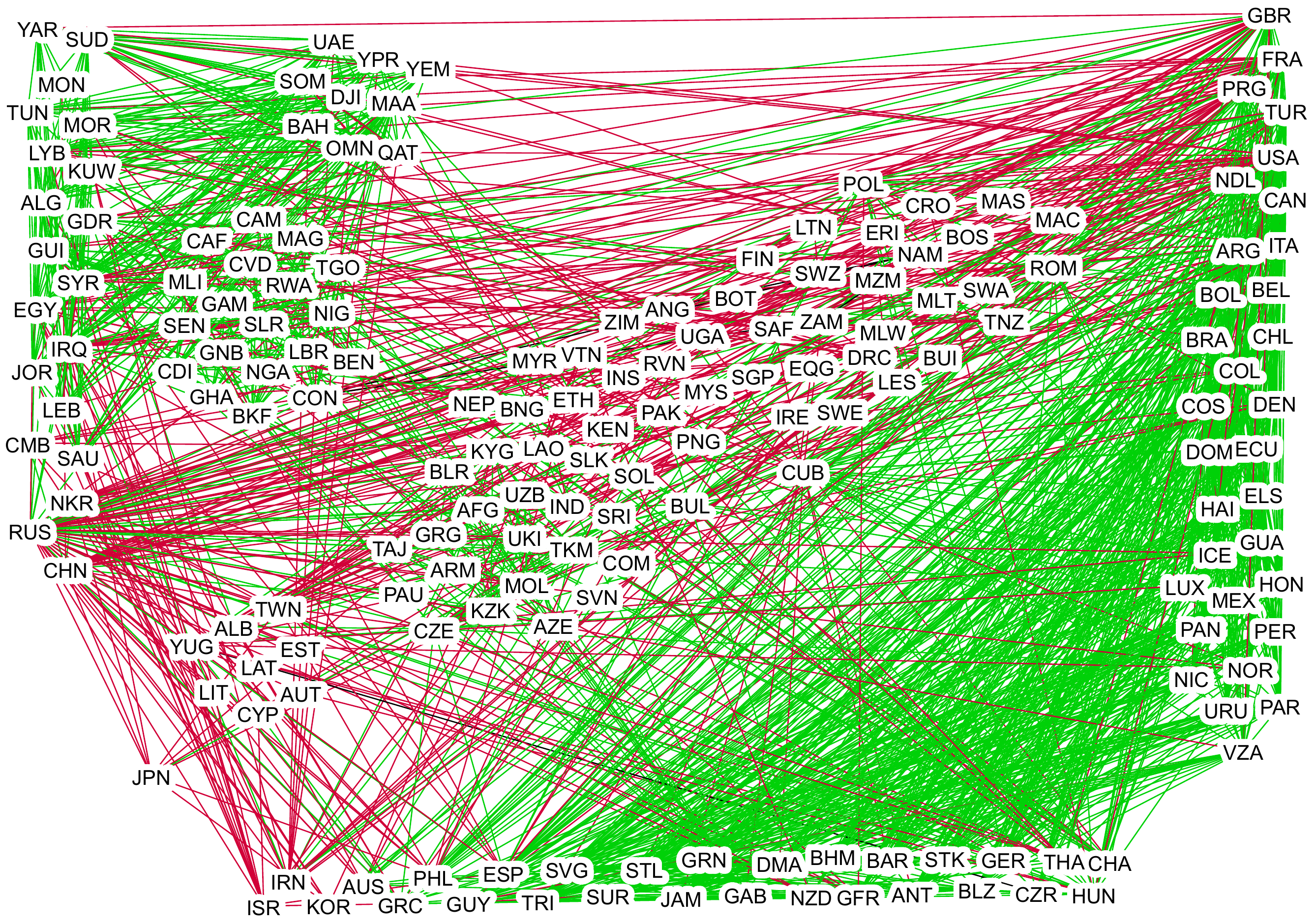}
	\caption{The partitioning of countries into groups A (right) and B (left) with most intra-group (inter-group) edges being positive (negative), the countries positioned in the bottom of the figure mostly belong to the same part as group A countries.}
	\label{4fig5.5}
\end{figure}

20 countries form another part, that we call group B, which opposes group A in over 40 time frames. Group B includes North Korea, Sudan, Tunisia, Morocco, Libya, Kuwait, Algeria, German Democratic Republic (East Germany), Guinea, Syria, Egypt, Iraq, Jordan, Lebanon, Russia, Saudi Arabia, Cambodia, Mongolia, China, and Yemen.

The optimal colours of the nodes associated with the remaining 102 countries change several times over the 1946-1999 period. Figure~\ref{4fig5.5} shows the partitioning of the countries in which groups A and B are positioned on the right and left sides respectively. The countries that mostly belong to the same part as group A countries are positioned at the bottom of Figure~\ref{4fig5.5}. The countries more inclined towards group B are positioned closer to the top left side of Figure~\ref{4fig5.5}.

\subsection{Computations}

For this dataset, the XOR model in Eq.\ \eqref{3eq4} provides the exact values of the frustration index in less than $0.1$ seconds on an ordinary computer with an Intel Core i5 7600 @ 3.50 GHz processor and 8.00 GB of RAM (as discussed in Chapter \ref{ch:3}).

%\clearpage
\section{Financial portfolios}\label{4ss:finance}\label{4s:d4}
%balance means more predictability  
%If the portfolio
%consists of negative edges and is balanced, then it is likely that the position has a hedging
%disposition within it

There are studies investigating financial networks of securities modelled by signed graphs \cite{harary_signed_2002,huffner_separator-based_2010, figueiredo2014maximum}. Harary et al.\ originally suggested analysing portfolios using structural balance theory \cite{harary_signed_2002}. They represented securities of a portfolio by nodes and the correlations between pairs of securities by signed edges \cite{harary_signed_2002}. They used $\pm0.2$ as thresholds for considering a signed edge between two securities of a portfolio. Simplifying a portfolio containing Dow Jones, London FTSE, German DAX, and Singapore STI to a signed graph with four nodes, they observed that the graph has remained in a state of balance from October 1995 to December 2000 \cite{harary_signed_2002}. H\"{u}ffner, Betzler, and Niedermeier considered portfolios containing 60-480 stocks and thresholds of $\pm0.325,\pm0.35,\pm0.375$ to evaluate the scalability of their algorithm for approximating the frustration index \cite{huffner_separator-based_2010}. Their dataset is also analysed in \cite{figueiredo2014maximum}.
%Analysing such large portfolios does not particularly have a practical application due to a phenomenon know as over-diversification \cite{??}.

\subsection{Datasets}

In this subsection, we consider well-known portfolios recommended by financial experts for having a low risk in most market conditions \cite{bogle2015}. These portfolios are known as lazy portfolios and usually contain a small number of \textit{well-diversified} securities \cite{bogle2015}. We consider 6 lazy portfolios each consisting of 5-11 securities. Table~\ref{4tab3} represents the six lazy portfolios and their securities.

\begin{table}[ht]
	\centering
	\caption{Six portfolios and their securities}
	\label{4tab3}
	\begin{tabular}{p{1.5cm}p{1.5cm}p{1.5cm}p{1.7cm}p{1.4cm}p{1.7cm}p{1.5cm}}
		\hline
		Portfolio & \urllink{http://tinyurl.com/y88fkv67}{Ivy portfolio (P1)} & \urllink{http://tinyurl.com/yc5scb9z}{Simple portfolio (P2)} & \urllink{http://tinyurl.com/y74k72e9}{Ultimate Buy \& Hold (P3)} & \urllink{http://tinyurl.com/y922s5f2}{Yale Endowment (P4)} & \urllink{http://tinyurl.com/ybtx7295}{Swensen's lazy portfolio (P5)} & \urllink{http://tinyurl.com/ybzalvzz}{Coffee House (P6)} \\ \hline
		Financial Expert&Mebane Faber&Larry Swedroe&Paul Merriman&David Swensen&David Swensen&Bill Schultheis\\ \hline
		VEIEX                                                                          &    & x  & x  & x  & x  &                        \\
		VGSIX                                                                          &    &    & x  & x  & x  & x                      \\
		VIPSX                                                                          &    & x  & x  & x  & x  &                        \\
		VTMGX                                                                          &    &    & x  & x  & x  &                        \\
		VIVAX                                                                          &    & x  & x  &    &    & x                      \\
		NAESX                                                                          &    & x  & x  &    &    & x                      \\
		EFV                                                                            &    & x  & x  &    &    &                        \\
		VFINX                                                                          &    &    & x  &    &    & x                      \\
		VFISX                                                                          &    &    & x  &    & x  &                        \\
		VISVX                                                                          &    &    & x  &    &    & x                      \\
		VTSMX                                                                          &    &    &    & x  & x  &                        \\
		IJS                                                                            &    & x  &    &    &    &                        \\
		TLT                                                                            &    &    &    & x  &    &                        \\
		VFITX                                                                          &    &    & x  &    &    &                        \\
		VBMFX                                                                          &    &    &    &    &    & x                      \\
		VGTSX                                                                          &    &    &    &    &    & x                      \\
		GSG                                                                            & x  &    &    &    &    &                        \\
		IEF                                                                            & x  &    &    &    &    &                        \\
		VEU                                                                            & x  &    &    &    &    &                        \\
		VNQ                                                                            & x  &    &    &    &    &                        \\
		VTI                                                                            & x  &    &    &    &    &                        \\ \hline
		$n$		&5 &6 &11 &6 &6 &7\\ \hline
	\end{tabular}
\end{table}

%The financial networks representing the lazy portfolios are generated by simply generating complete weighted graphs in which weights are the correlation coefficients between the securities. We use the correlation coefficient data of the securities that can be found on Portfolio Visualizer website \cite{Portfolio_visualizer}. We do not over-simplify the correlation coefficients by the arbitrary thresholding used in \cite{harary_signed_2002,huffner_separator-based_2010}. Instead we model the portfolio as a weighted signed graph and analyse it using the extended optimisation model for weighted signed graphs (\eqref{4eq21}).
%\urllink{http://tinyurl.com/y9byk8nn}{Bogleheads Three Fund}                     & 3   &   &  &  &  \\
%\urllink{http://tinyurl.com/y8koqrqs}{Harry Browne Permanent Portfolio}          & 4   &  &  &  &  \\
%\urllink{http://tinyurl.com/ybczam4z}{Bogleheads Four Funds}                     & 4   &  &    &   &   \\
%\urllink{http://tinyurl.com/y8uxh6fj}{Larry Swedroe Minimize FatTails Portfolio} & 4   &  &    &   &   \\
%\urllink{http://tinyurl.com/y7ln3evs}{Rick Ferri Core Four}                      & 4   &  &    &   &   \\ 
%\urllink{http://tinyurl.com/yapufyjr}{Bill Bernstein No Brainer}                 & 4   &  &    &   &   \\
%Table \ref{tab6} shows the securities in each of the six portfolios.
The signed networks representing the lazy portfolios are generated by considering prespecified thresholds as in \cite{harary_signed_2002,huffner_separator-based_2010}. We use the daily returns correlation data that can be found on the Portfolio Visualizer website \cite{Portfolio_visualizer} and thresholds of $\pm0.2$ similar to \cite{harary_signed_2002}. Correlation coefficients with an absolute value greater than $0.2$ are considered to draw signed edges between the securities with respect to the sign of correlation. 

\begin{figure}[ht]
	\subfloat[A weighted complete graph representing correlation coefficients, edge weights $\leq -0.2$ are shown on the edges]{\includegraphics[height=2.5in]{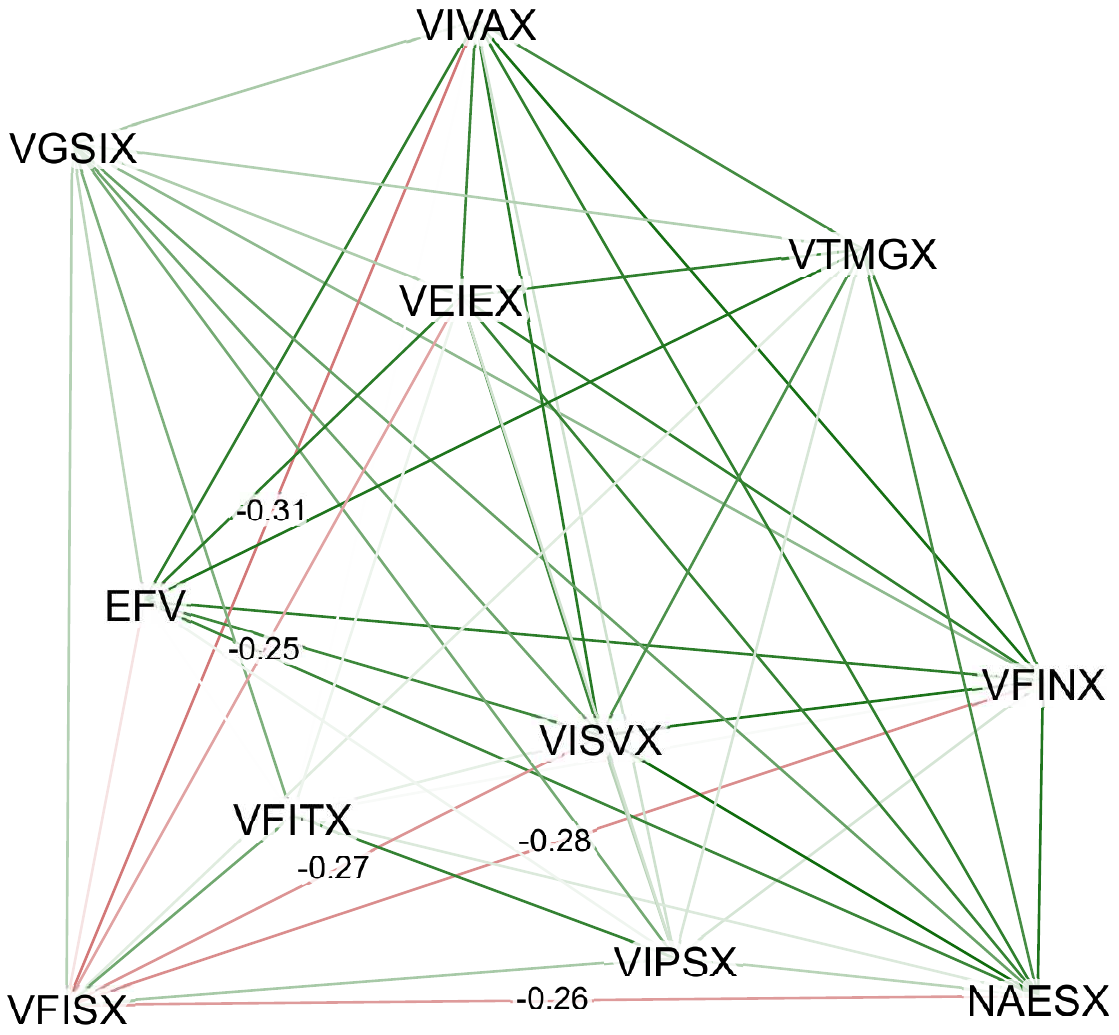}%
		\label{4fig6a}} 
	\hfil
	\subfloat[The portfolio signed graph produced by thresholding on $\pm0.2$, frustrated edges are shown by thick lines]{\includegraphics[height=2.5in]{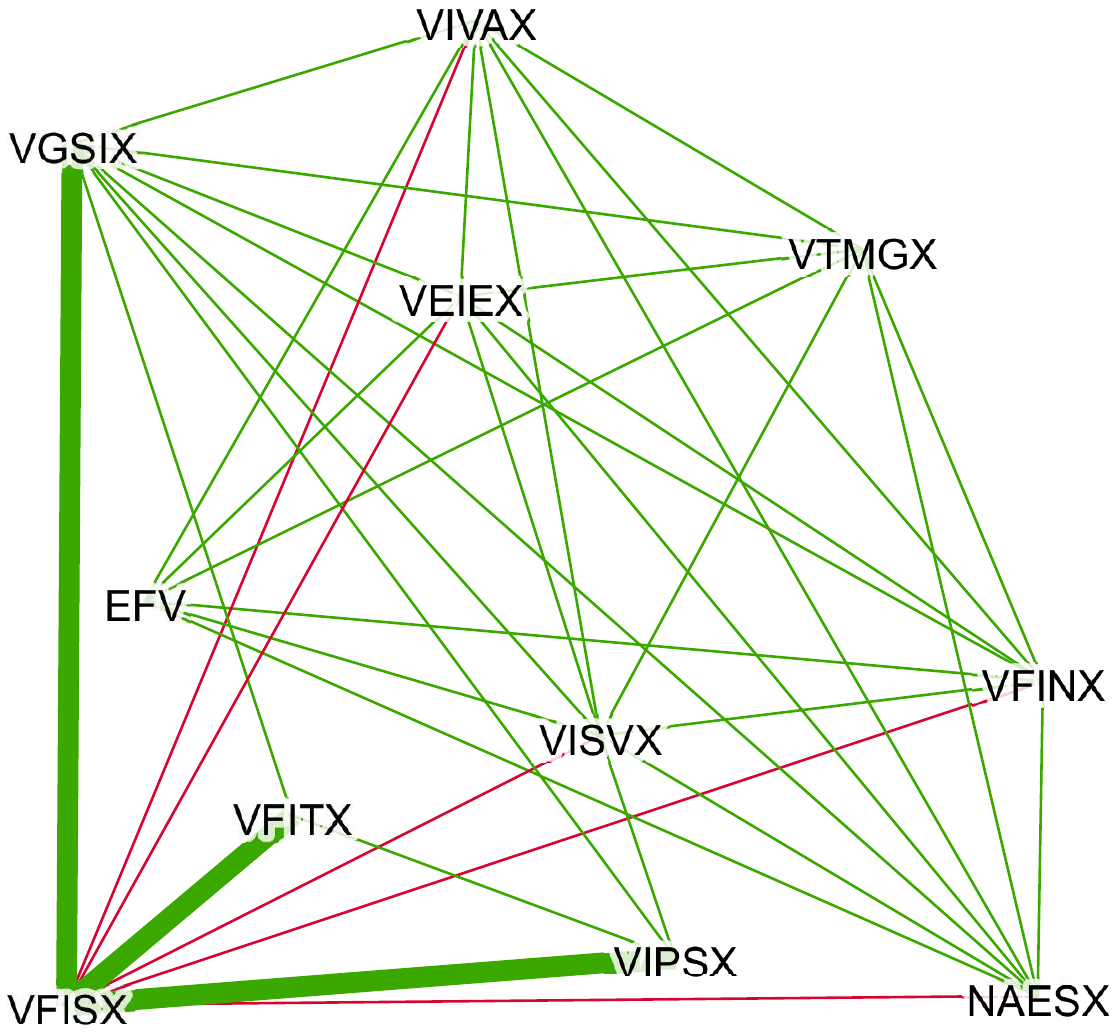}%
		\label{4fig6b}} 
	\hfil
	\caption{Portfolio P3 in 2016-10 (unbalanced) illustrated as (a) weighted and (b) signed networks using Gephi %(colour version online)
	}
	\label{4fig6}
\end{figure}

Figure \ref{4fig6} shows two networks of portfolio P3 based on the October 2016 data. The nodes represent 11 securities of the portfolio and the colours of the edges correspond to the correlations between the securities (green for positive and red for negative correlation). Lighter colours in Figure~\ref{4fig6} (a) represent smaller absolute values of correlation coefficient.

\subsection{Results}

\begin{figure}[ht]
	\centering
	\includegraphics[width=1\textwidth]{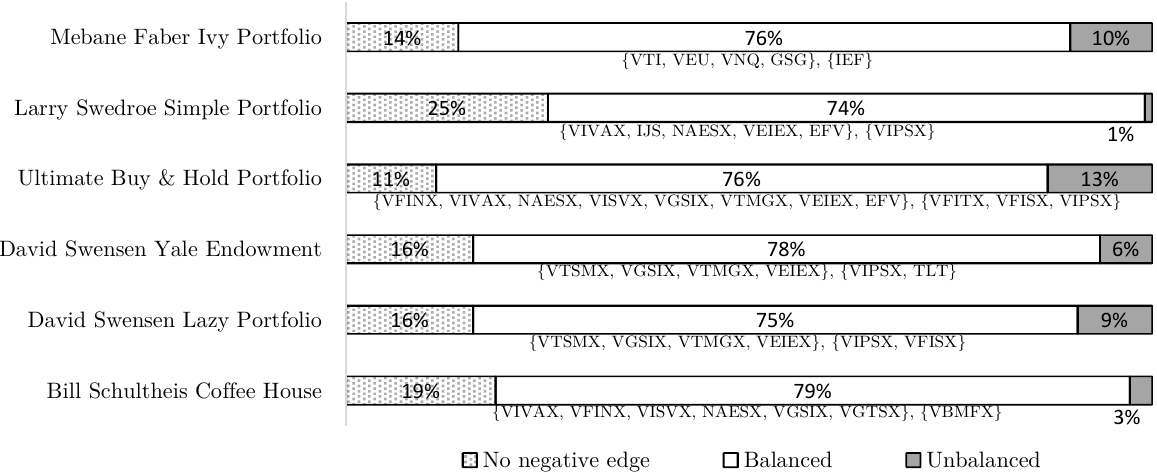}
	\caption{Frequencies of all-positive, balanced, and unbalanced networks over 108 monthly time frames}
	\label{4fig7}
\end{figure}

We analyse 108 monthly time frames for each of the six portfolios which correspond to the signed networks of each month within the 2008-2016 period. %In 11-25\% of the time frames, the portfolio signed networks do not have any negative edge. 
The signed networks obtained are totally balanced and have negative edges in a large number of time frames (74-79\%). In a relatively small number of time frames (1-13\%), the underlying network is unbalanced. Figure~\ref{4fig7} illustrates the results which are consistent with the findings of Harary et al.\ in \cite{harary_signed_2002} in terms of balanced states being dominant. 

%\FloatBarrier

%\FloatBarrier

\begin{figure}
	\centering
	\includegraphics[width=1\textwidth]{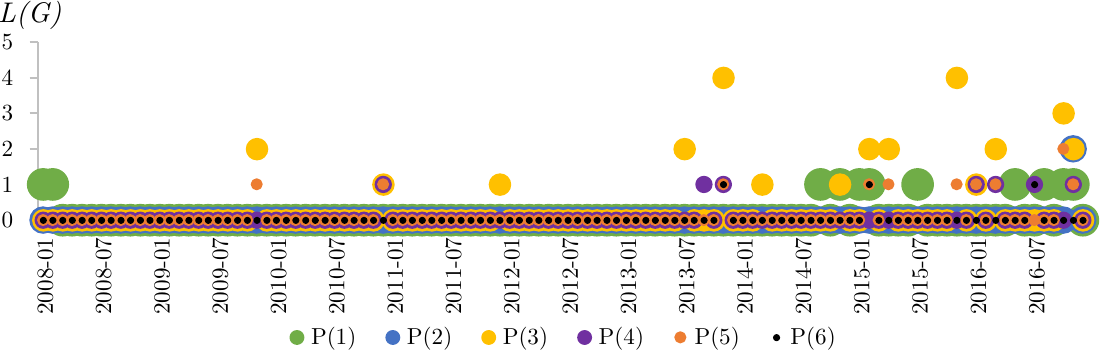}
	\caption{Frustration index of six portfolios over 108 monthly time frames %(colour version online)
	}
	\label{4fig9}
\end{figure}

More detailed results on balance states and frustration index of six portfolios over time are provided in Figure~\ref{4fig9}. We observe that there are some months when several portfolios have an unbalanced underlying signed graph (non-zero frustration index values). One may suggest that common securities explain this observation, but P(1) does not have any security in common with other portfolios which suggests otherwise. It can be observed from Figure~\ref{4fig9} that non-zero frustration index values are rather rare and usually very small.

%\begin{figure}
%	\centering
%	\includegraphics[width=1\textwidth]{fig9-eps-converted-to.pdf}
%	\caption{State of six portfolios over 108 monthly time frames %(colour version online)
%	}
%	\label{fig8}
%\end{figure}

\subsection{Optimal partitioning}

The optimal partitioning of each portfolio into two sub-portfolios (with positive correlations within and negative correlations in between) can be obtained from optimal node colours. The optimal partitioning remains mostly unchanged over balanced states among 108 time frames. Figure~\ref{4fig7} also shows the most common optimal partitioning of the securities for each portfolio. 

\subsection{Computations}

Regarding sensitivity of the results to the cut-off threshold, other thresholds (like $\pm0.1$ and $\pm0.3$) also lead to balanced states being dominant. Using thresholds of $\pm0.1$ leads to relatively more unbalanced states and less all-positive states, while thresholds of $\pm0.3$ have the opposite effect. Regarding computational performance for these small instances, a basic optimisation formulation of the problem with no speed-up technique (such as the model formulated in Eq.\ \eqref{2eq7}) would solve the portfolio instances in a reasonable time on an ordinary computer.

\section{Closely related problems from chemistry and physics}\label{4s:d56}

In this section, we briefly discuss two problems from chemistry and physics that are closely related to the frustration index of signed graphs. The parallels between these problems and signed graphs allow us to use the XOR model in Eq.\ \eqref{3eq4} to tackle the NP-hard computation of important measures for relatively large instances. We discuss computation of a chemical stability indicator for carbon molecules in Subsection \ref{4s:d5} and the optimal Hamiltonian of Ising models in Subsection \ref{4s:d6}.

\subsection{Bipartivity of fullerene graphs} \label{4ss:chemistry}\label{4s:d5}

\begin{figure}[ht]
	\centering
	\subfloat[Fullerene graph of C240]{\includegraphics[height=2.2in]{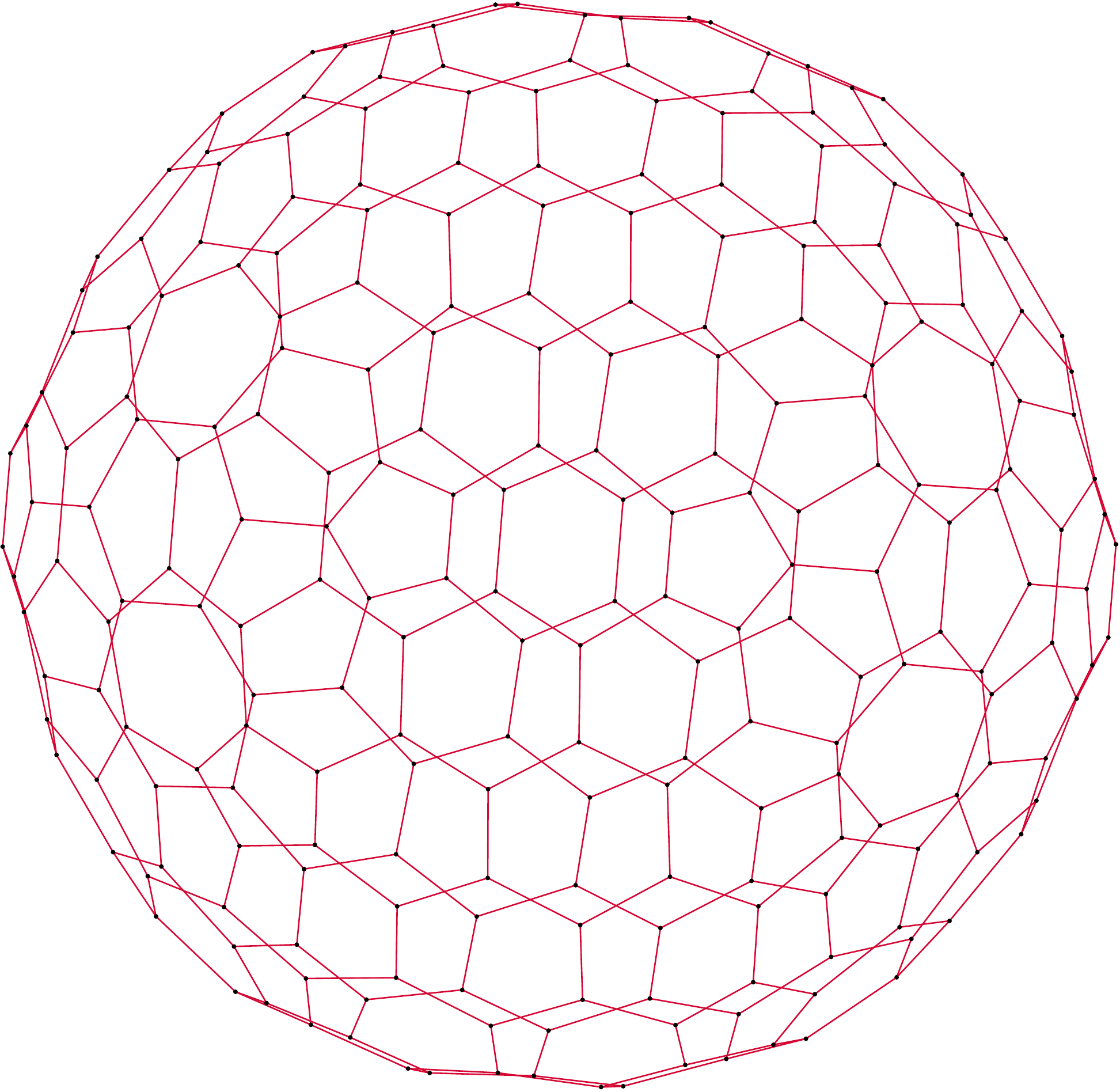}%
		\label{4fig10a}} 
	\hfil
	\subfloat[Fullerene graph of C240 made bipartite after removing 24 edges]{\includegraphics[height=2.2in]{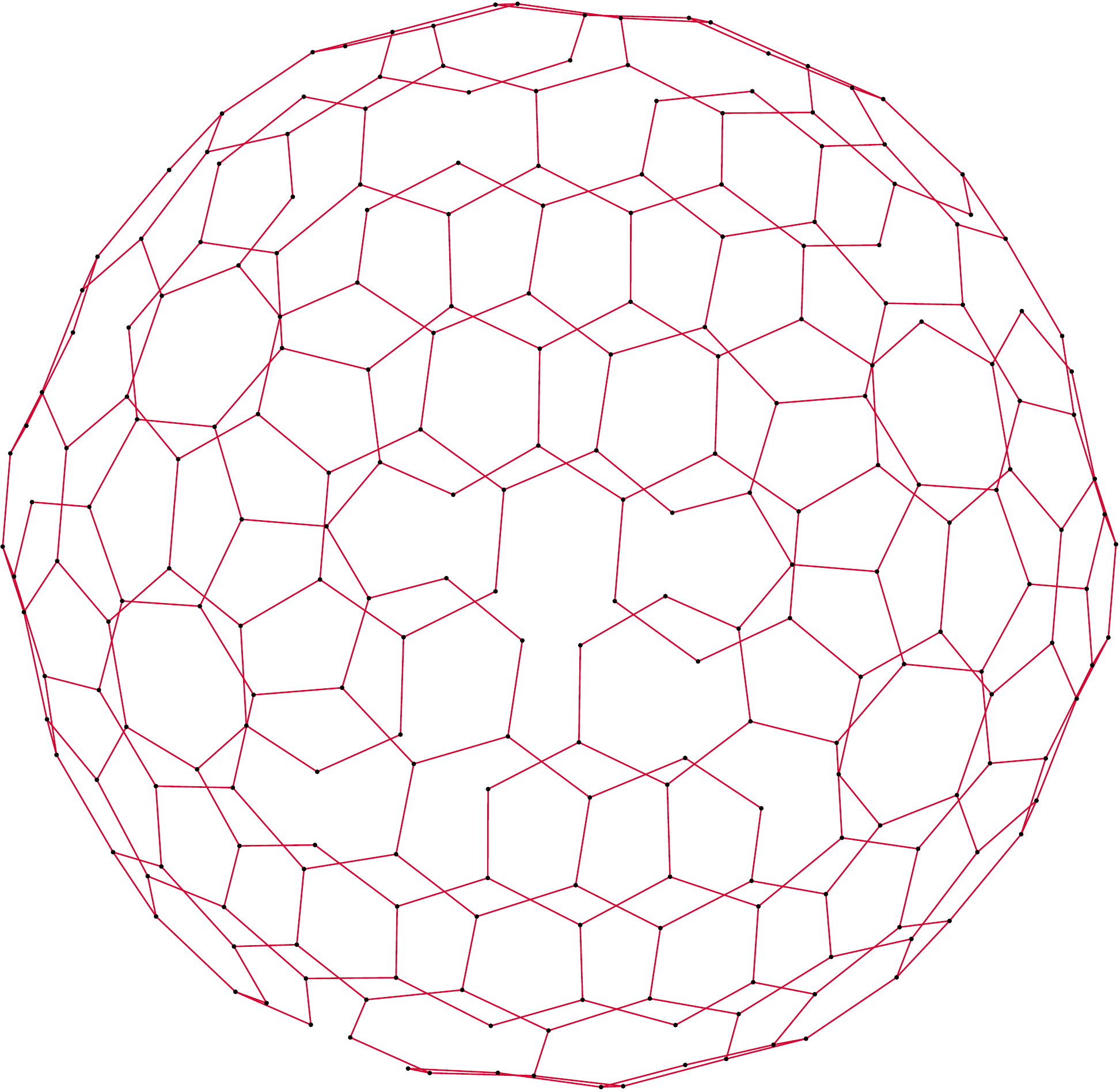}%
		\label{4fig10b}} 
	\hfil
	\subfloat[Fullerene graph of C2160]{\includegraphics[height=2.2in]{C2160-eps-converted-to.pdf}%
		\label{4fig10c}} 
	\hfil
	\subfloat[Fullerene graph of C6000]{\includegraphics[height=2.2in]{C6000-eps-converted-to.pdf}%
		\label{4fig10d}} 
	\caption{Several fullerene graphs represented as signed graphs in which all edges are negative %(colour version online)
	}
	\label{4fig10}
\end{figure}

Previous studies by Do{\v{s}}li{\'c} and associates suggest that graph bipartivity measures are potential indicators of chemical stability for carbon structures known as fullerenes \cite{doslic2005bipartivity, doslic_computing_2007}. The graphs representing fullerene molecular structure are called fullerene graphs where nodes and edges correspond to atoms and bonds of a molecule respectively.
Do{\v{s}}li{\'c} recommended the use of bipartivity measures in this context based on observing strong correlations between a bipartivity measure and several fullerene stability indicators. The correlations were evaluated on a set of eight experimentally verified fullerenes (produced in bulk quantities) with atom counts ranging between 60 and 84 \cite{doslic2005bipartivity}. 
Do{\v{s}}li{\'c} suggested using the \textit{spectral network bipartivity measure}, denoted by $\beta(G)$, which was originally proposed by Estrada et al.\ 
\cite{estrada2005}. 
This measure equals the proportion of even-length to total closed walks as formulated in 
\eqref{4eq5}
in which $\lambda_j$ ranges over eigenvalues of $|\textbf{A}|$ (the entrywise absolute value of adjacency matrix $\textbf{A}$). Note that $\beta(G)$ ranges between $0.5$ and $1$ and greater values represent more bipartivity. 
\begin{equation}\label{4eq5}
\beta(G)=\frac{\sum_{j=1}^{n} \cosh {\lambda_j}}{\sum_{j=1}^{n} e^{\lambda_j}}
\end{equation}
Two years later, Do{\v{s}}li{\'c} and Vuki{\v{c}}evi{\'c} suggested using the \textit{bipartite edge frustration} as a more intuitive measure of bipartivity to investigate the stability of fullerenes
\cite{doslic_computing_2007}. This measure equals the minimum number of edges that must be removed to make the network bipartite \cite{holme2003,yannakakis1981edge} and is closely related to the frustration index of signed graphs. Subfigure \ref{4fig10a} shows a graph that is made bipartite in Subfigure \ref{4fig10b} after removing 24 edges.
Do{\v{s}}li{\'c} and Vuki{\v{c}}evi{\'c} have observed no strong correlation between the bipartite edge frustration and $\beta(G)$ 
\cite{doslic_computing_2007}.
However, both measures have performed well in detecting the most stable fullerenes among all isomers with 60 and 70 atoms 
\cite{doslic_computing_2007}. 
More recently, Estrada et al.\  \cite{estrada2016} 
proposed \textit{spectral bipartivity index}, denoted as $b_s(G)$ and formulated in \eqref{4eq6},
as a bipartivity measure with computational advantages over $\beta(G)$. Note that $b_s(G)$ ranges between $0$ and $1$ and greater values represent more bipartivity.
%The upper bound is obtained if and only if the graph is bipartite (which has no odd-length closed walks) and the lower bound is reached by the complete graph (which asymptotically has an equal number of odd- and even-length closed walks).
\begin{equation}\label{4eq6}
b_s(G)=\frac{\sum_{j=1}^{n} e^{-\lambda_j}}{\sum_{j=1}^{n} e^{\lambda_j}}=\frac{\Tr(e^{-\textbf{A}})}{\Tr(e^{\textbf{A}})}
\end{equation}
%There are two other frustration-based measures of bipartivity suggested in \cite{holme2003}.Both measures are based on approximating $L(G)$ and we do not consider them.

\subsubsection{Relevance}
The bipartite edge frustration of a graph is equal to the frustration index of the signed graph, $G$, obtained by declaring all edges of the fullerene graph to be negative. Using this analogy, we provide some results on the bipartivity of large fullerene graphs. 
According to Do{\v{s}}li{\'c} et al., a motivation for using the bipartite edge frustration is exploring the range of atom counts for which there are no confirmed stable isomers yet %This measure has the advantage of exact computation over spectral indicators like $\beta(G), b_s(G)$.}
\cite{doslic_computing_2007}.
The least bipartite fullerene graphs represent the most stable fullerene isomers \cite{doslic2005bipartivity, doslic_computing_2007}.
Therefore, lower bipartivity (smaller values of $\beta(G)$, $b_s(G)$, and $F(G)$) can be interpreted as higher stability.

\subsubsection{Datasets}
We use the XOR model in Eq.\ \eqref{3eq4} to compute the bipartite edge frustration of several fullerene graphs with atom count ranging from 180 to 6000. The fullerene graph of C240 (molecule with 240 carbon atoms) and its bipartite subgraph are visualised in Subfigures \ref{4fig10a} -- \ref{4fig10b} followed by C2160 and C6000 fullerene graphs in Subfigures \ref{4fig10c} -- \ref{4fig10d}.

Among the 14 fullerene graphs we consider are the \textit{icosahedral fullerenes} that have the structure of a truncated icosahedron. It is conjectured that this family of fullerenes has the highest chemical stability among all fullerenes with $n$ atoms \cite{doslic_computing_2007, faria2012odd}.
% The algorithm is based on finding a spanning subgraph with no vertices of odd degree for the dual of the fullerene graph. This uses an $O(n^3)$ algorithm for minimum weight perfect matching
%Fowler et al.\ discuss that the connection between independence number and fullerene stability is not clear.
%Fowler PW, Daugherty S, Myrvold W, Independence number and fullerene stability,
%Chem Phys Lett 448:75–82, 2007
% independence number is a quadratic function of bipartite edge frustration
% Tomislav Došlić (2007): independence number has a significant stability predicting potential, but is much more difficult to compute

\subsubsection{Results}
%We use the binary linear model \eqref{eq4} to compute the bipartite edge frustration for giant fullerene molecules with up to 6000 carbon atoms.
The values of bipartivity measures for 14 fullerene graphs are computed in Table~\ref{4tab4} where we have also provided the normalised frustration index, $F(G)=1-2L(G)/m$, to compare the bipartivity of fullerenes with different atom counts. The closeness of $F(G)$, $\beta(G)$, and $b_s(G)$ values to 1 are consistent with fullerene graphs being almost bipartite (recall that these three measures take value 1 for a bipartite graph) \cite{doslic2005bipartivity}.
\begin{table}[ht]
	\centering
	\caption{Bipartivity measures computed for a range of large fullerene graphs}
	\label{4tab4}
	\begin{tabular}{llllll}
		\hline
		Fullerene graph & $m$  & $L(G)$ & $F(G)$  & $\beta(G)$ & $b_s(G)$ \\ \hline
		C180            & 270  & 18     & 0.86667 & 0.99765    & 0.99529  \\
		C240$^\dagger$  & 360  & 24     & 0.86667 & 0.99823    & 0.99647  \\
		C260            & 390  & 24     & 0.87692 & 0.99837    & 0.99674  \\
		C320            & 480  & 24     & 0.9     & 0.99867    & 0.99735  \\
		C500            & 750  & 30     & 0.92    & 0.99915    & 0.99830  \\
		C540$^\dagger$  & 810  & 36     & 0.91111 & 0.99921    & 0.99843  \\
		C720            & 1080 & 36     & 0.93333 & 0.99941    & 0.99882  \\
		C960$^\dagger$  & 1440 & 48     & 0.93333 & 0.99956    & 0.99912  \\
		C1500$^\dagger$ & 2250 & 60     & 0.94667 & 0.99972    & 0.99943  \\
		C2160$^\dagger$ & 3240 & 72     & 0.95556 & 0.99980    & 0.99961  \\
		C2940$^\dagger$ & 4410 & 84     & 0.96190 & 0.99986    & 0.99971  \\
		C3840$^\dagger$ & 5760 & 96     & 0.96667 & 0.99989    & 0.99978  \\
		C4860$^\dagger$ & 7290 & 108    & 0.97037 & 0.99991    & 0.99983  \\
		C6000$^\dagger$ & 9000 & 120    & 0.97333 & 0.99993    & 0.99986  \\ \hline
		\multicolumn{4}{l}{$\dagger$ icosahedral fullerene}
	\end{tabular}
\end{table}

Both spectral measures provide a monotone increase in the bipartivity values with respect to increase in atom count. However, $F(G)$ seems to provide distinctive values for icosahedral fullerenes. In particular for this set of fullerenes, we observe $F(C240)\not>F(C180)$, $F(C540)\not>F(C500)$, and $F(C960)\not>F(C720)$ which are consistent with the conjecture that icosahedral fullerenes are the most stable isomers \cite{doslic_computing_2007, faria2012odd}. 
%It is a conjecture that icosahedral fullerenes with $n$ vertices have a bipartite edge frustration of $L(G)=\sqrt{12n/5}$ \cite[Conjecture 13]{doslic_computing_2007}. The conjecture has not been tested due to the complexity involved in computing bipartite edge frustration \cite{doslic_computing_2007}. Our results for icosahedral fullerene in Table~\ref{4tab4} (marked with *) are in alignment with the conjecture.
\subsubsection{Computations}

Computing the bipartite edge frustration of a graph in general is computationally intractable and heuristic and approximation methods are often used instead \cite{holme2003}. For bipartite edge frustration of fullerene graphs which are planar; however, a polynomial time algorithm of complexity $\mathcal{O}(n^3)$ exists \cite{doslic_computing_2007}. Previous works suggest that this algorithm cannot process graphs as large as $n=240$ \cite{doslic_computing_2007}. Our computations for obtaining $L(G)$ of fullerene graphs with 180--2940 atoms take from split second to a few minutes. The solve times for computing $L(G)$ of C3840, C4860, and C6000 are 29.8, 68.1, and 97.5 minutes respectively.

As indicated by Table~\ref{4tab4} results, the XOR model in Eq.\ \eqref{3eq4} 
allows computing frustration-based measures of bipartivity in the range of atom counts for which there are no experimentally verified stable isomers yet. The performance of frustration-based fullerene stability indicators requires further research that is beyond our discussion in this chapter.

%\clearpage
\subsection{Ising models with $\pm1$ interactions}\label{4s:d6}

Closely related to the frustration index of signed graphs, are the ground-state properties of Ising models. The most simple and standard form of Ising models represents patterns of atomic magnets based on interactions among spins and their nearest neighbours. A key objective in Ising models with $\pm1$ interactions is finding the spin configurations with the minimum energy \cite{friedli2017statistical}. The standard nearest-neighbour Ising model with $\pm1$ interactions and no external magnetic field is explained in what follows.

Each spin is connected to its neighbours in a grid-shaped structure. Two connected spins have either an aligned or an unaligned coupling. The positive (negative) interaction between two spins represents a coupling constant of $J_{ij}=+1$ ($J_{ij}=-1$) alternatively called matched (mismatched) coupling. Under another terminology from physics, positive and negative edges are referred to as ferromagnetic bonds and anti-ferromagnetic bonds respectively \cite{zaslavsky2017}. Each spin can either take an upward or a downward configuration. We discuss finding a spin configuration for a given set of fixed coupling constants that minimises an energy function \cite{friedli2017statistical}.

Frustration arises if and only if a matched (mismatched) coupling has different (same) spin configurations on the endpoints. The energy of a spin configuration is calculated based on the Hamiltonian function: $H=-\sum_{ij} J_{ij} s_i s_j$ in which the sum $\sum_{ij}$ is over all the coupled spins. Note that $J_{ij}$ represent the couplings limited to $\pm1$ in the Ising model with the type of interactions relevant to this chapter. $s_1, s_2, \dots , s_n$ are the decision variables that take values $+1$ or $-1$ and represent upward/downward spin configurations. The Hamiltonian function of these Ising models is very similar to the energy function in \cite{facchetti_computing_2011} that is also used in other studies \cite{iacono_determining_2010, esmailian_mesoscopic_2014, ma_memetic_2015, ma_decomposition-based_2017, Wang2016}. The problem of minimising $H$ over all possible spin configurations is NP-hard for many structures \cite{liers2004computing}.

\begin{figure}[ht]
	\centering
	\subfloat[2D Ising model $50\times50$]{\includegraphics[height=2in]{grid2d-eps-converted-to.pdf}%
		\label{4fig11a}} 
	\hfil
	\centering
	\subfloat[3D Ising model $10\times10\times10$]{\includegraphics[height=2.2in]{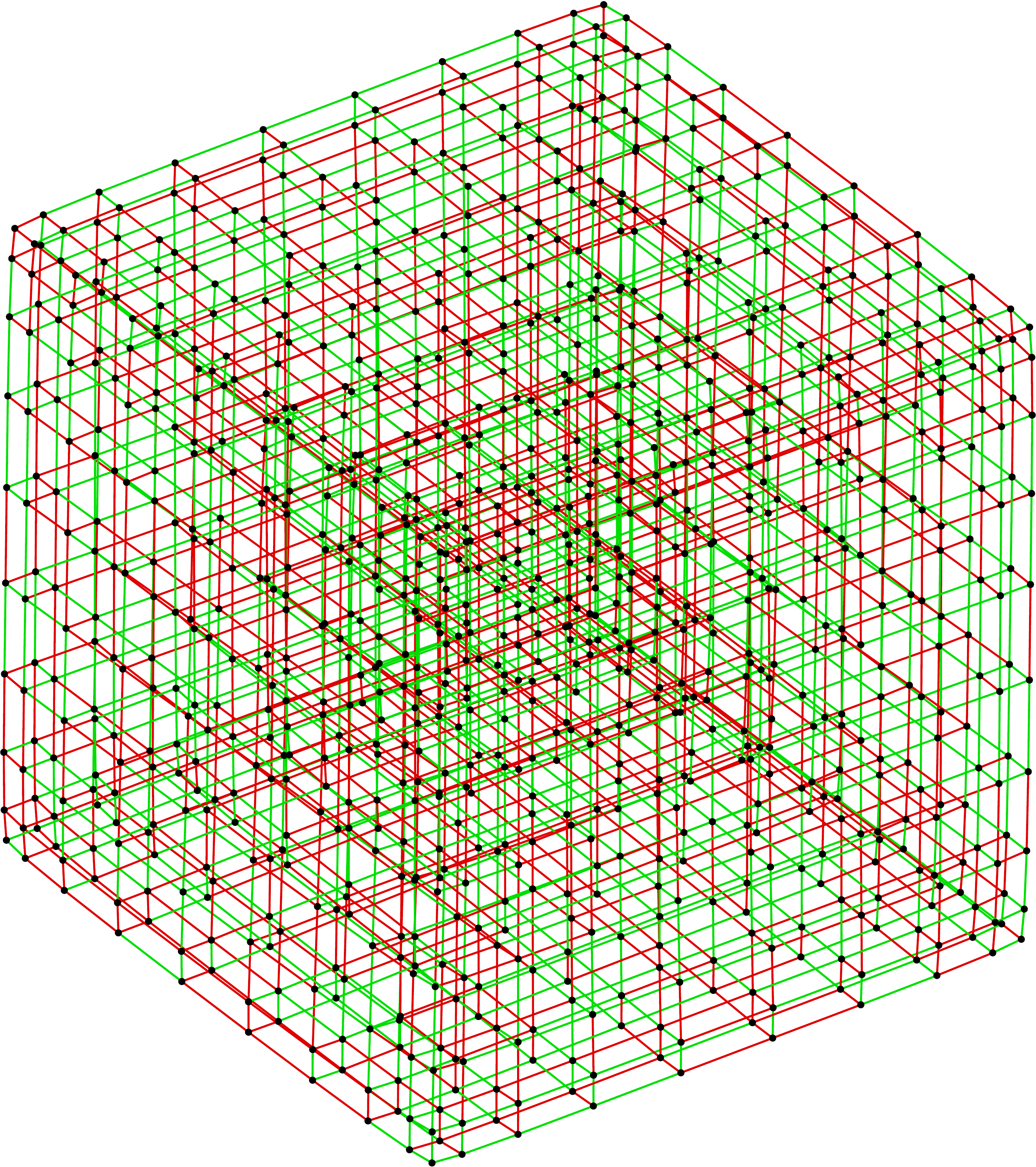}%
		\label{4fig11b}} 
	\hfil
	\caption{Signed graphs with two and three dimensional structure representing simple Ising models with $50\%$ unaligned couplings %(colour version online)
	}
	\label{4fig11}
\end{figure}
%\FloatBarrier

\subsubsection{Relevance}

In order to make a connection between Ising models and signed graphs, we represent spins by nodes and spin configurations by node colours. Signs on the edges represent coupling constants where matched and mismatched couplings between spins are modelled as positive and negative edges respectively.

If $X^*$ represents the optimal colouring leading to $L(G)$ for a given signed graph, the minimum value of the corresponding Hamiltonian function can be calculated by $H(X^*)=- \sum_{i,j} a_{ij}(2x_{i}-1)(2x_{j}-1)$. The minimum value of $H$ is obtained based on the optimal spin configuration associated with $X^*$. Alternatively, one may consider the fact that frustrated edges and non-frustrated (satisfied) edges contribute values $1$ and $-1$ to the Hamiltonian function respectively. For an Ising model with $m$ edges, this gives $H(X^*)=%(1)L(G) + (-1)(m - L(G))=
 2L(G)-m$ as the optimal Hamiltonian function value.

\subsubsection{Datasets}

\begin{figure}[ht]
	\centering
	\subfloat[Dimension 4]{\includegraphics[height=1.4in]{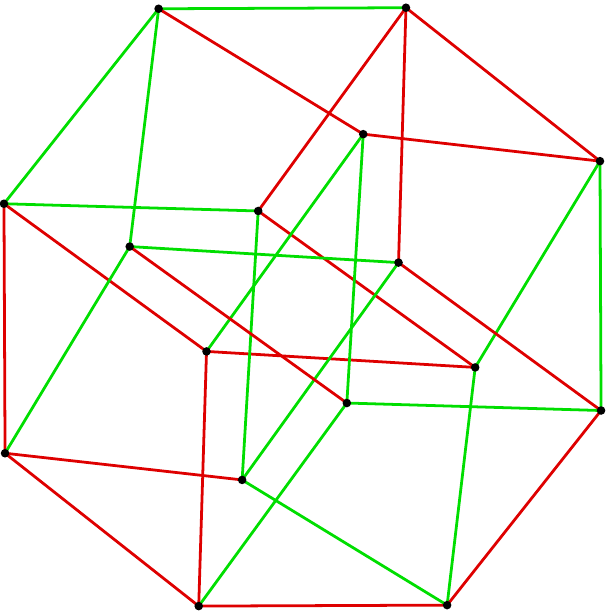}%
		\label{4fig12a}} 
	\hfil
	\subfloat[Dimension 5]{\includegraphics[height=1.4in]{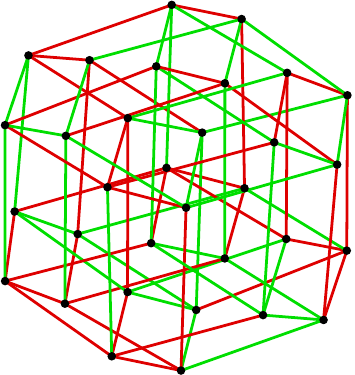}%
		\label{4fig12b}} 
	\hfil
	\subfloat[Dimension 6]{\includegraphics[height=1.4in]{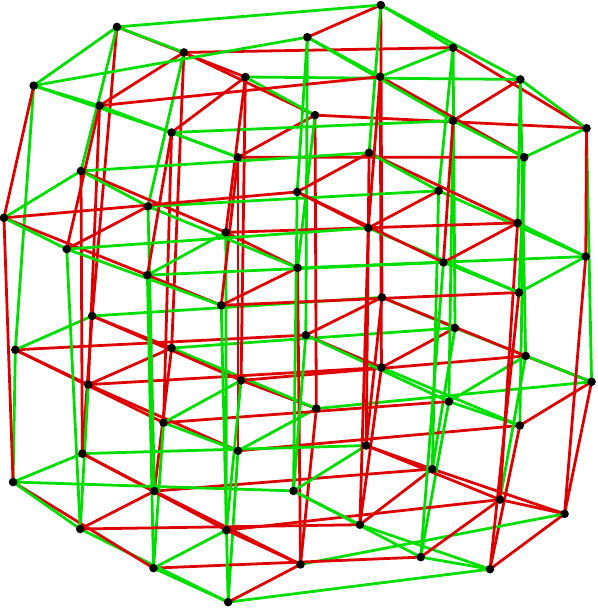}%
		\label{4fig12c}} 
	\hfil
	\subfloat[Dimension 7]{\includegraphics[height=1.7in]{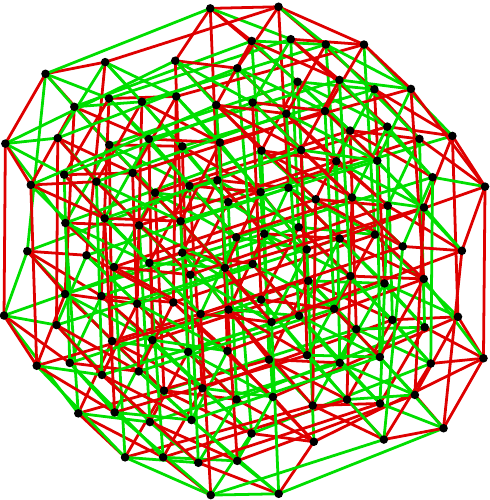}%
		\label{4fig12d}} 
	\hfil
		\subfloat[Dimension 8]{\includegraphics[height=1.7in]{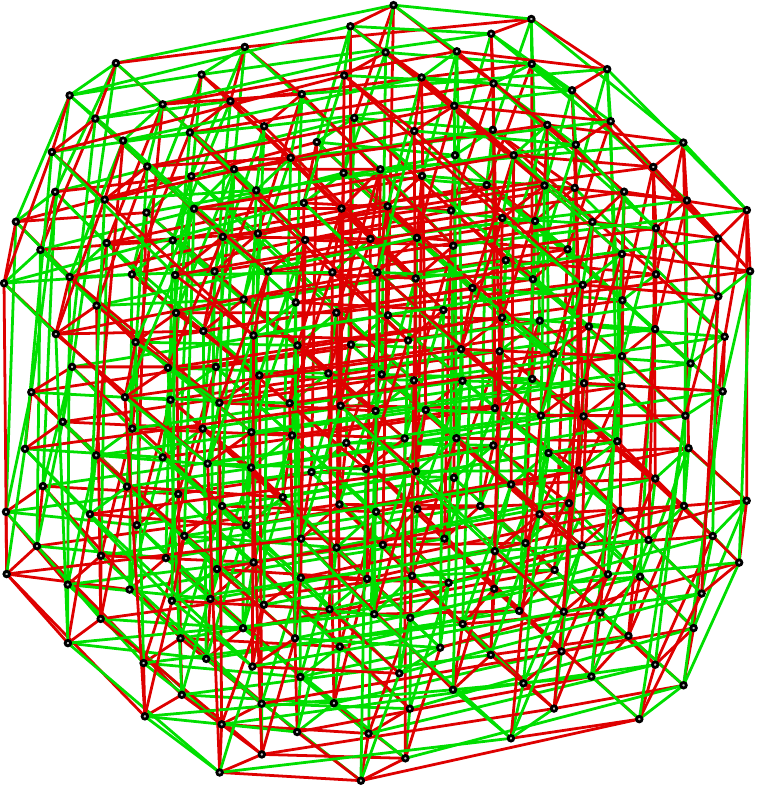}%
		\label{4fig12e}}
	\hfil 
	\caption{Five signed graphs with hypercube structure representing more structurally complex Ising models with $50\%$ unaligned couplings %(colour version online)
	}
	\label{4fig12}
\end{figure}

We use the XOR model in Eq.\ \eqref{3eq4} to compute the frustration index in Ising models of various grid size and dimension for several 2D and 3D grid structures and hypercubes. Figure~\ref{4fig11} illustrates a 2D and a 3D Ising model with $50\%$ unaligned couplings. Five hypercubes of dimension 4--8 with $50\%$ unaligned couplings are visualised in Figure~\ref{4fig12}.

\FloatBarrier

For each Ising model, we generate 10 grids and randomly assign $\pm1$ to the edges to achieve the pre-defined proportion of negative edges based on our experiment settings. For each Ising model with a specific dimension (Dim.) and grid size (Gri.), we consider three experiment settings with $m^-/m \in \{ 25\%, 50\%, 75\%\}$.

\subsubsection{Results}

Table \ref{4tab5} provides results on Ising models with a fixed dimension and grid size in each row. The mean and standard deviation of the frustration index values and the average solve time (in seconds) for each experiment setting are provided in Table~\ref{4tab5}.

\begin{table}[ht]
	\centering
	\caption{Frustration index values (and average solve times in seconds) for several Ising models}
	\label{4tab5}
	\begingroup
	\renewcommand{\arraystretch}{1.5}
	\begin{tabular}{p{0.8cm}p{0.8cm}p{2.5cm}p{2.5cm}p{2.5cm}}
		\hline
		&										       & $m^-/m=25\%$        & $m^-/m=50\%$       & $m^-/m=75\%$       \\ \cline{3-5} 
		{Dim., \quad Gri.} 							    &      {$n$, \quad \quad$m$}                   & $L(G)$ mean$\pm$SD (solve time)   & $L(G)$ mean$\pm$SD (solve time)  & $L(G)$ mean$\pm$SD (solve time)\\ \hline 
		2,    \quad                  50                      & 2500, 4900                                   & 691.1$\pm$12 \quad (28.2)  & 720.9$\pm$9.2 \quad (31.9) & 687.7$\pm$10.5 \quad (42.3)  \\ 
		2,                      100                     & 10000, 19800                                 & 2814.1$\pm$16 \quad (2452.3)     & 2938.2$\pm$22.3  \quad (2660.8)  & 2802.5$\pm$24.7 \quad (4685.2) \\ 
		2,                      150                     & 22500, 44700                                 & 6416.1$\pm$25.9  \quad (5256.2)  & 6698.5$\pm$55.3  \quad (5002.0)  & 6396.3$\pm$41.7 \quad (6761.0) \\ 
		2,                      200                     & 40000, 79600                                 & 11449$\pm$47     \quad (13140.6)  & 11930.3$\pm$58.9 \quad (12943.7)  & 11411.8$\pm$46  \quad (22720.4) \\ \hline 
		3,  \quad                    5                       & 125,         300                             & 51.4$\pm$1.7     \quad \quad (0.1)  & 52.4$\pm$2.5     \quad \quad (0.1)  & 51$\pm$3.2     \quad \quad (0.1)  \\  
		3,  \quad                    10                      & 1000,        2700                            & 491.5$\pm$7.5    \quad (82.9)  & 509.1$\pm$4      \quad (539.0)  & 491.6$\pm$7    \quad (96.4)  \\  
		3,   \quad                   15                      & 3375,       9450                             & 1762.1$\pm$16.1  \quad (8488.2)  & 1839.1$\pm$10.4  \quad (21384.3)  & 1761.1$\pm$14.2 \quad (9244.1) \\ \hline
		4,   \quad                   2                       & 16,             32                           & 5.6$\pm$0.8   \quad   \quad (0.1)  & 4.8$\pm$1   \quad   \quad    \quad (0.1)  & 5.6$\pm$0.8  \quad   \quad (0.1)  \\  
		5,   \quad                   2                       & 32,            80                            & 14.5$\pm$1.1   \quad   \quad (0.1)  & 15$\pm$1.2   \quad     \quad (0.1)  & 15$\pm$1.2   \quad   \quad (0.1)  \\  
		6,   \quad                   2                       & 64,           192                            & 38.8$\pm$2    \quad    \quad (0.1)  & 41$\pm$1.6    \quad    \quad (0.1)  & 38$\pm$2.2   \quad   \quad (0.1)  \\  
		7,   \quad                   2                       & 128,          448                            & 94.6$\pm$3.1   \quad   \quad (0.6)  & 99.6$\pm$3.2   \quad   \quad (1.6)  & 96$\pm$2.4   \quad   \quad (1.1)  \\  
		8,   \quad                   2                       & 256,         1024                            & 232.4$\pm$3.7    \quad (206.1)  & 245.8$\pm$3.6    \quad (4742.2)  & 231$\pm$4.7    \quad (543.87)  \\ \hline
	\end{tabular}
	\endgroup
\end{table}

%Intuitively, we expect more unbalanced cycles (and a greater frustration index) when $m^-/m = 50\%$ given that in the structures considered here all cycles have an even length. 
The results in Table~\ref{4tab5} show that in most cases the Ising model with $m^-/m = 50\%$ has the highest frustration index value (and therefore the highest optimal Hamiltonian value) among models with a fixed grid size and dimension. This can be explained by considering that in the structures investigated in Table~\ref{4tab5} all cycles have an even length. Therefore a higher number of negative cycles (each containing at least one frustrated edge) is obtained when the number of positive and negative edges are equal.

\subsubsection{Computations}

Hartmann and collaborators have suggested efficient algorithms for computing the ground-state properties in 3-dimensional Ising models with 1000 nodes \cite{hartmann2015matrix} improving their previous contributions in 1-, 2-, and 3-dimensional \cite{hartmann2014exact,hartmann2011ground,hartmann2013information} Ising models. Recently, they have used a method for solving binary optimisation models to compute the ground state of 3-dimensional Ising models containing up to $268^3$ nodes \cite{hartmann2016revisiting}.

While there are computational models for specialised Ising models based on the type of underlying structure \cite{hartmann2014exact,hartmann2011ground,hartmann2013information,hartmann2015matrix,hartmann2016revisiting}, the XOR model in Eq.\ \eqref{3eq4} can be used as a general purpose computational method for finding the ground state of Ising models with $\pm1$ interactions regardless of the underlying structure.

%\clearpage

\section{Conclusion of Chapter \ref{ch:4}} \label{4s:conclu}

%In this chapter, we discussed the applications of a recently developed computational method in six different disciplines where network models have signed edges. The present study applies mathematical programming techniques to tackle the NP-hard computation of a standard measure of balance suggested in late 1950's \cite{abelson_symbolic_1958,harary_measurement_1959} and known as the frustration index (line index of balance).

In this chapter, the frustration index is used for analysing a wide range of signed networks from sociology and political science (Section \ref{4s:d1}), biology (Section \ref{4s:d2}), international relations (Section \ref{4s:d3}), finance (Section \ref{4s:d4}), and chemistry and physics (Section \ref{4s:d56}) unifying the applications of a fundamental graph-theoretic measure. Our results contribute additional evidence that suggests many signed networks in sociology, biology, international relations, and finance exhibit a relatively low level of frustration which indicates that they are relatively close to the state of structural balance.

The numerical results also show the capabilities of the optimisation-based model in Eq.\ \ref{3eq4} in making new computations possible for large-scale signed networks with up to $10^5$ edges. The mismatch between exact optimisation results we provided on social and biological networks in Sections \ref{4s:d1} -- \ref{4s:d2} and those in the literature \cite{dasgupta_algorithmic_2007, huffner_separator-based_2010, iacono_determining_2010, facchetti_computing_2011, ma_memetic_2015, ma_decomposition-based_2017} suggests the necessity of using accurate computational methods in analysing signed networks. This essential consideration is more evident from our results on international relations networks in Section~\ref{4s:d3} where inaccurate computational methods in the literature \cite{patrick_doreian_structural_2015} have led to making a totally different inference with respect to the balance of signed international relations networks.

This chapter provides extensive results on financial portfolio networks in Section~\ref{4s:d4} confirming the observations of Harary et al.\ \cite{harary_signed_2002} on small portfolio networks being mostly in a totally balanced state. In Section~\ref{4s:d56}, we extended the applications of the frustration index to a fullerene stability indicator in Subsection \ref{4s:d5} and the Hamiltonian of Ising models in Subsection \ref{4s:d6}. It is hoped that these discussions pave they way for using exact optimisation models for more efficient and reliable computational analysis of signed networks, fullerene graphs and Ising models.

%In particular, the superiority of the model is evident in the numerical results with guaranteed solution quality provided for large-scale social, biological, and political signed networks, molecular graphs of fullerenes, and large-scale Ising models.

%\section*{Acknowledgement}
%The first author would like to show his gratitude to the Centre for eResearch at University of Auckland for providing the high performance computer and to Marco De La Pierre and Tomislav Do{\v{s}}li{\'c} for valuable discussions. 

%The author is immensely grateful to for providing insight and expertise that greatly assisted this research. 

% references section

%Instructions	%1. remove bbl file
				%2. set bibliographystyle to {imaiai} and run twice to get the reference list in the correct format
				%3. set bibliographystyle to {plain} and run twice to get the correct style for in-text reference

%\bibliographystyle{imaiai} 
%\bibliographystyle{plainnat}
%\bibliographystyle{plain}
%\bibliography{refs}

%\end{document}

% ====================================================
%
% ENDMATTER
%
% Appendices and bibliography 
% Pagination arabic, re-starts at 1
%
% ====================================================
\cleardoublepage
\chapter{Conclusion and Future Directions}
\cleardoublepage

	{In Chapter \ref{ch:1}, we discussed quantifying the answer to this simple question: is the enemy of an enemy a friend? We formalised the concept of a measure of partial balance, compared several measures on synthetic and real datasets, and investigated their axiomatic properties. We evaluated measures to be used in future work based on their properties which led to finding key axioms and desirable properties satisfied by a measure known as the frustration index. We recommended its usage in future work, although it requires intensive computation.
		
	The findings of Chapter \ref{ch:1} have a number of important implications for future investigation. Although we focused on partial balance, the findings may well have a bearing on link prediction and clustering in signed networks \cite{Gallier16}. Some other relevant topics of interest in signed networks are network dynamics \cite{tan_evolutionary_2016} and opinion dynamics \cite{li_voter_2015}. Effective methods of signed network structural analysis can contribute to these topics as well. 
	
	The intensive computations required for obtaining the frustration index encouraged us to focus in Chapter \ref{ch:2} on developing computational methods that exactly compute this measure for decent-sized graphs in a reasonable time. Our studies of this graph-theoretic measure revealed that while it has several applications in many fields, it was mostly approximated or estimated using heuristic methods. We also found out that the frustration index was almost never computed exactly in non-trivial examples because of the complexity in its computation which is closely related to classic NP-hard graph problems. We linearised a quadratic programming model to compute this measure exactly. We obtained numerical results on graphs with up to 3000 edges that showed most real-world social networks and some biological networks have small frustration index values which indicate that they are close to a state of structural balance.
	
	In Chapter \ref{ch:3}, we focused on reformulating the optimisation model we had developed for computing the frustration index. We suggested three new integer linear programming models that were mathematically equivalent, but had major differences in performance. We also took advantage of some structural properties in the networks to develop speed-up techniques. Our algorithms were shown to provide the global optimal solution and outperform all previous methods by orders of magnitude in solve time. We showed that exact values of the frustration index in signed graphs with up to 15000 edges can be efficiently computed using our suggested optimisation models on inexpensive hardware.
	
	Chapters \ref{ch:2} -- \ref{ch:3} have a number of important implications for future investigation. The optimisation models introduced can make network dynamics models more consistent with the theory of structural balance \cite{antal_dynamics_2005}. Many sign change simulation models that allow one change at a time use the number of balanced triads in the network as a criterion for transitioning towards balance. These models may result in stable states that are not balanced, like \textit{jammed states} and \textit{glassy states} \cite{marvel_energy_2009}. This contradicts not only the instability of unbalanced states, but the fundamental assumption that networks gradually move towards balance. Deploying decrease in the frustration index as the criterion, the above-mentioned states might be avoided resulting in a more realistic simulation of signed network dynamics that is consistent with structural balance theory and its assumptions.
	
	The efficient computational methods we developed encouraged us to explore the frustration index beyond its classic friend-enemy interpretation in the social context. In Chapter \ref{ch:4}, we investigated a range of applications from biology and chemistry to finance, international relations, and physics. This helped us unify the concept of signed graph frustration index whose practical applications can be found among mostly unanswered questions in several research areas. We discussed how the frustration index turns out to be a measure of distance to monotonicity in systems biology, a predictor of fullerene chemical stability, a measure of bi-polarisation in international relations, an indicator for well-diversified portfolios in finance, and a proxy for ground-state energy in some models of atomic magnets in physics. We used a high-performance computer to solve a wide range of instances involving graphs with up to 100000 edges concerning applications in several fields.
	
	While Chapter \ref{ch:4} provided an overview of the state-of-the-art numerical computations on signed graphs and the vast range of applications to which it can be applied, it is by no means an exhaustive survey on the applications of the frustration index. From a computational perspective, this thesis and some other recent studies \cite{Giscard2016, giscard2016general} call for more advanced computational models that put larger networks within the reach of exact analysis. As another future research direction, one may consider formulating edge-based measures of stability for directed signed networks based on theories involving directionality and signed ties \cite{leskovec_signed_2010,yap_why_2015}.

	}

%\printbibliography[title={Works cited}, heading=bibintoc]
%\bibliographystyle{abbrv}
\bibliographystyle{acm}
\bibliography{thesisrefs}

%\cleardoublepage % start afresh on a new page
%\setcounter{page}{1} % re-sets the page counter
%\appendixpage* % makes a page to mark beginning of appendices
% \input{appendix1} 

\end{document}